\documentclass[traditabstract]{aa}
\usepackage{txfonts}
\usepackage{booktabs}
\usepackage{url}
\usepackage{longtable}
\usepackage{supertabular}
\usepackage{rotating}

\usepackage{natbib}
\bibpunct{(}{)}{;}{a}{}{,}  
\usepackage{multirow}
\usepackage[table]{xcolor}
\definecolor{lightgray}{gray}{0.8}

\usepackage{lscape}
\usepackage{color}
\usepackage{amsmath}
\usepackage{float}

\usepackage[flushleft]{threeparttable}

\usepackage{adjustbox}

\usepackage{subcaption}

\def\hzrg{HzRG}
\def\hzrgs{HzRGs} 
\def\ujy{\,$\mu$Jy}

\def\mum{\,$\mu$m}
\def\msun{\,M$_{\odot}$}
\def\msunyr{\,M$_{\odot}$ yr$^{-1}$}

\def\nodata{...}

\def\herge{HeRG\'E}

\def\spitzer{\textit{Spitzer}}
\def\herschel{\textit{Herschel}}

\def\mrmoose{\textsc{Mr-Moose}}

\def\irs{IRS (16\,\mum)}
\def\mips1{MIPS (24\,\mum)}
\def\mipssju{MIPS (70\,\mum)}
\def\mipssex{MIPS (160\,\mum)}
\def\pacsb{PACS (70\,\mum)}
\def\pacsg{PACS (100\,\mum)}
\def\pacsr{PACS (160\,\mum)}
\def\spires{SPIRE (250\,\mum)}
\def\spirem{SPIRE (350\,\mum)}
\def\spirel{SPIRE (500\,\mum)}
\def\scubas{SCUBA (450\,\mum)}
\def\scuba{SCUBA (850\,\mum)}
\def\laboca{LABOCA (870\,\mum)}
\def\atca{ATCA (7\,mm)}

\def\liragn{L$^\mathrm{IR}_\mathrm{AGN}$}

\def\lirsf{L$^\mathrm{IR}_\mathrm{SF}$}

\def\lir{L$^\mathrm{IR}$}

\newcommand{\CIone}{[C{\small I}] $^3\mathrm{P}_1-^3\mathrm{P}_0$}

\def\l500{L$^{\rm 500 MHz}_{\rm ext}$}

\def\ltsima{$\; \buildrel < \over \sim \;$}
\def\simlt{\lower.5ex\hbox{\ltsima}} 
\def\gtsima{$\; \buildrel > \over \sim \;$}
\def\simgt{\lower.5ex\hbox{\gtsima}} 

\usepackage[maxfloats=50]{morefloats}

\begin{document}

\title{Massive galaxies on the road to quenching: ALMA observations of powerful high redshift radio galaxies}

\author{Theresa Falkendal \inst{1,2}
  \and Carlos De Breuck\inst{1} 
  \and Matthew D. Lehnert\inst{2}
  \and Guillaume Drouart\inst{3}
  \and Jo{\"e}l Vernet\inst{1}
  \and Bjorn Emonts\inst{4}
  \and Minju Lee\inst{5,6}
  \and Nicole P.H. Nesvadba\inst{7}
  \and Nick Seymour\inst{3}
  \and Matthieu B\'ethermin\inst{8}
  \and Sthabile Kolwa\inst{1}
  \and Bitten Gullberg\inst{9}
  \and Dominika Wylezalek \inst{1,10}}

\institute{European Southern Observatory, Karl-Schwarzchild-Str. 2, 85748 Garching, Germany
\and
Sorbonne Universit\'{e}, CNRS UMR 7095, Institut d'Astrophysique de Paris, 98bis bvd Arago, 75014, Paris, France
\and
International Center for Radio Astronomy Research, Curtin University, Perth WA 6845, Australia
\and 
National Radio Astronomy Observatory, 520 Edgemont Rd, Charlottesville, VA 22903, USA 
\and
Department of Astronomy, The University of Tokyo, 7-3-1 Hongo, Bunkyo-ku, Tokyo 133-0033, Japan
\and
National Observatory of Japan, 2-21-1 Osawa, Mitaka, Tokyo 181-0015, Japan
\and
Institut d'Astrophysique Spatiale, CNRS, Universit\'e Paris Sud, 91405 Orsay France
\and
Aix-Marseille Univ., CNRS, LAM, Laboratoire d’Astrophysique de Marseille, 13013 Marseille, France
\and
Centre for Extragalactic Astronomy, Department of Physics, Durham University, South Road, Durham DH1 3LE, UK
\and
Department of Physics and Astronomy, Johns Hopkins University, Bloomberg center, 3400 N. Charles St, Baltimore, MD, 21218, USA}

\date{accepted for publication, \aap, 10 July 2018}  
  
\abstract{We present 0\farcs3 (band 6) and 1\farcs5 (band 3) ALMA
observations of the (sub)millimeter dust continuum emission for 25
radio galaxies at $1<z<5.2$. Our survey reaches a rms flux density of
$\sim$50\,$\mu$Jy in band 6 (200-250\,GHz) and $\sim$20\,$\mu$Jy in band
3 (100-130\,GHz). This is an order of magnitude deeper than single-dish
850\,$\mu$m observations, and reaches fluxes where synchrotron and thermal
dust emission are expected to be of the same order of magnitude. Combining
our sensitive ALMA observations with low-resolution radio data from ATCA,
higher resolution VLA data, and infrared photometry from \herschel\ and
\spitzer, we have disentangled the synchrotron and thermal dust emission. We
determine the star-formation rates and AGN infrared luminosities
using our newly developed \underline{M}ulti-\underline{r}esolution and
\underline{m}ulti-\underline{o}bject/\underline{o}rigin \underline{s}pectral
\underline{e}nergy distribution fitting code (\mrmoose). We
find that synchrotron emission contributes substantially at
$\lambda\sim$1\,mm. Through our sensitive flux limits and accounting for
a contribution from synchrotron emission in the mm, we revise downward
the median star-formation rate by a factor of seven compared to previous
estimates based solely on \herschel\ and \spitzer\ data. The hosts of
these radio-loud AGN appear predominantly below the main sequence of
star-forming galaxies, indicating that the star formation in many of
the host galaxies has been quenched.  Future growth of the host galaxies
without substantial black hole mass growth will be needed to bring these
objects on the local relation between the supermassive black holes and
their host galaxies. Given the mismatch in the timescales of any star
formation that took place in the host galaxies and lifetime of the AGN,
we hypothesize that a key role is played by star formation in depleting
the gas before the action of the powerful radio jets quickly drives out
the remaining gas. This positive feedback loop of efficient star formation
rapidly consuming the gas coupled to the action of the radio jets in
removing the residual gas is how massive galaxies are rapidly quenched.}

\keywords{Galaxies: active -- Galaxies: evolution -- Galaxies: high
redshift -- Galaxies: jets -- Galaxies: star formation -- Galaxies: ISM}

\titlerunning{ALMA observations of quenched high redshift radio galaxies}
\authorrunning{Falkendal, De Breuck, Lehnert et al.}

\titlerunning{}
\authorrunning{T. Falkendal et al.}

\maketitle

\section{Introduction} 

The connection between active galactic nuclei (AGN), their host galaxies
and environments has been one of the central questions in extra-galactic
astrophysics for over 30 years \citep{Balick1982}. This contemporary
debate centers around two predominant issues concerning the influence of
AGN on their environment: Is their influence ``positive'' or ``negative'',
meaning do they either increase or decrease the star-formation efficiency
of their hosts? How do super-massive black holes (SMBHs) regulate
their own growth? These two questions are intertwined. When the SMBH
is active, it may well regulate its own growth while also enhancing or
inhibiting the stellar or baryonic mass growth of its host. The empirical,
approximately linear relationship between the mass of SMHB and both galaxy
bulge mass and the velocity dispersion \cite[e.g.][]{Magorrian1998,
Gebhardt2000, Ferrarese2000, Haring2004}, suggests that the growth of these
two components is concomitant along the observed relationship. However,
it is not clear if this relation is causal or if it simply reveals
a connection between galaxy-galaxy mergers and that the growth of galaxy
components are limited asymptotically \citep[central limit theorem,
e.g.,][]{Peng2007, Jahnke2011}.

SMBHs and host galaxies share several properties. Both SMBHs and galaxies
have exponential cut-offs at the high mass end of their co-moving space
densities \citep[e.g.,][]{Shankar2009, Ilbert2013, Kelly2013}.  The
population of both SMBHs and galaxies also exhibit mass downsizing whereby the
oldest, in the case of galaxies, and the most massive of SMBHs grew early
and rapidly \citep[e.g.,][]{Thomas2005, Thomas2010, Merloni2008}. However,
there is a mismatch in both the shape and co-moving number density between
galaxies and dark matter halos, especially at the low and high mass ends
of these functions \citep{Benson2003}. Because powerful AGN can have
a mechanical and radiative energy output similar to or exceeding that of the binding
energy of a massive galaxy and dark matter halo, AGN are thought to play a
key role in regulating the baryonic growth of galaxies.  Both observations
and simulations have suggested that there may be a positive trend between
the mean black hole accretion rate and star-formation rate
\citep[SFR; e.g.,][]{Delvecchio2015, McAlpine2017}, while the mean
SFR as a function of black hole accretion rate shows no correlation for
low luminosity sources \cite[e.g.,][]{Stanley2015, McAlpine2017}. One
should be cautious when interpreting both theoretical and observational
results in the definition of what exactly AGN feedback is and how AGN
affect their host galaxies to explain the properties of an ensemble of
galaxies \citep{scholtz18}. The strength and nature of AGN feedback --
the cycle whereby the SMBH regulates both its own growth and that of its host
-- depends on galaxy mass and morphology. For example, the most massive
elliptical galaxies are generally metal-rich and old, while less massive
lenticular galaxies, which make up the bulk of the early-type galaxy
population, have star formation histories that lasts significantly
longer \citep{Thomas2005, Thomas2010, emsellem2011, krajnovic2011}.
Clearly, if AGN feedback plays a crucial role in shaping the ensemble
of galaxies, its impact on massive dispersion dominated galaxies must
result in somewhat different characteristics in these galaxies compared
to rotationally-dominated and predominately less massive lenticular galaxies.

To gain a deeper understanding about how the growth of host galaxy and
the SMBH are intertwined, it is important to study the characteristics
of the star formation occurring in the host galaxies of actively
fueled black holes.  Within this context, powerful radio galaxies
generally, and high-redshift radio galaxies (\hzrgs) in particular,
are important test beds of our ideas on the physics underlying
AGN feedback. At almost all redshifts, powerful radio-loud AGN are
hosted by galaxies that are among the most massive \citep{Bithel1990,
Lehnert1992, vanBreugel1998, Rocca2004, Best2005}. Since we know that at
low redshift the star formation history of many of these galaxies was
brief, but intense \citep[e.g.,][]{Thomas2005,Tadhunter2011}, and they
lie at the exponential high mass tail of the stellar mass distribution
\citep[e.g.,][]{Seymour2007, Ilbert2013}, if AGN play an important role
in shaping massive galaxies, it is in these galaxies that this must be
most evident. Most importantly, \hzrgs\, are luminous sources not only
in the radio, but also throughout the mid-infrared (MIR) and
sub-mm continuum. This generally implies that they have high rates of
star-formation \citep{Archibald2001, Reuland2003}, and a rapid accretion
onto the SMBH. Due to the fact that the broad line region and rest-frame
ultraviolet-optical continuum emission from their accretion disks is
obscured, we can observe their stellar emission \citep{Seymour2007,
DeBreuck2010}. All of these arguments make \hzrgs\ important targets for
understanding the complex relationship between AGN and massive galaxies.

The substantial mechanical and radiative AGN luminosity and apparently
significant star-formation rates of \hzrgs\ leads to a quandary. If
AGN feedback effectively suppresses black hole and galaxy growth,
why do the host galaxies of AGN grow so rapidly \citep[mass doubling
times of $\sim$0.1-1\,Gyr,][]{Drouart2014}? This is the ``coordination
problem'', the apparent contradiction which is a paradoxical situation
where the strongest phase of energy and momentum injection into the
interstellar medium of the host galaxy by the AGN is not co-concomitant
with strong suppression of star formation \citep[see e.g.,][]{Drouart2014,
Drouart2016}. Models predict an offset between the fueling of the AGN and
the suppression of star formation because the timescales for fueling the
AGN is substantially shorter than the star-formation timescale within the
host. Could the impact of AGN feedback be to increase the star formation
efficiency in galaxies \citep[i.e., ``positive feedback'';][]{Silk2013,
kalfoutzou2017}?

The bulk of the bolometric output of \hzrgs\ is emitted in the infrared
\citep[IR; e.g.,][]{Miley2008}. Both AGN and star-formation (SF, dust
heated by stars) contribute energy to the dust continuum spectral
energy distribution; the AGN heats the dust to warm temperatures
($T\gtrsim$60\,K) emitting in the MIR, while the SF generally heats
the dust to lower temperatures ($T\lesssim$60\,K) and emits mainly in the
far-infrared (FIR). The \herschel\ space telescope provided
an opportunity to cover both sides of the peak of thermal emission
allowing us to disentangle the dust components heated by AGN and SF. In
our \herschel\ radio galaxy evolution (\herge) project, we constrained
spectral energy distributions (SEDs) for a sample of 70 radio galaxies at
1$<z<$5.2 \citep{Drouart2014}. While this led to substantially improved
estimates of the SFR, one limitation of \herschel\ data are their low
spatial resolution (e.g., 36\,\arcsec\ at 500\,\mum). As shown by several
arcsecond resolution follow-up observations, the dust continuum emission
often splits into several components, which are not necessarily coincident
with the AGN host galaxy \citep{DeBreuck2005, Ivison2008, Ivison2012,
Nesvadba2009, Emonts2014, Gullberg2016}. Sub-arcsecond resolution imaging
is therefore essential to separate the star formation occurring in the
AGN host galaxies from that occurring in the nearby companion galaxies.
We have therefore started a large systematic follow-up program
of our \herge\ sample with the Atacama Large Millimeter Array (ALMA).

As \hzrgs\ are, by selection, the brightest radio sources at each redshift,
synchrotron radiation may make a substantial contribution at (sub)mm
wavelengths. This was already discussed by \citet{Archibald2001},
who concluded the 850\,\mum\, fluxes of three sources in their sample
of 47 may be dominated by synchrotron emission. However, to make such
an assessment, one has to assume that the sub-mm fluxes are a straight
power-law extrapolation of the radio SED. While one may expect the spectra
to steepen at high frequencies due to aging of high energy electrons,
there are also suggestions that the SEDs in at least some \hzrgs\ remain
a power law with a constant exponent even at the highest observed radio
and/or mm frequencies \citep{Klamer2006}.

The high sensitivity of ALMA also allows us to reach flux density levels
more than an order-of-magnitude fainter than previous single-dish
observations with LABOCA or SCUBA. Reaching such depths implies that we
may reach flux density levels of the extrapolated synchrotron emission
in most of the sources in our sample. It is therefore essential to
disentangle the thermal dust and synchrotron components. To achieve this,
we adopt two strategies: (1) multi-frequency photometry covering the
range 10$<\nu_{\rm obs}$<200\,GHz, and (2) spatially resolving the radio
core and lobes. For this first strategy, we combine our ALMA data with
7\,mm and 3\,mm observations from the Australia Telescope Compact Array
\citep[ATCA;][]{Emonts2014}. For the second strategy, we use the available
radio maps from the Very Large Array \citep[VLA;][]{Carilli1997, Pentericci2000,
DeBreuck2010}, which have similar spatial resolution as our ALMA data. A
more detailed study of the physics of the high-frequency synchrotron
emission is deferred to a forthcoming paper. For now, we simply consider
the possibility that synchrotron component will impact our (sub)mm
observations and our estimates of the SFRs of the AGN host galaxy.

The three main SED components, synchrotron emission, AGN and SF heated
thermal dust emission constitute the SEDs of \hzrgs. Because each
component potentially makes a significant contribution to the over all
SED, it requires an analysis of photometry covering an order-of-magnitude
range in spatial scales -- sub arcsecond to 10s of arc seconds.
To this end, we developed the \underline{M}ulti-\underline{r}esolution and
\underline{m}ulti-\underline{o}bject/\underline{o}rigin \underline{s}pectral
\underline{e}nergy
distribution fitting procedure \mrmoose\ \citep{drouart18}.  This versatile
code allows us to isolate the SF heated dust emission \lirsf\ from the
two spectrally adjacent components -- the AGN heated dust component at
higher frequencies and the synchrotron emission at lower frequencies.
We revise the \lirsf\ downwards by a factor of many compared to
previous \herschel\ determinations.

This paper is structured as follows: after introducing the sample and
observations in \S2, we briefly describe our fitting code in \S3. The
overall results are described in \S4, with detailed descriptions of each
individual source given in the appendix. We discuss the implications of
our results in \S5, and summarize our conclusions in \S6.

\section{Observations and data reduction}

\subsection{Sample}

Our sample consists of 25 \hzrgs\ over the redshift range
1$<$$z$$<$5.2. This is a subsample of the parent \herge\ sample of 70
\hzrg\, a project dedicated to observed \hzrgs\, with \herschel\
\citep[described in detail in][]{Seymour2007, DeBreuck2010}. To
summarize, the parent samples sources were selected to have luminosities
at rest-frame 3\,GHz greater than 10$^{26}$\,W\,Hz$^{-1}$ and have
ultra-steep radio spectra ($\alpha$=-1.0; S$_\nu$$\propto$$\nu^\alpha$,
at $\nu_\text{obs}$$\sim$1.4~GHz). The parent sample has complete 12-band
3.6 to 850\,\mum\ photometry from \spitzer, \herschel\ SCUBA/LABOCA
\citep{DeBreuck2010, Drouart2014}. The subsample of 25 sources observed
with ALMA were chosen to be easily observable by ALMA, i.e., the southern
part of the parent sample and were grouped in such a way that multiple
sources share a phase calibrator to minimize the overheads to the extent possible.

\subsection{ALMA Observations}

ALMA Cycle 2 band 6 (and band 4 for source TN\,J2007-1316)
observations were carried out from June 2014 to September 2015
(Table~\ref{tab:observation}). We used four 1.875\,GHz spectral windows,
tuned to cover molecular lines at the specific redshift of each
source \citep{McMullin2007}. The data was calibrated in CASA (Common
Astronomy Software Application) with the supplied calibration script
(with exception of MRC\,2224-273, for which the provided script was
changed to correctly compensate for different averaging factor in one
of the spectral windows). Since a significant fraction of our sources
have a low SNR, we decided to optimize the sensitivity by using natural
weighting to construct images. For all sources, except TN\,J2007-1316,
atmospheric absorption noise was present in observations, therefore we
excluded the affected channels from the final images. The settings used in
our data reduction were: cell size of 0.06\,arcsec (roughly five times smaller
than the beam size), barycentric reference frame (BARY), and the mode
``mfs'' (multi-frequency synthesis emulation). For the brightest source
MRC\,0114-211, a phase self-calibration was done which decreased the RMS
noise in the final image from 87\,\ujy\ to 59\,\ujy.

The ALMA Cycle 3 band 3 (and band 4 for source MRC\,0943-242)
observations were conducted from March 2016 to September 2016
(Table~\ref{tab:observation}). We used four 1.875\,GHz spectral windows
tuned to include the \CIone\ line in one of the side bands. Just as
for the Cycle 2 observations the data were calibrated in CASA with
the calibration scripts provided by the observatory and the continuum
maps were produced in the same way, i.e., natural weighting, cell size of
0.28\,arcsec, BARY and mode mfs, to be consistent over the whole sample.

For the source, 4C\,23.56, we use the ALMA band 3 and 6 continuum data
presented in \citet{Lee2017}. Please see \citet{Lee2017} for details of
the observations and data reduction.
%
\begin{table*}
\rowcolors{1}{}{lightgray}
\begin{center}
\begin{tabular}{lcccccccccc}
\toprule
Name 			&  ALMA	& RMS			& Beam size	&  Antennas	& Central frequency & Integration time & Observations\\
\rowcolor{white}	& band	& [\ujy /beam]	& [arcsec]	&		& [GHz]		& [minutes] 			& date	\\
\midrule
\rowcolor{white}	MRC\,0037-258		&6 	&57	&0.26$\times$0.22		&42		&230	&6		&2014-08-28	\\
\rowcolor{lightgray}MRC\,0114-211		&6 	&59	&0.27$\times$0.22		&42		&230 	&6		&2014-08-28	\\	
\rowcolor{white}						&	&	&						&36		&		&183	&2016-04-17	\\
\rowcolor{white}		\multirow{-2}{*}{TN\,J0121+1320}	&	\multirow{-2}{*}{3}	&\multirow{-2}{*}{12} &\multirow{-2}{*}{1.57$\times$1.19}	&36	 &\multirow{-2}{*}{102}	&78	&2016-04-17	\\
\rowcolor{lightgray}					&	&	&						&34		&		&7		&2014-08-28	\\
\rowcolor{lightgray}	\multirow{-2}{*}{MRC\,0152-209}	&	\multirow{-2}{*}{6}&	\multirow{-2}{*}{58}&\multirow{-2}{*}{0.39$\times$0.32}		&35	&\multirow{-2}{*}{245} 	&7	&2014-07-21	\\	
\rowcolor{white}						&	&	&						&34		&	&7		&2014-08-28	\\
\rowcolor{white} 	\multirow{-2}{*}{MRC\,0156-252}	&	\multirow{-2}{*}{6}&	\multirow{-2}{*}{50}	&\multirow{-2}{*}{0.39$\times$0.32}		&35	&\multirow{-2}{*}{245}	&7	&2014-07-21	\\	
\rowcolor{lightgray}					&	&	&						&36		&		&86		&2016-03-08	\\
\rowcolor{lightgray}		\multirow{-2}{*}{TN\,J0205+2242}	&	\multirow{-2}{*}{3}	&\multirow{-2}{*}{17} &\multirow{-2}{*}{1.87$\times$1.37} &36 &\multirow{-2}{*}{97}	&121 &2016-04-17	\\
\rowcolor{white}						&	&	&						&34		&	&7 &2014-08-28	\\	
\rowcolor{white}	\multirow{-2}{*}{MRC\,0211-256}	&	\multirow{-2}{*}{6}&	\multirow{-2}{*}{52}	&\multirow{-2}{*}{0.39$\times$0.32}	&35		&\multirow{-2}{*}{245}	&7	&2014-07-21	\\
\rowcolor{lightgray}TXS\,0211-122		&6 	&51	&0.47$\times$0.25		&36		&250	&5	&2014-09-01	\\
\rowcolor{white}	MRC\,0251-273		&6 	&55	&0.61$\times$0.44		&35		&231	&3	&2015-06-13	\\  
\rowcolor{lightgray}MRC\,0324-228		&6 	&52	&0.57$\times$043		&35		&231	&3	&2015-06-13	\\
\rowcolor{white}	MRC\,0350-279		&6 	&56	&0.42$\times$0.28		&34		&245	&4	&2014-08-28	\\
\rowcolor{lightgray}MRC\,0406-244		&6 	&65	&0.41$\times$0.28		&34		&245	&4	&2014-08-28	\\  
\rowcolor{white}	PKS\,0529-549		&6 	&45	&0.42$\times$0.2		&34		&233	&5	&2014-09-02	\\
\rowcolor{lightgray}TN\,J0924-2201		&6	&79	&0.68$\times$0.58		&36		&243	&2	&2014-04-28	\\ 
\rowcolor{white}	MRC\,0943-242		&4	&13	&1.61$\times$1.03		&42		&133	&146			&2016-03-06	\\ 
\rowcolor{lightgray}					&	&	&						&34		&	&4	&2015-09-26	\\ 
\rowcolor{lightgray}					&	&	&						&36		&	&2	&2014-04-28	\\  
\rowcolor{lightgray}	\multirow{-3}{*}{MRC\,0943-242}	&	\multirow{-3}{*}{6}&	\multirow{-3}{*}{61}&\multirow{-3}{*}{0.47$\times$0.38}	&37		&\multirow{-3}{*}{243}	&3	&2015-06-12	\\
\rowcolor{white}						&	&	&						&34		&	&4	&2015-09-26	\\   
\rowcolor{white}						&	&	&						&36		&	&2	&2014-04-28	\\
\rowcolor{white}	\multirow{-3}{*}{MRC\,1017-220}	&	\multirow{-3}{*}{6}&	\multirow{-3}{*}{67}	&\multirow{-3}{*}{0.48$\times$0.36}	&37		&\multirow{-3}{*}{243}	&3	&2015-06-12	\\	
\rowcolor{lightgray}4C\,03.24			&3	&18	&1.92$\times$1.46		&42		&102	&101	&2016-03-06	\\
\rowcolor{white}	TN\,J1338+1942		&3	&12	&1.72$\times$1.36		&43		&92		&39	&2016-04-16	\\
\rowcolor{lightgray}TN\,J2007-1316		&4 	&44	&0.44$\times$0.34		&43		&150	&3	&2015-06-29	\\
\rowcolor{white}						&	&	&						&39		&	&3	&2015-07-19	\\
\rowcolor{white}	\multirow{-2}{*}{MRC\,2025-218}	&	\multirow{-2}{*}{6}&	\multirow{-2}{*}{46}	&\multirow{-2}{*}{0.29$\times$0.26}	&34		& \multirow{-2}{*}{232}	&3	&2014-08-18	\\ 
\rowcolor{lightgray}					&	&	&						&39		&	&3	&2015-07-19	\\			 
\rowcolor{lightgray}	\multirow{-2}{*}{MRC\,2048-272}	&	\multirow{-2}{*}{6}&	\multirow{-2}{*}{45}&\multirow{-2}{*}{0.27$\times$0.25}	&34		&\multirow{-2}{*}{232}	&3	&2014-08-18	\\ 
\rowcolor{white}						&	&	&						&39		&	&3	&2015-07-19	\\
\rowcolor{white}	\multirow{-2}{*}{MRC\,2104-242} 	&	\multirow{-2}{*}{6}&	\multirow{-2}{*}{43}&\multirow{-2}{*}{0.27$\times$0.27}	&34		&\multirow{-2}{*}{232}	&3	&2014-08-18	\\
\rowcolor{lightgray}4C\,23.56$^1$		&3	&16	&0.87$\times$0.60		&43		&106	&61		&2015-06-30	\\
\rowcolor{white}	4C\,23.56$^1$		&6	&81	&0.78$\times$0.68		&25		&265	&4		&2014-04-25	\\
\rowcolor{lightgray}4C\,19.71			&3	&13	&1.76$\times$1.59		&36		&103	&141	&2016-03-06	\\
\rowcolor{white}	MRC\,2224-273		&6 	&72	&0.41$\times$0.39		&33		&251	&4		&2014-07-28	\\
\bottomrule
\end{tabular}
\end{center}
\caption{Details about the ALMA observations. Several sources were
observed more than once because the first observation did not meet
the requested sensitivity and/or resolution. The additional data were
included in the final reduction when they improved the signal-to-noise
of the resulting image. $^1$The ALMA are from \citet{Lee2017}, see that
study for details concerning the reduction and observations.}
\label{tab:observation}
\end{table*}

\subsection{ATCA 7\,mm and 3\,mm data}

The ATCA observations of 7\,mm continuum in some of the sources were
conducted over 2009\,$-$\,2013.  These continuum data were part of a
survey to search for cold molecular CO(1--0) gas in high redshift radio
galaxies \citep{Emonts2014}. The data were obtained using the compact
hybrid H75, H168 and H214 array configurations, with maximum baselines
of 89, 192 and 247\,m, respectively. This resulted in synthesized
beams ranging from roughly 6$-$13\,arcsec. We used the Compact Array
Broadband Backend \citep[CABB;][]{Wilson2011} with an effective 2\,GHz
bandpass and 1\,MHz channels, centered on the redshifted CO(1--0)
line \citep[32$-$48\,GHz;][]{Emonts2014}. The data were calibrated
using the software package, MIRIAD \citep{Sault1995}. The observing
and data reduction strategy, as well as the basic data products, were
previously described in \citet{Emonts2011} and \citet{Emonts2014}. For
MRC\,0943-242, we also used the ATCA 3\,mm ATCA/CABB system on March
21 2012 in the H168 array configurations and on September 30 and Oct
1 2012 in the H214 array configuration to obtain an upper limit of
the radio continuum at 88.2\,GHz. This frequency corresponds to the
redshifted CO(3-2) line, which was not detected in our observations. We
used PKS\,1253-055 (March) and PKS\,0537-441 (Sept/Oct) for bandpass
calibration, PKS\,0919-260 for frequency gain calibration, and Mars
(March) and Uranus (Sept/Oct) to set the absolute flux levels. The total
on-source integration time was 5.1\,hrs. We used a standard observing
and data reduction strategy to calibrate these 3\,mm data, matching the
strategy of the 7\,mm data. The resulting synthesized beam of these
3\,mm data is 2.4$^{\prime\prime}$\,$\times$\,1.9$^{\prime\prime}$
(PA 82$^{\circ}$) after robust +1 weighting \citep{Briggs1995}. The rms
noise level of these 3\,mm data are 0.3\,mJy\,beam$^{-1}$.

\section{Mr-Moose}

\mrmoose\ is a new SED fitting code, developed with the goal of
being able to handle multi-resolution photometric data that have multiple
spatially-resolved detections in the same photometric band. The specific
motivation for developing this new code is to be able to make full use
of the information contained in deep and high resolution ALMA data. When
combining ALMA and radio interferometric (such as JVLA and ATCA) data
with previous low resolution data (such as \herschel\ or \spitzer) where
the beam is too large to resolve individual components, one needs a SED
fitting tool able to handle multiple components in order to make full
use of the interferometric data which often contain multiple resolved
components. \mrmoose\footnote{https://github.com/gdrouart/MrMoose} is
open source and the current version operates in MIR to radio wavelengths
\citep{drouart18}. The code relies on simple analytic models to describe
the underlying physical processes. It is up to the user to define which
data point should be fitted to which analytic models. Each photometric
data point can be associated to any number of models and combinations of
different models. It therefore \textsl{only} requires the user to make
educated guesses about the underlying physical processes responsible
for the observed flux and does not require the user to select only one
possible choice. The code fits simultaneously all pre-selected analytic
models and uses Bayesian statistics to find the most likely solution given
the set of observed fluxes. The Bayesian, Monte Carlo Markov Chain (MCMC)
approach provides marginalized posterior probability density functions
(PDF) for each of the free parameter.

In this paper, we model the MIR through radio spectral energy
distributions with three components: a component representing dust heated
by an AGN which we model as a power law with a slope, $\gamma$, and an
exponential cut-off at $\nu_\text{cut}$ \eqref{AGN_eq};  a component
representing dust heated by the young stellar population of the host
galaxy or companion which we model as a modified blackbody \eqref{BB_eq};
and a component representing synchrotron emission which we modeled
as a simple power law with constant slope, $\alpha$, with no cut-off
at high frequencies \eqref{sync_eq}. Table~\ref{tab:MrMoose_models}
summarizes the allowed range and the number of free parameters in the
three different models.

\begin{table*}[ht]
\begin{center}
\begin{tabular}{lccc}
\toprule
Model name& degree of freedom	& Free parameters	& Fixed parameters\\
\midrule
AGN heated dust	& 2 & N$_{\text{AGN}}$,\, $\gamma$\,[0,6],\, 	&$\lambda_{\text{cut}}$=33\mum\\
SF heated dust 	& 2 & N$_\text{BB}$, T\,[20,70K]& $\nu_0$=1.5 GHz, $\beta$=2.5\\
Synchrotron	 	& 2 & N$_\text{sync}$, $\alpha$\,[$-$4,0]	&\nodata\\
\bottomrule
\end{tabular}
\end{center}
\caption{List of the free parameters in our models and their allowed ranges.
N$_{\text{AGN}}$ is the normalization of the AGN power law with slope,
$\gamma$, and rest-frame cut-off wavelength, $\lambda_{\text{cut}}$.
The $\lambda_{\text{cut}}$ is fixed for all galaxies except 4C\,23.56. We
choose this particular galaxy to fit this parameter because it
is AGN-dominated (see Fig.~\ref{f_cut_fit}). N$_\text{BB}$ is the
normalization of the modified blackbody of temperature, T. N$_\text{sync}$
is the normalization of the synchrotron power-law with slope $\alpha$.}
\label{tab:MrMoose_models}
\end{table*}

\subsection{Analytic models}

In this paper, simple analytic models are used to fit each component
of the SED, instead of, using a (perhaps more realistic) physical
models because the limited number of data points available. As is often the case for high-redshift studies, we are limited
to broad band photometric data. So even though this sample of galaxies
has been observed with \herschel, \spitzer, LABOCA, and now ALMA, we
are still limited to $\sim$10 data points in the infrared. To properly
constrain more complex models, more data points are needed. We therefore
rely on less complicated, but empirically justifiable models, to fit
the data in a statistically robust way and to prevent the temptation
to over-interpret the physical processes underpinnings of our results
(e.g., the characteristics of the AGN torus which may be responsible
for reprocessing the emission from the accretion disk).

\subsubsection{AGN model}

\begin{figure}
\centering
\includegraphics[width=\linewidth]{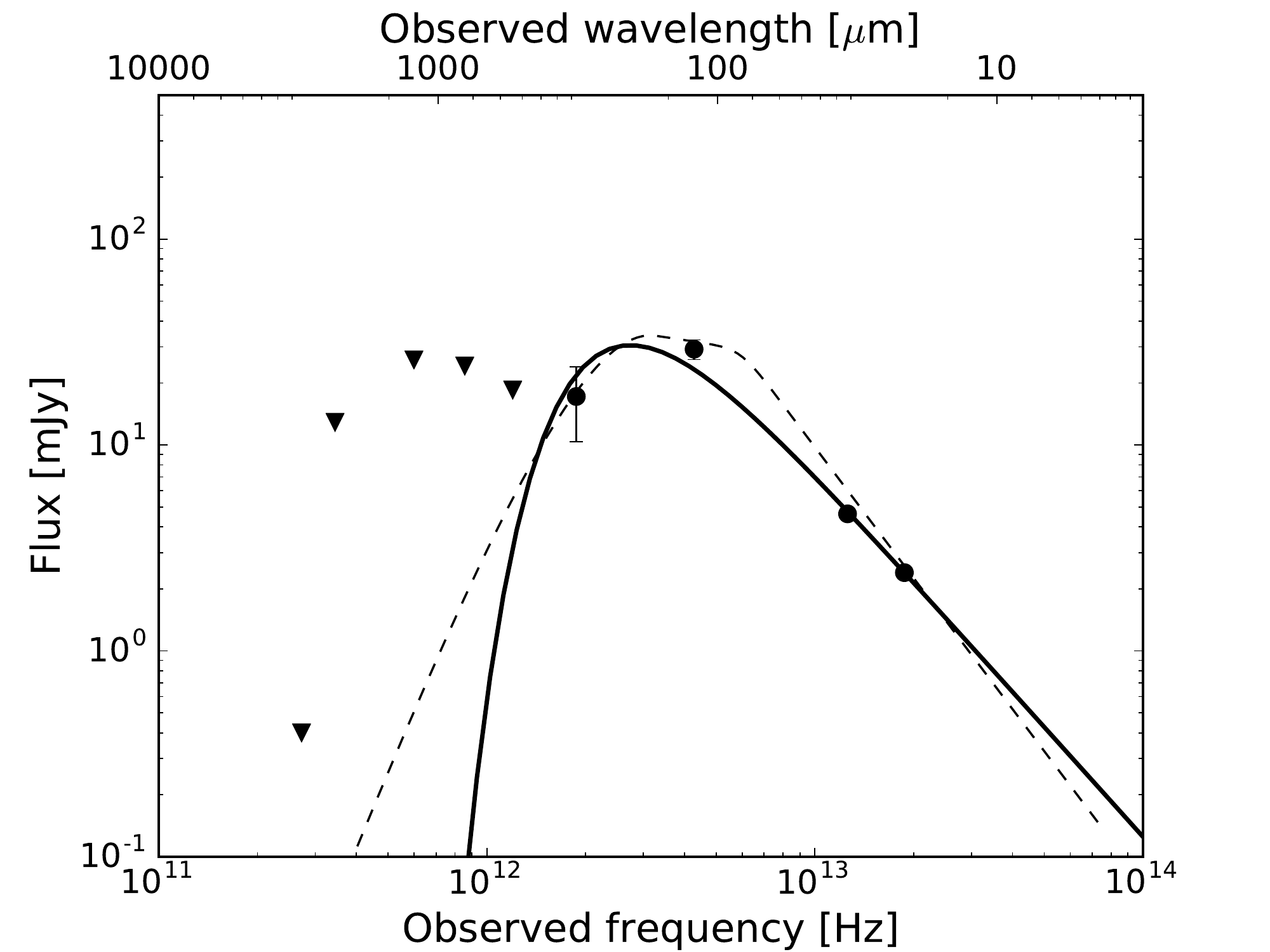}
\caption{Best fit of the SED of 4C\,23.56 when assuming that
the AGN is solely responsible for the mid-infrared emission
and allowing $\lambda_\text{cut}$ to be an additional free parameter.
The black solid line shows the best fit with $\lambda_\text{cut} =
33^{+3}_{-3}$\,\mum. The dashed line is the best fit AGN model
used in \cite{Drouart2014} which itself is the average AGN used in the
DecompIR SED fitting code \citep{Mullaney2011}. See Table~\ref{table_4C23}
for details about the photometric data.}
\label{f_cut_fit}
\end{figure}

The IR-luminosity of the source is modeled following the study of
\cite{Casey2012}. The mid- and far-infrared SEDs are fitted with a
combination of a simple power law with a low frequency exponential cut-off
and a single temperature modified blackbody. These two simple models
represents the dust heated by both the AGN and star-formation in the
galaxy respectively. However, in this paper, the two models to describe
the AGN and SF component are de-coupled and normalized individually to
fit the photometry. The functional form of the dust heated by the AGN is,

\begin{equation}
S_\text{AGN} = N_\text{AGN} \,\,  \nu^{-\gamma} \,\,e^{-(\nu_{\rm cut} / \nu)^2},
\label{AGN_eq}
\end{equation}

\noindent
where $\nu_{\rm cut}$ is the rest-frame frequency of the exponential
cut-off of the power law. The shape at longer wavelengths of the heated
dust by AGN is not well constrained because cold dust emission often
completely dominates at longer wavelengths making it very difficult
to determine the actual shape of the AGN emission. Therefore, we use a
simple exponential with a fixed rest-frame cut-off wavelength/frequency
at $\lambda_\text{cut}$ =33\,\mum\ or $\nu_{\rm cut}$=9.085\,THz. This
value was determined by letting $\lambda_\text{cut}$ be a free parameter
for one galaxy and finding the best fit value. The mid-infrared emission
of 4C\,23.56 appears to predominately due to warm dust emission from its
AGN and the far-infrared emission is very faint suggesting that it has
a very low star formation rate. Within our sample, this makes 4C\,23.56
the most obviously suitable choice for determining $\lambda_\text{cut}$
for this sample of \hzrgs. Fig.~\ref{f_cut_fit} shows the best fit with
$\lambda_\text{cut}$ as a free parameter. The chosen model for the warm
dust used in this paper has also been adopted in other studies. For
example, \cite{Younger2009} found good agreement when fitting the IR
emission of luminous high-redshift galaxies with a modified blackbody
paired with a power law component at short wavelengths.

\subsubsection{Star-formation model}

The far-infrared emission from dust heated by star-formation is fitted
by a simple single temperature modified blackbody (i.e., a ``graybody''),

\begin{equation}
S_\text{BB} = N_\text{BB} \left(1- e^{-(\nu / \nu_0 ) ^ \beta}\right) \,\,B_\nu\left(\nu,\text{T}\right),
\label{BB_eq}
\end{equation}

\noindent
where $B_\nu(\nu,\text{T})$ is a blackbody (BB) distribution, $\nu_0$
is the critical frequency where the source becomes optically thin
\citep[assumed to be fixed at $\nu_0=1.5$\,THz;][]{Conley2011} and
$\beta$ is the emissivity, which we fixed at $\beta=2.5$. The value of
the emissivity depends on the characteristics of the dust grain size,
composition, distribution and how efficiently the grains re-emit the
absorbed energy. It is common to assume a $\beta$ in the range 1-2
\citep{Hildebrand1983} but values of up to 2.5 at sub-mm wavelengths
have been found in the integrated SEDs of galaxies \citep{Galametz2012,
Cortese2012}. A value of $\beta=2.5$ has been adopted in this paper.
This is justified because sources with both LABOCA and ALMA detections
in the submm have SEDs that are better fit with $\beta$=2.5 compared to
models with $\beta \sim 1.5-2$. Whether or not this is physical is a
difficult question to answer since we do not know the precise composition
of the dust grains in these sources. Also, the emission from dust
in galaxies is most likely a combination of regions with different
temperatures and mixtures of grain size distributions and compositions.
This means that when a single modified blackbody is fitted to the total
dust emission from a galaxy the estimated value of beta is affected by
these different effects which may result in an increase in $\beta$.

\subsubsection{Synchrotron model}

The synchrotron emission is fitted with a single power law,

\begin{equation}
S_\text{sync} = N_\text{sync} \,\, \nu^\alpha, 
\label{sync_eq}
\end{equation}

\noindent
with a constant slope $\alpha$. Such a simple representation
is perhaps not a physical model in that it does not include the
possibility of a steepening or cut-off due to the aging of the electron
populations. At which frequency this happens is not constrained with
the data we have available for most of the sources. Therefore, it is
important to realize that with this simple power law, we are fitting
the maximum possible contribution from synchrotron to the radio and mm
frequencies. {For the sources with a good photometric data coverage}
in the high radio frequencies (i.e., $>$10\,GHz), there is no evidence
for a high frequency cut-off or steeping, in agreement with previous
work \citep{Klamer2006,Emonts2011}.

\subsection{Fitting procedure}

\mrmoose\ fits a pre-selected number of models to the rest-frame
photometry of a galaxy. To do this, it uses Bayesian parameter estimation
to find posterior distributions of the free parameters which are
determined based on the prior distribution (uniform) and the likelihood
function. \mrmoose\ uses the Monte Carlo Markov Chain (MCMC) core provided
in the Python package, \textit{Emcee} \citep{Foreman-Mackey2013}. The
best fit model is determined by minimizing the likelihood ($\chi^2$);
a parameter of the goodness of the fit, calculated by comparing the
observed data with the model values for the combination of all models at
the same time for each photometric band. The parameter space is explored
by ``walkers'' taking random steps. Each walker makes a chain of random
steps with each new step being only dependent on the previous step in
the sequence. The new value of a parameter after each step is accepted
if the $\chi^2$ is lower than the likelihood of the previous step in the
chain, or rejected if it is higher, in which case, the previous value is
retained. The process continues resulting in a random walk. The parameter
space is thus explored during these "walks" and from the combination
of the chain of each individual walker, posterior probability density
functions (PDF) are estimated. These PDFs are used to find the best fit
values and the uncertainties of each parameter.

The likelihood function, $\chi^2$, is calculated as described in
\cite{Sawicki2012}. One particular aspect of calculating the likelihood
function in this way, is to emphasize that it treatments upper limits in
a continuous way. There is no sharp upper cut to the allowed value of
the modeled data can be for data points reported as upper limits. The
upper limits are included in a continuous way, and the modeled value
can also go above the 3$\sigma$ upper limit of the observed data,
but in that case the $\chi^2$ increase rapidly when the model starts
to over predict the flux of the upper limit. In the case where all of
the observed photometric data are detections, the likelihood function
reduces to the classical expression of $\chi^2$. We refer the reader
to the \textit{Appendix: The maximum likelihood formalism for SED
fitting with upper limits} in \citep{Sawicki2012} for details about the
derivation of the maximum-likelihood. We also refer the reader to the
paper \citep{drouart18} for a more detailed description of \mrmoose.

\subsection{Setting up the SED for fitting}

For each source, input files need to be specified individually. This is
necessary because the code can fit a large number of possible models to
each data set.  Depending on the complexity of the source, if there are
several individual resolved components, the user can launch the code with
any number of models adapted to the specific nature of each source. If
all the photometric data is unresolved then the multi-component part of
the code is not applicable and only a single combination of synchrotron
emission, modified blackbody and power-law component is needed.

It is up to the user to assign which data point belongs to which model or
set of models, and thus, requires some knowledge of the source properties
and what underling processes contribute to the observed flux. The code is not
made to be applied blindly to a large sample of galaxies. Even
though the process of assigning analytic models to each photometric data
point may sound subjective, we actually let the code decide between
multiple models as illustrated in Table~\ref{tab:example_0114}. In
case of doubt, we provide the code with many flexible options. For the
sources with spatially-separated detections in the same band, it is easy
to connect them to corresponding detections at other frequencies. To
show how the set up is done, we take MRC\,0114-211 as an example. This
source has unresolved \spitzer\ IRS through to LABOCA data, resolved
ALMA band 6 data with two spatially resolved continuum components, unresolved
ALMA band 3, ATCA and VLA L data and resolved VLA C- and X-band data
with two resolved components. The two radio components
 coincide with the two detections in ALMA. The FIR data are unresolved
and it is unclear if this is the combined flux from the two individual
sources detected in ALMA or just the flux from the host galaxy
with no contribution from the companion. For the two detections in
ALMA it is unclear if these are two thermal dust emitting objects or the
high frequency end of the synchrotron emission. The unresolved radio data
is the total flux from both synchrotron components. In this situation,
since we cannot decide what is the nature of the two ALMA components
and therefore allow contributions to their flux from both synchrotron
emission and the modified blackbody model. We refer the reader to
the full list of components and how they are assigned to various models
which is shown in Table~\ref{tab:example_0114}. The code then determines 
what contributes to each component, not the astrophysicist.

\begin{table}[ht]
\centering
\begin{adjustbox}{max width=0.48\textwidth}
\begin{tabular}{lll}
\toprule
Photometric band   & Assigned model   & Fitted model\\
\midrule
\irs 			& AGN, BB1, BB2 & AGN \\       
\mips1			& AGN, BB1, BB2 & AGN\\      
\pacsb			& AGN, BB1, BB2 & AGN, BB2 \\       
\pacsr			& AGN, BB1, BB2 & AGN, BB2  \\       
\spires			& AGN, BB1, BB2 	& AGN, BB2  \\       
\spirem			& AGN, BB1, BB2 & BB2 \\ 
\spirel			& AGN, BB1, BB2 & BB2 \\ 
\laboca			& AGN, BB1, BB2 & BB2 \\ 
ALMA 6 east comp.&  BB1, Sync2	& Sync2\\
ALMA 6 west comp.&  BB2, Sync1	& BB2\\
ALMA  3			& BB1, BB2, Sync1, Sync2 & Sync1, Sync2 \\ 
ATCA (7mm)		& Sync1, Sync2	& Sync1, Sync2 \\ 
VLA X west comp.	& Sync2 			& Sync2 \\ 
VLA X east comp.	& Sync1 			& Sync1 \\
VLA C west comp.	& Sync2			& Sync2\\  
VLA C east comp.	& Sync1			& Sync1 \\
VLA L			& Sync1, Sync2	&Sync1, Sync2 \\
\bottomrule                          
\end{tabular}
\end{adjustbox}
\caption{The list of models assigned component to each
photometric band for MRC\,0114-211. The last column lists the model components
that dominate each band as determined through SED fitting
with \mrmoose. Models that contribute less than 1\% to the total flux in
a particular band for the best fitting model are not listed.}
\label{tab:example_0114}
\end{table}

The best fit is determined by minimizing $\chi^2$, which is calculated
by fitting models to each spatial and photometric data point. The MCMC
attempts to find the most likely solution considering all the data
at the same time. For example, as we already outlined, in the case of
MRC\,0114-211, two black bodies and two synchrotron models were assigned to
the spatially resolved ALMA band 6 data points and what came out of the
fitting procedure is that the eastern component is consistent with being
dominated by synchrotron emission and the western is dominated by thermal dust emission from
the host galaxy (Sect.~\ref{MRC0114} and Table~\ref{tab:example_0114}).

\section{Results of the SED fitting with Mr-Moose}

\begin{table*}[!ht]
	\caption{Characteristics of the sources}
	\begin{center}
\begin{tabular*}{\linewidth}{c @{\extracolsep{\fill}} lcccccccc}
\toprule
Name & \multicolumn{2}{c}{Position} 		 &ALMA 	&Flux		& RMS  		\\
	  &	R.A.(J2000.0)   &	Dec.(J2000.0)    &band	& [mJy]    & [ $\mu$Jy/beam]    \\
\midrule
MRC\,0037-258 		& 00:39:56.44 	&$-$25:34:31.00 &6	&0.36 $\pm$0.08		&57 \\
TN\,J0121+1320		& 01:21:42.73	&$+$13:20:58.00	&3	&0.19 $\pm$0.01		&12	\\
MRC\,0156-252(N)		& 01:58:33.66	&$-$24:59:31.01 &6	&0.97 $\pm$0.58 		&49 \\
MRC\,0156-252(S) 	& 01:58:33.45 	&$-$24:59:32.12 &6	&0.63 $\pm$0.39 		&50 \\
TN\,J0205+2242		& 02:05:10.69	&$+$22:42:50.40	&3	&<0.05				&17 \\
MRC\,0211-256 		& 02:13:30.53 	&$-$25:25:20.81 &6	&0.67 $\pm$0.09 		&52 \\
TXS\,0211-211 		& 02:14:17.38	&$-$11:58:46.89 &6	&0.31 $\pm$0.08		&51 \\
MRC\,251-273 		& 02:53:16.68	&$-$27:09:11.94 &6	&0.35 $\pm$0.07		&55 \\
MRC\,0324-228		& 03:27:04:54	&$-$22:39:42.10 &6	&<0.21 		 		&52 \\
MRC\,0350-279 		& 03:52:51.60	&$-$27:49:22.60 &6	&<0.19  				&56 \\
MRC\,0406-244 		& 04:08:51.48 	&$-$24:18:16.47 &6	&0.50 $\pm$0.10		&65 \\
TN\,J0924-2201 		& 09:24:19.90	&$-$22:01:42.30	&6	&0.88 $\pm$0.70		&79	\\
MRC\,0943-242(Y) 	& 09:45:32.77 	&$-$24.28.49.29 &6	&0.84$\pm$0.40		&61 \\
MRC\,0943-242(O) 	& 09:45:32.22 	&$-$24.28.55.06 &6	&2.46 $\pm$0.15		&61 \\
MRC\,0943-242(T) 	& 09:45:32.39 	&$-$24.28.54.05 &6	&1.33 $\pm$0.12		&61 \\
MRC\,0943-242(F) 	& 09:45:32.44	&$-$24.28.52.55 &6	&0.42$\pm$0.96		&61 \\
MRC\,1017-220 		& 10:19:49.02 	&$-$22.19.59.86 &6	&0.52 $\pm$0.11		&67 \\
4C\,03.24(S)			& 12:45:38.36	&$+$03:23:20.70	&3	&0.71$\pm$0.14		&11 \\
4C\,03.24(H)			& 12:45:38.36	&$+$03:23:20.70	&3	&0.08$\pm$0.07		&11 \\
TN\,J1338+1942		& 13:38:25.98	&$+$19:42:31.00	&3	&0.18$\pm$0.09		&10 \\
TN\,J2007-1316 		& 20:07:53.26	&$-$13:16:43.60 &4	&<0.25 				&44 \\
MRC\,2025-218 		& 20:27:59.48 	&$-$21:40:56.90 &6	&<0.12 				&46 \\
MRC\,2048-272 		& 20:51:03.59  	&$-$27:03:02.50 &6	&<0.17 				&45 \\
MRC\,2104-242		& 21:06:58.28 	&$-$24:05:09.10 &6	&<0.12				&43 \\
4C\,23.56			& 21:07:14.84	&$+$23.31.44.91	&3	&0.23$\pm$0.05		&16 \\		
4C\,23.56			& 21:07:14.84	&$+$23.31.44.91	&6	&<0.4				&81 \\		
4C\,19.71(H)			& 21:44:07.45	&$+$19:29:14.60	&3	&0.07$\pm$0.05		&13 \\
4C\,19.71(N)			& 21:44:07.49	&$+$19.29.18.99	&3	&0.29$\pm$0.53		&13 \\
4C\,19.71(S)			& 21:44:07.53	&$+$19.29.10.78	&3	&0.11$\pm$0.05		&13 \\
MRC\,2224-273 		& 22:27:43.28	&$-$27.05.01.67 &6	&0.22$\pm$0.06 		&58 \\
\midrule
Multiple components$^*$\\
\midrule
	MRC\,0114-211(S) &01:16:51.40 &	$-$20.52.06.98 	&6	&9.82 $\pm$0.38		&87 \\
	MRC\,0114-211(N) &01:16:51.44 &	$-$20.52.06.96 	&6	&2.21 $\pm$0.31		&87 \\
	MRC\,0152-209(S) &01:54:55.76 &	$-$20.40.26.96 	&6	&2.30 $\pm$0.10		&58 \\
	MRC\,0152-209(N) &01:54:55.74 &	$-$20.40.26.59 	&6	&1.75 $\pm$0.13		&58 \\
	PKS\,0529-549(E) &05:30:25.53 &	$-$54.54.23.30 	&6	&0.37$\pm$ 0.07		&46 \\
	PKS\,0529-549(W) &05:30:25.44 &	$-$54.54.23.21	&6	&1.33 $\pm$0.16		&46 \\
\bottomrule
\end{tabular*}
\end{center}
\begin{tablenotes}
\small
\item Integrated fluxes are determined over regions where the
signal has a significance $>$1.5$\sigma$. Upper limits are 3$\sigma$
above the noise at the IRAC position of each source over an area of
one ALMA beam. The noise for each source is estimated by measuring
the root-mean-square of the pixels in the non-primary beam corrected
images. The peak flux is the deconvolved value from a single Gaussian fit.
The signal-to-noise estimates are from the image and do not include the
uncertainties in the flux calibration.  The characteristics of various
components of MRC\,0943-242 are indicated as ``Y'' for Yggdrasil,
``T'' for Thor, ``O'' for Odin, and ``F'' for Freja \citep[see][for
details]{Gullberg2016}. Other notations are ``N'' for the northern
component, ``H'' indicating the host galaxy, ``S'' for the southern
component, ``E'' for the eastern component, and ``W'' for the western
component. \\ * Sources with multiple components for which the individual
components are not robustly separated in the ALMA continuum images. The
flux of these sources is deblended by fitting two Gaussian profiles using
CASA 4.5.0. The flux is determined by integrating the fit, peak values are
deconvolved with the beam, the position is the flux center of the fit,
and noise level is estimated from the RMS of the uncorrected primary
beam image. For MRC\,0152-209, the flux in the ALMA continuum data
is divided between a northern and southern component \citep[see][for
details]{emonts2015a}.
\end{tablenotes}
\label{ALMA_flux}
\end{table*}

Combining \spitzer, \herschel, SCUBA/LABOCA, ALMA, ATCA and VLA data,
we fit the FIR--radio SED with \mrmoose\ to derive the IR luminosities of
both the SF and AGN components in our sources. Importantly, we are able
to disentangle the contribution of synchrotron at $\sim$1\,mm, which
can otherwise masquerade as thermal dust emission.  The contribution
of synchrotron to the long wavelength thermal dust emission, if not
well-constrained, can lead to a general over-estimate of \lirsf\ and thus
the SFR \citep[see also][]{Archibald2001}. The focus of this paper is to disentangle the emission
of cold (assumed to be heated by young stars) and warm (assumed to
be heated by the AGN) dust emission from individual components and to
separate the emission from nearby objects and radio hot spots/lobes
by identifying independent emission components. We did not include the
\spitzer\ IRAC bands since these can be dominated by stellar photospheric
emission and emission from PAH bands. No models for photospheric
emission from stars are included in the version of the SED fitting code,
\mrmoose, we used in this paper.

\subsection{SF and AGN IR luminosities}
\label{subsec:IR_lum}

\begin{table*}
\begin{threeparttable}
	\caption{Integrated AGN and SF luminosities.\label{tab:SED_results}}
\begin{tabular}{lcllcccccc}
\toprule
  Name 		&redshift	& \liragn & \lirsf	&	SFR		& LAS	& size$^*$& Stellar mass	& Temp.	& \lirsf / {L$^{\rm IR}_{\rm SB}$}$^{**}$\\
			&		& [$10^{12}$\,L$_\odot$]	&	 [10$^{12}\,$L$_\odot$]$^1$ &	[M$_\odot$\,yr$^{-1}$]&[arcsec] &[kpc]& [log($M_*/M_{\odot}$)]&[K]	&\\
\midrule
MRC\,0037-258    	&1.100 	&0.93$_{-0.06}^{+0.10}$ 	&<2.17	&<249		&27.6$^a$	&231.7	&11.56$^g$		&uncons. 		&\nodata\\
MRC\,0114-211    	&1.410	&2.00$_{-0.04}^{+0.05}$	&1.09$_{-0.28}^{+0.28}$	&126		&<2$^a$		&<17		&11.39$^g$		&40$^{+1}_{-13}$	&0.47\\
TN\,J0121+1320		&3.516	&<2.67	&5.43$_{-2.11}^{+2.32}$	&626		&0.3$^b$	&2.2		&11.02$^g$		&53$^{+9}_{-9}$		&0.72\\
MRC\,0152-209 (H)	&1.920 	&9.57$_{-1.06}^{+1.30}$	&15.82$_{-2.36}^{+1.85}$	&1817		&2.2$^c$	&19.0		&11.76$^g$		&69$^{+4}_{-4}$		&0.89\\
MRC\,0152-209 (C)	&1.920	&\nodata &0.94$_{-0.42}^{+0.88}$	&108		&\nodata	&\nodata	&\nodata		&28$^{+7}_{-5}$	&\nodata\\
MRC\,0156-252		&2.016	&10.08$_{-1.58}^{+1.54}$	&<1.99	&<228		&8.3$^d$	&71.2		&12.05$^g$		&uncons.		&\nodata\\
TN\,J0205+2242		&3.506	&<2.70	&<0.74	&<84		&2.7$^b$	&20.2		&10.82$^g$		&uncons.		&\nodata\\
MRC\,0211-256  		&1.300	&0.61$_{-0.08}^{+0.02}$	&1.02$_{-0.15}^{+0.18}$	&117		&2.4$^a$	&20.6		&<11.54$^g$		&36$^{+4}_{-4}$		&0.57\\
TXS\,0211-122    	&2.340	&10.75$_{-0.70}^{+0.98}$	&0.71$_{-0.55}^{+0.89}$	&81			&17.0$^d$	&142.5	&<11.16$^g$		&53$^{+17}_{-21}$ &>0.16\\
MRC\,0251-273    	&3.160	&9.98$_{-2.74}^{+2.95}$	&0.69$_{-0.52}^{+1.05}$	&79			& 3.9$^a$	&30.3		&10.96$^g$		&47$^{+19}_{-19}$ &>0.11\\
MRC\,0324-228    	&1.894	&7.76$_{-0.70}^{+0.82}$	&<0.86	&<98		& 9.6$^a$	&81.1		&10.7(8)$^f$	&uncons.		&<0.15\\
MRC\,0350-279      	&1.900	&0.66$_{-0.10}^{+0.24}$	&<0.77	&<88		& 1.2$^a$	&10.4		&<11.00$^g$		&uncons.		&\nodata\\
MRC\,0406-244      	&2.427	&9.40$_{-1.26}^{+1.41}$	&<1.63	&<186		& 10.0$^d$	&83.2		&11.1(3)$^f$	&uncons.		&<0.22\\
PKS\,0529-549       &2.575	&5.88$_{-0.83}^{+0.83}$	&8.86$_{-1.50}^{+1.66}$	&1018		& 3$^i$		&24.7		&11.46$^g$		&66$^{+5}_{-4}$		&0.84\\
TN\,J0924-2201		&5.195	&<14.98	&1.00$_{-0.71}^{+1.34}$	&142		& 1.2$^b$ 	&7.6		&11.10$^g$		&uncons.		&>0.23\\
MRC\,0943-242 (H)	&2.923	&6.39$_{-1.35}^{+1.57}$	&0.36$_{-0.29}^{+0.23}$	&41			& 3.9$^d$	&31.0		&11.3(4)$^f$	&52$^{+18}_{-22}$	&0.03\\
MRC\,0943-242 (C)	&2.923	&\nodata &6.49$_{-1.94}^{+1.41}$	&747		& \nodata	&\nodata	&\nodata	&42$^{+3}_{-4}$	&\nodata\\
MRC\,1017-220      	&1.768	&2.01$_{-0.09}^{+0.24}$	&<1.75	&<201		& <0.2$^c$	&<1.7		&<11.70$^g$		&uncons.		&\nodata\\
4C\,03.24			&3.570	&16.48$_{-4.04}^{+3.23}$	&1.23$_{-1.13}^{+2.08}$	&142		&6.0$^e$	&44.7		&<11.27$^g$		&59$^{+15}_{-20}$	&0.26\\
TN\,J1338+1942		&4.110	&11.24$_{-2.83}^{+3.84}$	&4.02$_{-1.63}^{+2.6}$	&461		&5.2$^c$	&36.6 	&11.04$^g$		&46$^{+9}_{-9}$		&0.59\\
TN\,J2007-1316     	&3.840	&5.13$_{-0.60}^{+1.35}$	&2.19$_{-1.63}^{+2.33}$	&251		& 3.5$^b$	&25.3		&11.9(0)$^f$	&59$^{+14}_{17}$	&0.35\\
MRC\,2025-218       &2.630	&1.16$_{-0.24}^{+0.75}$	&<0.36	&<41		& 5.1$^d$	&41.7		&<11.62$^g$		&uncons.		&\nodata\\
MRC\,2048-272     	&2.060	&<0.18	&<0.65	&<74		& 6.8$^c$	&58.2		&11.47$^g$		&uncons. 		&\nodata\\
MRC\,2104-242      	&2.491	&9.56$_{-1.58}^{+1.52}$ 	&<0.38	&<43		& 23.7$^c$	&196.2	&11.0(0.6)$^f$	&uncons.		&<0.07\\
4C\,23.56		   	&2.483	&27.58$_{-1.24}^{+1.60}$	&<0.49	&<56		& 53.0$^d$	&439.1	&<11.59$^g$		&uncons.		&\nodata\\
4C\,19.71		   	&3.592	&10.91$_{-3.74}^{+6.47}$	&0.74$_{-0.55}^{+1.52}$	&84			& 23.3$^d$	&172.0	&<11.13$^g$		&44$^{+24}_{-17}$	&0.17\\
MRC\,2224-273     	&1.679	&1.70$_{-0.15}^{+0.37}$	&1.21$_{-0.55}^{+0.72}$	&138		& <0.2$^a$	&<1.7		&11.41$^g$		&61$^{+11}_{-12}$	&0.55\\
\bottomrule
\end{tabular}
\begin{tablenotes}
\small
\item \liragn\ is the integrated AGN luminosity. \lirsf\ is the
integrated SF luminosity over the wavelengths 8--1000\,\mum\ in the
rest-frame assuming $\beta=2.5$ (\S~\ref{subsec:IR_lum}). Upper-limits
in \lirsf\ where estimated assuming $\beta=2.5$ and T=50\,K
(Sec.~\ref{subsec:IR_lum}). The star-formation rates, SFR, are calculated
from the \lirsf\ via the conversion given in \cite{Kennicutt1998}
but we scaled these values from the original Salpeter IMF assumed in
\cite{Kennicutt1998} to a Kroupa IMF by dividing by a factor of
1.5.\,LAS is the largest angular size of the radio source as given in
the references indicated by the superscript: $^a$\citet{Kapahi1998},
$^b$\citet{DeBreuck2000}, $^c$\citet{Pentericci2000},
$^d$\citet{Carilli1997}, $^e$\citet{vanOjik1996}, $^i$size extracted
from original map. (*) The physical size of radio source at 1.4\,GHz
(LAS) at the given redshift. The stellar masses are taken from the
studies indicated by the superscript: $^g$\citet{DeBreuck2010},
$^f$\citet{Drouart2016}. Both of these studies use a Kroupa 2001
IMF. Some values of the stellar mass have two estimates.  The value is
the parentheses replaces the last digits to give the alternative value
(see \S~\ref{subsec:stellar_masses}). For example, the possible stellar
masses for MRC\,2104-242 are log($M_*/M_{\odot}$)=11.0 or 10.6. (**) The
difference between the total luminosity of heated dust by star-formation,
\lirsf (this paper) and the total luminosity of dust heated by starbursts
L$^{\rm IR}_{\rm SB}$ \citep{Drouart2014}. Two of sources contain
additional sources in the ALMA images and for these two sources, we
indicate the host galaxy and companions as ``H'' and ``C'', respectively.
\end{tablenotes}
\end{threeparttable}
\end{table*}

From our well-sampled SEDs we estimate the total IR luminosity of the AGN
and SF component (\liragn\ and \lirsf). We estimate the total IR
luminosity as the integrated flux density over rest-frame 8-1000\,\mum\
continuum emission. The flux densities are derived from the best fit of
the analytic models, \eqref{AGN_eq} and \eqref{BB_eq} for the AGN and SF
component respectively. To determine the total integrated luminosity,
the analytic models need to be well constrained. For several sources,
this is not the case for the SF component. Either because there are only
upper limits in the FIR and ALMA bands (e.g., Fig.~\ref{fig_0350}), the
measured ALMA flux is not consistent with originating from pure dust
emission (e.g., Fig.~\ref{fig_0156}) or there is only one detection in
the FIR (e.g., Fig. \ref{fig_0924}) which is not enough to constrain
a model with two free parameters.  In these cases, only an upper limit of
\lirsf\ can be estimated. This is done by scaling a modified BB (with
fixed $\beta = 2.5$ and T=$50$\,K) to the ALMA data point.

The estimated upper limit of the \lirsf\ is dependent on the exact values
of $\beta$ and T that are assumed. A flatter slope, $\beta=1.5-2$,
will lower the inferred IR luminosity. Assuming a higher (or lower)
temperature then 50\,K for a fixed $\beta$ will increase (or decrease)
the integrated IR luminosity. We illustrate these dependencies in
Fig.~\ref{fig:Lum_IR}. In our sample, eight sources are not detected in our
ALMA data: TN\,J0205+2242, MRC\,0324-228, MRC\,0350-279, TN\,J2007-1316,
MRC\,2025-218, MRC\,2048-272, MRC\,2104-242 and 4C\,23.56. Four sources
have ALMA fluxes which are consistent with being dominated by synchrotron
emission: MRC\,0037-258, MRC\,0152-209, MRC\,0406-244 and MRC\,1017-220,
The SFR for these sources are determined by scaling a modified BB to
the ALMA detection and are thus very conservative upper limits (i.e.,
they could be much lower).

In the case of only upper-limits in the MIR, it is not possible
to constrain the AGN contribution to the SED. This is the case
for four sources: TN\,J0121+1320, TN\,J0205+2242, TN\,J0924-2201
and MRC\,2048-272. For these four galaxies, \liragn\ is given as a
upper-limit and have been estimated by scaling the analytic AGN model,
\eqref{AGN_eq}, to the \spitzer\ IRS~16\,\mum\ upper-limit with a fixed
slope of $\gamma$=2. The best fit values and upper-limits of \liragn\
and \lirsf\ are given in Table~\ref{tab:SED_results}.

\subsection{Calculating uncertainties of the integrated IR luminosity}

The estimated total integrated luminosity results from our fitting
the SEDs. Each combination of the parameters affects the total infrared
luminosity in a unique way. There are degeneracies between parameters,
meaning that several combinations of parameters can give the same
integrated luminosity. Therefore the standard uncertainty estimates,
such as quadratically summing the uncertainties for the luminosities
of each individual component, is not an accurate reflection of the true
uncertainty. In order to estimate accurately the total luminosity and its
associated uncertainty, we performed an after-the-fit post-processing
calculation. Because each Monte Carlo chain contains all the required
information about each fitted parameter, we build the marginalized
distribution for the integrated luminosity for each step and each
walker after convergence by integrating the model in the defined
wavelength limits (8-1000\,\mum\, in the rest-frame in our case). From
this distribution we are therefore able to derive the percentile values
that are listed in Table~\ref{tab:SED_results}.

\subsection{High frequency synchrotron}\label{subsec:synch}

The synchrotron emission is modeled by assuming a power-law with
a constant slope without any steepening or cut-off at high
frequencies. This means that we are estimating the maximal
possible contribution from individual synchrotron components out to
frequencies where the low frequency tail of the cold dust emission
and synchrotron possibly overlap. Through the use of already published
VLA L, C and X-band data \citep{Carilli1997, Kapahi1998, Condon1998,
Pentericci2000, DeBreuck2010, Broderick2007} as well as ATCA 7\,mm and
ALMA band 3 for a subset of our sample, we were able to determine
whether or not the detections in ALMA are likely to be the high
frequency extrapolation of the synchrotron emission or low frequency
thermal emission from dust.

There are in total 13 individually resolved detections in ten ALMA maps
which have been found though the SED fitting procedure to most likely
be dominated by synchrotron emission. For four sources, MRC\,0037-258,
MRC\,0156-252 (both components), MRC\,0406-244, and MRC\,1017-220,
the total ALMA flux is consistent with being dominated by synchrotron
emission. For these sources their \lirsf\ and SFR are only given as
upper limits (Figs.~\ref{fig:Lum_IR}, \ref{fig:MS}, \ref{fig:sSFR_z},
and \ref{fig:SFR_LAS}). In two sources, MRC\,0114-211 and PKS\,0529-549,
one out of two ALMA components are dominated by synchrotron emission and
the other detection appears dominated by thermal dust emission. For three
sources, MRC\,0943-242, 4C\,23.56 and 4C\,19.71, both the synchrotron
lobes and host galaxies are detected in ALMA band 3 or 4. The best fit of
these sources is consistent with the lobes being dominated by synchrotron
emission.  The detections of the host galaxies for MRC\,0943-242 and
4C\,19.71 are consistent with thermal dust emission, while for 4C\,23.56,
the modified blackbody component is completely unconstrained.

\subsection{IR luminosity model comparison}
\label{subsec:IRcomparison}

The IR luminosities were calculated by integrating the flux density of
the two analytic models used in this analysis (see Eqs.~\eqref{AGN_eq}
and \eqref{BB_eq}). This procedure differs from the previous SED fitting
work on the parent sample where starburst templates and an average
AGN model were used to fit the SED \citep{Drouart2014}. To investigate
how these different approaches may influence our results, we now make
a direct comparison of our results with those of \citeauthor{Drouart2014}.

\begin{figure}
\centering
\begin{subfigure}{\linewidth}
\includegraphics[width=\linewidth]{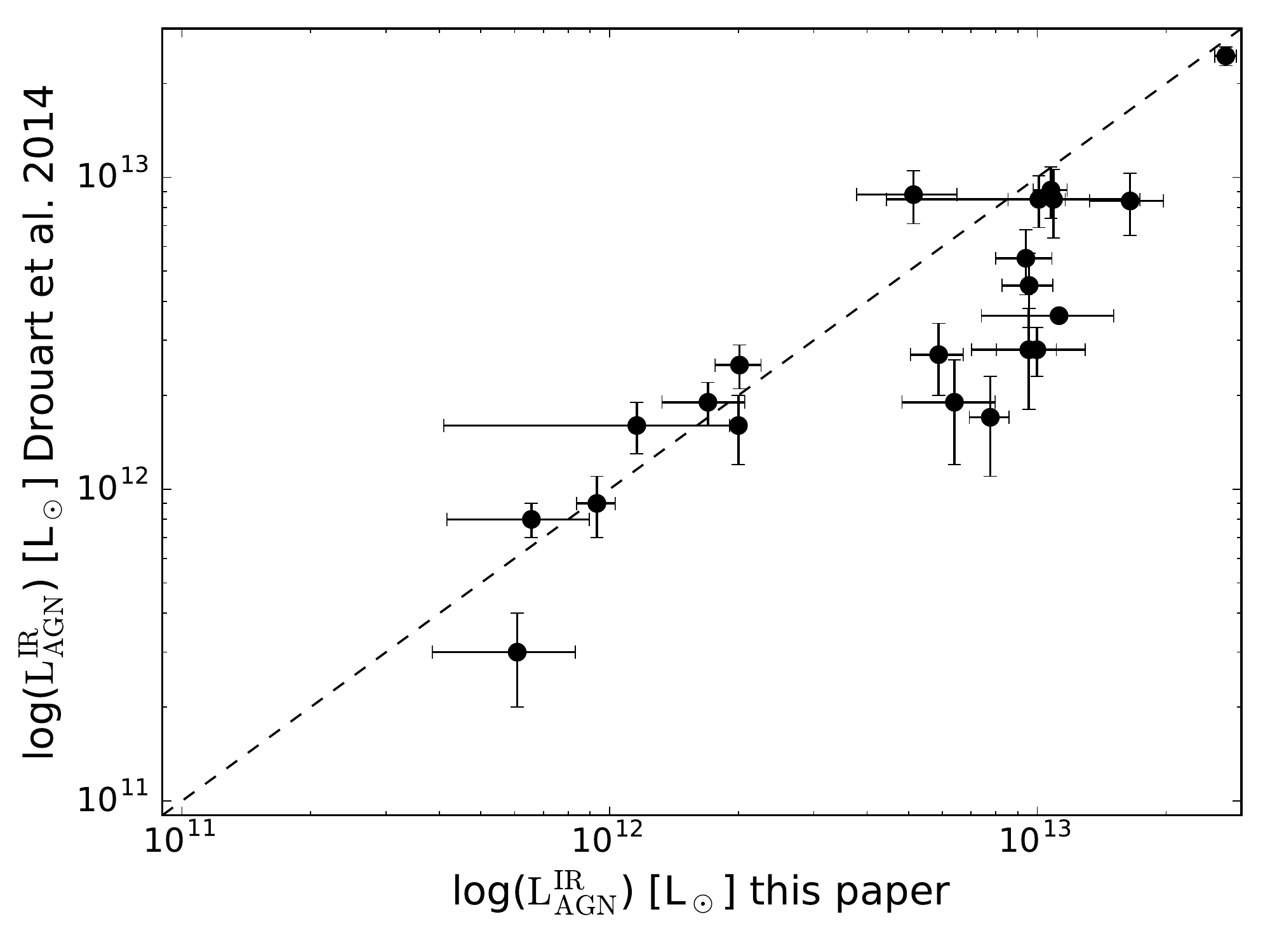}
\end{subfigure}
\begin{subfigure}{\linewidth}
\includegraphics[width=\linewidth]{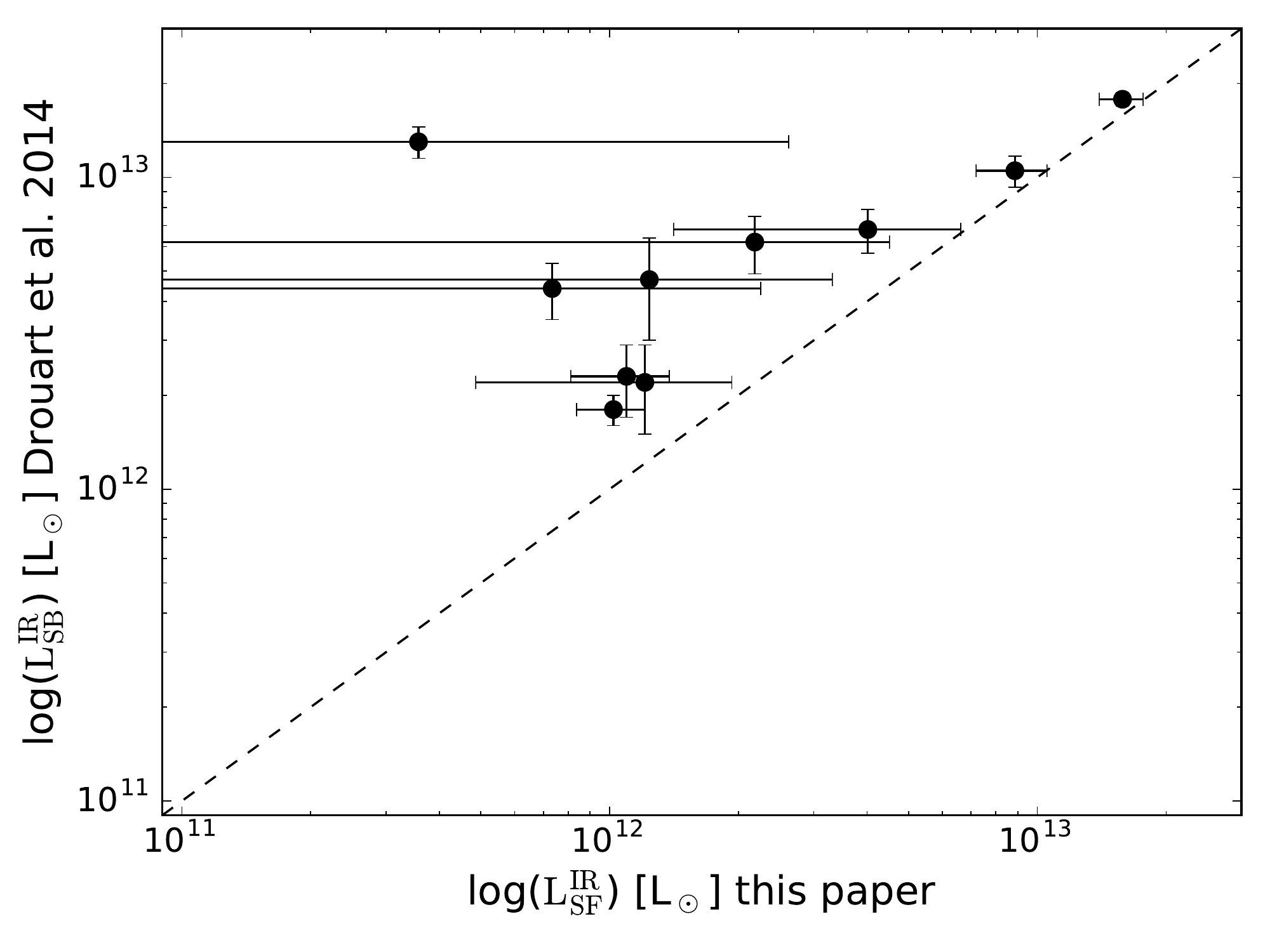}
\end{subfigure}
\caption{Comparison between the integrated IR AGN and SF
luminosities computed in this paper compared to the results of
\cite{Drouart2014}. \textit{(top)} A comparison between the infrared
luminosity of the AGN and \textit{(bottom)} a comparison between
the infrared luminosity of the starburst or star forming component.
In both panels, the dashed line is the one-to-one relationship between
luminosities.}
\label{fig:one-to-one}
\end{figure}

The study of \citeauthor{Drouart2014} used the AGN model implemented
in DecompIR \citep[see Fig.~\ref{f_cut_fit}; ][]{Mullaney2011} The
simple empirically motivated model implemented in this paper does
not deviate much form the average AGN model of \citet{Mullaney2011}.
A direct comparison of the total integrated AGN luminosities (Fig. \ref{fig:one-to-one}), \liragn\ suggests a modest, $\sim$20\%,
offset in the median of \liragn\ in between the two approaches.

\begin{table}[ht]
\begin{center}
\begin{tabular*}{\linewidth}{c @{\extracolsep{\fill}} lccc}
\toprule
Quality&Number&L$_\mathrm{SF}^\mathrm{IR}$/L$_\mathrm{SB}^\mathrm{IR}$\\
\midrule
Constrained L$^\mathrm{IR}$ & 11 & 0.49\\
\lirsf upper limit & 3 & <0.15\\
	\\[-1em]
L$_\mathrm{SB}^\mathrm{IR}$ upper limit & 3 & >0.17\\
	\\[-1em]
All sources & 17 & $\sim$0.38\\
\bottomrule
\end{tabular*}
\caption{The average ratio of IR luminosities of this paper (denoted
as L$^\mathrm{IR}_\mathrm{SF}$) and those from \cite[][denoted as
L$^\mathrm{IR}_\mathrm{SB}$]{Drouart2014}. The ratio is given for three
cases: the L$^\mathrm{IR}$ estimated for sources with detections in both
ALMA and Herschel and thus the infrared luminosities are constrained in
both studies (Constrained L$^\mathrm{IR}$); when the infrared luminosity
is given as an upper limit in this paper but is given as a detection in
\citeauthor{Drouart2014} (\lirsf\ upper limit); and where L$^\mathrm{IR}$
for the source is constrained but L$^\mathrm{IR}_\mathrm{SB}$ is an
upper limit in \citeauthor{Drouart2014} (L$_\mathrm{SB}^\mathrm{IR}$
upper limit).}
\label{tab:SF_comparison}
\end{center}
\end{table}

\begin{figure*}
\centering
\begin{subfigure}{.4\textwidth}
\includegraphics[width=\linewidth]{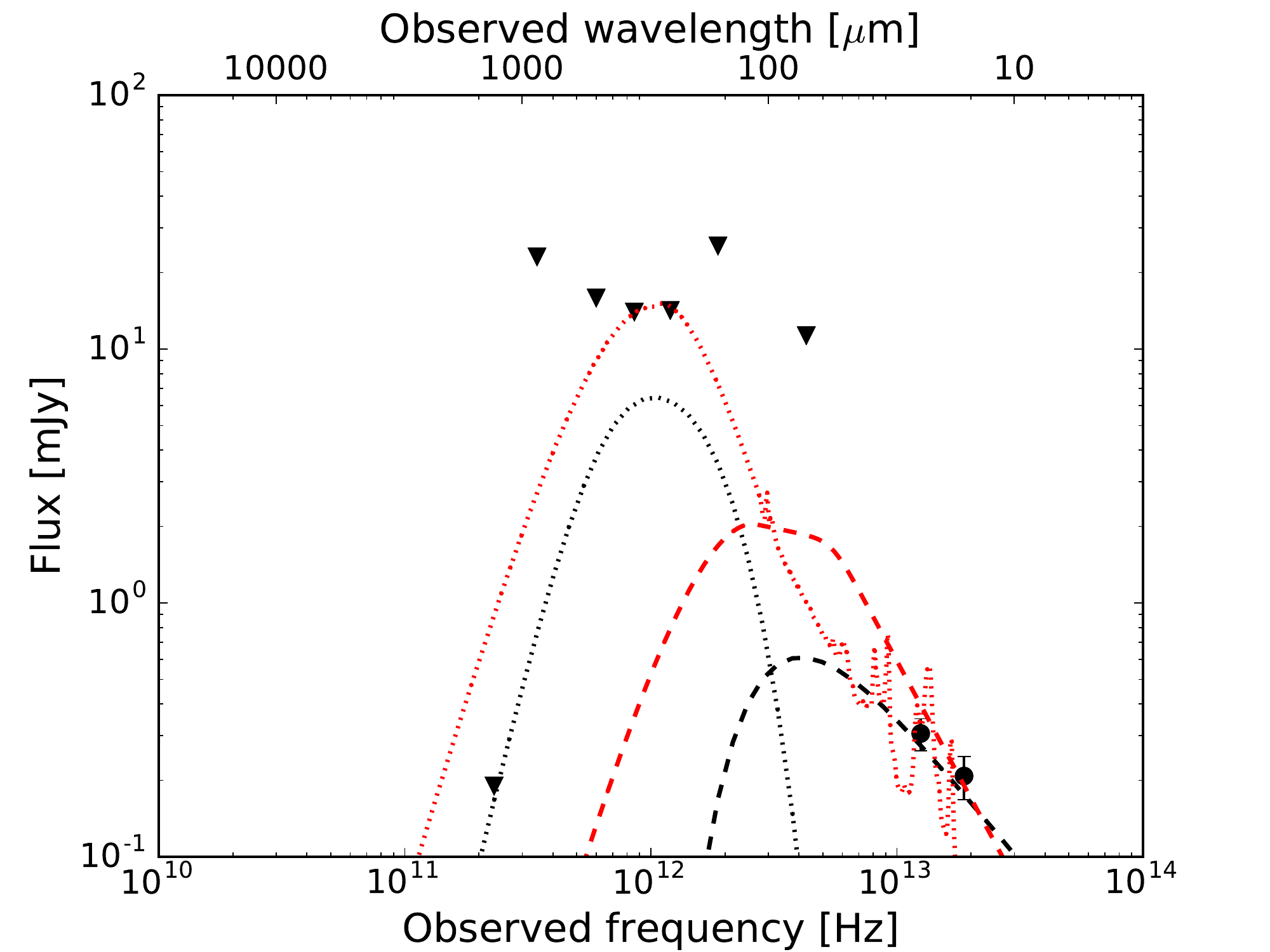}
\end{subfigure}
\begin{subfigure}{.4\textwidth}
\includegraphics[width=\linewidth]{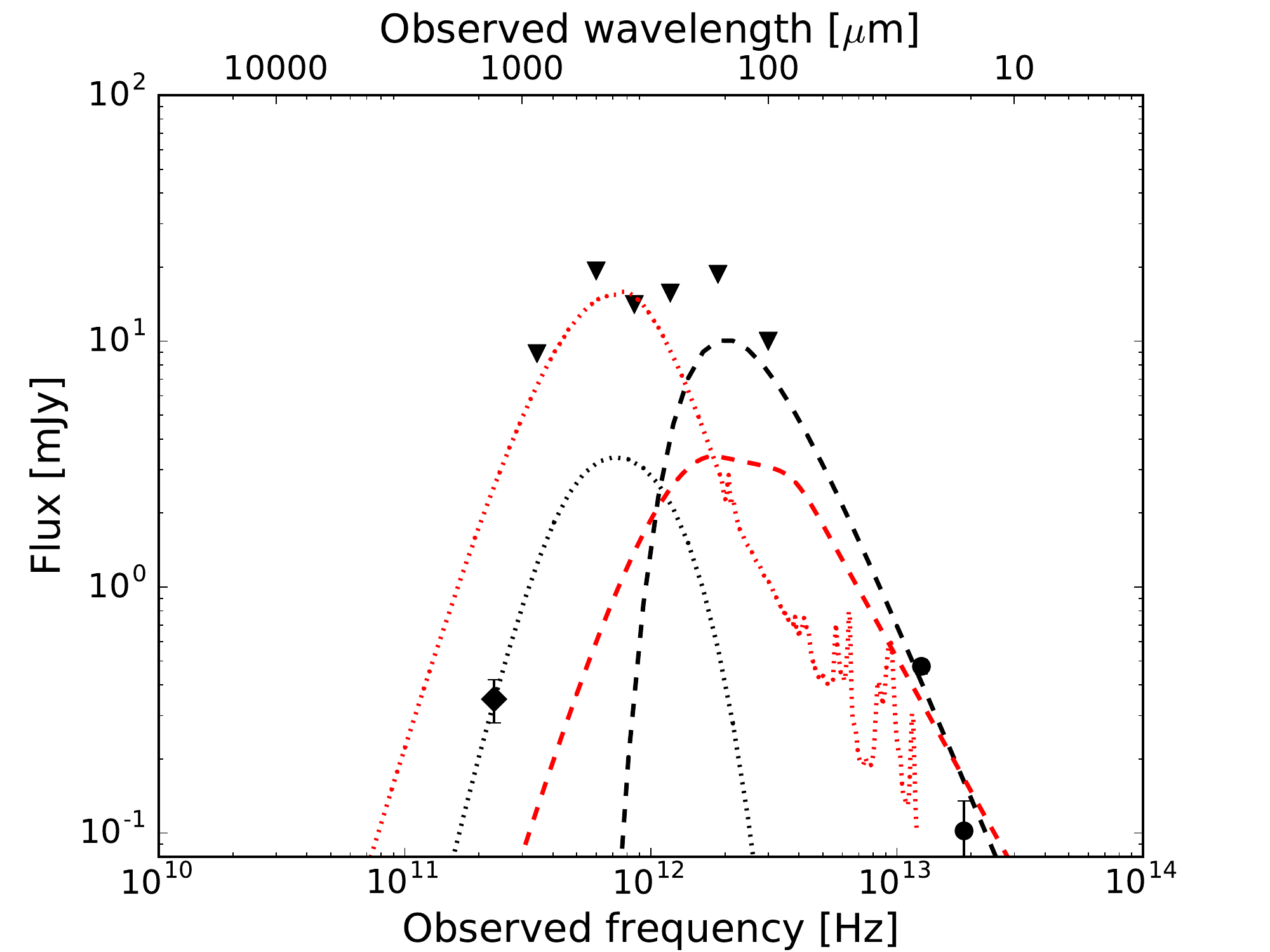}
\end{subfigure}
\begin{subfigure}{.4\textwidth}
\includegraphics[width=\linewidth]{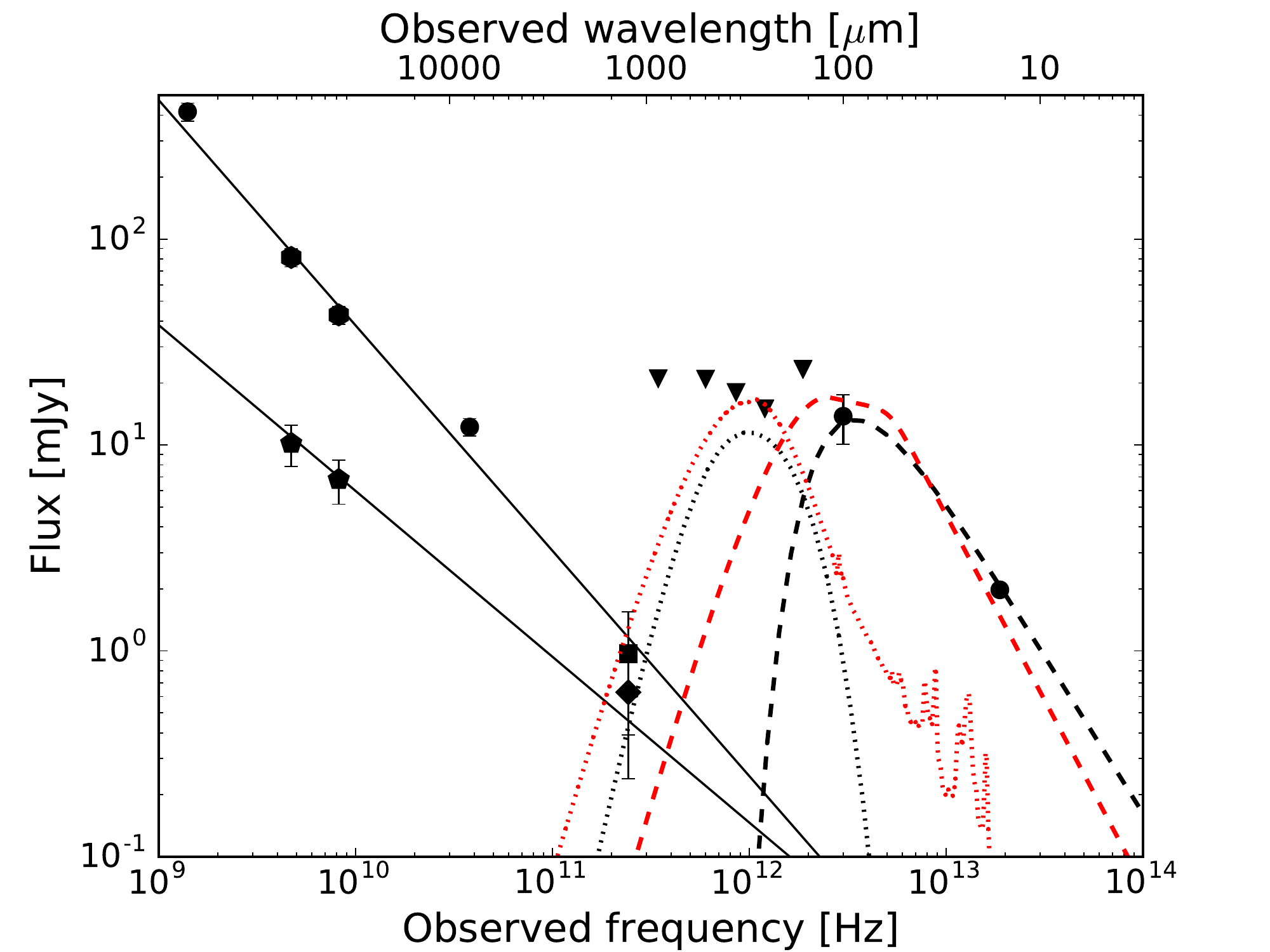}
\end{subfigure}
\begin{subfigure}{.4\textwidth}
\includegraphics[width=\linewidth]{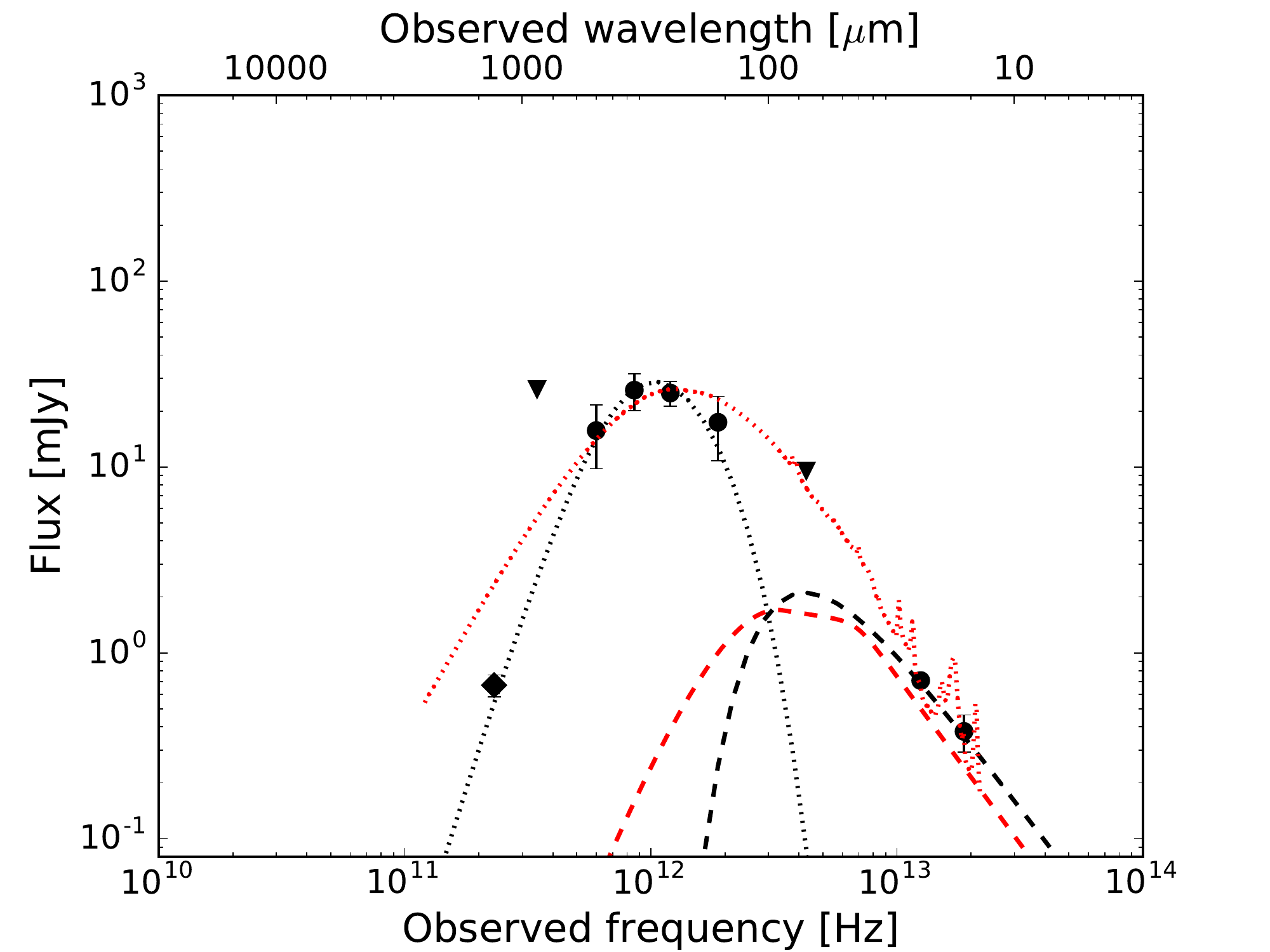}
\end{subfigure}
\caption{Model comparison between the SF and AGN models of this paper and
\cite{Drouart2014} for galaxies MRC\,0350-279 \textit{(top left)},
MRC\,0251-273 \textit{(top right)}, MRC\,0156-252 \textit{(bottom
left)} and MRC\,0211-256 \textit{(bottom right)}. In each plot,
the black solid lines represent the synchrotron emission, dotted
lines indicate the FIR thermal emission due to star-formation, and
dash-dotted indicates the best-fit MIR emission due to the AGN as
determined from the best fits to the photometry for each galaxy. The
red lines with the same styles represent the same components as fitted
in \citeauthor{Drouart2014} A synchrotron power-law was not fit in the
analysis of \citeauthor{Drouart2014}}
\label{fig:SF_model_comp}
\end{figure*}

We compare the star-forming models in this paper and those from
\citeauthor{Drouart2014} The differences are predominately due to having
data in the rest-frame submm which is more sensitive, has higher spatial
resolution, and covers a wider wavelength range. As can be seen from
the SED shapes of the starburst/star-forming components, they differ
both in the presence of PAH features in the \citeauthor{Drouart2014}
models, and sometimes in the overall shape depending on which starburst
template is used (Fig. \ref{fig:SF_model_comp}). The PAH features do
not contribute significantly to the total integrated \lirsf. However,
in general when it comes to SED fitting, the choice of template
can change the integrated \lir\ by up to a factor of four, as mentioned
in Sect~4.4 of \cite{Drouart2014}. The factor of four is estimated in
the case one cannot discriminate between the most extreme starburst
templates. Even in the less extreme case, there is still a factor of
$\sim$2 between the sets of templates that are consistent with the same
data points. The SED parameters that influence this difference include
the assumed dust temperature, opacity and emissivity \citep[see also
Fig.~8 of][who show how different templates can give the same integrated
IR luminosity]{Casey2018}. The bottom right panel of figure \ref{fig:SF_model_comp} illustrates
the importance of having at least one sensitive measurement on the
Rayleigh-Jeans side of the emission peak, even when there are multiple
detections near the peak of the SED. In this HzRG MRC\,0211-256,
the observed ALMA band 6 flux is 10$\times$ lower than the
predicted flux in the \cite{Drouart2014} model.

We identify four general categories where the SED fits lead to significant
differences in \lirsf\, between the two studies: (1) only upper limits
in the FIR; (2) one detected ALMA component and upper limits in the
rest of the FIR bands; (3) two spatially resolved detections in the
ALMA bands; and (4) when the FIR consists mainly of detections. It is
clear that there are cases where the differences are not significant
(e.g., MRC\,0156-252). However, in some cases, isolating a sub-component
in high resolution, 0\farcs3, sensitive ALMA imaging leads to a
significantly lower \lirsf\ which simply is not possible using only
the low-resolution \herschel\ data (e.g., MRC\,0251-273). There are also
cases where significantly deeper ALMA data ($\ga$10$\times$ deeper than
any previous submm/mm observations) still does not detect any emission
(e.g., MRC\,0350-279). Accordingly, our limits on \lirsf\ are also much
more stringent, but formally still consistent with the shallower upper
limits of \citeauthor{Drouart2014} Furthermore, we also include a more
robust extrapolation of the synchrotron component due to now including
the ALMA and ATCA $\sim$90\,GHz data.

As mentioned above, it is important to note that choice of template
can change the \lirsf\, by $\sim$2 even with good photometric coverage of
the peak of the thermal dust emission. In our sample, six sources have
$\geq$3 FIR detections from \herschel\, and LABOCA. Out of these six,
three have good agreement in the infrared luminosity, having ratios
of 0.72--0.89, with \citeauthor{Drouart2014}. One is an ALMA source
with multiple components, MRC\,0943-242, which explains the large
difference in \lirsf. Only two sources, MRC\,0211-256 and 4C\,03.24,
have poor agreement due to differences between a modified blackbody
and the starburst templates used in \cite{Drouart2014}. Considering
that \citeauthor{Drouart2014} found a potential difference of $\sim$2
within the templates used in their study, it is to be expected that we
are finding a factor of two to three difference compared to their results for a
few of our sources, especially when we also include an additional ALMA
data point that constrains the Rayleigh-Jeans side of the emission peak.

Quantitatively, we find that our estimated far-infrared
luminosities of the component due to star formation are
only a fraction of those found by \citeauthor{Drouart2014}
(Fig~\ref{fig:one-to-one} and Tab.~\ref{tab:SF_comparison}).  If we
only include detections, we find that our estimates of \lirsf\ are
only $\sim$50\% of those estimated in \citeauthor{Drouart2014}
If we also include detections and upper limits of \lirsf\ in either
of the two papers, then our estimates are only $\sim$40\% of those
in \citeauthor[][(Table~\ref{fig:SF_model_comp})]{Drouart2014}. We
discuss the implications of these significantly lower \lirsf\ in
Sects.~\ref{sec:RGsSMBHs} and \ref{sec:MS}. If we compare all the
sources together and estimate the median IR luminosity including upper
limits of the overlapping 25 sources in both studies, we find our IR
luminosities are a factor $\sim$7 lower.\footnote{For this comparison,
we used a Kaplan-Meier estimator \citep{feigelson85}.}

\subsection{Notes on the Stellar Masses}
\label{subsec:stellar_masses}

Given the importance of stellar masses in our analysis, we briefly discuss
the nature of the mass estimates we are utilizing. All stellar masses
used in this paper are based on those estimated in \citet{Seymour2007,
DeBreuck2010, Drouart2016}. Our stellar masses are based on 6-band
\spitzer\ photometry covering 3.6--24\,\mum, augmented with near-IR
imaging.  The AGN in our sample may contribute flux to the optical to
mid-IR photometry used to determine the stellar masses. AGN emission
contributes from both direct and scattered continuum (dominating
at $\lambda_\mathrm{rest}<$1\mum), and dust emission from the torus
(dominating at $\lambda_\mathrm{rest}$>5\mum). Our sample is composed
of Type-2 AGN where the direct AGN contribution is obscured by the dusty
torus. One exception, MRC\,2025-218, has a SED which is consistent
with AGN-dominated continuum emission and thus, although it is detected
in the photometry used to estimate masses, we assume its stellar mass
is an upper limit. \spitzer\ photometry used in \citet{Seymour2007}
and \citet{DeBreuck2010} allowed them to extrapolate the hot dust
emission from the AGN down to rest-frame 1--2\,\mum\ where the old
stellar population peaks. The stellar masses of objects where this hot
dust contribution \textit{may} dominate are conservatively listed as
upper limits.

In the remaining objects, \citet{DeBreuck2010} derived the stellar
masses assuming a maximally old stellar population. While such
estimates are reasonable, they may slightly over-estimate the masses
\citep[for a detailed discussion, see][]{Seymour2007}. The remedy this,
\citet{Drouart2016} combined the \spitzer\ data with existing optical
and near-IR photometry on a sub-sample to perform a multi-component SED
fitting through population synthesis. In cases of overlap, we use the
stellar masses derived by \citet{Drouart2016}. The paper from which each
mass estimate is taken is indicated in Table~\ref{tab:SED_results}.

\begin{figure*}[ht]
\centering
\includegraphics[width=\linewidth]{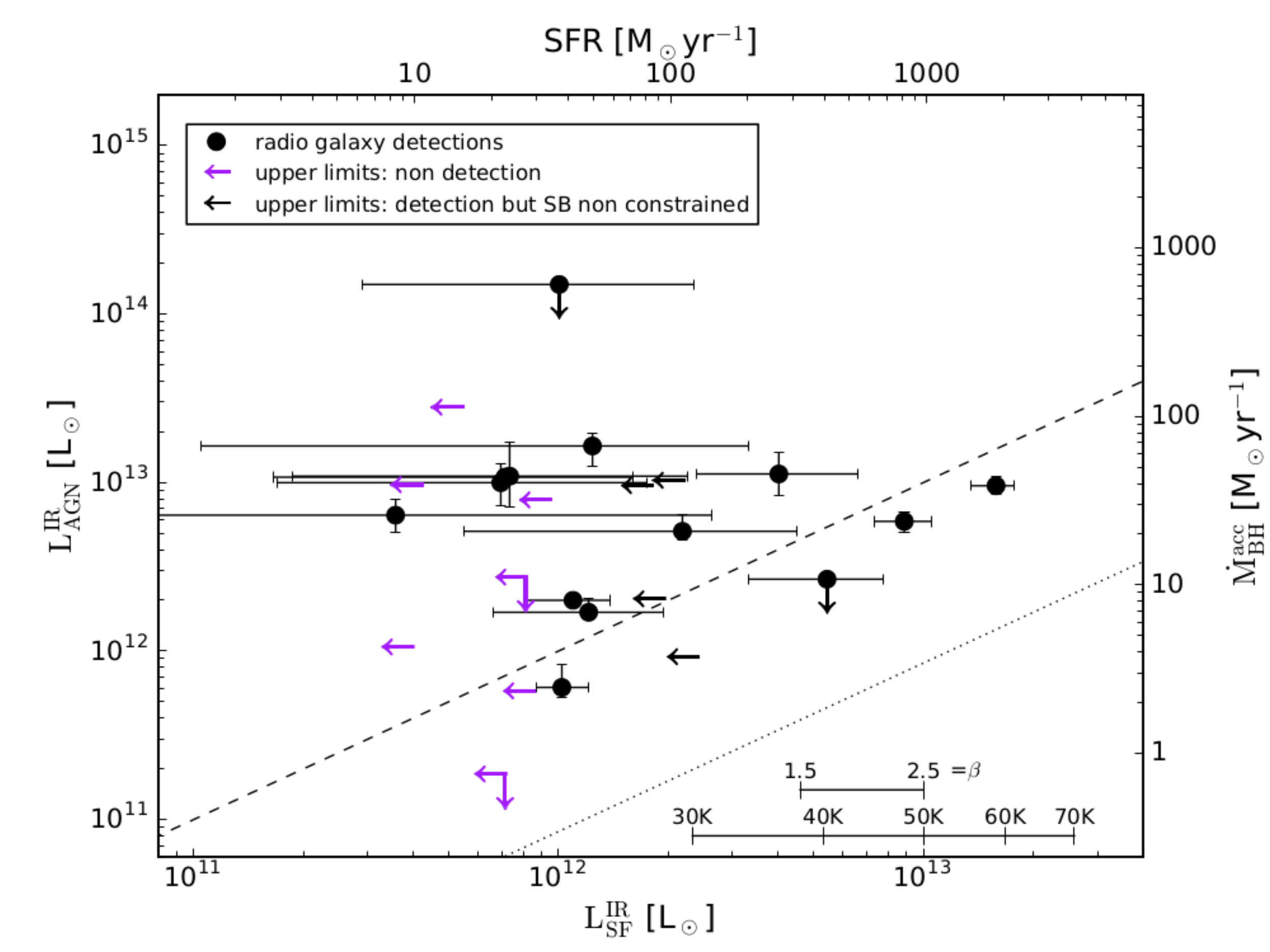}
\caption{Estimated AGN luminosity, L$_\mathrm{AGN}^\mathrm{IR}$,
versus the SF luminosity, L$_\mathrm{SF}^\mathrm{IR}$.  The dashed
black line indicates the values where L$_\mathrm{AGN}^\mathrm{IR}$
= L$_\mathrm{SF}^\mathrm{IR}$, while the dotted black line
indicates parallel growth of the stellar mass and black hole mass,
$\dot{\rm M}_\mathrm{BH}^\mathrm{acc}$=0.002$\times$SFR. Filled black
circles are galaxies detected with ALMA and with constrained SF and
AGN luminosity estimates. Purple arrows are sources with upper limit in the ALMA
band. The upper limits of L$_\mathrm{SF}^\mathrm{IR}$ are approximated
by scaling a modified blackbody to the 3-$\sigma$ upper limit estimated
assuming $\beta$=2.5 and T=50\,K. We assumed $\beta$=2.5
for all of our fits and 50\,K is approximately the medium temperature of
our best fits (Table~\ref{tab:SED_results}). Black arrows
indicate galaxies which are detected in ALMA but where the observed ALMA
flux(es) are consistent with an extrapolation of the lower frequency
synchrotron emission implying little or no contribution from thermal dust
emission. We indicate with bars in the lower right corner of the plot,
how the upper limits of the \lirsf\ would shift if one of the fixed
parameters, $\beta$ or T, are changed with respect to our assumed values
of $\beta=2.5$ and T=50\,K.}
\label{fig:Lum_IR}
\end{figure*}

\section{Relationship between radio galaxies and their supermassive
black holes}
\label{sec:RGsSMBHs}

We can now compare the relative growth rates of both galaxies and their central supermassive black holes, using estimates of the star-formation rates and the mass accretion rates, respectively. Such
an investigation addresses the question of how galaxies and black
holes evolved to the black hole mass-bulge mass relation we
observe locally \citep{Magorrian1998, Gebhardt2000, Ferrarese2000,
Haring2004}. We have already discussed this issue for radio galaxies
in \citet{Drouart2014}, so we only briefly highlight how our results
re-enforce the conclusions of that paper.
The differences between our approach and the one of \citet{Drouart2014} are detailed in Section \ref{subsec:IRcomparison}.
Other than these differences, we follow the analysis of \citeauthor{Drouart2014} quite closely, that is, we use the same
conversion factors between IR luminosities and SFR and AGN accretion
rates, and a very similar SED to determine the AGN luminosities (see
Fig.~\ref{f_cut_fit} and \ref{fig:one-to-one} for a direct comparison).

\subsection{Determining growth rates of star-formation and SMBH}

To estimate the star formation rates of galaxies, we use the conversion
factor from \citet{Kennicutt1998}:
\begin{equation}
{\rm SFR} = 1.72 \times 10^{-10} \times {\rm L}_{\rm SF}^{\rm IR}, 
\end{equation}
using a \cite{Salpeter1955} IMF. For consistency when comparing the SFR
with stellar masses from \cite{Drouart2014} and \cite{DeBreuck2010},
we divide these estimates by a factor of 1.5 to convert from a
Salpeter IMF to a Kroupa IMF. Our new data and fitting have resulted
in significantly lower star-formation rate estimates for some sources
compared to \citeauthor{Drouart2014}. If we only consider the 25
sources in common, the median star-formation rate as estimated by
\citeauthor{Drouart2014} is 760\,\msunyr. The SFR for our sample span from
about 40 to $\sim$2000\,\msunyr\ with a median\footnote{the medians were
estimated using a Kaplan-Meier estimator \citep{feigelson85}.} value
of 110\,\msunyr (Table~\ref{tab:SED_results}). Our star-formation rates
are seven times lower than previously estimated by \citeauthor{Drouart2014}
for the same sources.

To convert \liragn\ to black hole accretion rate $\dot{\rm M}^{\rm
acc}_{\rm BH}$, we follow \cite{Drouart2014}: 
\begin{equation} 
\kappa ^{\rm Bol}_{\rm AGN} \times  {\rm L}^{\rm IR}_{\rm AGN} = \epsilon \dot{\rm
M}^{\rm acc}_{\rm BH} c^2,
\end{equation} 
where the efficiency factor $\epsilon =0.1$ and the bolometric correction factor $\kappa^{\rm Bol}_{\rm AGN} = 6$. We refer to \cite{Drouart2014} for a more detailed discussion. A large fraction of the sample lies above the one-to-one line between
the \lirsf\ and \liragn, showing that the FIR emission from the SF
component is generally much weaker than the luminosity of the AGN. This is
consistent with other samples of powerful AGN \citep[e.g.,][]{Netzer2014,
Netzer2016, Stanley2015}

\subsection{Relative growth rates of galaxies and SMBHs}

What are the relative growth rates of the stellar and black hole
mass? To make this comparison, we simply scale the IR luminosities
of each component as just described. If the galaxies evolve
along the local relation, we would expect the accretion rate,
$\dot{\mathrm{M}}_\mathrm{BH}^\mathrm{acc}$, to be about $\sim$0.2\% of the SFR
\citep[we chose 0.2\% to be consistent with \citeauthor{Drouart2014}
and is within the uncertainty of estimates in the literature at the time;
see][]{KH13}. Of course, there are many assumptions that must be made in
order to use these relations and one should be aware that the empirical
relation is really between integrated IR luminosities, \lirsf\ and \liragn,
with scaling factors.  Nevertheless, we find that our sample lies more
than an order of magnitude above the local parallel growth relation of
0.2\% (Fig.~\ref{fig:Lum_IR}). This suggests that the black holes can
become overly massive relative to their host galaxies if the accretion
time spans the same time scale as the star formation. In fact, given that
the host galaxies are already massive, it is likely that this implies
that the SMBHs are overly massive at the epoch they are observed.

In \citet{Drouart2014}, we argued that for the growth of the host
galaxy and SMBH to ultimately be consistent with the local relation, the
on-going star formation would have to last about a factor of eight longer
than the observed level AGN activity. Shifting L$_\text{SF}^\text{IR}$
downwards by about a factor of seven, now implies that the star
formation must last over a factor of 50 longer. If the lifetime of the
radio loud phase is $\sim$25\,Myrs \citep[][]{martini01, schmidt17},
this would suggest that the star formation has to last more than
a Gyr. Since we predominately have upper limits for the star
formation rates of the majority of the galaxies, this appears unlikely.
There is evidence at high redshift that perhaps SMBHs are already overly
massive compared to their host galaxies, where overly massive means that
they do not have the local value of the black hole mass to spheroidal
mass ratio \citep[e.g.,][but see \citealt{willott17}]{nesvadba11, wang13,
willott15, trakhtenbrot15, shao17, vayner17}.  Thus, the time required
for the stellar mass to ``catch up'' to the mass of the SMBH is actually
much longer than we have estimated here. Our new results therefore
exacerbate the problem already discussed in \citeauthor{Drouart2014} that
it appears difficult for the mass ratio of the SMBH and the spheroidal
component of the radio galaxies to fall on the local relation through
star formation. We caution however that the average black hole accretion
rates over longer time scales of star formation are not well constrained
by the relatively instantaneous estimates provided here and in Drouart
et al. \citep{hickox14, Stanley2015, volonteri15a}.

\subsection{Keeping up with rapid SMBH growth}
The host galaxies of \hzrgs\ need to catch up with the growth
of the SMBH, as they appear to be already overly massive. In order
to end up on the local mass relationship, the stellar component needs
to grow through a mechanism that does not fuel substantially the supermassive black hole.
In the following sections, we discuss the possibility of growth by
mergers as a way to explain how the sample of high-z galaxies in
our study can evolve on to the local relationship.

\subsubsection{Growth through major mergers}

High redshift powerful radio galaxies like the ones studied here are
found in environments which are over-dense \citep[e.g.,][]{wylezalek13,
hatch14, dannerbauer14, cooke15, cooke16, noirot16, noirot18}.
In such environments, mergers are likely an important mode of galaxy
growth. However, a few caveats must be kept in mind when considering
galaxy mergers as the mechanism allowing galaxies and SMBHs of
powerful radio galaxies to evolve onto the local mass relation. The
first requirement is that mergers do not bring substantial amounts of
gas to grow the SMBH significantly compared to the mass of the accreted
stars. Major mergers, which may increase the stellar mass considerably,
would have to be gas poor galaxies as major mergers can carry
gas efficiently to small scales (kpc-scales) through dissipation which
may lead to significant black hole growth. Generally, massive galaxies
at high redshift, those that would constitute major mergers for radio
galaxies, are gas-rich \citep[e.g.,][]{bolatto15, noble17, emonts18}. So
unless massive galaxies within the over-dense environments of radio
galaxies are particularly gas poor \citep[see e.g.,][]{Emonts2014,
Lee2017, dannerbauer17, emonts18} then major mergers do not appear
to be particularly favored for growing the stellar content of radio
galaxies. Having said that, the quenching time of moderately massive
galaxies in clusters is likely a small fraction of the Hubble time
\citep[e.g.,][]{muzzin12, foltz18} but with reduced efficiency with
increasing redshift \citep{nantais16, nantais17}. The second significant
problem with major mergers as the driver of the stellar growth is that the
merging galaxy likely also contains a supermassive black hole.  In the early
universe, the merger partner may have a black hole that is massive
relative to the mass of its host \citep[e.g.,][]{willott15}. After
the merger has advanced to the coalescence stage of the merger,
which occurs in a few dynamical times of the most massive galaxy,
the black holes will merge in less than a Hubble time \citep[$\la$1\,Gyr
for M$_{\star}\sim$10$^{11}$\,M$_{\sun}$, which is approximately the
stellar masses of our galaxies;][]{berczik06, merritt07}  A final,
but perhaps less important limitation in such a picture is that the
relative velocities of the merging galaxies should be relatively low,
of-order the internal dynamical velocity of the stars in the most massive
galaxy. Thus, relative low speed encounters are favored for efficient
merging.  In the over-densities surrounding the high redshift radio
galaxies, the dispersion of the most massive galaxies in the potential
appears to be high \citep{kuiper11, noirot18}.

\subsubsection{Growth through minor mergers}

Minor mergers may be an effective way to allow the mass of old stellar
populations to grow in massive galaxies without fueling significant SMBH
growth.  There are several pieces of evidence that suggest hypothesizing
that minor mergers contributed significantly to the stellar mass growth of
massive galaxies. High resolution imaging suggests that there may be low
mass galaxies in the surroundings of some radio galaxies \citep{Miley2006,
Seymour2012}. So the potential merging sources are close at hand. Beyond
just the necessary association of low mass galaxies, there are a number
of lines of evidence that support the notion that massive early-type
galaxies grew substantially through minor mergers. Some of these
are: \textit{(1)} the size evolution of massive galaxies in the early
Universe to the present may be driven principally through minor mergers
\citep[e.g.,][]{Daddi2005, vanDokkum2008, Delaye2014, vulcani16, hill17};
\textit{(2)} the change in the mass and luminosity function of galaxies
with redshift and as a function of environment \citep[e.g.,][]{Ilbert2013,
sarron17}; \textit{(3)} the elemental abundance ratios, abundance
gradients, and age gradients in the outer regions of local massive
spheroids are consistent with accreting galaxies with a range of masses,
perhaps predominately low mass, which had their star formation truncated
early in their growth \citep{Huang2013, Greene2013, barosa16}; and
\textit{(4)} the mass of massive early-type galaxies grew by about a
factor of four over the last $\sim$10\,Gyr \citep[e.g.,][]{vanDokkum2010,
Ilbert2013}.

Interestingly, \citet{hill17} identified the epoch at which the stellar
growth of very massive galaxies, M$_{\star}\ga$10$^{11.5}$\,M$_{\sun}$,
transition from growing substantially through star formation to one
where mergers dominate the stellar mass growth. This is at the low
redshift end of the objects in our study but is overall consistent
with the quenching we observe. Moreover, at the average redshift we
are observing our sample, they also find a factor of three increase in the
stellar mass, which again is similar to what is needed to close the gap
between the mass of the supermassive black holes and the host galaxies
\citep[see also][]{vanDokkum2010, vulcani16}. Both the fossil record in
nearby massive galaxies and their \textit{in situ} evolution suggest that
minor mergers played a role in their stellar mass growth and physical
properties \citep[e.g.,][]{Greene2013, hilz13, laporte13, hirschmann15}.

\subsubsection{A comparison with BCGs and X-ray-selected AGNs}

\noindent Radio galaxies lie in over-densities and have been suggested
to be progenitors of brightest cluster galaxies \citep[BCGs;][]{hatch14}, a
more relevant comparison is not with the general properties of massive
galaxies but the stellar mass growth of BCGs. Statistical samples of
BCGs are limited to redshifts lower than about 1 which is lower than the
median redshift of our sample, z$\sim$2.4. Results from these studies
suggest that BCG typically grew by about a factor of two over the last
8-10 Gyrs \citep{a-s98, bellstedt16}.  Overall, these growth rates are
consistent with semi-analytic models which indicate that, since about
z$\sim$1-1.5, BCGs grew by about a factor of 2-4 \citep{deLucia07,
tonini12}. However, any theoretical result explaining the growth of BCGs
is sensitive to the treatment of dynamical friction and tidal stripping
through galaxy-galaxy interactions as galaxies move through the cluster
potential \citep[e.g.,][]{shankar15}. We conclude that if some of the
galaxies in our sample are destined to become BCGs, then our overall
current understanding of the stellar growth of these massive clusters
galaxies is consistent with closing the difference in relative masses
of the supermassive blackholes and their host galaxies.

Other samples, such as X-ray selected AGN, show a range of relative growth
rates of SMBH and host galaxies. Some studies, like those that select
star forming galaxies and then investigate their black hole accretion
rates (using amount of X-ray emission observed above that expected
that due to galaxy stellar populations) and SFR, find that black holes
and galaxies are growing in lock-step \citep[e.g,][]{Delvecchio2015}.
Similarly, some studies of X-ray selected AGN with a wide range of AGN
bolometric luminosities that black holes and galaxies grow in lock-step
\citep[z$\sim$2, e.g.,][]{mullaney12}. But such results are not found
universally. \citet{Netzer2016}, again for an X-ray selected sample
of AGN, but now at somewhat higher redshifts than previous studies,
find that SMBHs are growing more rapidly on average than their hosts.
\citet{cisternas11} find that there is no evolution in the black
hole-to-galaxy mass ratio out to z$\sim$1, except perhaps for high mass
black holes where black holes are overly massive relative to their host
galaxies. Could the variety of results be simply due to the mass of both
the galaxy and the SMBH \citep{cisternas11}?

\begin{figure*}[!ht]
\includegraphics[width=18.5cm]{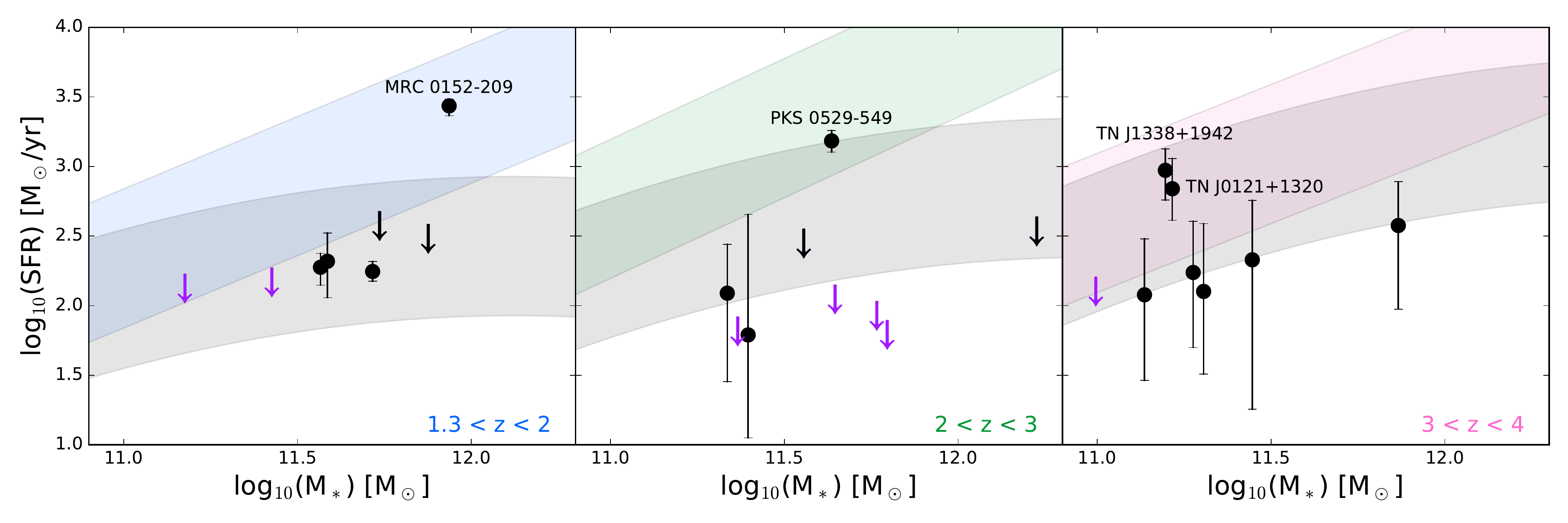}
\caption{Relationship between the SFR and stellar mass for different
redshift bins. The colored shaded regions shows the MS of
\citep{Santini2017} for each respective redshift bins with a 0.3\,dex scatter around the MS. The gray shaded regions shows the MS
with a 0.3\,dex scatter from \citet{Schreiber2015}, with a turnover
at higher stellar masses. Our stellar masses have been scaled from
a Kroupa to a Salpeter IMF to be consistent with the IMF used to
make the MS in this comparison. In the highest redshift bin, 3$<z<$4,
two sources, TN\,J1338+1942 and TNJ0924-2201, have been added to
the right most panel despite having redshifts outside of the range
used to construct the MS (z=4.110 and 5.195 respectively). Given
the redshift dependence on the normalization of the MS, these
galaxies may lie relatively lower in comparison with the mean
relation of a MS derived using galaxies over a more appropriate,
higher redshift range (see Fig.~\ref{fig:sSFR_z}).}
\label{fig:MS}
\end{figure*}

\section{SFR and Main Sequence Comparison: On the road to quenching}
\label{sec:MS}

The main sequence of star-forming galaxies -- the empirical relation
between the star-formation rate and stellar mass with a slope of
approximately 1 and a scatter of about a factor of two -- has been
studied extensively both theoretically and observationally \citep[the
MS, e.g.,][]{Brinchmann2004, Noeske2007, Elbaz2007, Daddi2007,
Santini2009, Santini2017, Peng2010, Whitaker2012, stark13, Speagle2014,
Lehnert15, dave16, LeeB2017, LF18, davidzon18}.  The relation is
observed over a wide redshift range, z$\sim$0-7, and its normalization
increases with increasing redshift. The MS has often been parameterized
as a simple power-law but there has also been evidence that the MS
flattens at higher stellar masses \citep[e.g][]{Whitaker2014,
Lee2015, Tasca2015, Schreiber2015, Tomczak2016, LeeB2017}. The slope
of the MS and the turnover mass in the case of a flattening MS
depend on the sample selection, redshift range and the technique
used to determine the stellar mass and SFR. The estimated turnover
stellar mass is $\sim$10$^{10}$-$10^{10.5}$\,M$_{\odot}$
\citep[e.g.,][]{Schreiber2015}. The MS can be used to study and
classify galaxies according to their relative SFR, where large
deviation from the MS suggests that galaxies are ``starbursts'' if
they lie above the MS, or quenched, if they fall below the MS. The
definition of whether a galaxy is a starburst or is quenched varies
in literature, but is often taken as either offset by more than three
times the scatter or a factor of ten above or below the mean relation
\citep[e.g.,][]{Rodighiero2011}. We now compare our sample of \hzrg
with the MS and discuss what the implications are now that we include
spatially-resolved FIR ALMA data and account for possible contamination
from synchrotron emission.

\subsection{Comparison with the MS and impact of submm spatial resolution}

\noindent Our sample of radio galaxies has a wide range of relative SFR
compared to galaxies which lie along the MS \citep[Fig.~\ref{fig:MS};
][]{Schreiber2015, Santini2017}. Unfortunately, the number of
possible MS parameters such as slope, zero-point, and whether or not
the MS was fitted with a turnover at high stellar masses makes a direct
comparison with our results challenging. To make matters worse, there
are additional potential differences in the methods and wavelength
range used to estimate stellar masses and SFR, the range of stellar
masses studied, and, for consistency with the broad redshift range of
our sample, the redshift range that any individual study covered. For
example, \citet{Santini2017} redshift span of 1.3$<z<$6 and stellar mass
range,$\sim$10$^{7.5}$-10$^{11}$\,M$_{\odot}$. Our masses are generally
higher than their upper mass limit. \citet{Schreiber2015} span stellar
masses from $\sim$10$^{9}$ to 10$^{11.3}$\,M$_{\odot}$, which makes their
mass range comparable to the radio galaxies in our sample. Unfortunately,
their redshift range is rather limited, 0.5$\leq z \leq$2.5, for making a
robust comparison with our sample difficult. Of course, these limitations
are purely observational depending on the depth and area covered by the
surveys from which these results are derived.
Our HzRG sample consists of the rarest, most massive galaxies
selected from all sky radio surveys.
HzRGs thus allows us to extend MS studies to the
most massive end of the galaxy mass distribution, and as now discussed, our comparison
suggests that the HzRGs in our sample fall either on or below the
star-forming MS (Fig.~\ref{fig:MS}).

We first compare our results with the MS from \cite{Schreiber2015}
(gray regions in Fig.~\ref{fig:MS}), covering 0.5$<z<$4 and for
M$_*$=10$^{11.33}$\,\msun, close to the mean stellar mass of
10$^{11.35}$\,\msun\ of our sample.  In comparison to \citet{Schreiber2015},
who fits a turnover in the MS at high stellar masses, we find two
sources lie above, 14 sources lie within the 0.3\,dex scatter of the
MS, and nine lie below the MS. We note however, that three of the sources
that lie along the MS have only upper limits in their estimated
SFR. In addition, two sources have uncertainties in their SFR estimate,
which give them a significant probability of lying below the MS.
All of these sources could well lie below the MS. Thus, in comparison
with the results of \citet{Schreiber2015}, we find that the galaxies
in our sample generally fall below the main sequence.

Since \citeauthor{Schreiber2015} fit their data in the SFR-stellar mass
plane with a function that has a turnover, comparing
our results with a study that does not allow for such a turnover, may
result in a change in how we characterize our results. We therefore also compare our results with the MS from \citet{Santini2017}, which does not include such a turnover (colored regions in Fig.~\ref{fig:MS}). We again find that our HzRG sample significantly lies below the MS. None of our
sources lie above the MS of \citet{Santini2017}, six sources lie along
the MS, five sources lie at at the lower $\pm$0.3\,dex boundary, and six sources lie below. The remaining seven sources have upper limits in SFR, and may well lie below the MS of \citet{Santini2017}. At any rate, the comparison
with both \citet{Schreiber2015} and \citet{Santini2017} suggest that
many of our sample galaxies lie below the MS and are consistent with
being quenched.

Three out of the four galaxies with the highest SFR in our
sample and which lie on or above the MS, TN\,J0121+1320, MRC\,0152-209,
PKS\,0529-549, and TN\,J1338-1942, have morphological features in the
ALMA continuum suggesting that perhaps they might be mergers (we indicate
these four sources by name in Fig.~\ref{fig:MS} to highlight the fact
that they are high in the SFR-stellar mass plane compared to the other
galaxies in our sample). The Dragonfly galaxy, MRC\,0152-209, has two
clearly interacting mm components. Indeed, for this source there is strong
evidence that the dynamics of this system are consistent with a strong
interaction of three components with significant mass transfer between them
\citep{emonts2015a, emonts2015b}. Both TN\,J0121+1320 and PKS\,0529-549
have elongated shapes which are different from the synthesized beam,
which in analogy with the Dragonfly galaxy, is consistent with them
being mergers. We would need (sub-)mm line kinematics and continuum
morphology at higher resolution to know definitively whether or not they
are mergers. TN\,J1338-1942 is unresolved in our data and, within this
criterion, is not consistent with being a merger.  

\begin{figure}[!t]
\centering
\includegraphics[width=\linewidth]{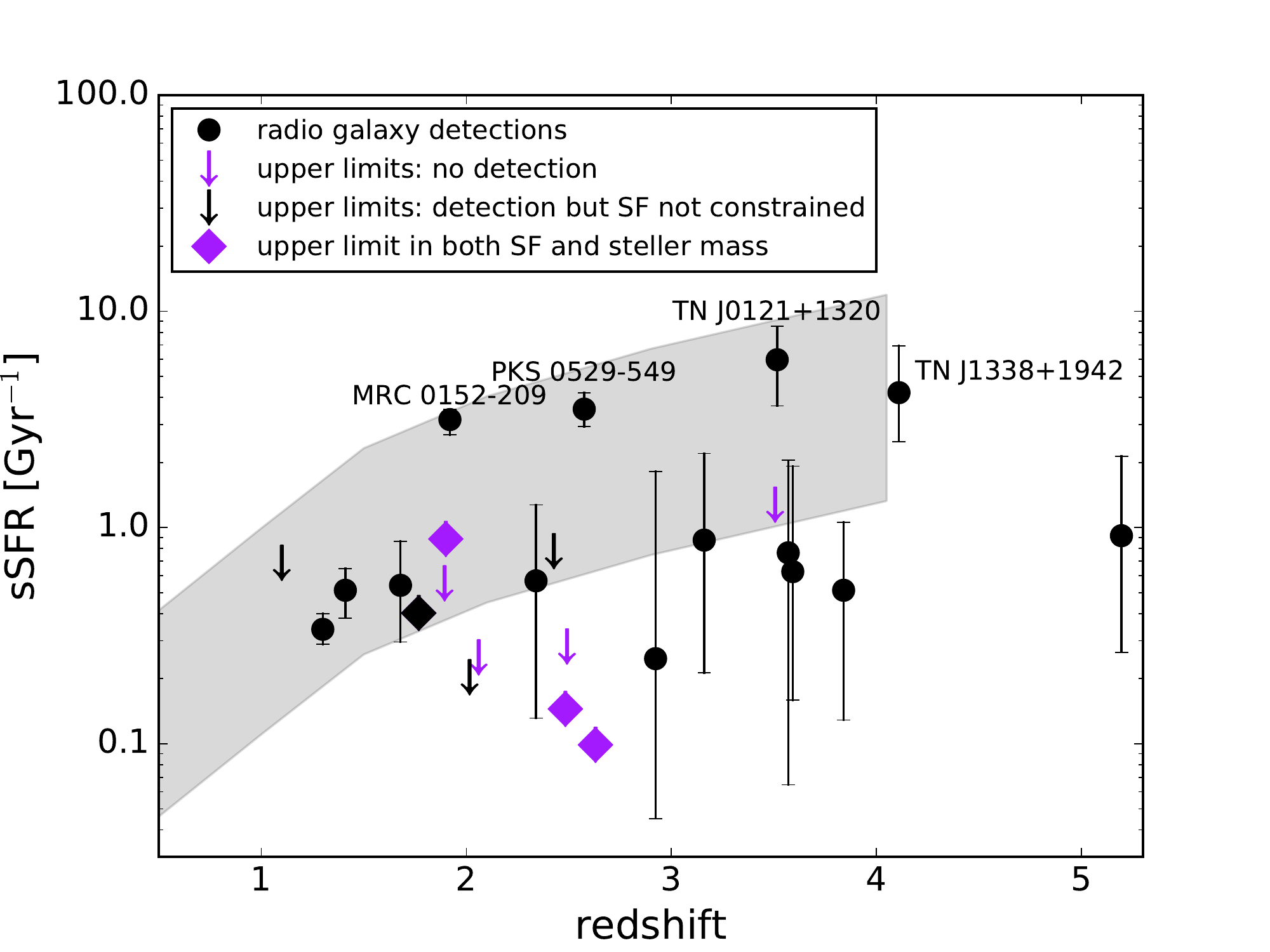}
\caption{Specific star formation rate, sSFR (Gyr$^{-1}$), as
a function of redshift. The black filled circles, black arrows and
purple arrows represent the radio galaxies of our sample that are
detected, detected but with unconstrained SF component and undetected
with ALMA, respectively. The diamonds indicate sources which have both
upper limits on the stellar mass and the SFR. The shaded region shows
the galaxy main sequence \citep[adapted from][]{Schreiber2015} for
M$_*$=10$^{11.33}$\,\msun. We only indicate the redshift range 0.5$<z<$4
in the shaded region as z=4 is the redshift limit of the objects studied
in \citeauthor{Schreiber2015}. We do not extrapolate to higher redshifts
\citep{stark13}.}
\label{fig:sSFR_z}
\end{figure}

We can also compare the sSFR our sample with that of the
ensemble of star-forming galaxies as a function of redshift
(Fig.~\ref{fig:sSFR_z}). Just as with our comparison with main sequences
at various redshifts, we find that a large fraction of our sample falls
below the main sequence and its evolution. To be conservative, we made
this comparison with the MS evolution estimated in \citet{Schreiber2015},
which includes a turnover in their fitting of the MS. Even though the
galaxies are massive and lie above the mass at which the turnover starts
and the MS flattens, our galaxies generally lie below. Of course, if
we used main sequences that do not allow for a turnover in the stellar
mass-SFR relation, then the difference between our sources and the
evolution of the MS at constant mass would be even more dramatic. We note
that since the slope of the MS is about one and if there is no flattening
in the MS at high stellar masses, then the exact mass used for comparison
with the radio galaxies is not particularly important.

Our results are different from previous studies. In previous studies
massive high redshift radio galaxies have often been associated with
high BH accretion rate combined with high SFR because they are extremely
luminous in both the mid-IR \citep{Ogle2006, Seymour2007, DeBreuck2010}
and the sub-mm \citep{Archibald2001, Reuland2003, Stevens2003}.
Not surprisingly given the sensitive upper limits or finding that
synchrotron emission may dominate the emission at mm wavelengths,
our star formation rate estimates are lower than those found by other
studies. On the other hand, our results on the AGN-heated dust luminosity are consistent with previous studies, which is not surprising given that we are using the same or similar data. Our finding that HzRGs fall mainly along or below the MS is therefore new. Most of the sources lying below the MS relation are non-detections in the ALMA band, and often not detected in the SPIRE bands either. These galaxies have very low SFR and are an order-of-magnitude weaker then what was previously estimated. These galaxies are not star forming, they are on their way to being quenched.

\subsection{Comparison with X-ray selected high redshift
AGNs}
\label{subsec:AGNcomp}

\noindent Naturally, given the importance of understanding the relation between the
growth of the stellar population of galaxies and the influence of AGN, it
is useful to compare our results for the star-formation rates of powerful
high redshift radio galaxies with those of other types AGN. Within the
perspective of the overall population of AGN, it is not clear exactly what
the role of the AGN is in quenching the star formation. We find that many
of our sources do not currently have significant star formation. Our
upper limits, constraining the median SFR to be $\sim$100\,\msunyr ,
are also consistent with little or no star formation, especially for
those sources with upper limits in the \herschel\ bands and (likely)
only synchrotron emission in the mm. As we already discussed, the typical
lifetime of AGN is of-order a few 10\,Myrs. The IR and sub-mm thermal dust
emission probes star formation, past and present, over timescales of 100\,Myrs or more \citep[i.e., comparable to a few internal dynamical times
of the galaxies;][]{lehnert96, boquien14, boquien16}.  These differing
timescales implies that unless the AGN are significantly longer lived
than we currently understand them to be, the luminous or mechanical output
from AGN is unlikely to be the sole mechanism for shutting down star
formation in these galaxies.

\begin{figure}
\centering
\includegraphics[width=\linewidth]{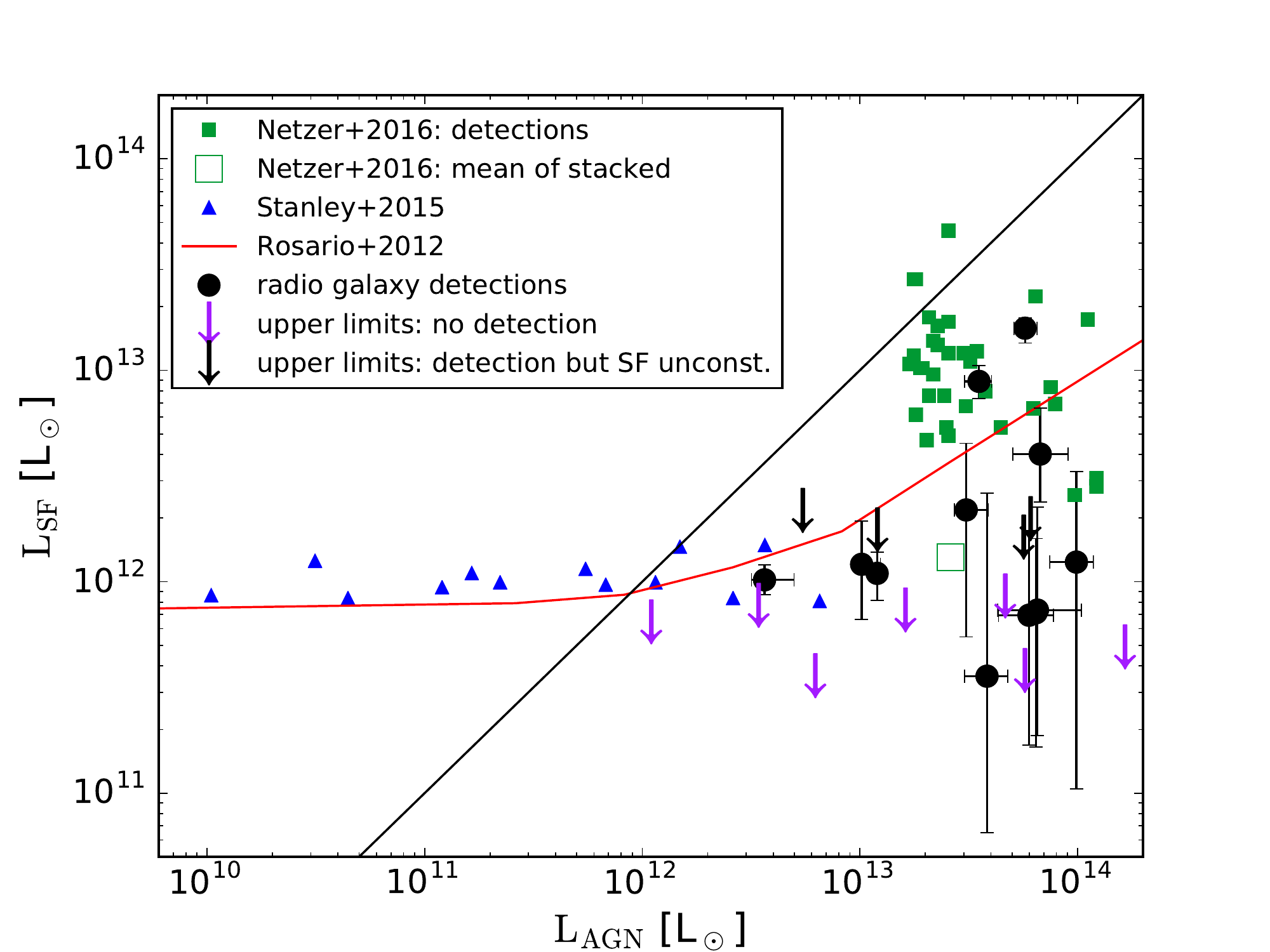}
\caption{Comparison of our results in the L$_\mathrm{SF}$--L$_\mathrm{AGN}$
plane to other studies. \liragn\ for our sample is scaled by a factor
six to estimate the bolometric luminosity.  Black circles indicate
sources detected with ALMA. Black downward pointing arrows indicate
sources detected in ALMA where the sub-mm flux is consistent with being
dominated by synchrotron emission. Purple downward pointing arrows
indicate sources not detected with ALMA. The X-ray selected sources
from \cite{Netzer2016} detected with \herschel\ over the redshift range
$z$=2-3.5 are indicated with green squares;  the large open green square
is the SFR of the undetected sources which were stacked to estimate
their median SFR. Blue triangles indicate X-ray selected sources from
\cite{Stanley2015} over the redshift range, 1.8$<z<$2.1. The red
line is the fit from \cite{Rosario2012} to the mean relation for AGNs
over the redshift range 1.5$<z<$2.5 scaled up by a factor of two
\citep[see][for details]{Netzer2016}. We indicate when the luminosity
due to star formation equals that of the AGN with a black solid line.}
\label{fig:Lum_AGN_SF}
\end{figure}

We can perhaps understand this by comparing the results of studies of the AGN influence on the star formation within their
hosts. In Fig.~\ref{fig:Lum_AGN_SF}, we show such a comparison focusing
on X-ray selected AGN from various studies \citep[based on a similar
figure in ][but see also \citealt{Rosario2012, Stanley2015, stanley17}
for earlier or different renditions of the same plot]{Netzer2016}. Our
galaxies lie at the upper end of the distribution of AGN luminosities
and their total infrared luminosities are dominated by the emission from
the AGN they host. Generically, our host galaxies are forming stars at
a much lower rate on average than the X-ray selected AGN, at least for
those that have well determined star formation rates. \citet{Netzer2016}
found that if they stacked the X-ray selected AGN without detections in
the infrared results in a much lower mean IR luminosity due to young
stars. This stacked luminosity and the implied SFR is comparable to
our sample. These final results may suggest that the key parameter in
the difference in differential growth rates of actively fueled SMBHs is
in fact is the luminosity of the AGN. All of our sources and those of
\citeauthor{Netzer2016} are among the most luminous at their respective
redshifts.

Such an hypothesis would explain a wide range of results. If we focus on
studies that sample a wide range of AGN bolometric or IR luminosities and
estimated SFRs, the evidence points to both the AGN and galaxies growing
in lock step \citep{mullaney12, rovilos12, Delvecchio2015} and there is
generally no widespread evidence for quenching \citep[e.g.,][]{harrison12,
rosario13, stanley17}. However, one has to be cautious in these simple
relations between galaxies being on the main sequence and the impact of
AGN on their star formation rates. The effect of AGN on star formation
rates may be subtle and may also influence the control sample
of galaxies without active black holes. \citet{scholtz18} make the
interesting point that the star-formation rate distribution at constant
mass of galaxies with or without AGN is similar, 
and this result actually agrees with simulations. They suggest that the
agreement between AGN and non-AGN in lying on the same relation in the
SFR-stellar mass plane is because the impact of AGN is evident in both
samples, and it is this effect that is driving the slope of decreasing
sSFR with increasing mass \citep[see also][]{mainieri11}. So the effect
of AGN feedback is subtle, broadening distributions and not necessarily
correlating with AGN power/luminosity as might be naively expected.
In a study similar to ours, \citet{mullaney15} used ALMA observations of
X-ray selected AGN, finding that the estimated SFRs were lower relative
to previous findings.  This reduction lead to AGN hosts having a different
distribution than the star forming galaxy population, meaning the AGN
population of star forming galaxies had a broader, and perhaps even offset
distribution of star formation rates \citep{rovilos12, rosario15}.

\subsection{A significant synchrotron contribution in submm}

\noindent Our high resolution ALMA data allowed us to do component separation such
that the contribution of synchrotron to the mm emission is now clear,
originating from radio lobes and the nuclei. For nine out of 25 sources, 
the best-fit SEDs of specific spatial components are consistent with a power-law extrapolation from radio frequencies. These spatial
components are likely dominated by synchrotron emission. The total flux
of five galaxies, not just the lobes, are associated with pure synchrotron
emission and are consistent with no contribution from thermal dust
emission. For two host galaxies, which have two spatially-resolved
emission components in the ALMA band, the majority of the mm-flux is
due synchrotron with only a minor contribution from thermal dust emission.

It is surprising to find our high frequency data is consistent with
an extrapolation extrapolation of the synchrotron intensities in the
radio \citep{Jaffe1973, carilli91, Blundell2006}. The energy loss
though aging of the electrons and inverse Compton scattering should
steepen the observed spectrum at higher frequencies. But what we see in
a few sources is that the synchrotron spectrum continues straight out
to high frequencies \citep[e.g.,][]{gopal-krishna01}. One possible
explanation is that the electrons are continuously accelerated within the
lobes, perhaps through strong oblique shocks \citep[e.g.,][]{summerlin12}.
The details and analdysis of the radio spectra including ALMA synchrotron
detections will be presented in a subsequent study.

\subsection{Star formation and radio source sizes}
\begin{figure}
\centering
\includegraphics[width=\linewidth]{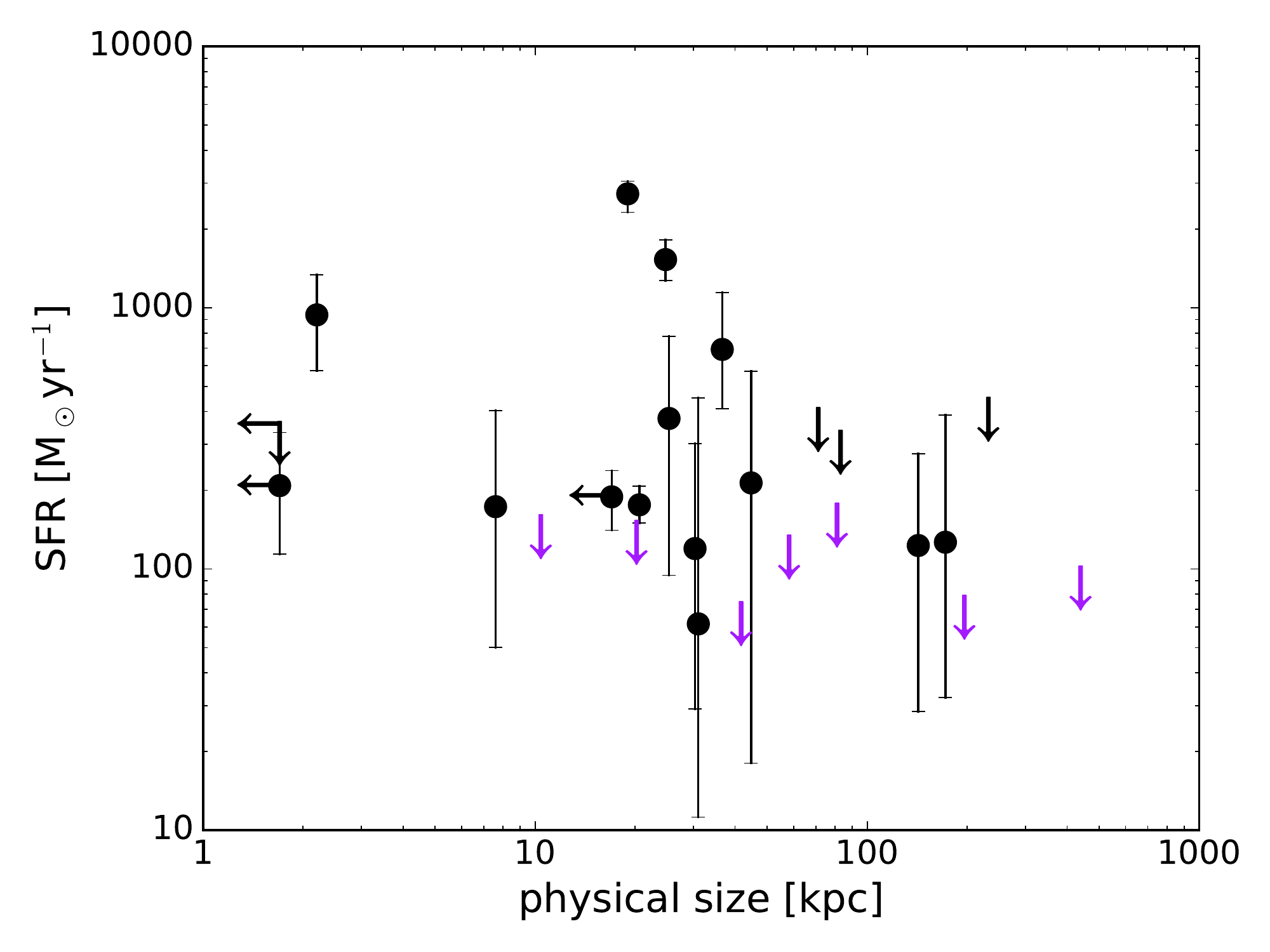}
\caption{Star formation rate, SFR in \msunyr , versus the projected
physical size of the radio emission in kpc, calculated using
the largest angular size (LAS) of the radio emission at 1.4\,GHz
\citep{DeBreuck2010}. Black filled circles are detections, black
arrows detections but with unconstrained SF component and purple arrows
non-detections (see text for details).}
\label{fig:SFR_LAS}
\end{figure}

\noindent The spatial separation of the emission from the radio lobes -- the
projected distance between the two lobes or between the core and lobes --
can provide useful constraints on the nature of the AGN. The separation,
if the sources expand at constant velocity, can be used as a proxy for
the age of the radio source \citep[e.g.,][]{carilli91}. Larger sources
correspond to older sources. A possible test of a scenario where the AGN
is the agent driving the quenching of star formation is that galaxies
with the lowest star formation rates have the largest radio sources.
In fact, we do see a possible trend between the largest angular size of
a radio source and the predominance of upper limits to the star formation
rates (Fig.~\ref{fig:SFR_LAS}). This trend is robust in the sense that high 
significance estimates of the SFR all appear below largest angular
sizes of $\sim$3\,arcsec. At the typical redshifts of our sources, this
corresponds to approximately 10-15\,kpc in radius, or about the expected
extent of the interstellar medium of a massive galaxy. We do not have
a sufficient number of sources or sensitive enough upper-limits to say
that there is a correlation.  Such a correlation would be interesting, as it would support
a simple model of propagating jets that may enhance star formation
when they are still confined within the host galaxy, but then quickly
preventing further star formation after they break out.

Timescale arguments may also complicate the comparison between radio size and SFR.
The duration of AGN activity
is thought be around 10$^7$ to 10$^8$\,yrs \citep[e.g.,][]{martini01,
hopkins05, schmidt17}.  The star formation likely lasts much longer. Other than observing a proclivity for the upper-limits in the SFR to be
associated with radio sources larger than the host galaxy proper, it may
not be that surprising that we do not see a very clear correlation. There
are several factors that may influence the rate at which radio jets expand
into the surrounding media. Perhaps the most important is the environment
into which the jets are expanding. For jets which propagate outwards into
denser environments, the radio source may be confined and can either
simply expand more slowly, or can decelerate \citep{Shabala2017}. There
may not be a simple linear relationship for individual sources and they
may not all take the same amount of time to reach the same size. There
are also of course projection and beaming effects to consider. Since we
observe a wide range of SFRs, radio morphologies, and lobe asymmetries in
this sample \citep{DeBreuck2010}, it is likely that the characteristics
of the surrounding environments of the individual radio sources play a
significant role in determining the scatter in this relation. Moreover,
given that the star formation is expected to occur on time scales longer
than the duration of the AGN activity perhaps not seeing a clear cut
correlation is not unexpected, regardless of the processes
that can influence radio source sizes and morphologies. Other effects
may play a role since the galaxies in our sample with high SFRs, have
morphologies consistent with being mergers (Sect.~\ref{sec:MS}).

\section{Possible interpretation of our results}

Our interpretation of our results is that: \textit{(i)} the quenching
we observe for the majority of the sample is naturally explained
by our galaxies predominately being at the high mass end of the
galaxy population \citep{Ilbert2013} and having very massive SMBHs
\citep{nesvadba11}. We know that such galaxies in the local universe
must be quenched rapidly and soon after their most substantial period of
growth \citep[e.g.,][]{Thomas2005, Thomas2010}.  They also likely grew
subsequently, after they are quenched, through the accretion of relatively
low mass, likely gas-poor galaxies \citep[e.g.,][]{Greene2013}. Thus,
we can say that their quenching is mainly a result of the fact that they
are already massive only a few Gyrs after the Big Bang; and, \textit{(ii)}
given the fact that some of galaxies in the sample of high redshift radio
galaxies have intense star formation and young stars \citep[][Man et
al. 2018, in prep.]{dey97}, it is likely that star formation played a
key role in quenching the host galaxies of distant radio galaxies. This
is a natural conclusion, given that the AGN are likely only luminous for
10s of Myrs \citep{schmidt17}, which is generally shorter than we would
expect bursts of star formation to last. Other samples of AGN with
lower host masses have substantial outflows \citep[e.g.,][]{harrison16b}, and
are not generally quenched.

It is difficult to isolate the impact of mass and environment
on the rate and timing of quenching. Mass quenching is more important at
earlier times in the evolution of galaxies and may be more important in
denser regions \citep[e.g.,][]{Peng2010, muzzin12, Lee2015, darvish16,
kawinwanichakij17, darvish18}.  And in the case of powerful radio
galaxies, which lie in over-dense environments, both gas-rich and gas-poor
mergers likely play an important role in both the growth of the stellar
mass and the black holes.  \citet{volonteri15} suggest that in the merger
phase, the AGN dominates the bolometric luminosity but the accretion
can be very stochastic \citep[see also][]{gabor13, degraf17}. It appears
that the galaxies in our sample with the highest star-formation rates 
all host very powerful AGN, and are potentially all advanced mergers,
consistent with this picture. In fact, PKS\,0529-549, which
has one of the highest SFRs of all the galaxies in our sample, has a
modest gas fraction of about 15\%, a high star-formation efficiency
(SFR/molecular gas mass), and has been transforming its gas into stars
rapidly (Man et al., in prep.). The star formation efficiencies in the other
radio galaxies with high SFRs also appear extreme (10-100\,Gyr$^{-1}$;
Man et al.). But of course, that does not explain our results in
themselves.  \citet{dubois15}, in a study using numerical simulations
of the relative growth of SMBHs and their host galaxies, found that
star formation may regulate the black hole accretion rate. During the
most rapid, gas-rich phase of the growth of massive galaxies, it may be
that a larger fraction of the gas in the ISM is not available to fuel
the SMBHs, but is consumed via star formation \citep[see][]{degraf17}.
As the gas fractions decline, the relative power of the AGN compared
to that of the star formation increases, resulting in an increased star
formation efficiency. Concomitantly, the increased star formation rate
can then disperse the dense gas making it easier for the jets to drive
vigorous and efficient outflows \citep{nesvadba06, nesvadba17}.

So we suggest that mass is the primary difference in the characteristics
of our sample of galaxies compared to other samples of AGN \citep[see
also][]{mainieri11, stanley17}. This difference makes it then much
easier for the AGN to dominate the bolometric output as the SMBHs in
massive galaxies may be ``overly'' massive for their host masses. Even
relatively low accretion rates would lead to powerful AGN emission. The
mechanical and radiative output from the young massive stars, heats the
interstellar medium, changes its phase distribution, and puffs it up.
In such a state, the coupling between the mechanical energy of the jet
and heated and expanded interstellar medium of the galaxy, would be high
\citep{biernacki18}. This leads to a positive feedback loop at the end
of the most rapid growth phase of massive galaxies where the residual
gas of star formation is rapidly removed through the action of the AGN,
leading to the rapid session of star-formation. The galaxy is on the
road to being quenched.

\section{Conclusions} 

With 0\farcs3 resolution mm ALMA data, and utilizing the new SED
fitting tool, \mrmoose, we have estimated \lirsf\ and \liragn.  We have
disentangled the IR luminosity into components heated by the stellar
populations and that heated by the luminous AGN that reside in these
host galaxies.  With our deep ALMA observations, we reach depths at
which we expect the thermal dust and synchrotron contributions to the
mm emission to be of the same order-of-magnitude. This high sensitivity
is the reason why it is essential to disentangle both spectrally and
spatially the contribution from the high frequency synchrotron emission
at (sub-)mm wavelengths. From a study of 25 powerful high redshift radio
galaxies, our main conclusions are:

\begin{itemize}

\item We find that the SFRs are lower than previously estimated, having
a median which is $\sim$7 times lower \citep[cf.][]{Drouart2014}. This
is a result of having deep ALMA data enabling us to estimate robust
upper limits, to disentangle several emission components, and determine
if the mm emission represents the high frequency end of the power-law
extrapolation of the radio emission. We interpret any flux density
estimate that is consistent with the power-law extension from the radio
to the (sub-)mm as synchrotron emission, and not as dust heated by
young stars.

\item Unlike many studies of high redshift AGN, we find that a large
fraction of our sources do not lie within the scatter of the relation
between the stellar mass and the star-formation rate of star forming
galaxies (the ``main sequence''). Since the deviation of the radio
galaxies with upper limits does not meet the general criterion for
quenched galaxies, we interpret such galaxies as being ``on their way
to being quenched''.

\item Finding lower star formation rates generally exacerbates the
problem of differential growth between the black hole and the host
galaxy discussed by \citet{Drouart2014}. This favors a scenario where
the host galaxies ultimately grow by dissipationless merging which
adds additional stellar mass without increasing significantly the mass
of the black hole. We find that at the typical specific star-formation
rates we estimate for the host galaxies, the host galaxy needs $\sim$50
times longer than the time over which the SMBH is active to ``catch up''
to having the local ratio of black hole mass to stellar mass.

\item There is no clear relation between the radio size and SFR, although
we note that the upper limits on the star formation rate tend to be
in sources where the LAS is larger than $\sim$3 arc seconds. However,
given that the star formation is expected to occur on time scales of the
same order or longer than the duration of the AGN activity (10$^7$ to
10$^8$\,yrs) perhaps not finding a clear cut correlation is not completely
surprising even if the radio jets are driving the quenching we observe.

\end{itemize}

From these results, we suggest that mass is the primary difference in
the characteristics of our sample of galaxies compared to other samples
of AGN. The host galaxies of our sample of radio galaxies are generally
more massive and likely harbor high mass black holes than AGN selected
using other methods. Hosting more massive black holes means that it is
also easier for the AGN to dominate the bolometric output of the galaxy,
even if it forming stars vigorously. In order to comprehend how the host
galaxies are quenching so rapidly, we hypothesize a positive feedback
loop between the AGN and star formation. The star formation leads to
rapid gas depletion and its intense energy input heats the remaining
gas allowing the interaction between the gas and jets to be efficient.
The AGN then blows away the residual gas at the end of the star formation
episode.  It is in this way perhaps that the host galaxies of powerful
high redshift radio galaxies are ``on the road to being quenched".

\begin{acknowledgements}
M.D.L. wishes to thank the ESO visitors program for its continued
support and Marta Volonteri and Gary Mamon for fascinating discussions.
T.F. acknowledges financial support from ESO and INSU/CNRS for this
work. T.F., C. De B. and M. D. L. thank Joe Silk for interesting
suggestions and analysis and an anonymous referee for
their constructive criticisms and suggestions. BG acknowledge support from the ERC Advanced Programme DUSTYGAL (321334) and STFC (ST/P0000541/1). This paper makes
use of the following ALMA data: ADS/JAO.ALMA\#2012.1.00039.S,
ADS/JAO.ALMA\#2013.1.00521.S, ADS/JAO.ALMA\#2015.1.00530.S, and
ADS/JAO.ALMA\#2012.1.00242.S. ALMA is a partnership of ESO (representing
its member states), NSF (USA) and NINS (Japan), together with NRC
(Canada), NSC and ASIAA (Taiwan), and KASI (Republic of Korea), in
cooperation with the Republic of Chile. The Joint ALMA Observatory is
operated by ESO, AUI/NRAO and NAOJ. The Australia Telescope is funded
by the Commonwealth of Australia for operation as a National Facility
managed by CSIRO.  The authors wish to express their sincerest thank
you to the staff of the CSIRO for their assistance in conducting these
observations and the program committee for their generous allocation
of time and continuing support for our research. The National Radio
Astronomy Observatory is a facility of the National Science Foundation
operated under cooperative agreement by Associated Universities, Inc.
\end{acknowledgements}

\bibliographystyle{aa}
\bibliography{references}

\clearpage
\newpage

\begin{appendix}

\section{SED fits}

\subsection{MRC\,0037-258}

\begin{figure}
\includegraphics[scale=0.52]{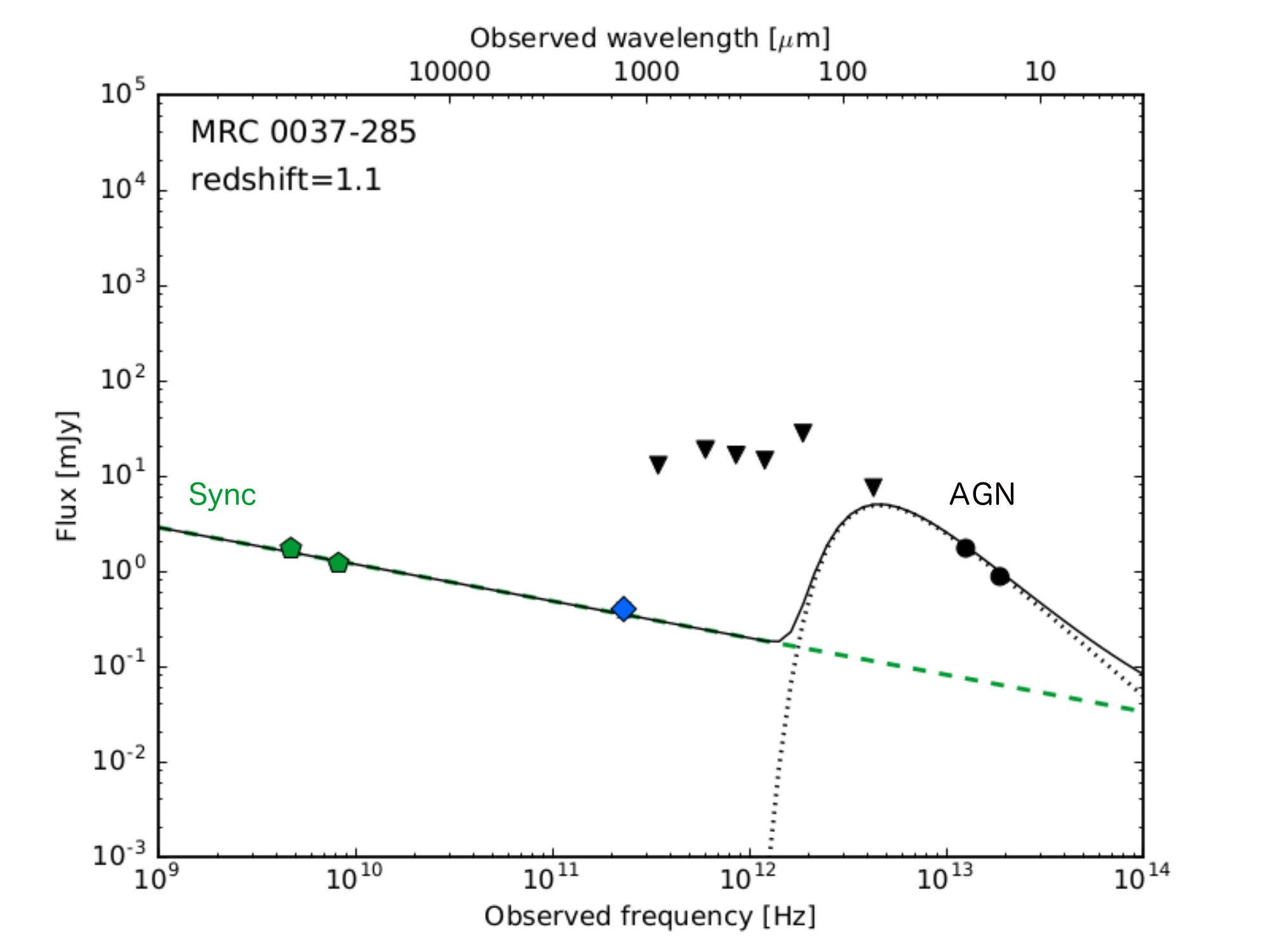}
\caption{SED of \textbf{MRC\,0037-258}. Black solid line represents the best
fit total model, green dashed line is synchrotron power-law of the
radio core, dotted line indicates the AGN contribution. The colored data
points indicate the data which have sub-arcsec resolution and black points indicate data of
low spatial resolution. Green pentagons represent the fluxes from the
radio core and the blue diamond indicates ALMA band 6 detection. Filled
black circles indicate detections (>$3\sigma$) and downward pointing
triangles indicate the $3\sigma$ upper limits (Table~\ref{table_0037}).}
\label{fig_0037}
\end{figure}

\begin{figure}
\centering
\includegraphics[scale=0.35]{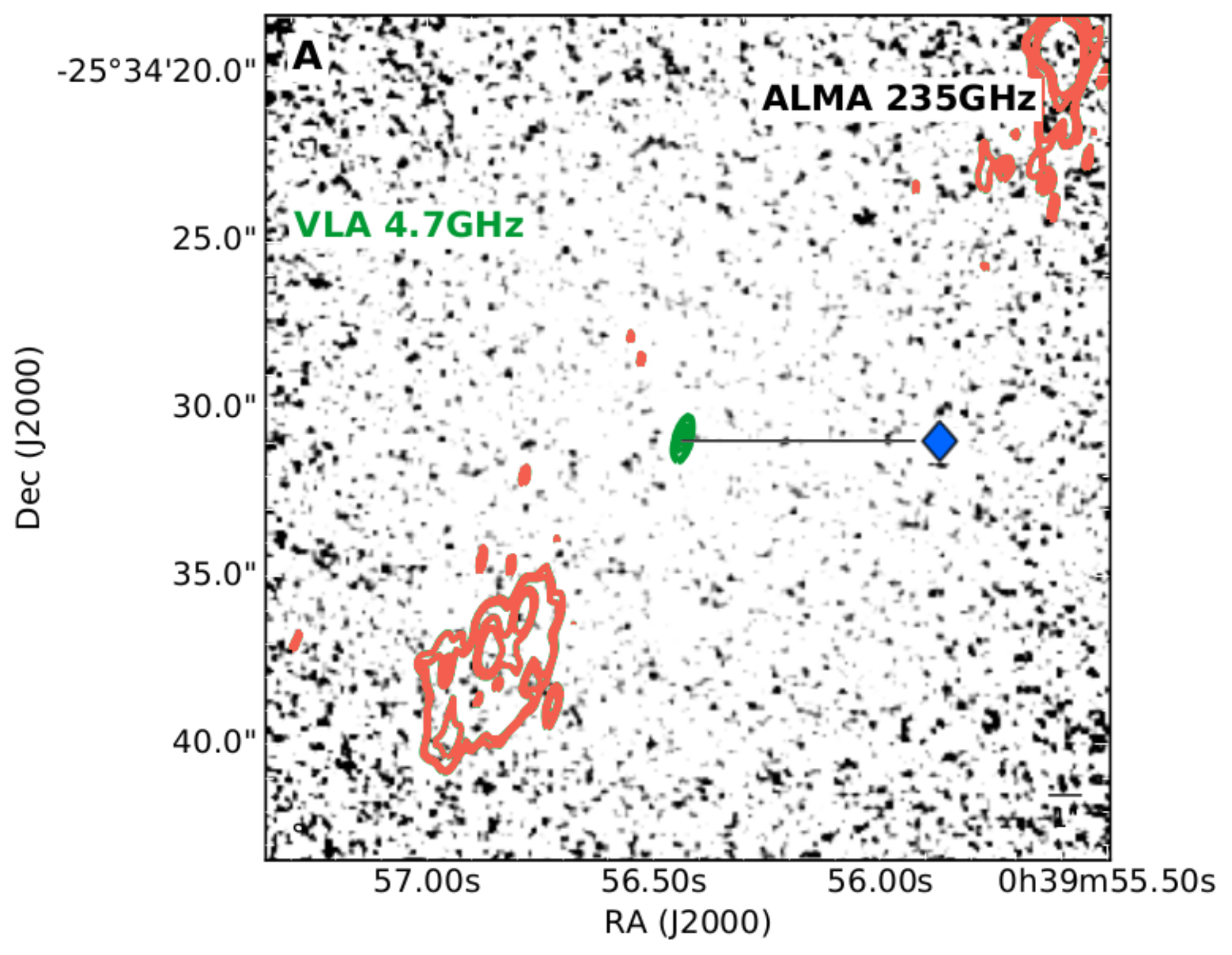} \\
\includegraphics[scale=0.35]{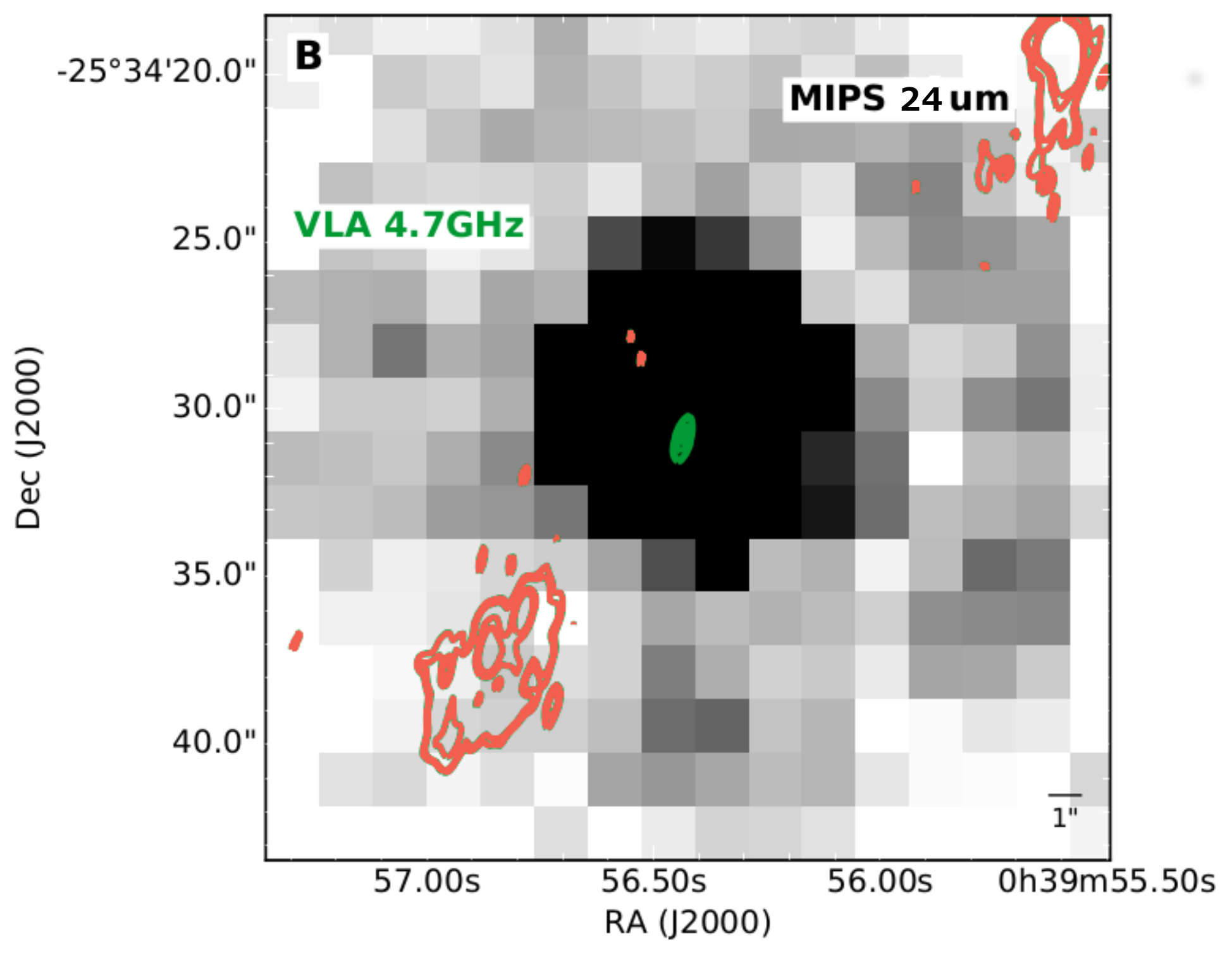}
\caption{\textit{Panel A:} continuum map of ALMA band 6 in
grayscale with overplotted VLA C contours (the levels are
$3\sigma$, $\sqrt{2}\times3\sigma$, $3\sqrt{2}\times3\sigma$ and
$5\sqrt{2}\times3\sigma$ ($\sigma=46\,\mu$Jy)). The blue diamond indicates
the ALMA detection and is the same marker used in Fig. \ref{fig_0037}. The
green contours show the portion of the VLA data that is used in the SED
fit, the red contours are excluded in the fit. \textit{Panel B:} MIPS 24\,\mum\ continuum map with the same VLA C
contours overplotted.}
\label{map_0037}
\end{figure}

MRC\,0037-258 is detected with a single continuum component which
coincides with the position of the radio core (Fig.~\ref{map_0037}). SED
fitting with \mrmoose\ is done with three components, a synchrotron power law
for the radio core (the lobes are excluded), a modified BB and an AGN
component. The VLA data are assigned to only the synchrotron component
while the ALMA flux density is assigned to both the synchrotron and
modified BB. The PACS, SPIRE, MIPS and IRS data are fitted to the
modified BB and AGN components. In the best fit model, the ALMA flux is
dominated by synchrotron emission (Fig.~\ref{fig_0037}). This is because
the $3\sigma$ upper limits in the FIR are not given a high weight and
there is a solution where the synchrotron power law can account for
all of the observed flux. 

The plot of the marginalized distribution of each free parameter of the fit have been omitted from the paper, but are available online\footnote{http://www.eso.org/\%7Ecdebreuc/sed\_paper/} for each 25 sources.

\begin{table}[!ht]
\centering
\begin{threeparttable}
\caption{Data for MRC\,0037-258 (z=1.10) }
\label{table_0037}
\centering
\begin{tabular}{lcc}
\toprule
Photometric band                   & Flux{[}mJy{]}   & Ref. \\
\midrule
\irs      	&     0.877 $\pm$ 0.1 		& A\\
\mips1 	&   1.740  $\pm$ 0.039 		& A\\ 
\pacsb 	&    <7.5       				& B \\
\pacsr  	&  <28.0        				& B \\
\spires 	& <14.6       				& B \\
\spirem 	& <6.6       				& B \\
\spirel 	& <18.9          				& B \\
\laboca  	& <12.9      				& B \\
ALMA 6		&0.4$\pm$  0.08			& this paper\\
VLA X$^c$   & 1.20  $\pm$ 0.12$^a$    	& A \\
VLA C$^c$   &  1.73  $\pm$ 0.17$^a$	& A \\
\bottomrule                          
\end{tabular}
\begin{tablenotes}
\small
\item \textbf{Notes} ($c$) Radio core, ($a$) Flux estimated using AIPS from original radio map, convolved to the resolution of the VLA C band.
\item \textbf{References.} (A) \cite{DeBreuck2010}, (B) \cite{Drouart2014}
\end{tablenotes}
\end{threeparttable}
\end{table}


\clearpage
\newpage
\subsection{MRC\,0114-211}
\label{MRC0114}
\begin{figure}
\includegraphics[scale=0.52]{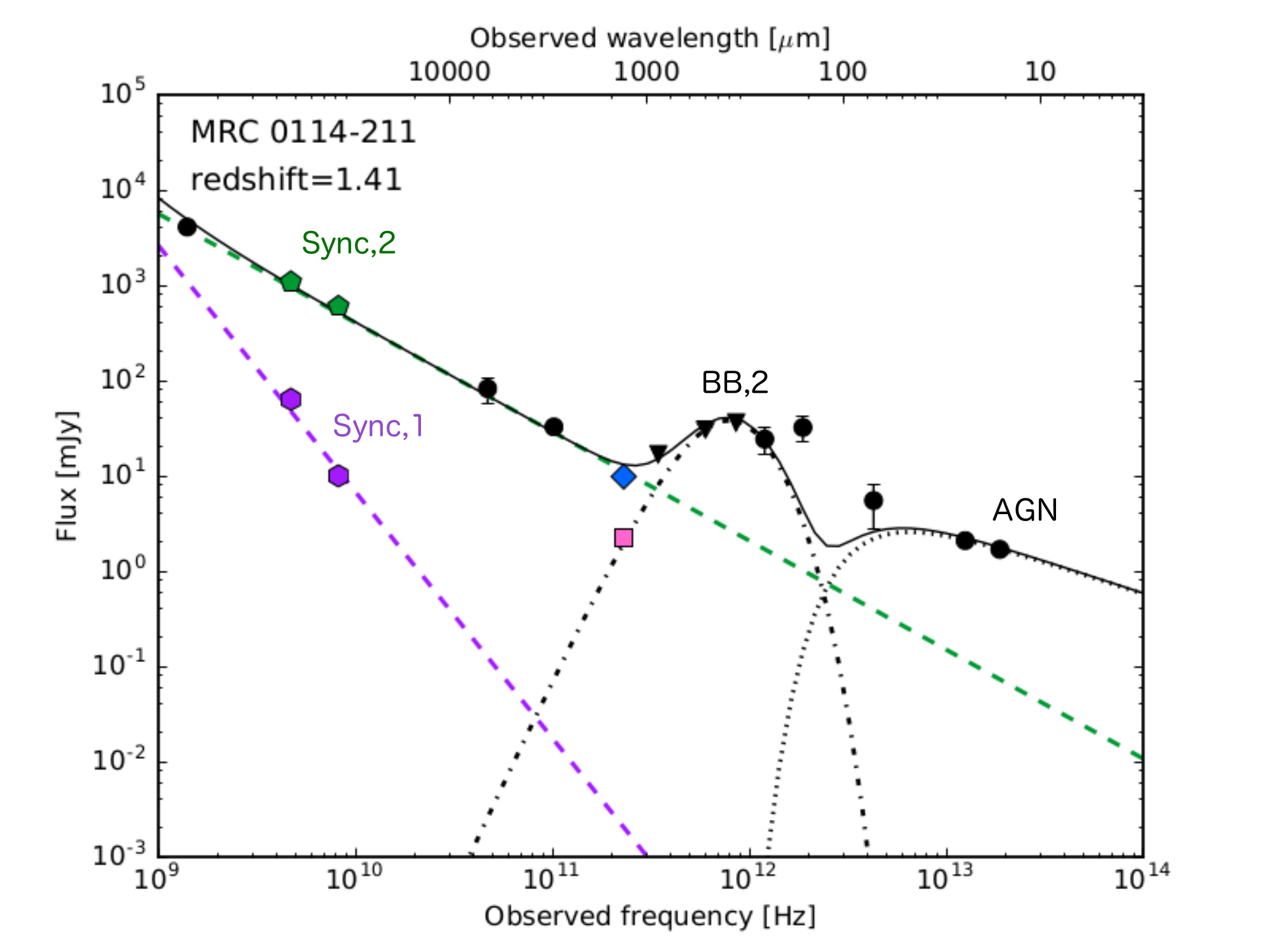}
\caption{SED of \textbf{MRC\,0114-211}. Black solid line shows best fit
total model, green and purple dashed lines are the western and eastern synchrotron
lobes, respectively.  The black dash-dotted line is one blackbody
component associated to one of the ALMA band 6 detections. Black dotted line
indicates the AGN component. The colored points indicate data with sub-arcsec
resolution and black ones indicate the data with low spatial resolution. Green
pentagons indicate the western radio lobe, purple hexagons indicate the eastern radio lobe,
the blue diamond indicates one of the ALMA band 6 detection, and the
magenta square is the second ALMA detection. Filled black circles indicate
detections (>$3\sigma$) and downward pointing triangles the $3\sigma$ upper
limits (Table~\ref{table_0114}).}
\label{fig_0114}
\end{figure}

\begin{figure}
\centering
\includegraphics[scale=0.35]{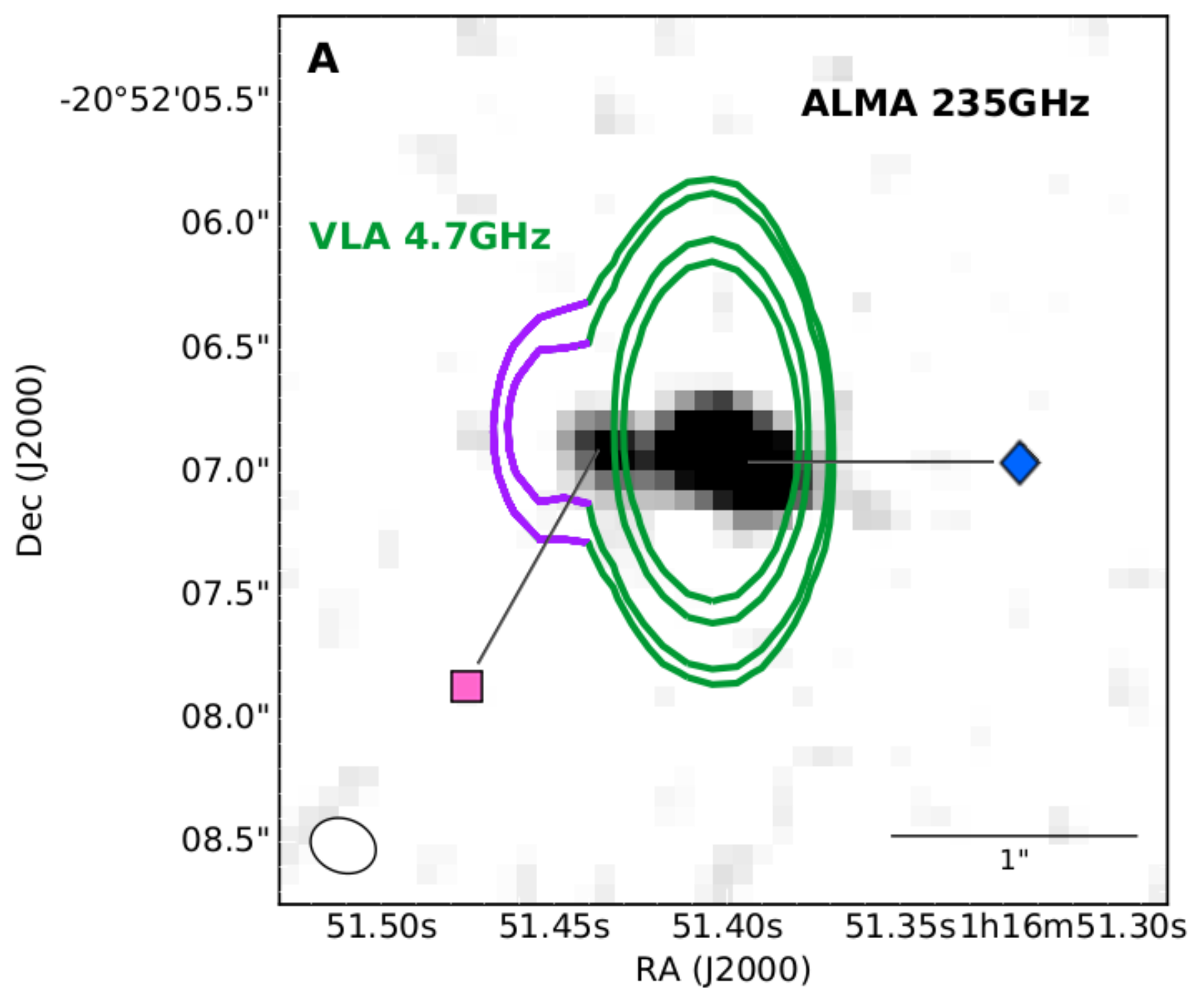} \\
\includegraphics[scale=0.35]{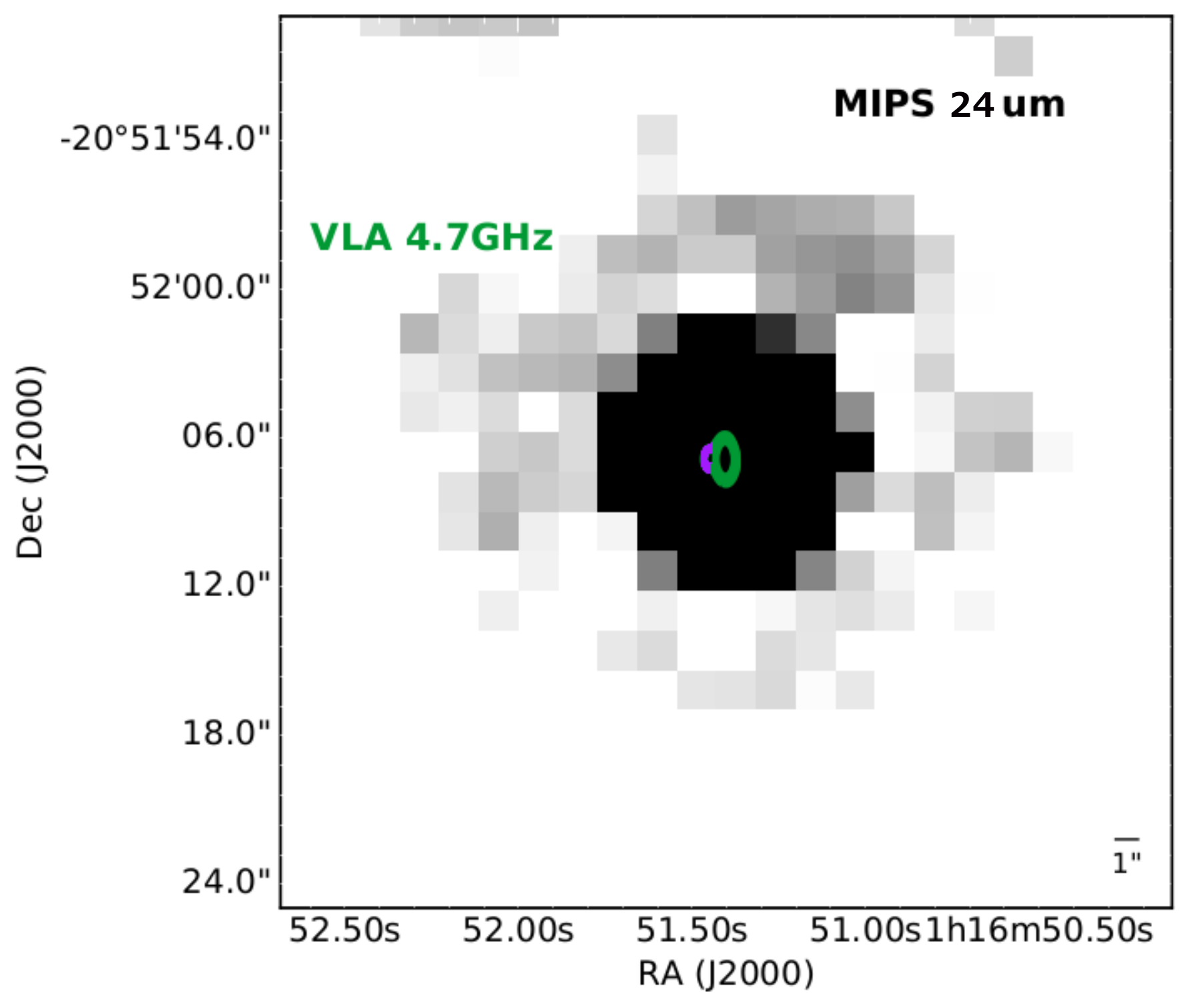}
\caption{\textit{Panel A:} continuum map of ALMA band 6 with VLA C contours 
	overlaid (levels are as Fig.~\ref{map_0037}; $\sigma=71\,\mu$Jy). The blue and pink markers indicate the two different components detected with ALMA and correspond to the same data points as Fig. \ref{fig_0114}. The purple and green contours show the two components of the VLA data and correspond to the markers of the same colors as in the SED fit. \textit{Panel B:} MIPS 24\,\mum\ continuum map with VLA 4.7\,GHz contours overlaid. We note that the scale of the MIPS image is five times larger than the image displayed in panel A.}
    	\label{map_0114}
\end{figure}

MRC\,0114-211 has two detected continuum components in ALMA band 6. Both
detections coincide with the two lower-frequency radio components
(Fig. \ref{map_0114}). The SED fitting is done with five components,
two synchrotron components (eastern and western radio components), two
modified BB (one for each ALMA detection) and one AGN component. The
western radio component (detected in VLA bands C and X) is assigned
to the western ALMA detection and the same setup is also applied to
the eastern radio and ALMA components. The VLA band X, ATCA 7\,mm, and
ALMA band 3 detections do not resolve the individual components and are
only considered for fitting the total radio flux (the combination of the
western and eastern synchrotron power-law components). The two ALMA band 6
points are both fitted with a combination of a synchrotron power-law and
a modified BB. The PACS. SPIRE, MIPS and IRS data points are fitted to
the combination of the two modified black bodies and a AGN component. The
best fit model is where the brighter western component in ALMA band 6
is pure synchrotron and the eastern component is dominated by thermal
emission (Fig.~\ref{fig_0114}). 

\begin{table}[ht]
\begin{threeparttable}
\caption{Data for MRC\,0114-211 (z=1.41) }
\label{table_0114}
\centering
\begin{tabular}{lcc}
\toprule
Photometric band                   & Flux{[}mJy{]}   & Ref. \\
\midrule
\irs 		&1.690   $\pm$    0.1		& A \\       
\mips1	& 2.09$\pm$        0.04	& A\\      
\pacsb	&5.5 $\pm$        2.7		& B \\       
\pacsr	& 32.2$\pm$        9.4		& B \\       
\spires	& 24.3   $\pm$      7.5 	& B \\       
\spirem	& <36.2         			& B \\ 
\spirel	& <30.8                 		& B \\ 
\laboca	& <16.8                     		& B \\ 
ALMA 6$^e$	&2.21 $\pm$    0.31 	& this paper\\
ALMA 6$^w$	&9.82 $\pm$ 0.9	& this paper\\
ALMA  3	& 32.6  $\pm$        3.2       & this paper \\ 
ATCA (7mm)	& 82.5    $\pm$    24.7      	& this paper \\ 
VLA X$^w$	& 599.9    $\pm$      59$^a$       & A \\ 
VLA X$^e$	& 9.99      $\pm$      0.99$^a$      & A \\
VLA C$^w$	& 1084.5     $\pm$     108$^a$     & A \\  
VLA C$^e$	& 63.7      $\pm$    6.37$^a$   & A \\
VLA L		& 4091.6    $\pm$      409 	& C \\
\bottomrule                          
\end{tabular}
     \begin{tablenotes}
      \small
      \item \textbf{Notes}  ($e$) East component,  ($w$) west component, ($a$) flux estimated using AIPS from original radio map, convolved to the resolution of the VLA C band.
      \item \textbf{References.} (A) \cite{DeBreuck2010}, (B) \cite{Drouart2014}, (C) \cite{Condon1998}.
    \end{tablenotes}
\end{threeparttable}
\end{table}


\clearpage
\newpage
\subsection{TN\,J0121+1320}
TN\,J0121+1320 has one continuum detection in the ALMA band 3
observations which coincides marginally with the compact radio component
(Fig.~\ref{map_0121}). SED fitting is done with three components, a single
synchrotron power law, a modified BB, and an AGN component. The single
component in radio VLA C and X bands is assigned to a synchrotron
power-law, the ALMA detection is assigned to both a synchrotron and a
modified BB model. The PACS, SPIRE, MIPS and IRS data points are fitted
with a combination of the modified BB component and an AGN component. The
best fit solution implies that the ALMA emission is due to thermal dust
emssion but the AGN component is unconstrained as there are only upper
limits in the mid-infrared. 

\begin{figure}
\includegraphics[scale=0.52]{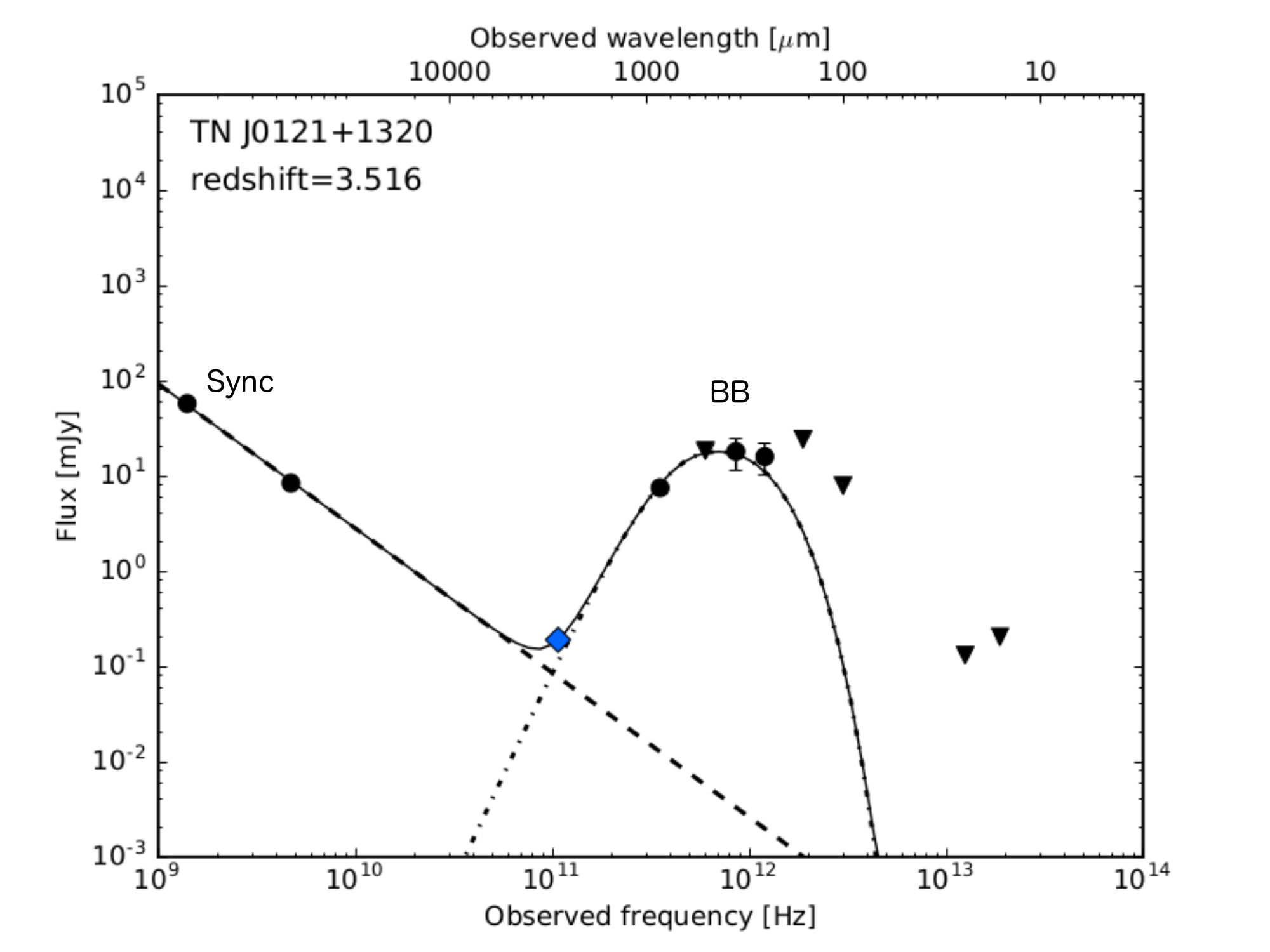}
\caption{SED of \textbf{TN\,J0121+1320}. Black solid line shows the best
fit total emission model, black dashed line represents the
synchrotron emission and the black dash-dotted lines represents the black
body emission. The colored data point indicate the data with sub-arcsec
resolution and black ones indicated data of low spatial resolution. The
blue diamond indicates the ALMA band 3 detection. Filled black circles
indicate detections (>$3\sigma$) and downward pointing triangles the
$3\sigma$ upper limits (Table~\ref{table_0121}).}
\label{fig_0121}
\end{figure}

\begin{figure}
\centering
\includegraphics[scale=0.35]{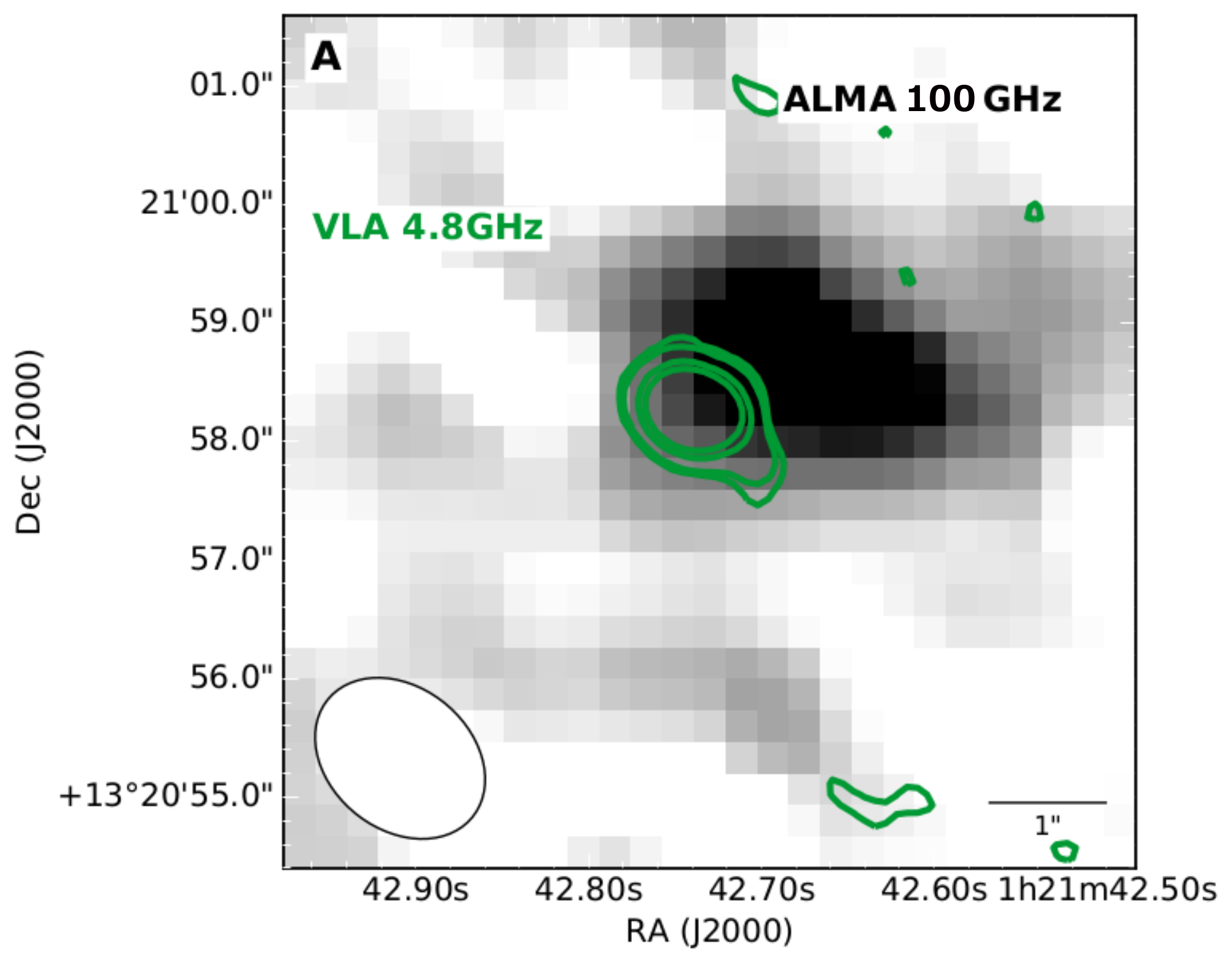} \\
\includegraphics[scale=0.35]{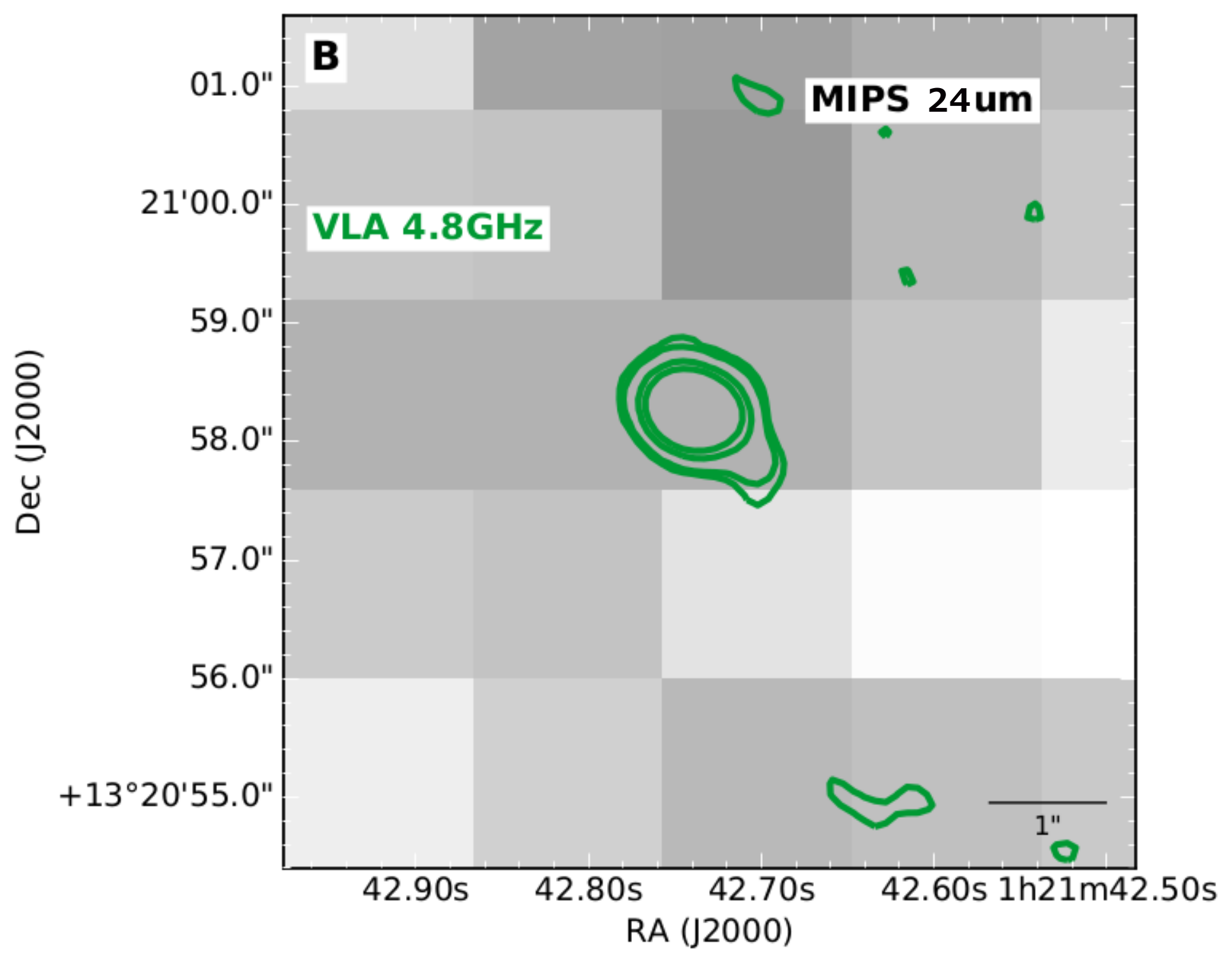}
\caption{\textit{Panel A:} continuum map of ALMA band 3 with overlaid VLA C contours (levels are as Fig. \ref{map_0037}, $\sigma=70$\ujy) \textit{Panel B: } MIPS 16\,\mum\ continuum map.}
    	\label{map_0121}
\end{figure}

\begin{table}[ht]
\begin{threeparttable}
\caption{Data for TN\,J0121+1320 (z=3.516) }
\label{table_0121}
\centering
\begin{tabular}{lcc}
\toprule
Photometric band                   & Flux{[}mJy{]}   & Ref. \\
\midrule
\irs 		&<0.204	& A \\       
\mips1	&<0.131	& A\\      
\pacsg	&<7.9	& B \\       
\pacsr	&<24.2	& B \\       
\spires	& 15.9  $\pm$     5.7 	& B \\       
\spirem	& 18.0$\pm$  6.6	& B \\ 
\spirel	& <18.4               	& B \\ 
\scuba	& 4.7 $\pm$ 1.0        	& B \\ 
ALMA 3	&0.19 $\pm$ 0.012$^a$	& this paper\\
VLA C	& 8.4 $\pm$ 0.5$^a$   	& C \\
VLA L	& 57.3 $\pm$ 2.7 		& D \\
\bottomrule                          
\end{tabular}
     \begin{tablenotes}
      \small
      \item \textbf{Notes}($a$) flux estimated using AIPS from original radio map, convolved to the resolution of the VLA C band.
      \item \textbf{References.} (A) \cite{DeBreuck2010}, (B) \cite{Drouart2014}, (C)  \cite{DeBreuck2000} (D) \cite{Condon1998}.
    \end{tablenotes}
\end{threeparttable}
\end{table}

 
\clearpage
\newpage
\subsection{MRC\,0152-209}
\begin{figure}
\includegraphics[scale=0.52]{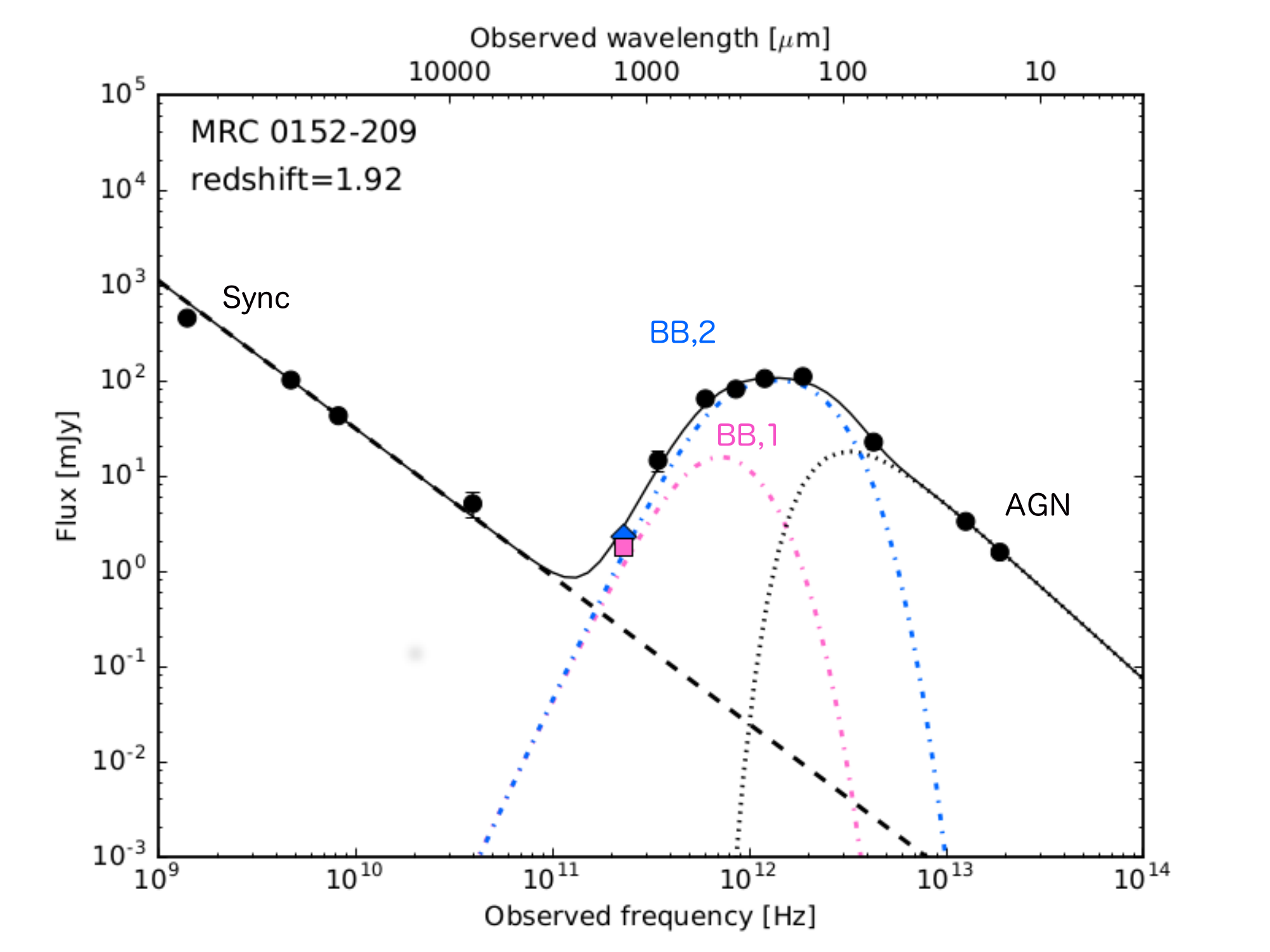}
\caption{SED of \textbf{MRC\,0152-209}. Black solid line shows best fit
total model, black dashed line represents the total synchrotron emission. The
magenta and blue dashed-dotted line represent the two black bodies of the northern and
southern ALMA 6 detections, respectively. Black dotted line indicates the
AGN component. The colored data points indicate data with sub-arcsec resolution
and black ones indicate data of lower spatial resolution. The magenta square
and blue diamond indicates the ALMA band 6 detection of the northern and
southern components, respectively. Filled black circles indicate detections
(>$3\sigma$) and downward pointing triangles the $3\sigma$ upper limits (Table~\ref{table_0152}).}
\label{fig_0152}
\end{figure}

\begin{figure}
\centering
\includegraphics[scale=0.35]{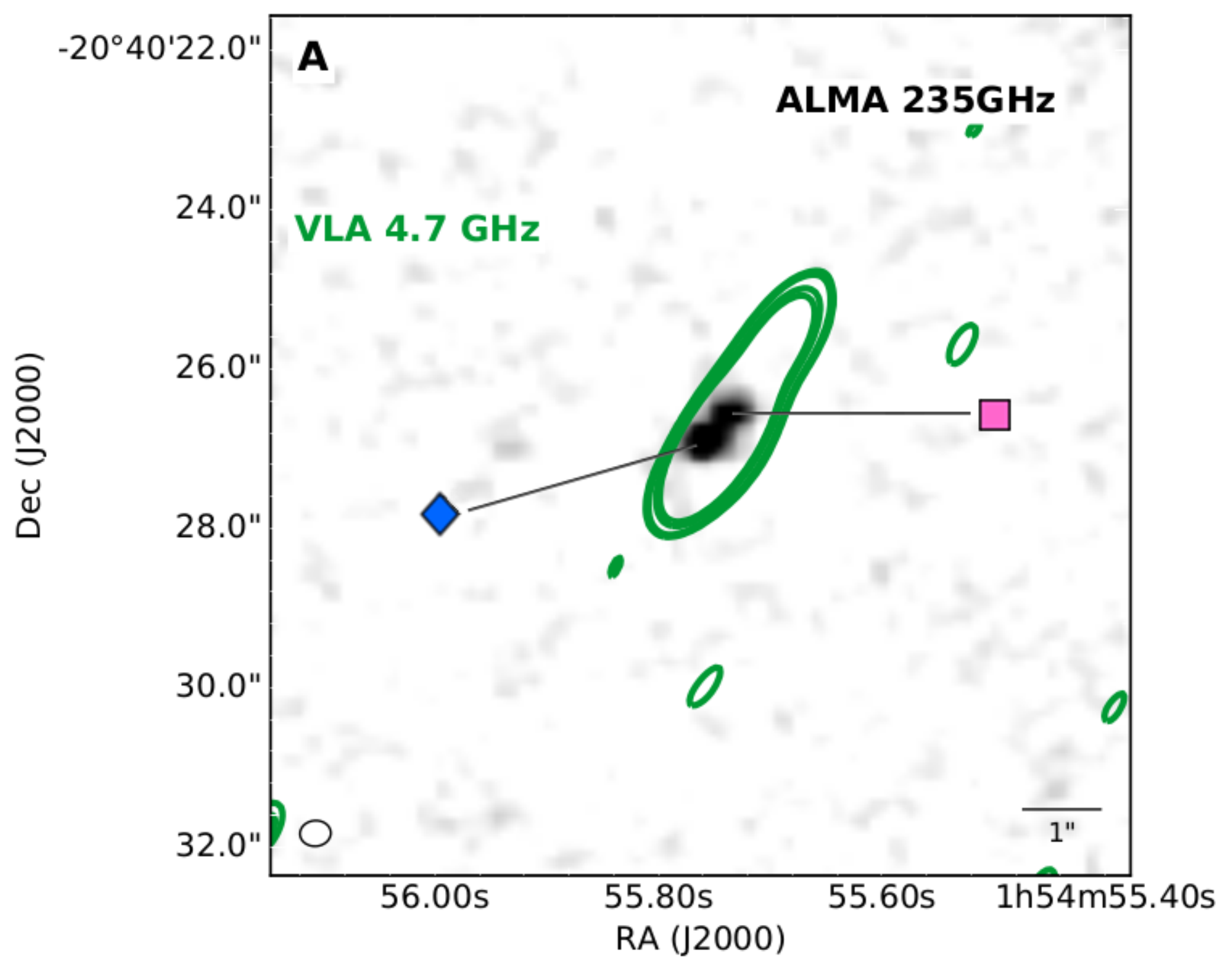} \\
\includegraphics[scale=0.35]{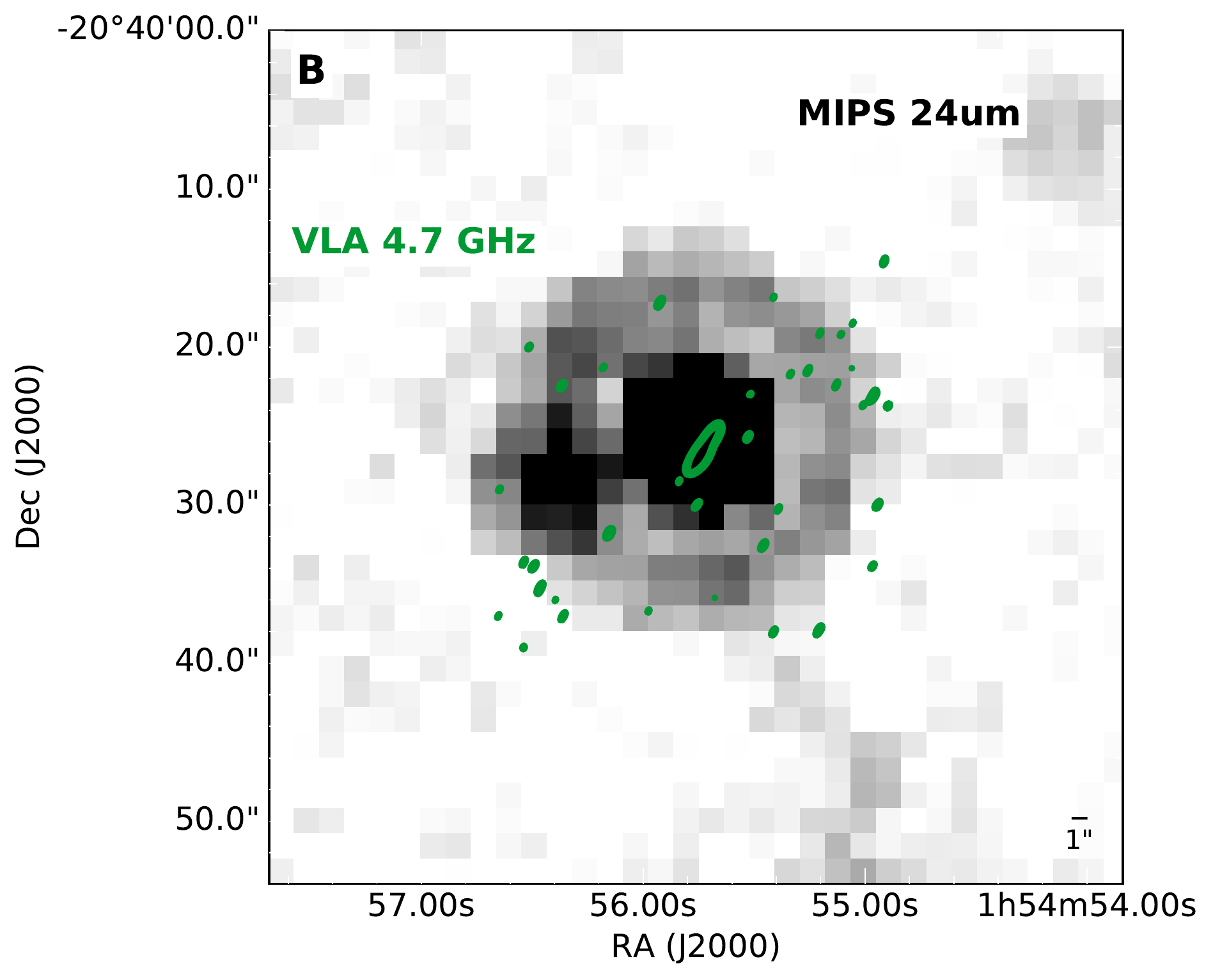}
	\caption{\textit{Panel A:} continuum map of ALMA band 6 with overlaid VLA C contours (levels are as Fig. \ref{map_0037}, $\sigma=56\,\mu$Jy). The blue and pink markers show the two ALMA detections and correspond the same markers used in Fig. \ref{fig_0152}. \textit{Panel B:} MIPS 24\,\mum\ continuum map and we note that the scale of the MIPS image is 5 times larger.}
\label{map_0152}
\end{figure}

MRC\,0152-209 has two detected continuum components in the ALMA
observations and both of which coincide with lower frequency radio detections
(Fig.~\ref{map_0152}). SED fitting is done with four components, one
total synchrotron, two modified BB and one AGN component. The VLA data
is fitted to one synchrotron power-law since it is not possible to
resolve individual components in any of the observed bands. The two ALMA band
6 detections are fitted to the same total synchrotron power-law and
two individual modified BB for each ALMA detection. The LABOCA, PACS,
SPIRE, MIPS and IRS data points are all fitted with a combination of
the two individual modified BB and an AGN component. The best fit model
is one where both of the ALMA detections are dominated by dust
emission. The radio slope is too steep to be able to account for the
observed ALMA continuum. It is unclear wherever the northern or southern dust
component corresponds to the host galaxy, but since the southern component
(indicated by the blue BB line in Fig. \ref{fig_0152}) is brighter, this
component is more likely to be the host galaxy and the northern emission,
is an interacting companion. Though this is not enough to determine which of
the component host the radio-loud AGN. In \cite{emonts2015a} they argue
that the NW component is the host, which might be more correct. If this
is the case, then the SFR would be one order of magnitude lower and the
galaxy would be on the MS, instead of lying above the MS relation. Though
the morphology of the source strongly suggest it is a merger, a SFR$\sim$1000\,M$_\odot$yr$^{-1}$ is reasonable for a merging system. 

\begin{table}
\begin{threeparttable}
\caption{Data for MRC\,0152-209 (z=1.92) }
\label{table_0152}
\centering
\begin{tabular}{lcc}
\toprule
Photometric band                   & Flux{[}mJy{]}   & Ref. \\
\midrule
\irs		&  1.580   $\pm$    0.1	& A \\   
\mips1	&  3.32    $\pm$    0.133 	& A \\
\pacsb  	& 22.6 $\pm$       3.5   	& B \\
\pacsr  	& 110.2  $\pm$     9.8   	& B \\
\spires  	& 105.0 $\pm$      8.6   	& B \\
\spirem  	& 81.3   $\pm$     7.3   	& B \\
\spirel  	& 64.4   $\pm$     6.8   	& B \\
\laboca  	& 14.5   $\pm$     3.3   	& B \\
ALMA 6$^s$& 2.30   $\pm$     0.23	& this paper, C\\
ALMA 6$^n$ &1.75     $\pm$   0.175	& this paper, C\\
ATCA (7\,mm)	&5.1 $\pm$1.5	& D	\\
VLA X  	& 42.75  $\pm$ 0.03$^a$	& E \\	
VLA C  	& 101.44 $\pm$   0.08	& E \\
VLA L  	& 453.10  $\pm$   45.30 	& F \\
\bottomrule                          
\end{tabular}
     \begin{tablenotes}
      \small
      \item \textbf{Notes}  ($s$) South companion, ($n$) north companion, likely to be the host as discussed in \cite{emonts2015a}, ($a$) flux estimated using AIPS from original radio map.
      \item \textbf{References.} (A) \cite{DeBreuck2010}, (B) \cite{Drouart2014}, (C) \citep{emonts2015a} , (D) \cite{Emonts2011_dragon} (E) \cite{Carilli1997}, (F) \cite{Condon1998}.
    \end{tablenotes}
\end{threeparttable}
\end{table}


\clearpage
\newpage

\subsection{MRC\,0156-252}

\begin{figure}
\includegraphics[scale=0.52]{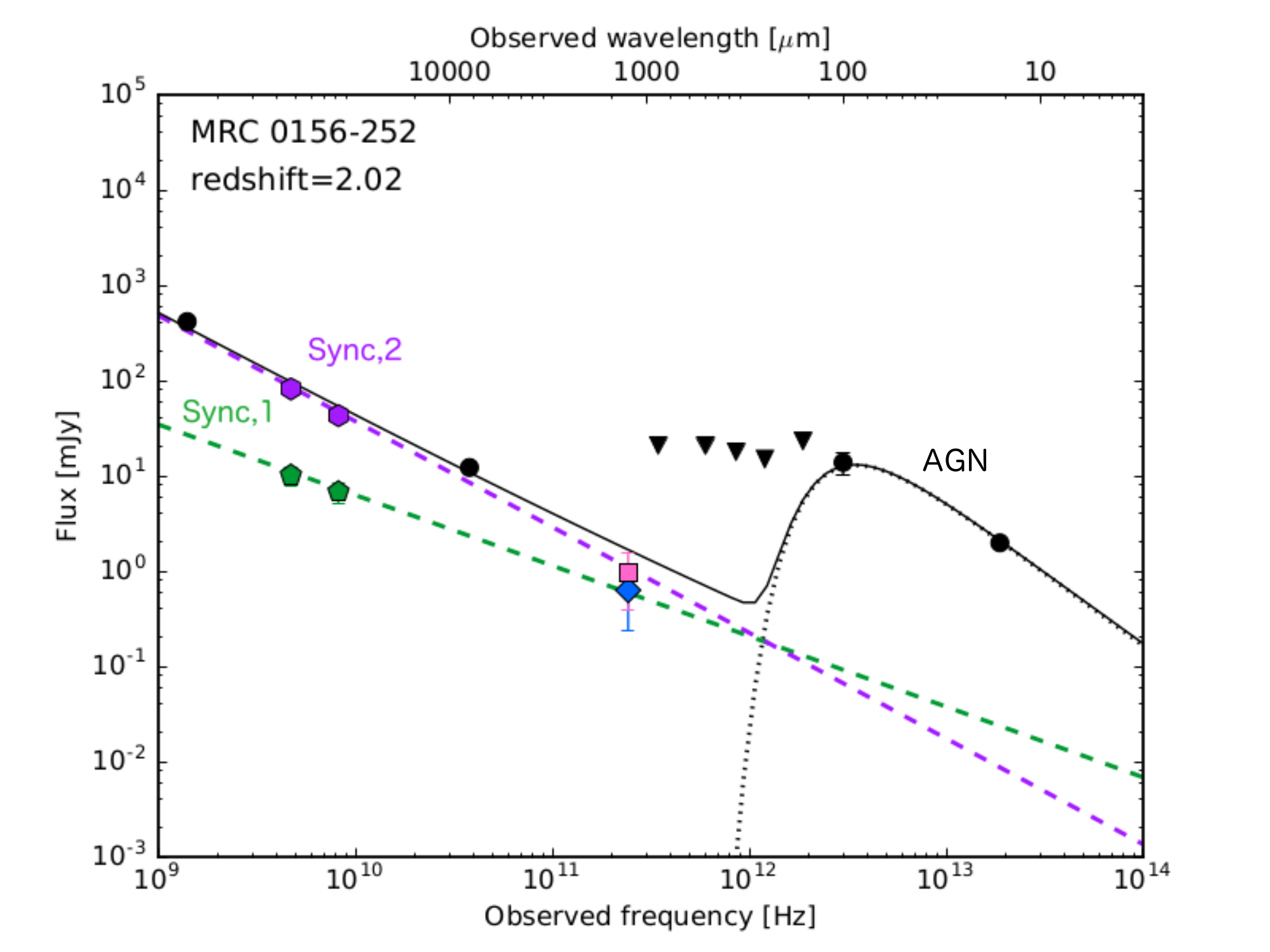}
\caption{The SED of \textbf{MRC\,0156-252}. Black solid line shows the best fit
total model, green and purple dashed line represents the northern lobe and synchrotron
emission from the core, respectively. Black dotted line indicates the AGN component. The
colored points are data which have sub-arcsec resolution and black ones
indicate data of lower resolution. Green pentagons represent the fluxes of the radio core, 
purple hexagons are the fluxes of the northern radio lobe, the blue diamond indicates the ALMA
band 6 detection which coincides with the radio core and the magenta square
is the second ALMA detection of the northern radio lobe. Filled
black circles indicate detections (>$3\sigma$) and downward pointing triangles
$3\sigma$ upper limits (Table~\ref{table_0156}).
}

\label{fig_0156}
\end{figure}

MRC\,0156-252 has two detected continuum detections, where one detection
coincides with the radio core and the other with the northern radio
lobe (Fig.~\ref{fig_0156}). SED fitting is done with five components,
two synchrotron power-laws (one for the core, one for the northern lobe,
the southern lobe is excluded from the fit), two modified BB for the two
ALMA detections, and one AGN component. The VLA L and ATCA 7\,mm bands are
fitted to the total synchrotron emission since these data do not resolve
the individual radio components, while the VLA C and X band fluxes are
fitted with two individual synchrotron power-laws. The two ALMA band 6
flux points are fitted to a combination of the individual synchrotron
power-law and modified black bodies. The LABOCA, PACS, SPIRE, MIPS and
IRS data points are fitted by a combination of a modified BB and an AGN
component. The best fit model is one where both the ALMA detections are
dominated by synchrotron emission. This is because there are only upper
limits in the FIR and they do not require the modified BB to be very
bright (Fig.~\ref{fig_0156}). 

\begin{figure}
\centering
\includegraphics[scale=0.35]{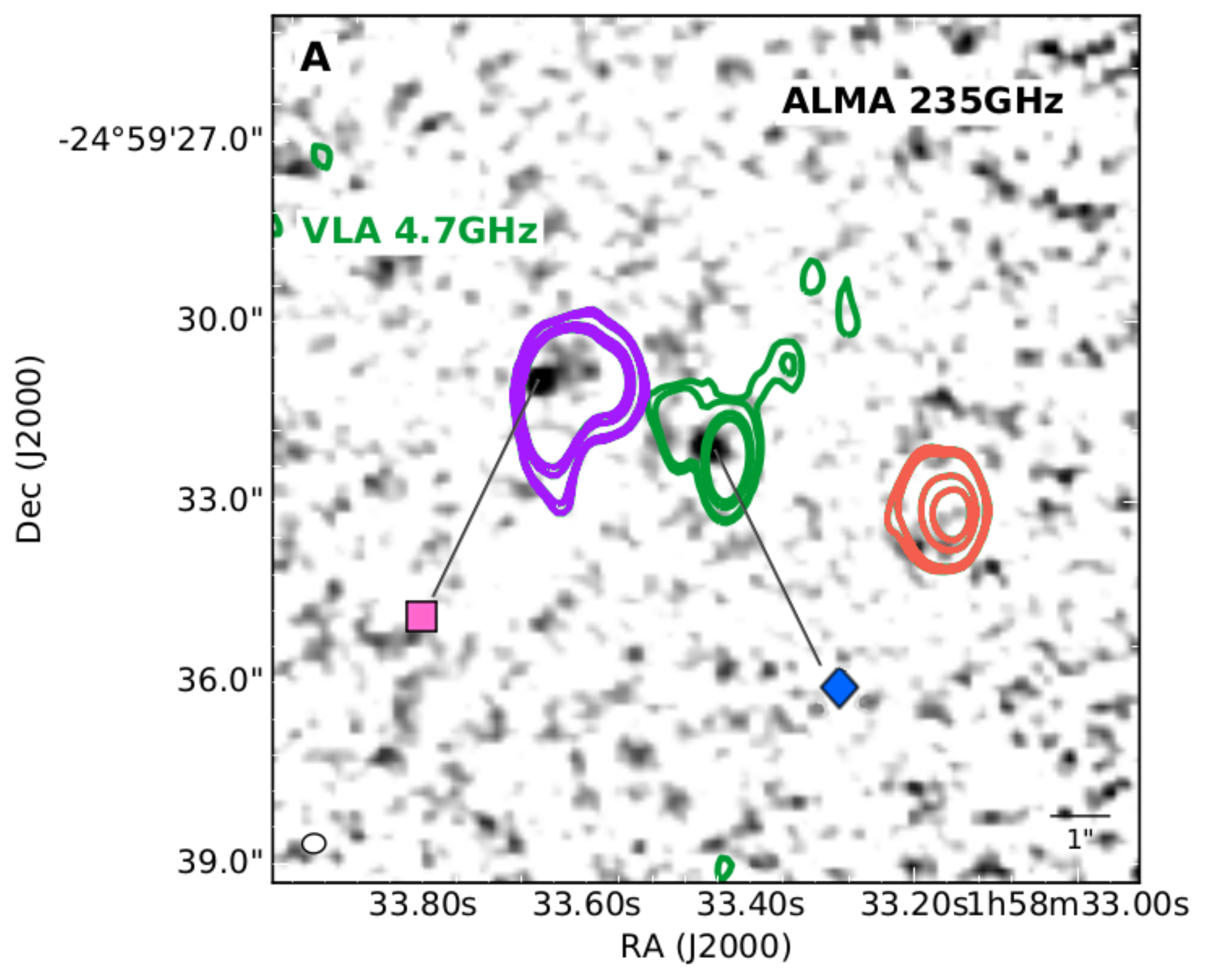} \\
\includegraphics[scale=0.35]{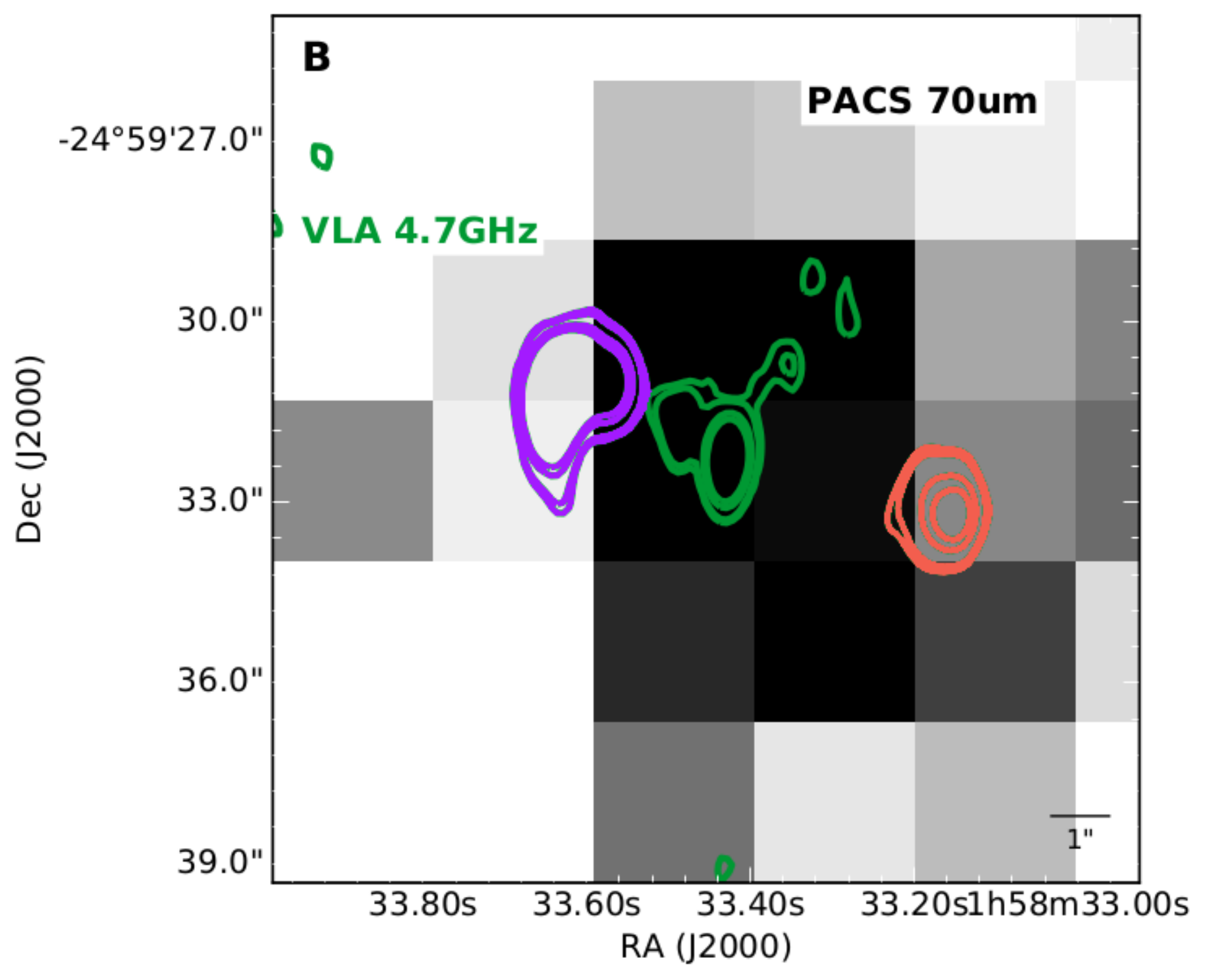}
	\caption{\textit{Panel A:} continuum map of ALMA band 6 with overlaid VLA C contours (levels are as Fig. \ref{map_0037}, $\sigma=96\,\mu$Jy). The blue and pink markers show the two ALMA detections and correspond to the same markers used for the SED in Fig \ref{fig_0156}. The core and northern radio lobes are color coded in the same colors as the flux markers in Fig. \ref{fig_0152}, the flux of the southern lobe (red radio contours) was not used in the SED fit.  \textit{Panel B:} PACS 70\,\mum\ continuum map with radio contours overlaid.}
    	\label{map_0156}
\end{figure}

\begin{table}
\begin{threeparttable}
\caption{Data for MRC\,0156-252 (z=2.02) }
\label{table_0156}
\centering
\begin{tabular}{lcc}
\toprule
Photometric band                   & Flux{[}mJy{]}   & Ref. \\
\midrule
\irs		& 1.980  $\pm$      0.1		& A \\
\pacsb	& 13.8   $\pm$      3.7   		& A \\
\pacsr	& <23.3   					& B \\
\spires	& <15.0          				& B \\
\spirem	& <18.0             			& B \\
\spirel	& <20.9             			& B \\
\laboca	& <21.0             			& C \\
ALMA 6$^h$	& 0.63   $\pm$  0.39		& this paper\\
ALMA 6$^n$	& 0.97    $\pm$ 0.58		& this paper \\
ATCA (7\,mm)	& 12.23 $\pm$   1.2   		& this paper \\
VLA X$^c$	& 6.8  $\pm$ 1.64 $^a$ 		& D \\
VLA X$^n$	& 42.84 $\pm$  4.2$^a$   		& D \\
VLA C$^c$	& 10.19  $\pm$ 2.32  $^a$	& D \\
VLA C$^n$	& 81.48  $\pm$  8.1 $^a$  	& D \\
VLA L		& 415.9  $\pm$  41.59		& E \\
\bottomrule                          
\end{tabular}
     \begin{tablenotes}
      \small
      \item \textbf{Notes} ($h$) ALMA detection at host location, ($c$) Synchrotron core, ($n$) North synchrotron lobe, ($a$) flux estimated using AIPS from original radio map, convolved to the resolution of the VLA C band.
      \item \textbf{References.} (A) \cite{DeBreuck2010}, (B) \cite{Drouart2014}, (C) \cite{Archibald2001}, (D) \cite{Kapahi1998}, (E) \cite{Condon1998}.
    \end{tablenotes}
\end{threeparttable}
\end{table}
%

\clearpage
\newpage
\subsection{TN\,J0205+2242}
\begin{figure}
\includegraphics[scale=0.52]{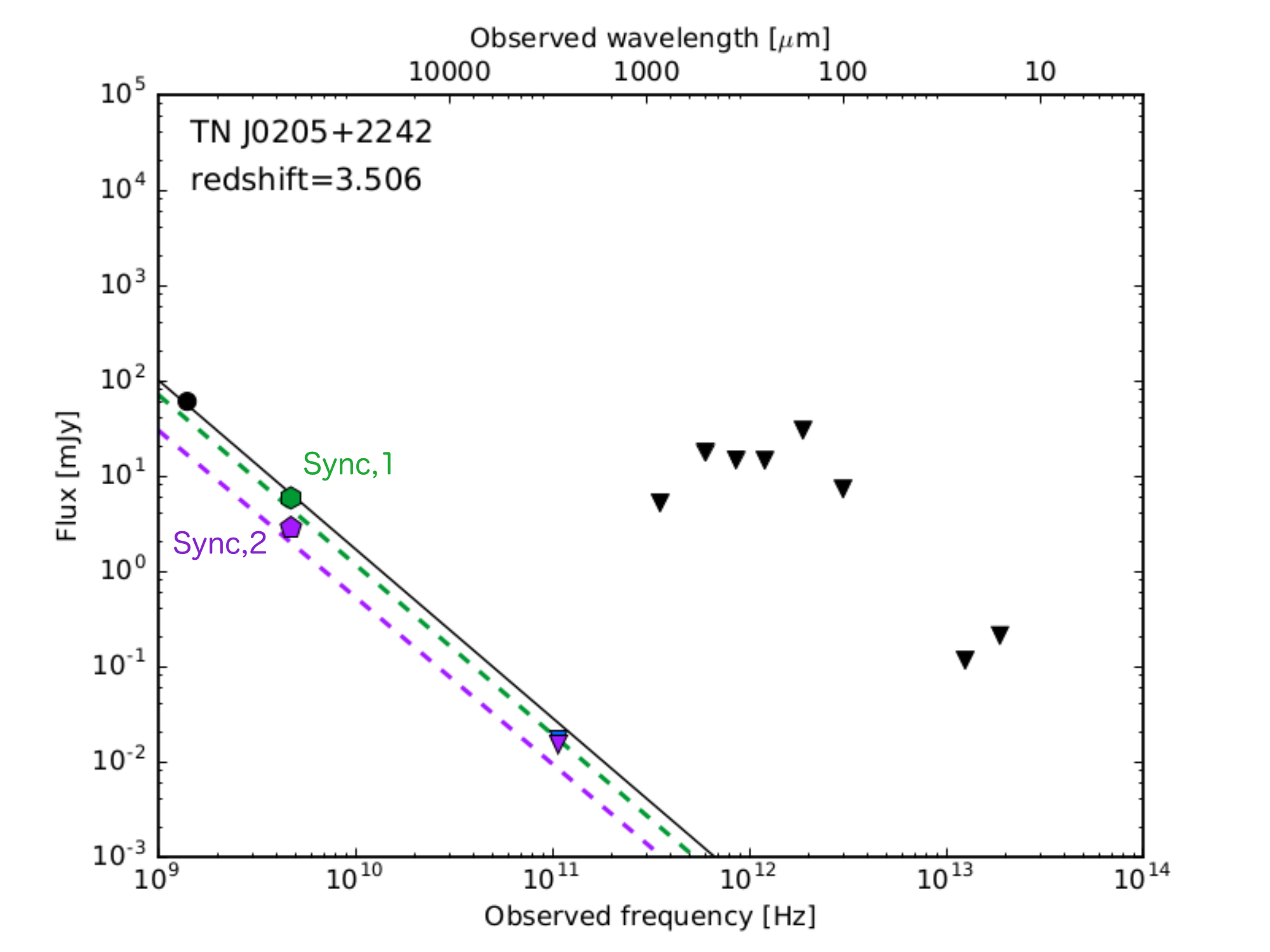}
\caption{SED of \textbf{TN\,J0205+2242}. Black solid line shows
best fit total model, green and purple dashed line is the north and
south synchrotron, respectively. The colored data points are sub-arcsec
resolution data and black ones indicated data of low resolution. The blue
and purple downward pointing triangles indicates the ALMA band 3 upper limits at
the location of the two radio lobes, where the blue triangle is the upper
limit of the host galaxy and the north synchrotron component. Filled
black circles indicate detections (>$3\sigma$) and downward pointing triangles
the $3\sigma$ upper limits (Table~\ref{table_0205}).}
\label{fig_0205}
\end{figure}

\begin{figure}
	\centering
      	\includegraphics[scale=0.35]{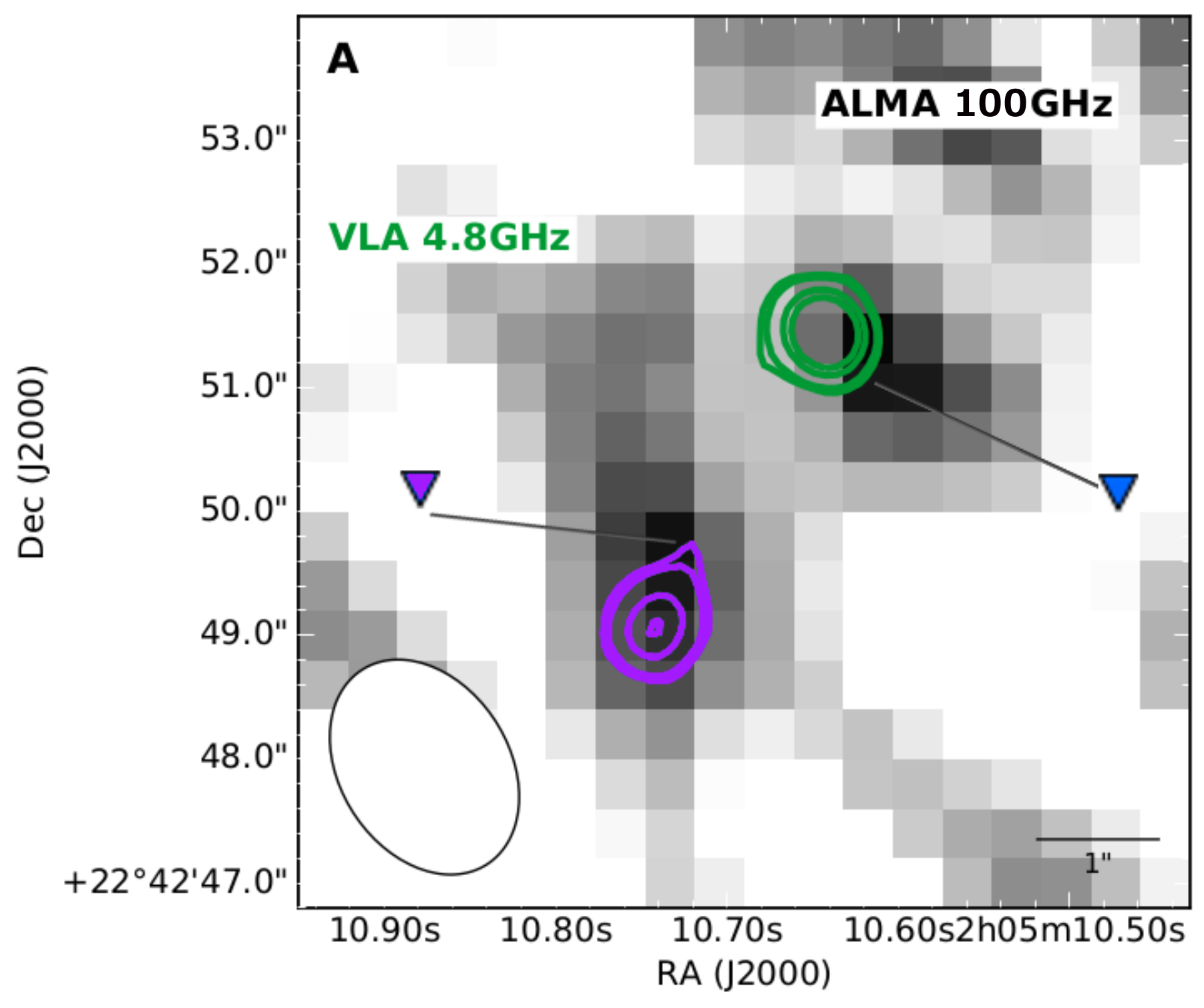}
\\
     	\includegraphics[scale=0.35]{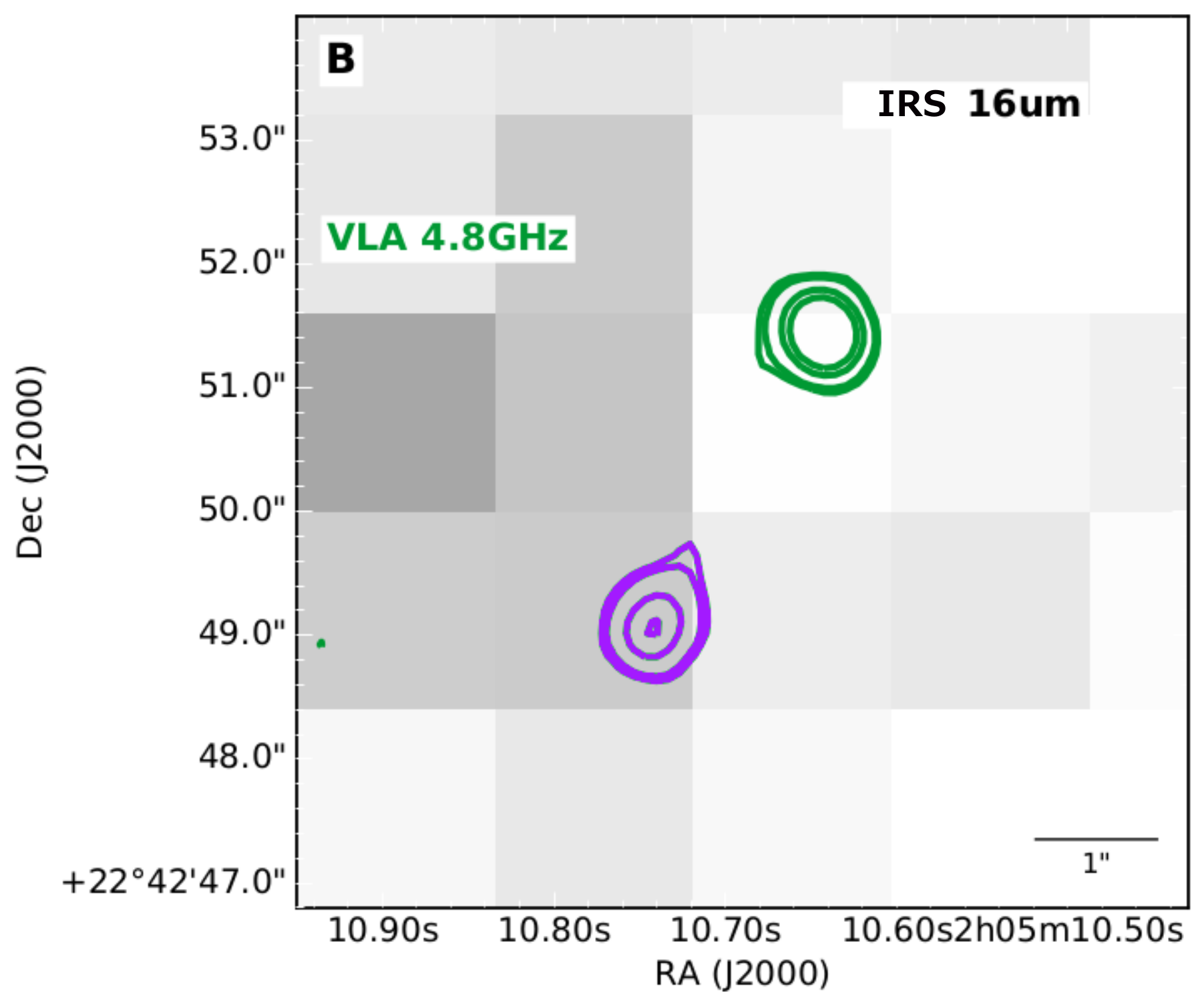}
 	\caption{\textit{Panel A:} continuum map of ALMA band 3, overlaid VLA C contours (levels are as Fig. \ref{map_0037}, $\sigma=85\,\mu$Jy). Blue and purple markers indicate the two different upper limits with ALMA and correspond to the same data points as Fig. \ref{fig_0205}. The green and purple show the two components of the VLA data and correspond to the markers of the same colors as in the SED fit. \textit{Panel B: } IRS 16\,\mum\ continuum map.}
    	\label{map_0205}
\end{figure}

TN\,J0205+2242 has no continuum detection (Fig.~\ref{map_0205}). SED fitting with \mrmoose\ is done
with five models, two synchrotron power-laws (northern and southern radio component), two modified BB and one AGN component. The northern radio
component (detected in VLA bands C and X) is assigned to
the northern ALMA upper limit and the same setup is also applied
to the southern radio and ALMA upper limit. The VLA band
X detection does not resolve
the individual components and are only considered for fitting
the total radio flux (the combination of the northern and southern
synchrotron power-law components). The two ALMA band 3 upper limits are both fitted with a combination of a synchrotron
power-law and a modified BB. The SCUBA, PACS, SPIRE and
IRS data points are fitted to the combination of the two modified
black bodies and a AGN component. The best fit only constrain the two
synchrotron models, both the SF and AGN dust components are unconstrained,
due to the fact that all FIR and MIR data points are upper limits
(Fig.~\ref{fig_0205}). 

\begin{table}
\begin{threeparttable}
\caption{Data for TN\,J0205+2242 (z=3.506) }
\label{table_0205}
\centering
\begin{tabular}{lcc}
\toprule
Photometric band                   & Flux{[}mJy{]}   & Ref. \\
\midrule
\irs		& <0.211  			& A \\
\pacsb	& <0.116    		& A \\
\pacsr	& <7.3   			& B \\
\spires	& <30.2          		& B \\
\spirem	& <14.6             	& B \\
\spirel	& <14.7             	& B \\
\scuba	& <5.2             		& C \\
ALMA 3$^n$	& <0.051       	& this paper\\
ALMA 3$^s$	& <0.045       	& this paper \\
VLA C$^n$	& 5.78  $\pm$ 0.59  $^a$		&  D\\
VLA C$^s$	& 2.87  $\pm$  0.28 $^a$  	&  D\\
VLA L		& 60.4  $\pm$  2.8			& E \\
\bottomrule                          
\end{tabular}
     \begin{tablenotes}
      \small
      \item \textbf{Notes} ($n$) North synchrotron component, ($s$) South synchrotron component, ($a$) flux estimated using AIPS from original radio map, convolved to the resolution of the VLA C band.
      \item \textbf{References.} (A) \cite{DeBreuck2010}, (B) \cite{Drouart2014}, (C) \cite{Reuland2003}, (D) \cite{DeBreuck2000}, (E) \cite{Condon1998}.
    \end{tablenotes}
\end{threeparttable}
\end{table}


\clearpage
\newpage
\subsection{MRC\,0211-256}
\begin{figure}
	\includegraphics[scale=0.52]{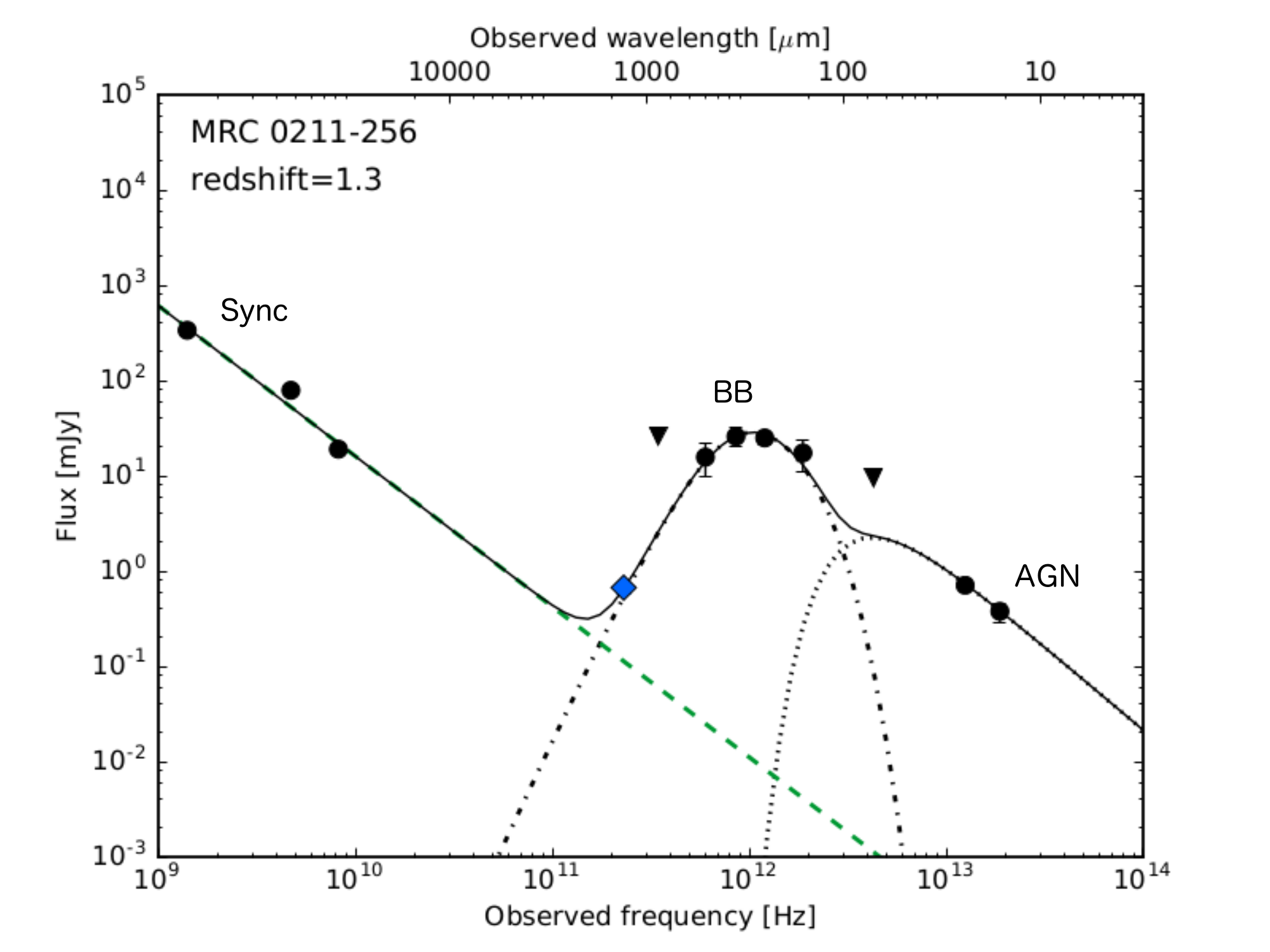}
	\caption{SED of {MRC\,0211-256}. Black solid line shows best fit total model, dashed line is total synchrotron, dash-dotted line is the blackbody component and the dotted line indicates the AGN component. The colored data point is sub-arcsec resolution data and black ones indicated data of low resolution. Blue diamond indicates ALMA band 6 detection. Filled black circles indicate detections (>$3\sigma$) and downward pointing triangles the $3\sigma$ upper limits (Table~\ref{table_0211}).}
	\label{fig_0211}
\end{figure}

\begin{figure}
	\centering
      	\includegraphics[scale=0.35]{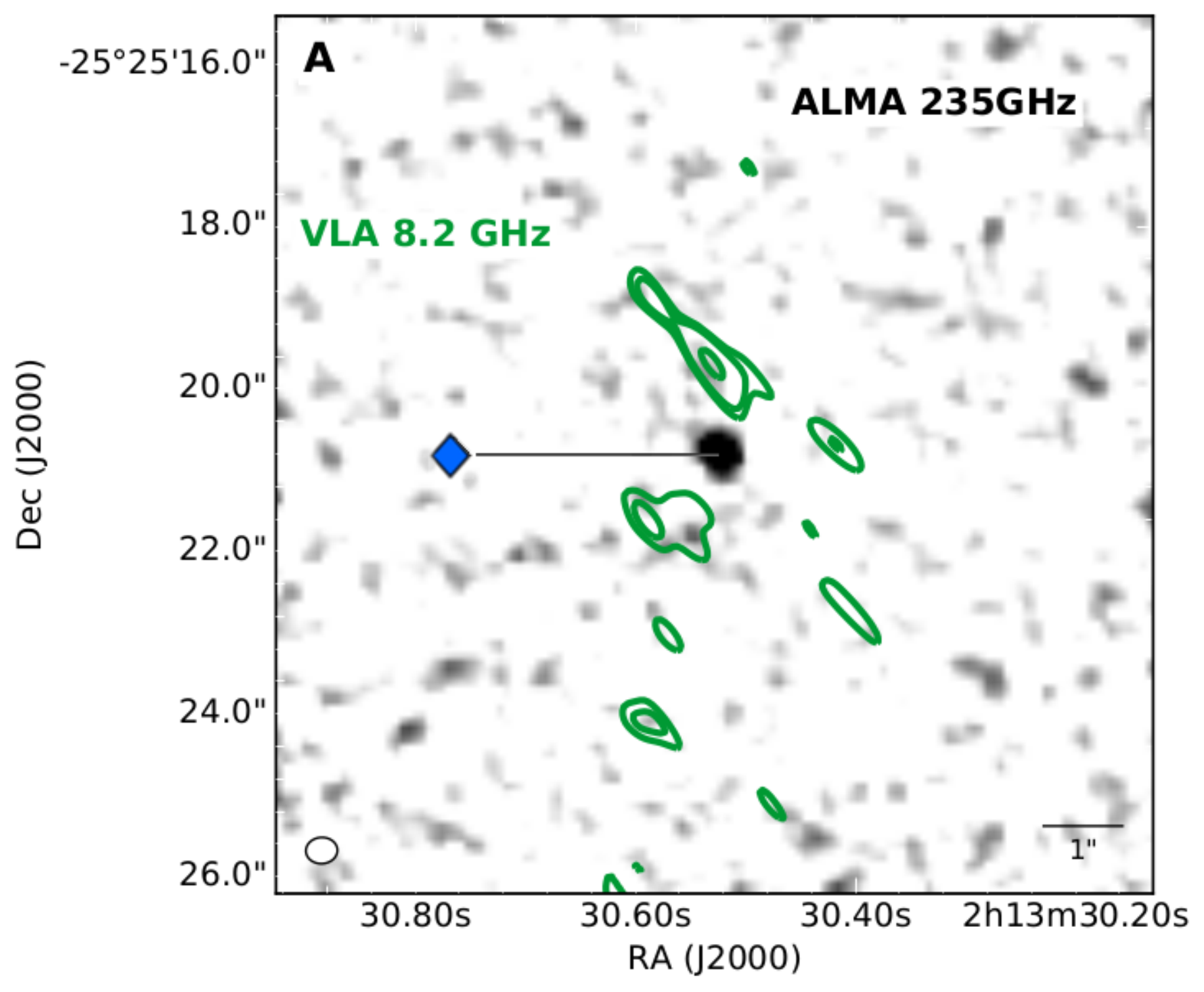}
\\
     	\includegraphics[scale=0.35]{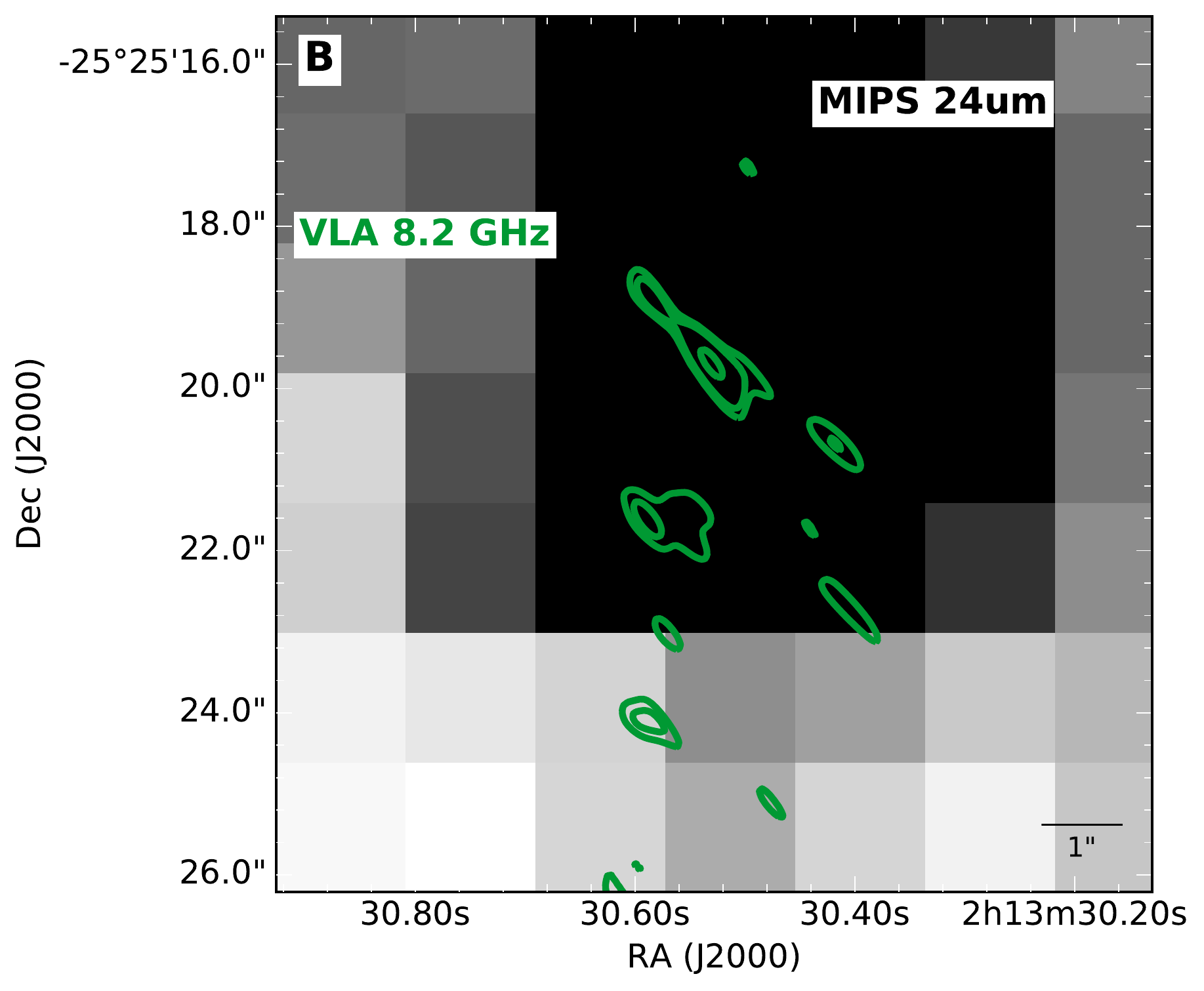}
 	\caption{\textit{Panel A:} continuum map of ALMA band 6 with overlaid VLA C contours (levels are as Fig. \ref{map_0037}, $\sigma=86\,\mu$Jy). The blue diamond indicates
the ALMA detection and is the same marker used in Fig. \ref{fig_0211}. The green contours show the portion of the VLA data that is
used in the SED fit. \textit{Panel B:} MIPS 24\,\mum\ continuum map.}
    	\label{map_0211}
\end{figure}

MRC\,0211-256 has one single continuum detection, which does not coincide
with the radio emission (Fig. \ref{map_0211}). SED fitting with MrMosse is
done with three components, one synchrotron power-law, one modified BB
and one AGN component. VLA L, X, and C bands are only considered for fitting
the total radio flux since only the total integrated flux for band L
and X, C are reported in  Condon et al. (1998) and Kapahi et al. (1998). The ALMA detection is assigned to both the synchrotron power-law and a modified BB component. The LABOCA, SPIRE, PACS, MIPS and IRS data points are
fitted to the combination of the modified BB and a AGN component. The best fit model gives a solution where the ALMA band 6 flux is dominated by thermal dust emission (Fig.~\ref{fig_0211}). 

\begin{table}
\begin{threeparttable}
\caption{Data for MRC\,0211-256 (z=1.3) }
\label{table_0211}
\centering
\begin{tabular}{lcc}
\toprule
Photometric band                   & Flux{[}mJy{]}   & Ref. \\
\midrule
\irs		&         0.378    $\pm$       0.08		& A \\
\mips1	&         0.710    $\pm$       0.03		& A \\
\pacsb	&         <9.5    					& B \\
\pacsr	&         17.4     $\pm$       6.6 		& B \\
\spires	&	         25.0     $\pm$       3.8 	& B \\
\spirem	&         25.9     $\pm$       5.8 		& B \\
\spirel	&         15.7     $\pm$       5.9 		& B \\
\laboca	&         <26.1       				& B \\
ALMA 6	&	0.67  $\pm$  0.09			& this paper\\
VLA X	&         19.04    $\pm$      2.03	$^a$	& C \\
VLA C	&         79      $\pm$        7.9 		& C \\
VLA L	&        337      $\pm$          33.7		& D \\
\bottomrule                          
\end{tabular}
     \begin{tablenotes}
      \small
      \item \textbf{Notes}  ($a$) Flux estimated using AIPS from original radio map.
      \item \textbf{References.} (A) \cite{DeBreuck2010}, (B) \cite{Drouart2014}, (C)  \cite{Carilli1997}, (D) \cite{Condon1998}.
    \end{tablenotes}
\end{threeparttable}
\end{table}


\clearpage
\newpage
\subsection{TXS\,0211-122}
\begin{figure}
	\includegraphics[scale=0.52]{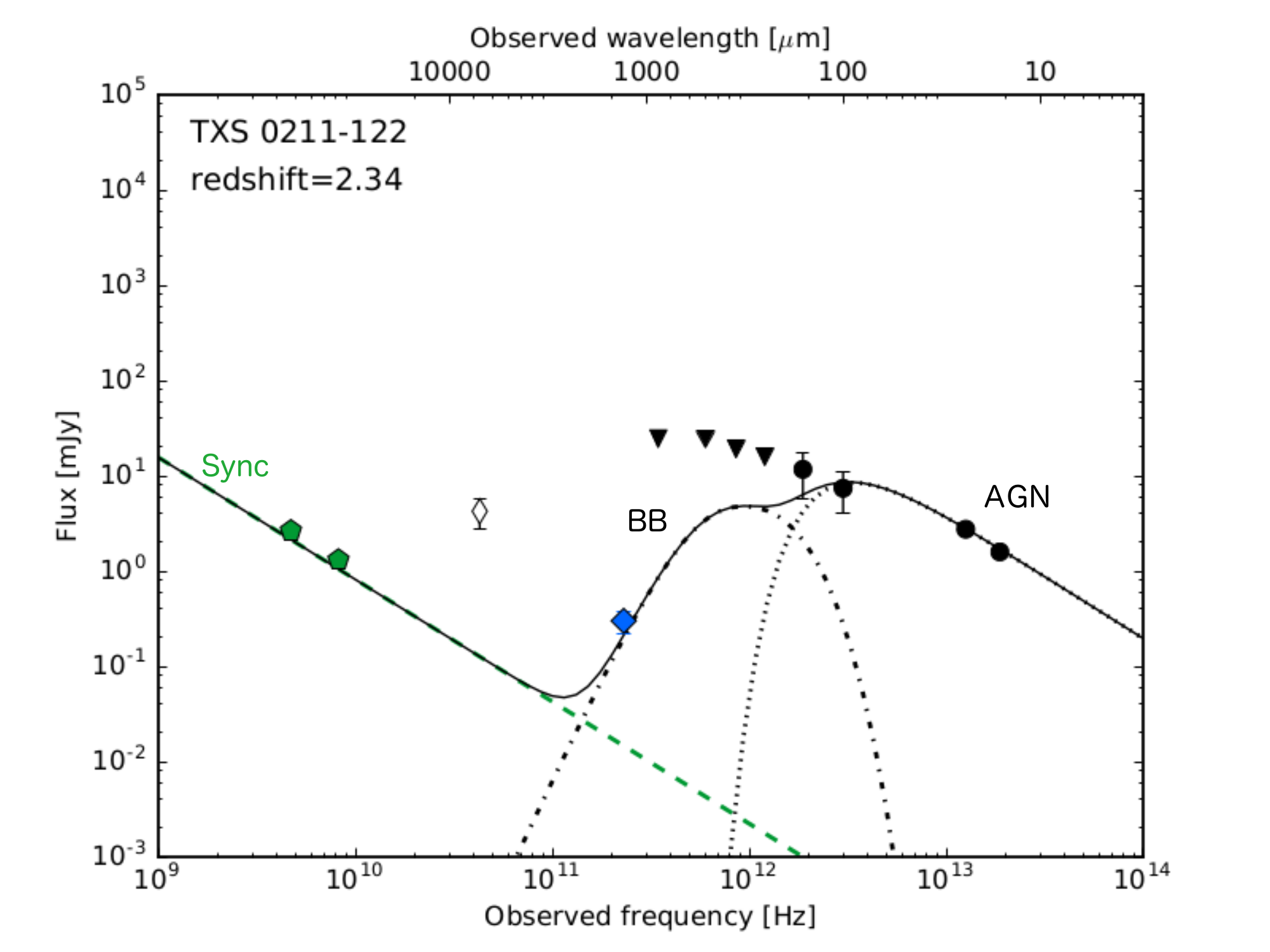}
	\caption{SED of \textbf{TXS\,0211-122}. Black solid line shows best fit total model, green dashed line is total synchrotron, black dash-dotted line is the blackbody component, black dotted line indicates the AGN component. The colored data points are sub-arcsec resolution data and black ones indicate data of low resolution. Green pentagons are the radio core and the blue diamond indicates ALMA band 6 detection. Filled black circles indicate detections (>$3\sigma$) and downward pointing triangles the $3\sigma$ upper limits (Table~\ref{table_T0211}). The open diamond shows available ATCA data but only plotted as a reference and was not used in the SED fit.}
	\label{fig_T0211}
\end{figure}

\begin{figure}
	\centering
      	\includegraphics[scale=0.35]{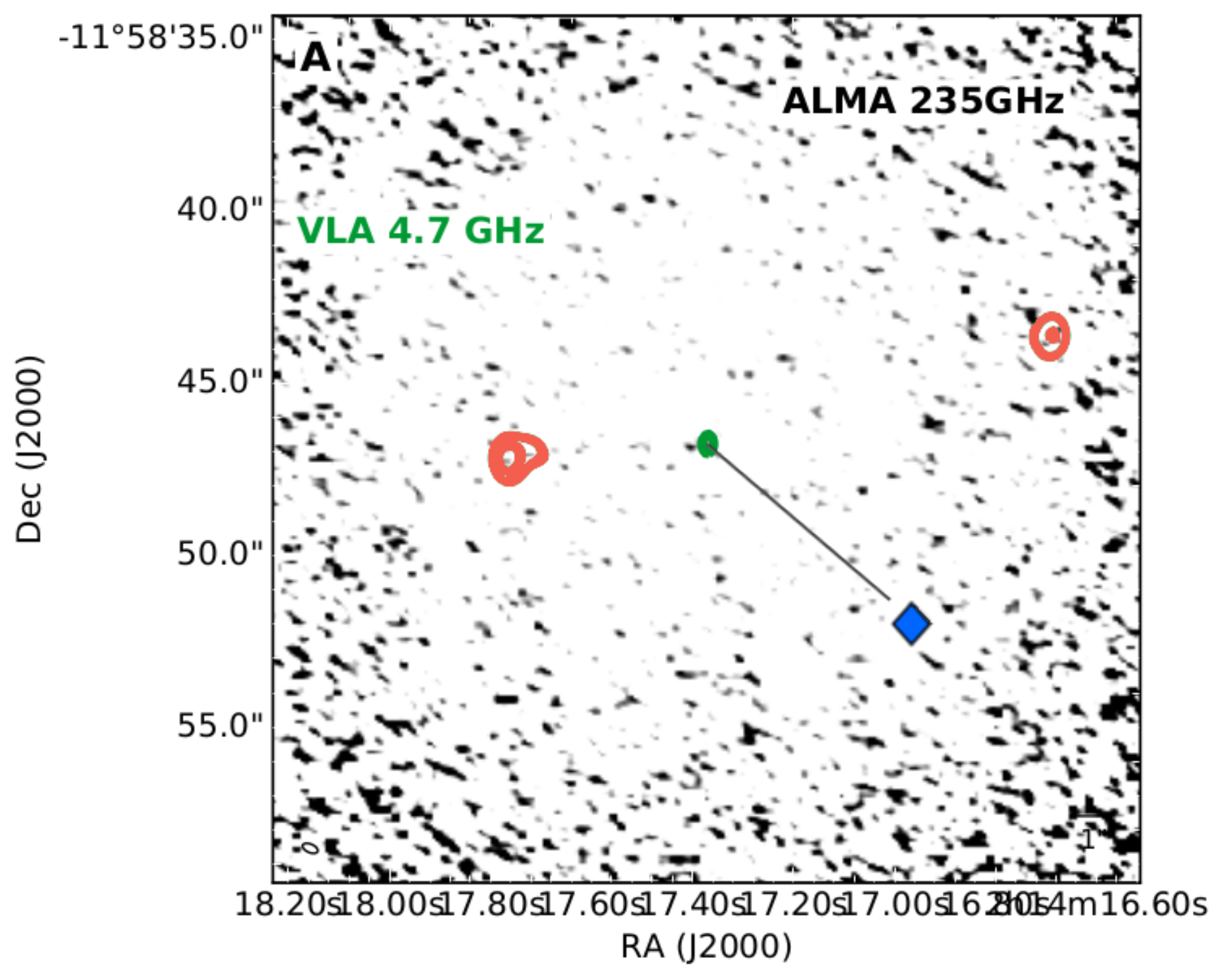}
\\
     	\includegraphics[scale=0.35]{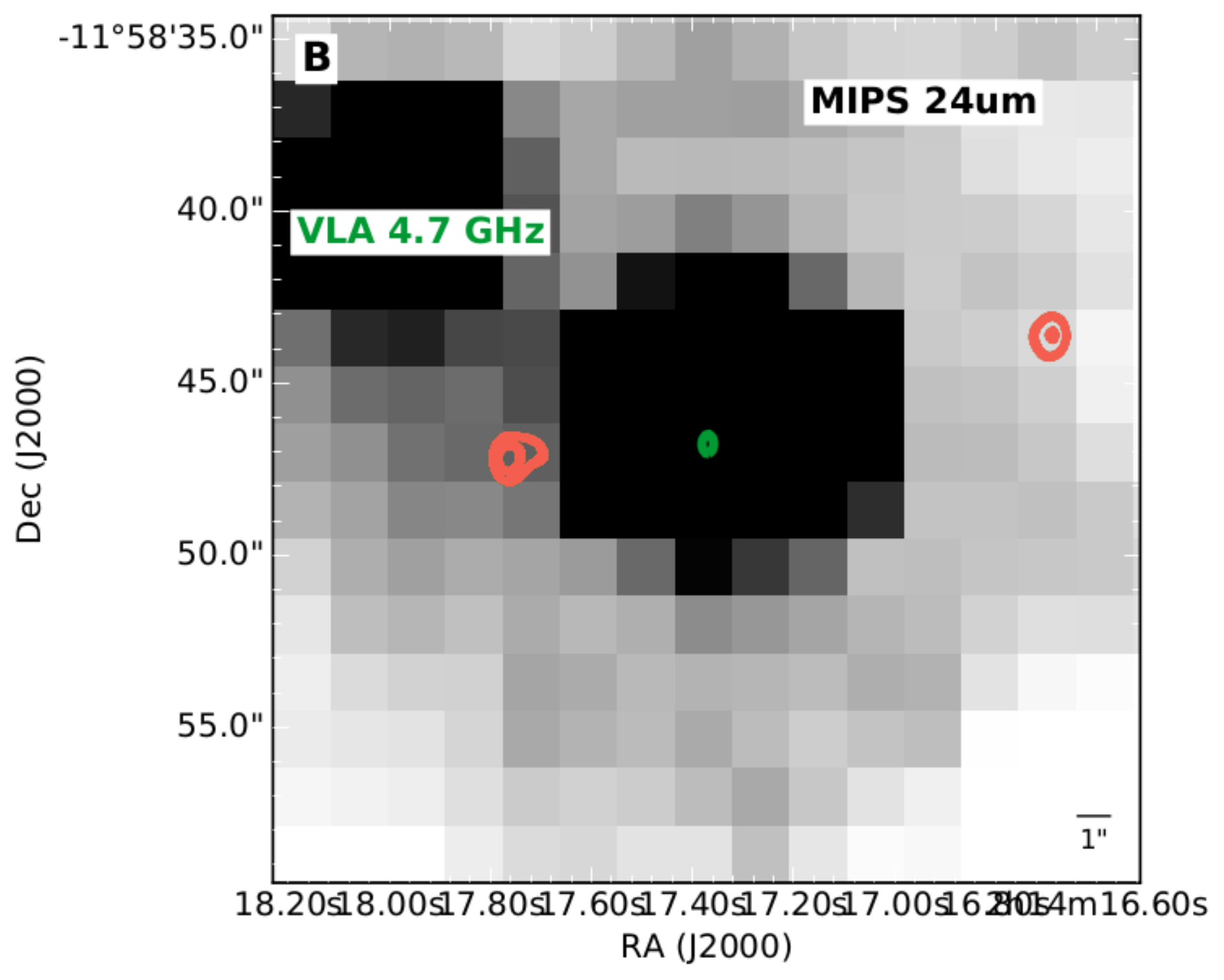}
 	\caption{\textit{Panel A:} continuum map of ALMA band 6 with overlaid VLA C contours (levels are as Fig. \ref{map_0037}, $\sigma=53\,\mu$Jy). The blue diamond indicates the ALMA detection and is the same marker used in Fig. \ref{fig_T0211}. The green contours show the portion of the VLA data that is used in the SED fit, the red contours are excluded in the fit. \textit{Panel B:} MIPS 24\,\mum\ continuum map.}
    	\label{map_T0211}
\end{figure}

TXS\,0211-122 have one single continuum detection, which coincide with
the radio core (Fig. \ref{map_T0211}). SED fitting with \mrmoose\ is done
with three components, one synchrotron power-law (for the radio core,
the two lobes are excluded in the fit), one modified BB and one AGN
component. The VLA data is fitted to the synchrotron component, the ALMA
data point is assigned to both the synchrotron power-law and a modified
BB. The LABOCA, PACS, SPIRE, MIPS and IRS are fitted to the combination of
the modified BB and a AGN component. The best fit model gives a solution
where the ALMA band 6 flux is dominated by emission from heated dust
(Fig. \ref{fig_T0211}). The reason why the PACS 160\,\mum\ data point is
not well fitted is due to the fact that the AGN power-law have a fixed
exponential cut off at 33\,\mum\ rest frame and the ALMA point puts hard
constraints on the normalization parameter of the modified BB. 

\begin{table}
\begin{threeparttable}
\caption{Data for TXS\,0211-122 (z=2.34) }
\label{table_T0211}
\centering
\begin{tabular}{lcc}
\toprule
Photometric band                   & Flux{[}mJy{]}   & Ref. \\
\midrule
\irs		& 1.590    $\pm$    0.22	& A \\
\mips1	& 2.75   $\pm$      0.04 	& A \\
\pacsg	& 7.4    $\pm$      3.4  	& B \\
\pacsr	& 11.7   $\pm$      5.9  	& B \\  
\spires	& <15.9  				& B \\
\spirem	& <19.2      			& B \\
\spirel	& <24.5      			& B \\
\laboca	& <24.6     			& B \\
ALMA 6 	& 0.30$\pm$   0.08 		& this paper\\
VLA X$^c$& 1.31 $\pm$        0.13$^a$	   	& C\\
VLA C$^c$& 2.66    $\pm$   0.26 $^a$	  	& C \\
\midrule
ATCA (7mm)$^*$	&	4.25$\pm$1.5	3 & this paper\\
\bottomrule                          
\end{tabular}
     \begin{tablenotes}
      \small
      \item \textbf{Notes} ($c$) Radio core ($a$) Flux estimated using AIPS from original radio map, convolved to the resolution of the VLA C band (*) data not used in SED fitting.
      \item \textbf{References.} (A) \cite{DeBreuck2010}, (B) \cite{Drouart2014}, (C) \cite{Carilli1997}.
    \end{tablenotes}
\end{threeparttable}
\end{table}


\clearpage
\newpage
\subsection{MRC\,0251-273}
\begin{figure}
	\includegraphics[scale=0.52]{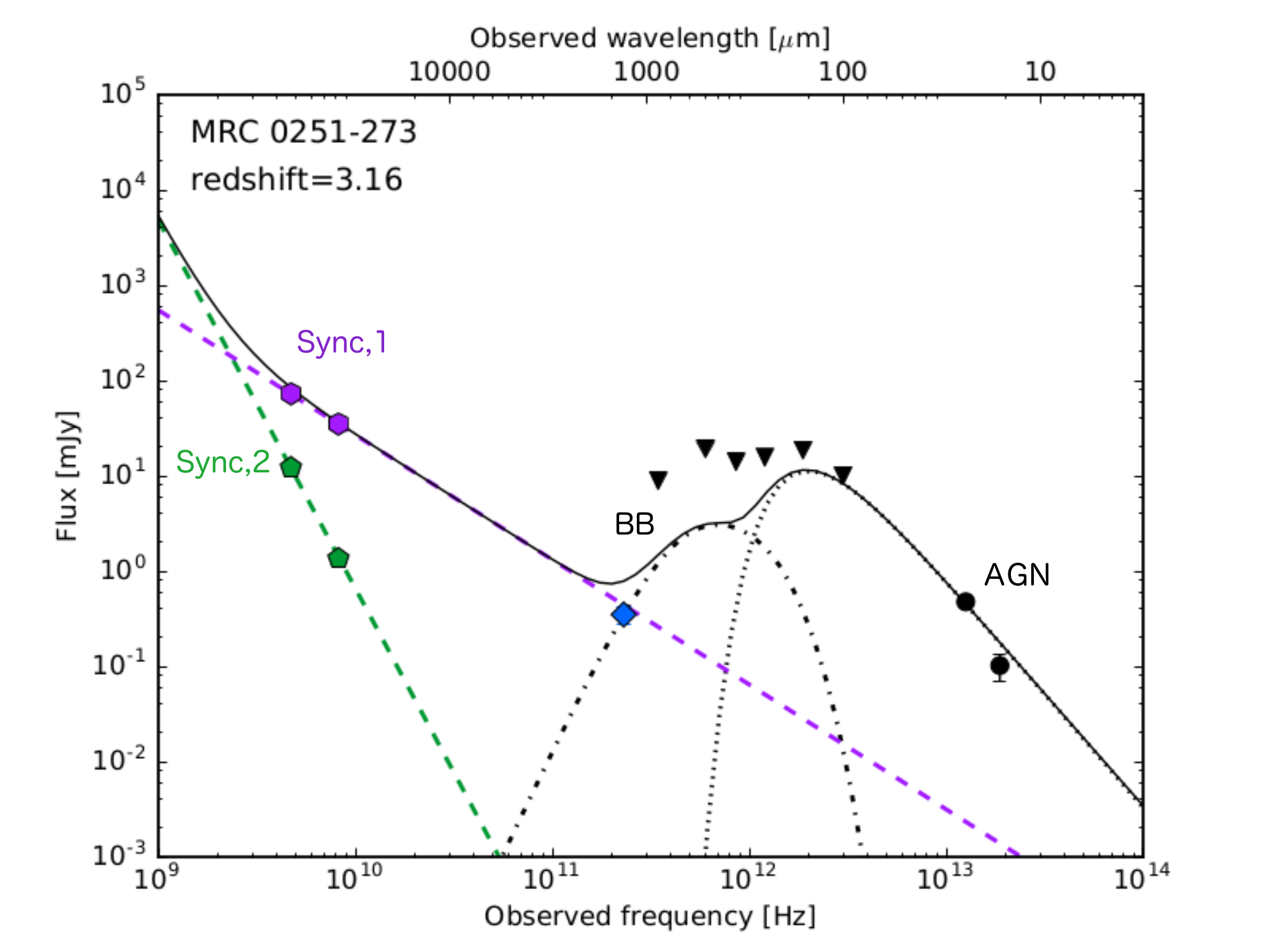}
	\caption{SED of \textbf{MRC\,0251-273}. Black solid line shows best fit total model, green and purple dashed line is north and south synchrotron lobes, respectively. Black dashed-dotted line shows the blackbody and the black dotted line indicates the AGN component. The colored data points are sub-arcsec resolution data and black ones indicate data of low resolution. Green pentagons are north synchrotron, purple hexagons are the south radio component and the blue diamond indicates the ALMA band 6 detection. Filled black circles indicate detections (>$3\sigma$) and downward pointing triangles the $3\sigma$ upper limits (Table \ref{table_0251}).}
	\label{fig_0251}
\end{figure}

\begin{figure}
	\centering
      	\includegraphics[scale=0.35]{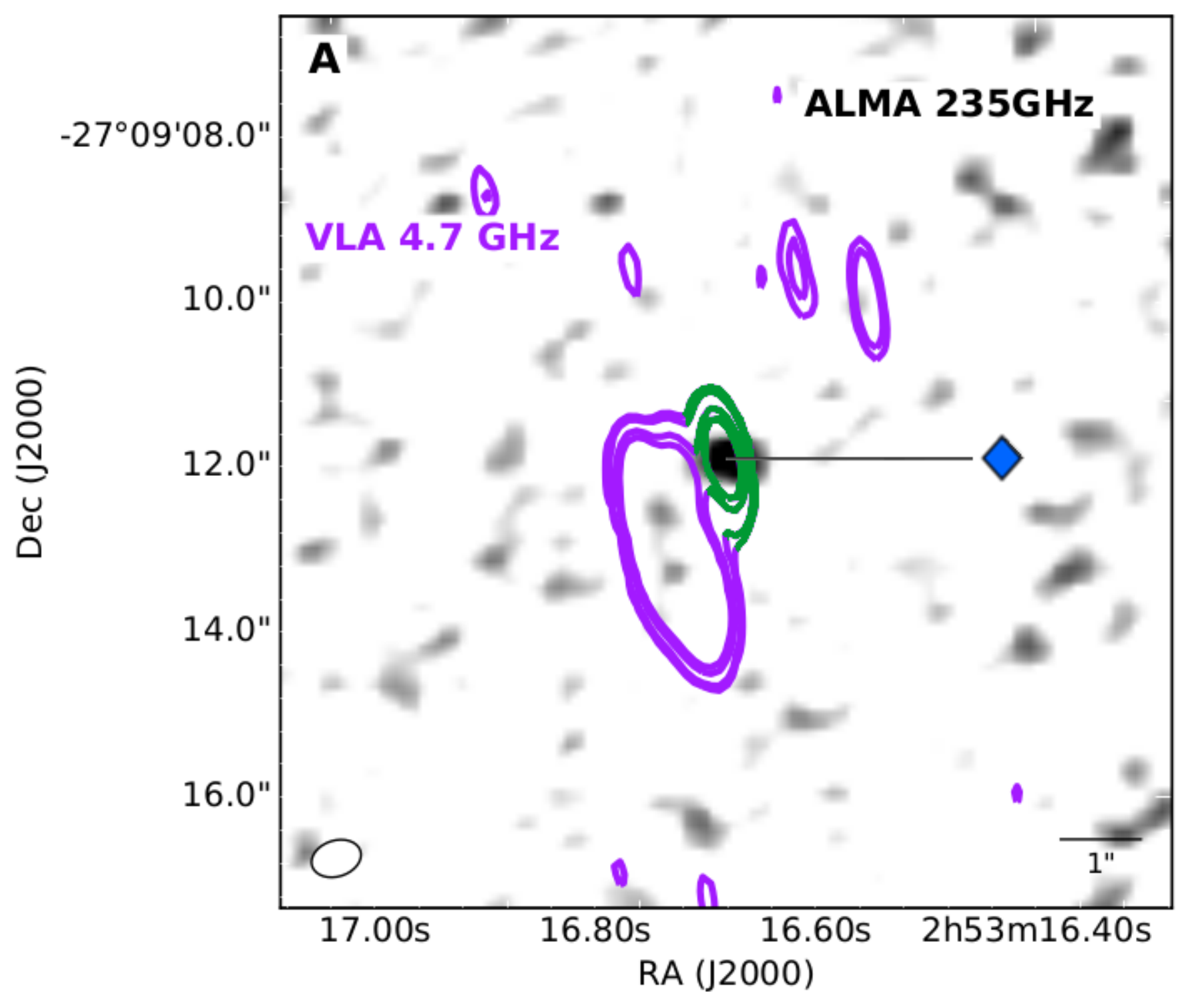}
\\
     	\includegraphics[scale=0.35]{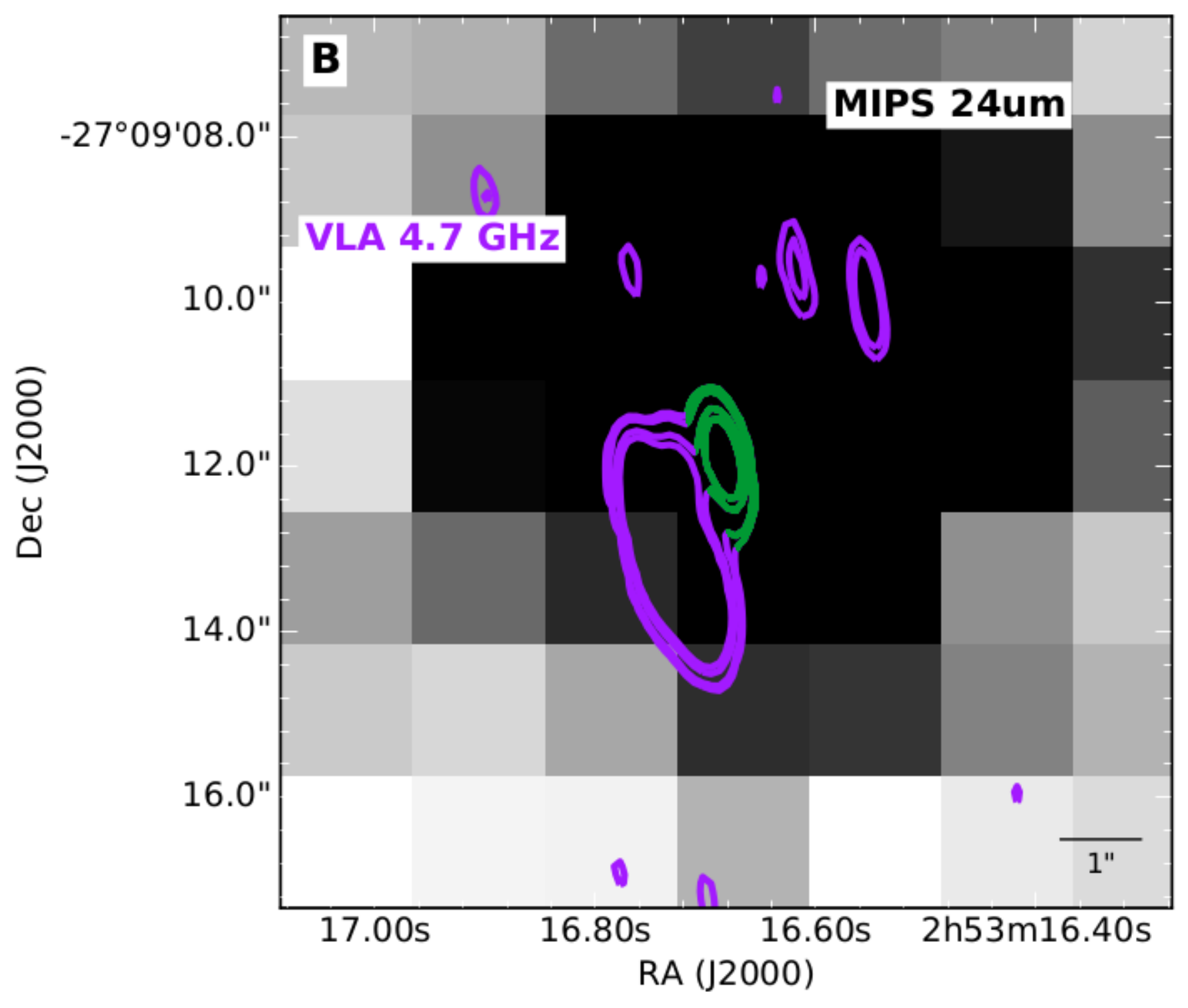}
 	\caption{\textit{Panel A:} continuum map of ALMA band 6 with overlaid VLA C contours (levels are as Fig.~\ref{map_0037}, $\sigma=48\,\mu$Jy). The blue diamond indicates the ALMA detection and is the same marker style used in Fig. \ref{fig_0251}. Green contours show the two components of the
VLA data and correspond to the markers of the same colors as
in the SED fit. Panel B: MIPS. \textit{Panel B:} MIPS 24\,\mum\ continuum map}
    	\label{map_0251}
\end{figure}

MRC\,0251-273 has one continuum detection which coincides with one
of the two radio components (Fig. \ref{map_0251}). SED fitting with
\mrmoose\ is done with four components, two synchrotron (northern and
southern radio component), one modified BB and one AGN component. The
northern radio component (detected in VLA bands C and X) is assigned
to the ALMA detection and fitted with with an individual power-law. The
southern radio component is only assigned to the VLA bands C and X. The
ALMA detection is associated to the northern synchrotron power-law and
a modified BB. The SCUBA, SPIRE, PACS, MIPS and IRS data are fitted
to the combination of the modified BB and a AGN component. The best
fit model gives that the ALMA detection is dominated by dust emission
(Fig. \ref{fig_0251}). The slope of the northern synchrotron component
is very steep and this can be due to not properly extracted photometry
because of blending between the two components. 

\begin{table}
\begin{threeparttable}
\caption{Data for MRC\,0251-273 (z=3.16) }
\label{table_0251}
\centering
\begin{tabular}{lcc}
\toprule
Photometric band                   & Flux{[}mJy{]}   & Ref. \\
\midrule
\irs		& 0.102   $\pm$     0.033	& A \\
\mips1	& 0.476   $\pm$     0.033  	& A \\
\pacsg	& <10.0        	&  B \\
\pacsb	& <18.7           &  B \\
\spires	& <15.7           &  B \\
\spirem	& <14.1             &  B \\
\spirel	& <19.3             &  B \\
\scuba	& <8.9               & C \\
ALMA 6	&	0.35 $\pm$   0.07 	& this paper\\
VLA X$^n$	& 1.35   $\pm$    0.14 $^a$  	& A \\
VLA X$^s$	& 34.86  $\pm$      0.35   $^a$ 	& A \\
VLA C$^n$	& 12.22    $\pm$   0.12     $^a$	& A \\
VLA C$^s$	& 72.36 $\pm$      0.72    $^a$		& A \\
\bottomrule                          
\end{tabular}
     \begin{tablenotes}
      \small
      \item \textbf{Notes}  ($n$) North radio component, ($s$) south radio component, ($a$) flux estimated using AIPS from original radio map, convolved to the resolution of the VLA C band.
      \item \textbf{References.} (A) \cite{DeBreuck2010}, (B) \cite{Drouart2014}, (C) \cite{Carilli1997}.
    \end{tablenotes}
\end{threeparttable}
\end{table}


\clearpage
\newpage
\subsection{MRC\,0324-228}
MRC\,0324-228 has no continuum detection with ALMA
(Fig. \ref{map_0324}). SED fitting with \mrmoose\ is done with four
components, two synchrotron power-law (northern and southern radio
component), one modified BB and one AGN component. The northern radio
component (detected in VLA bands C and X) is assigned to an individual
synchrotron power-law and the same setup is also applied for the southern
radio component. The VLA band L and ATCA 7\,mm data do not resolve the
individual components and are only considered for fitting the total radio
flux (the combination of the northern and southern synchrotron power-law
components). The ALMA band 6 detection is also assigned to the total radio
flux and to a modified BB. The LABOCA, SIPRE, PACS, MIPS and IRS data are
fitted to the combination of the modified BB and a AGN component. The best
fit model does not constrain the modified BB due to only upper limits in
the FIR. Without an intermediate data point between ALMA and ATCA 7\,mm it
is not possible to determine where the steepening occurs, but the upper
limit in ALMA is consistent with a continued synchrotron slope. 

\begin{figure}
	\includegraphics[scale=0.52]{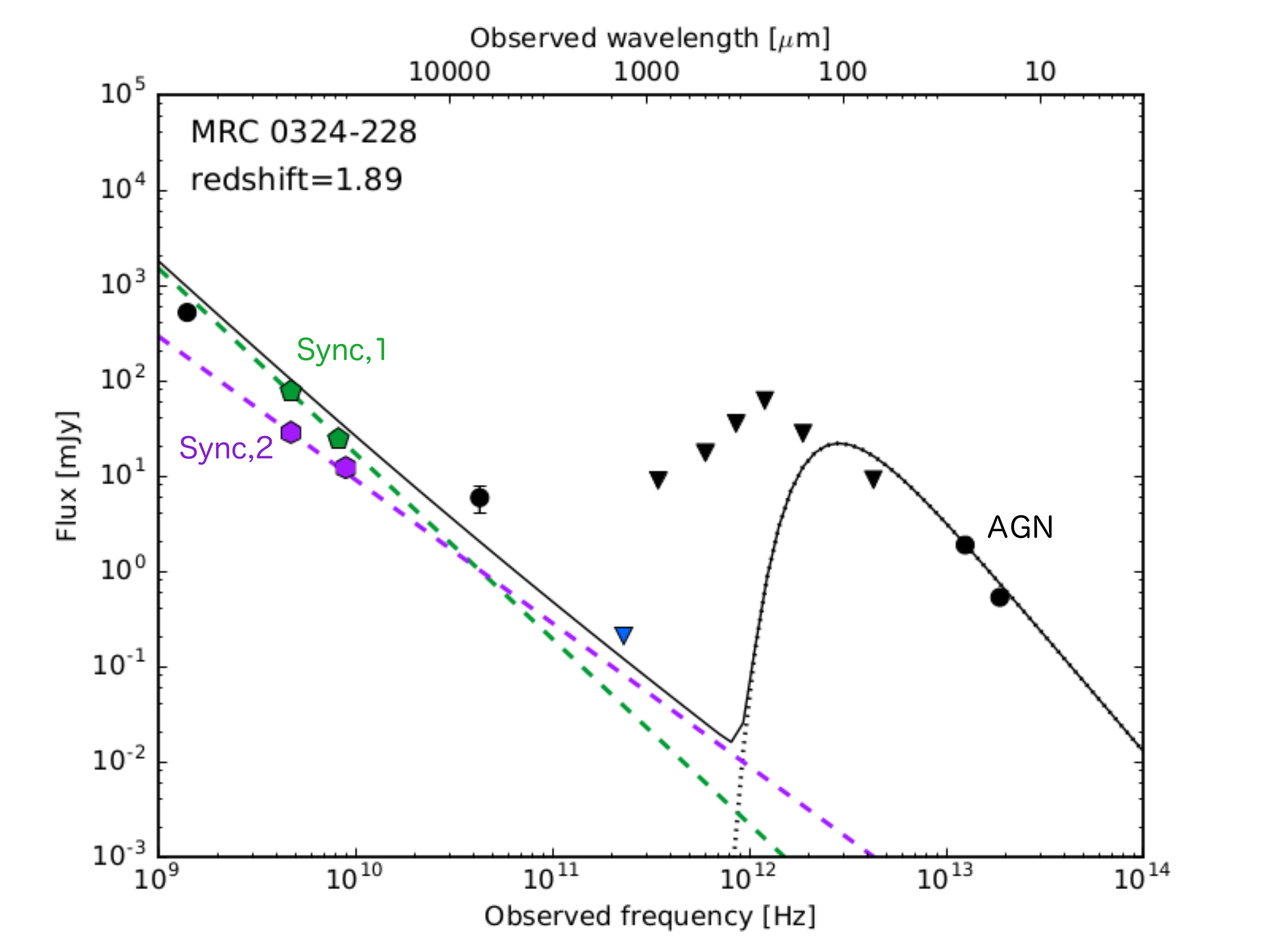}
	\caption{SED of \textbf{MRC\,0324-228}. Black solid line shows best fit total model, green and purple dashed line is north and south synchrotron lobes, respectively. Black dotted line indicates the AGN component. The colored data points are sub-arcsec resolution data and black ones indicate data of low resolution. Green pentagons are north synchrotron, purple hexagons are the south radio component and the blue triangle indicates the ALMA band 6 3$\sigma$ upper limit. Filled black circles indicate detections (>$3\sigma$) and downward pointing triangles the $3\sigma$ upper limits (Table~\ref{table_0324}). }
	\label{fig_0324}
\end{figure}

\begin{figure}
	\centering
      	\includegraphics[scale=0.35]{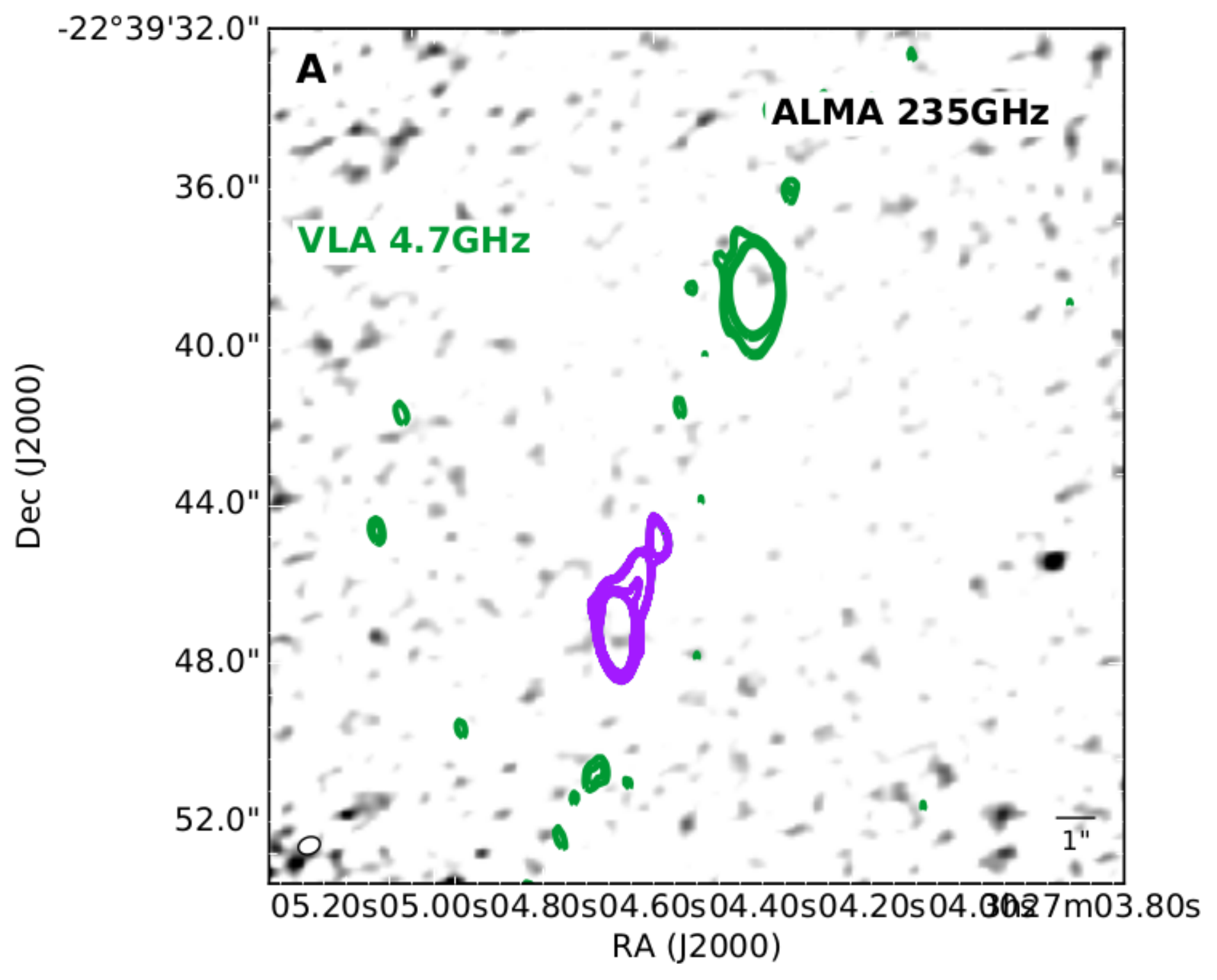}
\\
     	\includegraphics[scale=0.35]{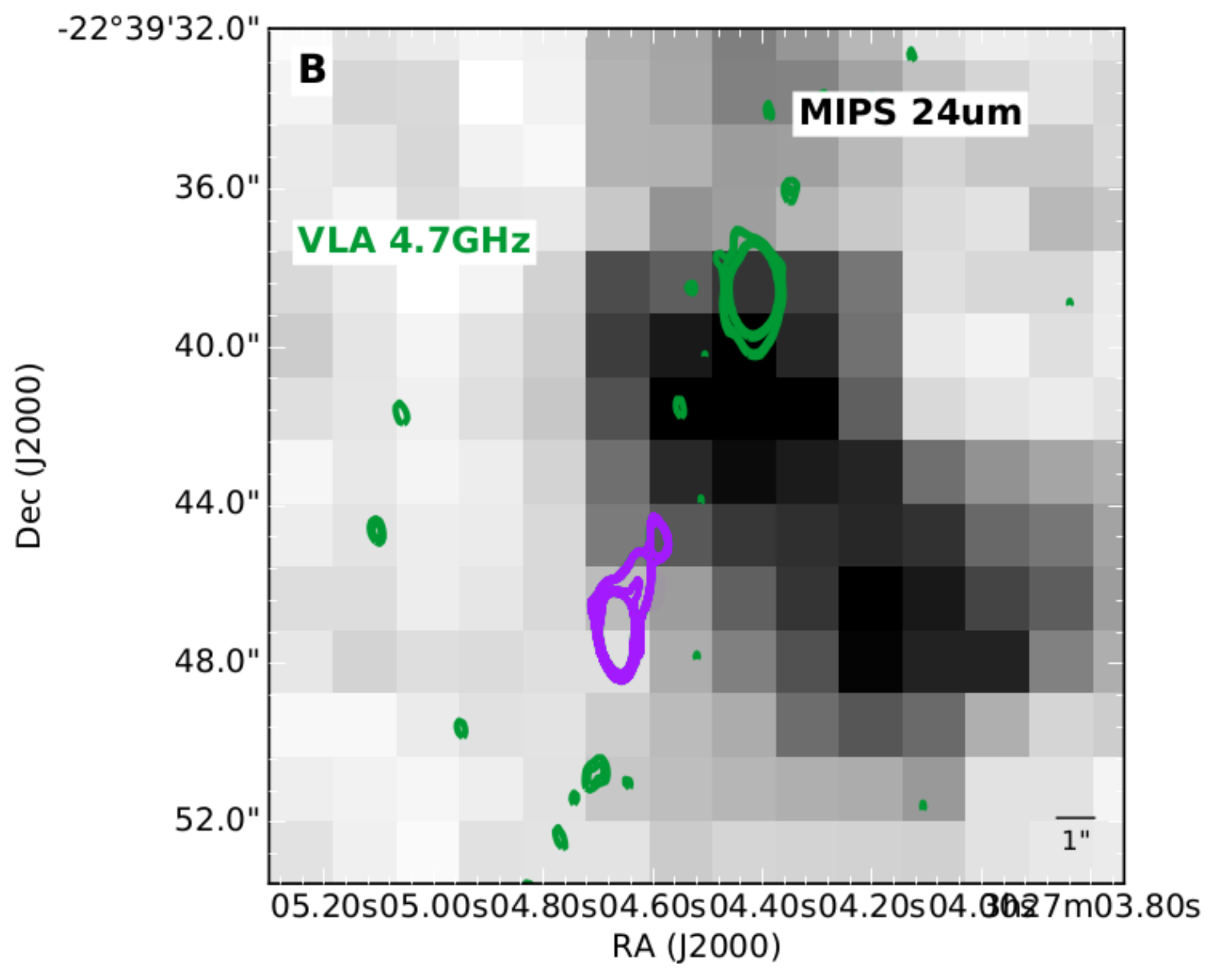}
 	\caption{\textit{Panel A:} continuum map of ALMA band 6 with overlaid VLA C (levels are as Fig. \ref{map_0037}, $\sigma=53\,\mu$Jy). The green and purple contours show the two components of the
VLA data and correspond to the markers of the same colors as
in the SED fit (Fig. \ref{fig_0324}).  \textit{Panel B: }MIPS 24\,\mum\ continuum map.}
    	\label{map_0324}
\end{figure}

\begin{table}
\begin{threeparttable}
\caption{Data for MRC\,0324-228 (z=1.89) }
\label{table_0324}
\centering
\begin{tabular}{lcc}
\toprule
Photometric band                   & Flux{[}mJy{]}   & Ref. \\
\midrule
\irs 		& 0.530    $\pm$   0.054		& A \\
\mips1	& 1.880    $\pm$   0.035		& A \\ 
\pacsb	&   <9.1         				& B \\
\pacsr	&   27.9   $\pm$     5.4        		& B \\   
\spires	&   61.8    $\pm$    6.7           		& B \\
\spirem	&   35.5    $\pm$    5.9           		& B \\
\spirel	&  17.5   $\pm$     7.4           		& B \\
\laboca	&  <9.0                      				& B \\
ALMA 6 	& <0.21 						& this paper\\
\atca		&   5.9     $\pm$     1.77          	& this paper \\
VLA X$^n$ 	& 24.54  $\pm$    0.24$^a$ 	        	& A \\
VLA X$^s$	&   12.22   $\pm$   0.12$^a$ 	        	& A \\ 
VLA C$^n$	&   77.86    $\pm$  0.77$^a$    	& A \\
VLA C$^s$	&   28.70   $\pm$   0.28$^a$	        	& A \\ 
VLA L		&   518.7   $\pm$    51     	        	& C \\
\bottomrule                          
\end{tabular}
     \begin{tablenotes}
      \small
      \item \textbf{Notes}  ($n$) North radio component, ($s$) south radio component, ($a$) flux estimated using AIPS from original radio map, convolved to the resolution of the VLA C band.
      \item \textbf{References.} (A) \cite{DeBreuck2010}, (B) \cite{Drouart2014}, (C) \cite{Condon1998}.
    \end{tablenotes}
\end{threeparttable}
\end{table}


\clearpage
\newpage
\subsection{MRC\,0350-279}
\begin{figure}
	\includegraphics[scale=0.52]{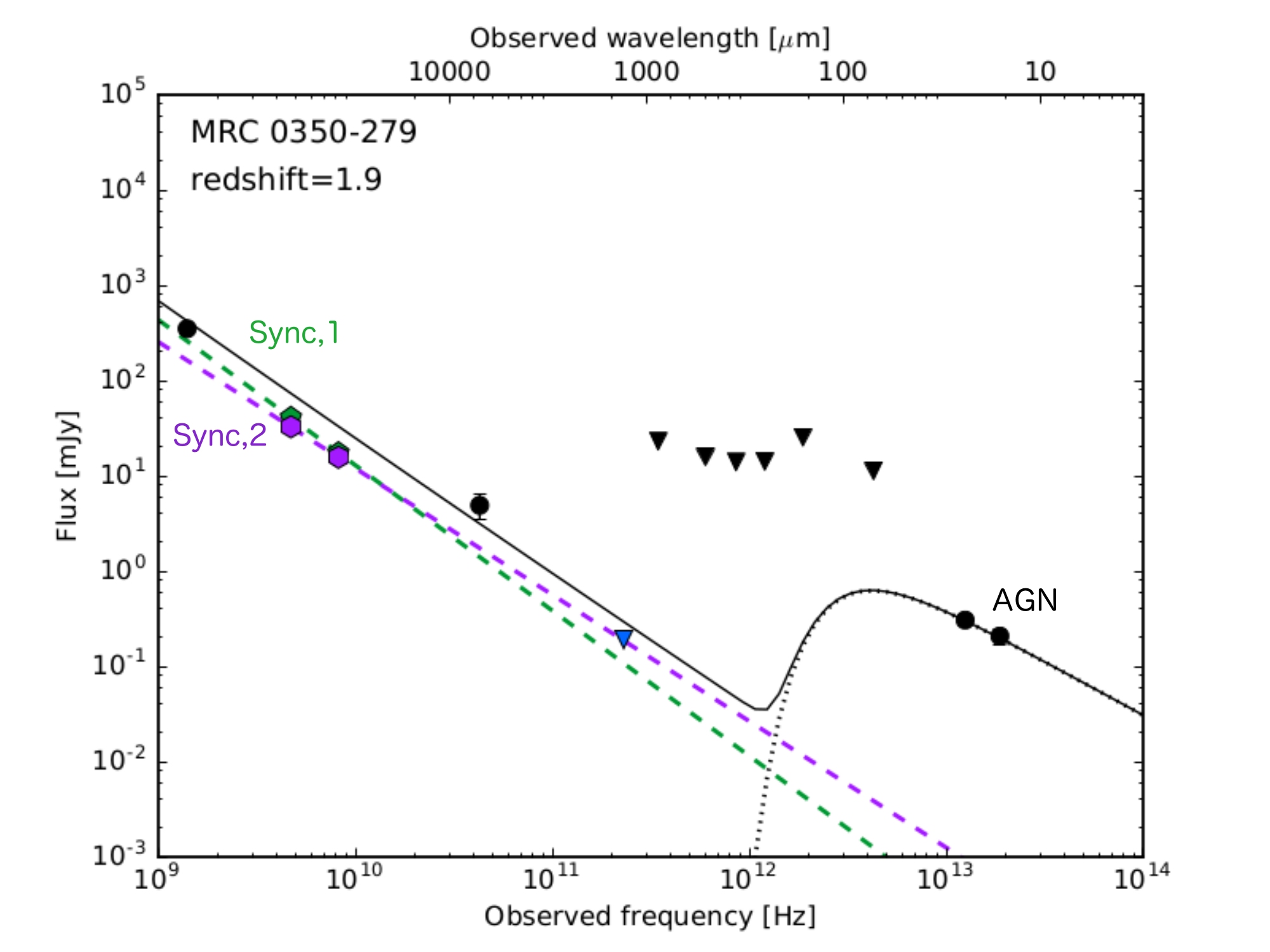}
	\caption{SED of \textbf{MRC\,0350-279}. Black solid line shows best fit total model, green and purple dashed line are the north and south synchrotron lobes, respectively. Black dotted line indicates the AGN component. The colored data points are sub-arcsec resolution data and black ones indicate data of low resolution. Green pentagons and purple hexagons are the north and south radio component, respectively and the blue triangle indicates the ALMA band 6 3$\sigma$ upper limit. Filled black circles indicate detections (>$3\sigma$) and downward pointing triangles the $3\sigma$ upper limits (Table \ref{table_0350}).}
	\label{fig_0350}
\end{figure}

\begin{figure}
	\centering
      	\includegraphics[scale=0.35]{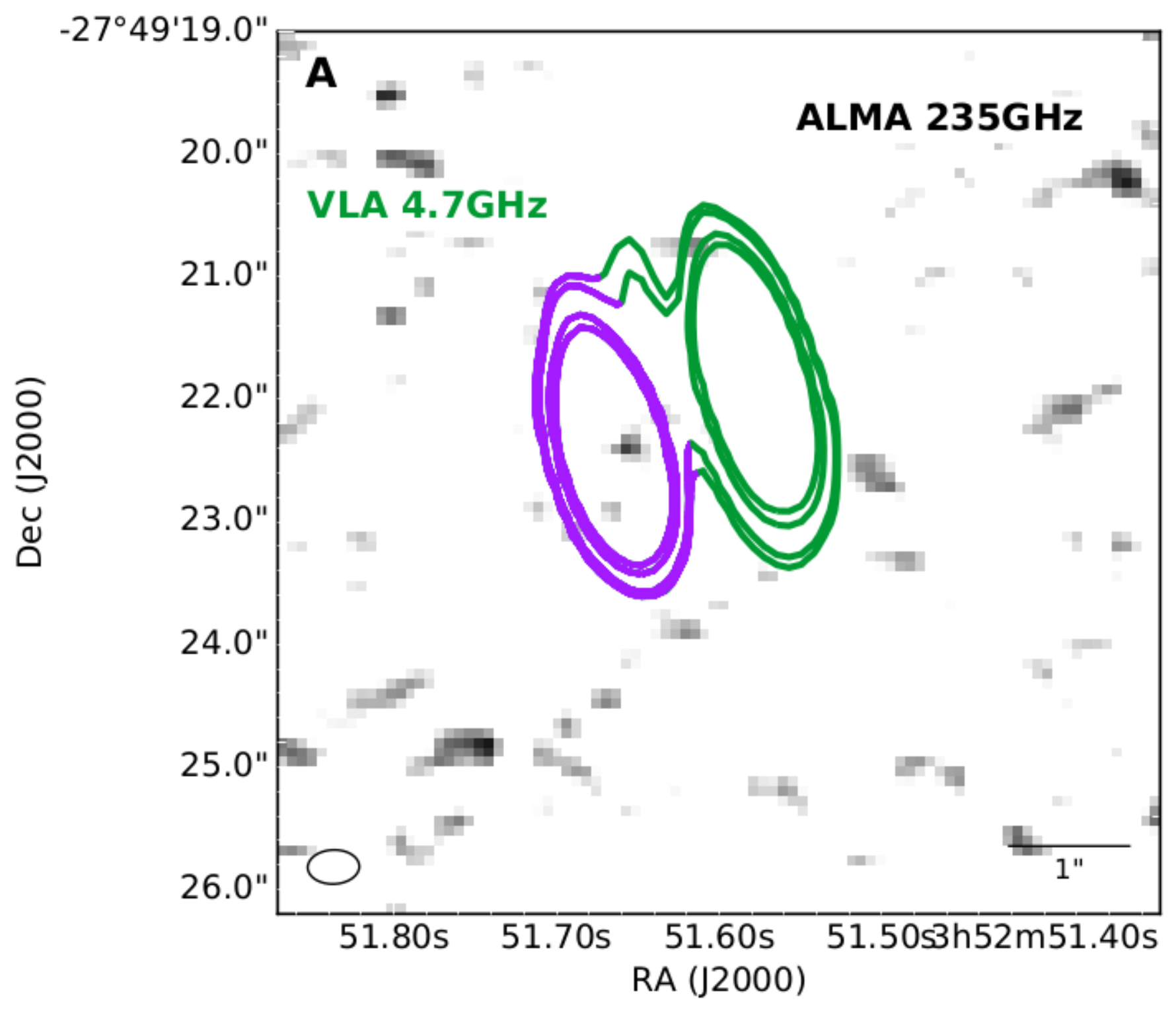}
\\
     	\includegraphics[scale=0.35]{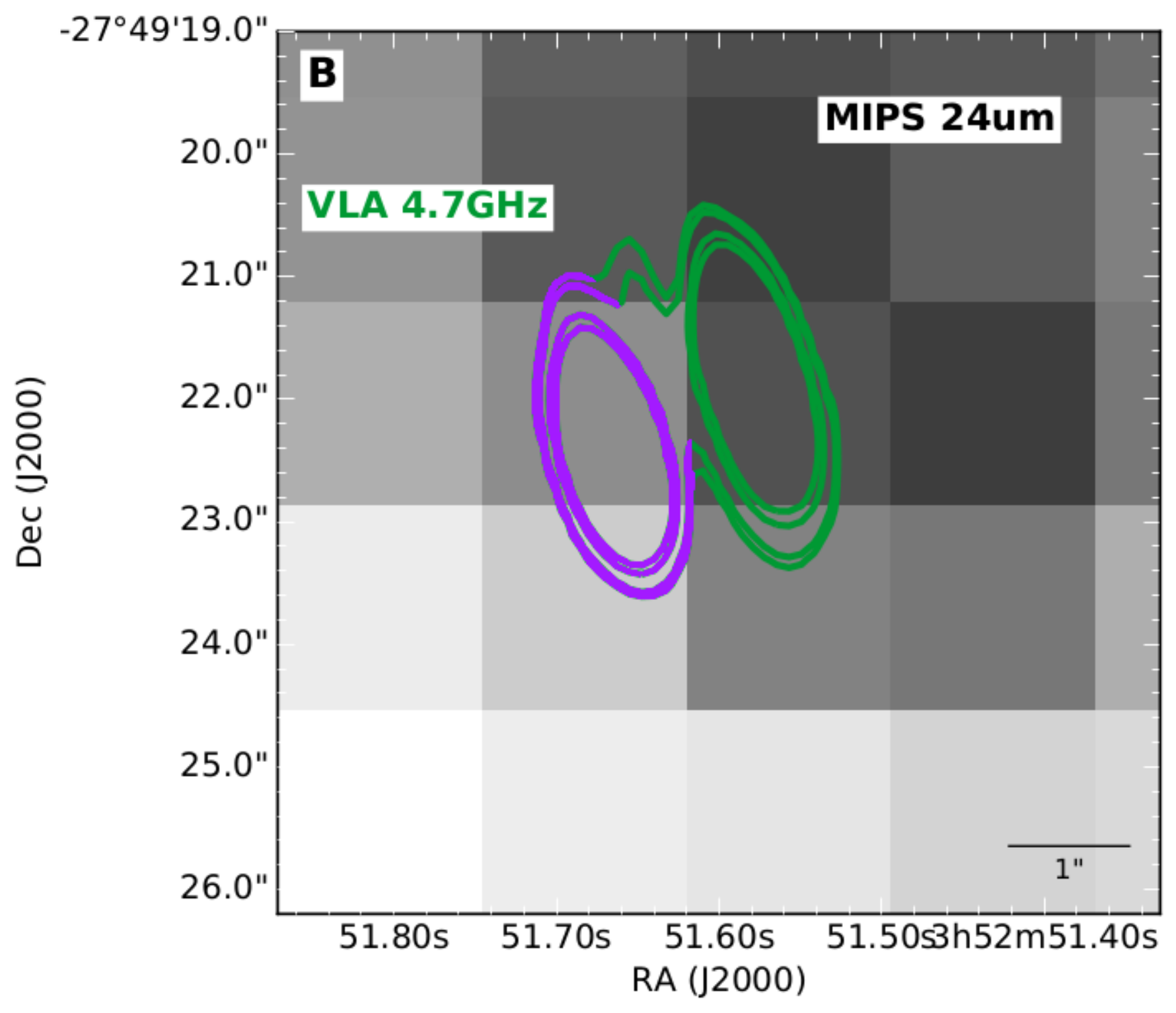}
	\caption{\textit{Panel A:} continuum map of ALMA band 6 with overlaid VLA C contours (levels are as Fig. \ref{map_0037},$\sigma=58\,\mu$Jy). The purple and green contours show the two components of the VLA data and correspond to the markers of the same colors as in the SED fit. \textit{Panel B: } MIPS 24\,\mum\ continuum map.}
    	\label{map_0350}
\end{figure}

MRC\,0350-279 has no continuum detection with ALMA (Fig. \ref{map_0350}). SED fitting with \mrmoose\ is done with four components two synchrotron power-law (northern and southern radio component), one modified BB and one AGN component. The northern radio component (detected in VLA bands C and X) is assigned to an individual synchrotron power-law and the same setup is also applied for the southern radio component. The VLA band L and ATCA 7\,mm data do not resolve the individual components and are only considered for fitting
the total radio flux (the combination of the northern and southern
synchrotron power-law components) The ALMA band 6 upper limit is assigned to the total radio flux and a modified BB. The LABOCA, SIPRE, PACS, MIPS and IRS data are fitted to the combination of the modified BB and a AGN component. The best fit model gives a solution where the ALMA detection is dominated by synchrotron, but the combined models is slightly over predict the 3$\sigma$ upper limit at 235\,GHz (Fig. \ref{fig_0350}). 

\begin{table}
\begin{threeparttable}
\caption{Data for MRC\,0350-279 (z=1.90) }
\label{table_0350}
\centering
\begin{tabular}{lcc}
\toprule
Photometric band                   & Flux{[}mJy{]}   & Ref. \\
\midrule
\irs		& 0.208    $\pm$    0.04			& A \\	
\mips1	& 0.306    $\pm$    0.044			& A \\
\pacsb	& <11.3      						& B \\
\pacsr	& <25.5      						& B \\      
\spires	& <14.2            						& B \\
\spirem	& <14.0            						& B \\
\spirel	& <15.9            						& B \\
\laboca	& <23.1            						& B \\
ALMA 6	& <0.19							& this paper\\
\atca		& 4.9      $\pm$        1.47      		& this paper \\
VLA X$^n$	& 17.48   $\pm$    0.17   $^a$ 		& A \\
VLA X$^s$	& 15.61   $\pm$    0.15    $^a$		& A \\
VLA C$^n$	& 41.05    $\pm$    0.41     $^a$		& A \\
VLA C$^s$	& 32.67   $\pm$    0.32    $^a$		& A \\
VLA L		& 350.3    $\pm$    35         			& C \\
\bottomrule                          
\end{tabular}
     \begin{tablenotes}
      \small
      \item \textbf{Notes}  ($n$) North radio component, ($a$) south radio component, ($a$) flux estimated using AIPS from original radio map, convolved to the resolution of the VLA C band.
      \item \textbf{References.} (A) \cite{DeBreuck2010}, (B) \cite{Drouart2014}, (C) \cite{Condon1998}.
    \end{tablenotes}
\end{threeparttable}
\end{table}


\clearpage
\newpage
\subsection{MRC\,0406-244}
\begin{figure}
	\includegraphics[scale=0.52]{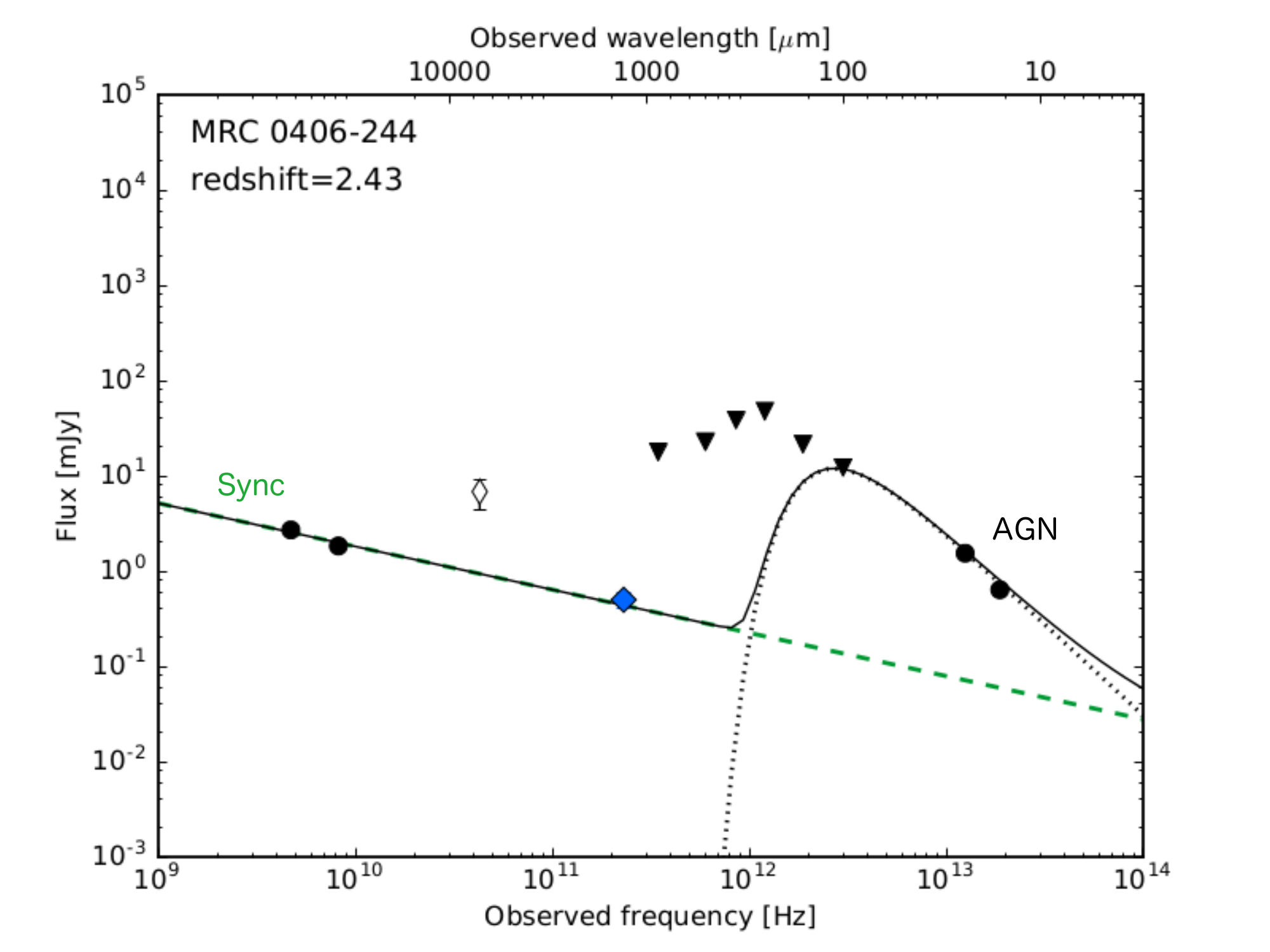}
	\caption{SED of \textbf{MRC\,0406-244}. Black solid line shows best fit total model, green dashed line is the synchrotron core and the black dotted line indicates the AGN component. The colored data points are sub-arcsec resolution data and black ones indicate data of low resolution. Green pentagons are the radio core and the blue diamond indicates the ALMA band 6 detection. Filled black circles indicate detections (>$3\sigma$) and downward pointing triangles the $3\sigma$ upper limits (Table \ref{table_0406}). The open diamond indicate available ATCA data but only plotted as a reference and was not used in the SED fit.}
	\label{fig_0406}
\end{figure}

\begin{figure}
	\centering
      	\includegraphics[scale=0.35]{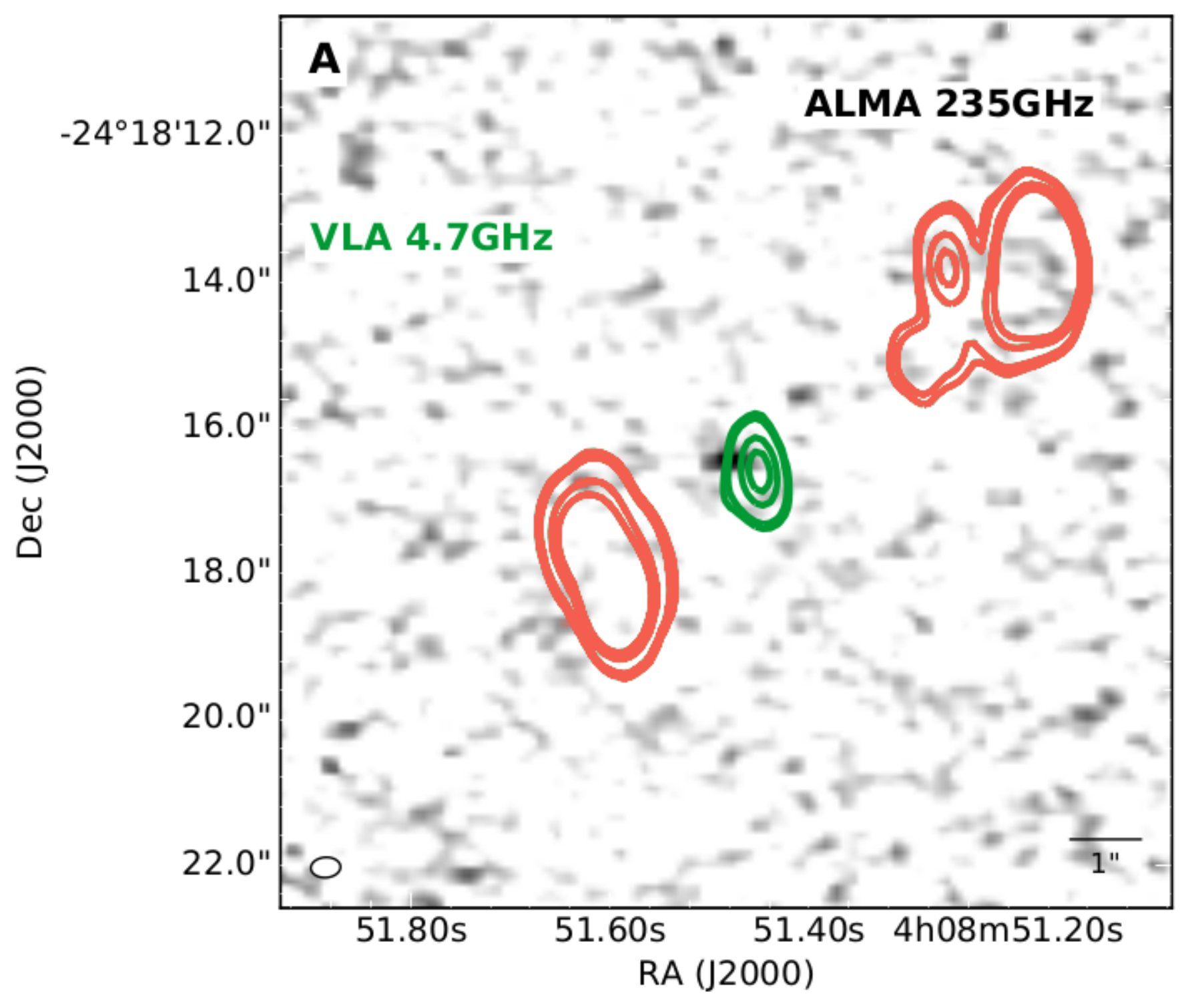}
\\
     	\includegraphics[scale=0.35]{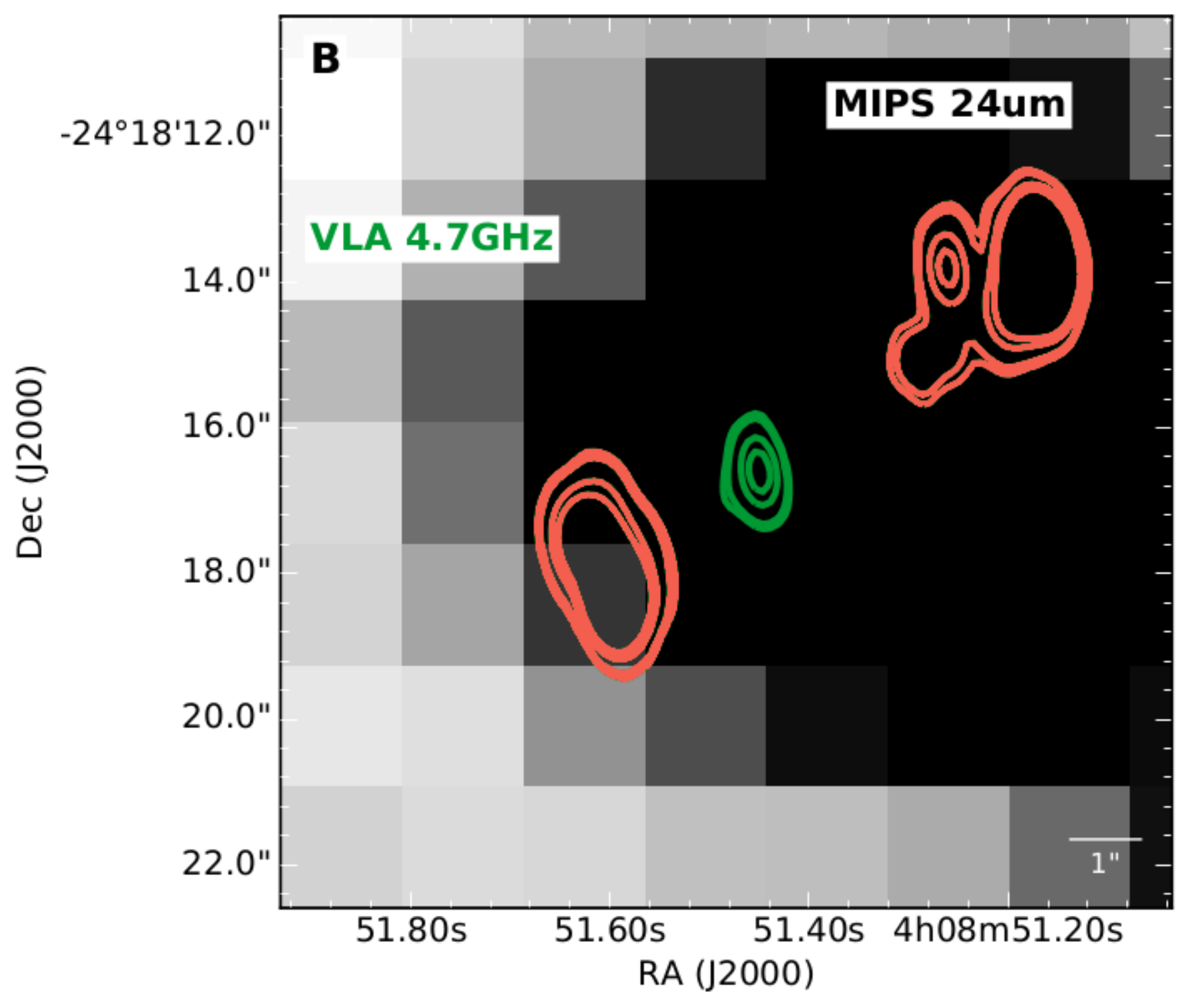}
 	\caption{\textit{Panel A:} continuum map of ALMA band 6 with overlaid VLA C contours (levels are as Fig. \ref{map_0037}, $\sigma=92\,\mu$Jy). The green contours show the portion of the VLA data that is used in the SED fit, the red contours are excluded in the fit. \textit{Panel B: } MIPS 24\,\mum\ continuum map.}
    	\label{map_0406}
\end{figure}

MRC\,0406-244 has one single continuum detection which coincides with the radio core (Fig. \ref{map_0406}). SED fitting with \mrmoose\ is done with three components, one synchrotron (of the radio core, the two lobes are excluded in the fit), one modified BB and one AGN component. The VLA data is fitted to the synchrotron power-law, the ALMA detection is assigned to both the synchrotron component and a modified BB. The LABOCA, SPIRE, PACS, MIPS and IRS is fitted to combination of the modified BB and a AGN component. The best SED fits favors a solution where the ALMA detection is dominated by synchrotron emission, due to the many upper limits in FIR and that there is no intermediate data point which could conclude whether the synchrotron is steepening at higher frequencies. The modified BB is completely unconstrained (Fig. \ref{fig_0406}). 

\begin{table}
\begin{threeparttable}
\caption{Data for MRC\,0406-244 (z=2.43) }
\label{table_0406}
\centering
\begin{tabular}{lcc}
\toprule
Photometric band                   & Flux{[}mJy{]}   & Ref. \\
\midrule
\irs		& 0.637    $\pm$     0.086			& A \\
\mips1	& 1.540    $\pm$     0.04  			& A \\
\pacsg	& <12.3                  				& B \\
\pacsr	& 21.5     $\pm$     7.9			& B \\   
\spires	& 47.6    $\pm$      5.6  			& B \\
\spirem	& 38.7    $\pm$      5.3   			& B \\
\spirel	& 22.8    $\pm$      5.9   			& B \\
\laboca	& <17.8          					& B \\
ALMA 6 	&0.5     $\pm$      0.1				& this paper\\
VLA X$^c$	& 1.84    $\pm$     0.18$^a$		& A \\
VLA C$^c$	& 2.70    $\pm$     0.21$^a$		& A \\
\midrule
ATCA (7mm)$^*$	&6.82$\pm$2.43	&this paper\\
\bottomrule                          
\end{tabular}
     \begin{tablenotes}
      \small
      \item \textbf{Notes}($c$) Radio core, ($a$) Flux estimated using AIPS from original radio map, convolved to the resolution of the VLA C band (*) data not used in fitting.
      \item \textbf{References.} (A) \cite{DeBreuck2010}, (B) \cite{Drouart2014}.
    \end{tablenotes}
\end{threeparttable}
\end{table}


\clearpage
\newpage
\subsection{PKS\,0529-549}
\begin{figure}
\includegraphics[scale=0.52]{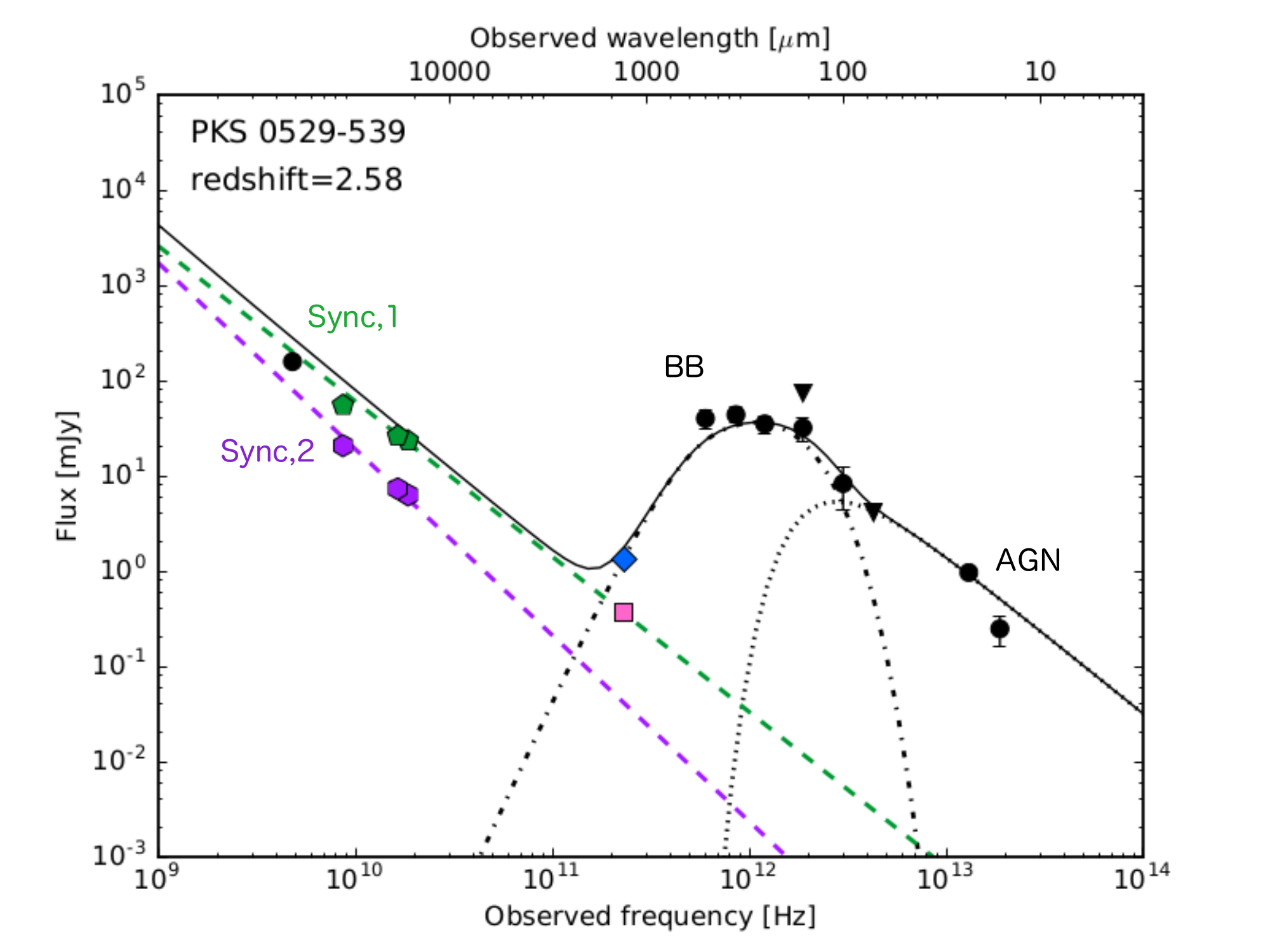}
\caption{SED of \textbf{PKS\,0529-549}. The black solid line shows best fit summed model, green and purple dashed lines are eastern and western synchrotron lobes, respectively. Blue dashed-dotted line shows the scaled BB assigned to the west ALMA detection. The colored data points indicate data with sub-arcsec resolution and black ones indicate data with lower resolution. Green pentagons indicate the location of the eastern synchrotron emission, purple hexagons represent the western radio component, the blue diamond and pink square represents the west and east ALMA detection, respectively. Filled black circles indicate detections (>$3\sigma$) and downward pointing triangles the $3\sigma$ upper limits (Table~\ref{table_0529}).}
	\label{fig_0529}
\end{figure}

\begin{figure}
\centering
\includegraphics[scale=0.35]{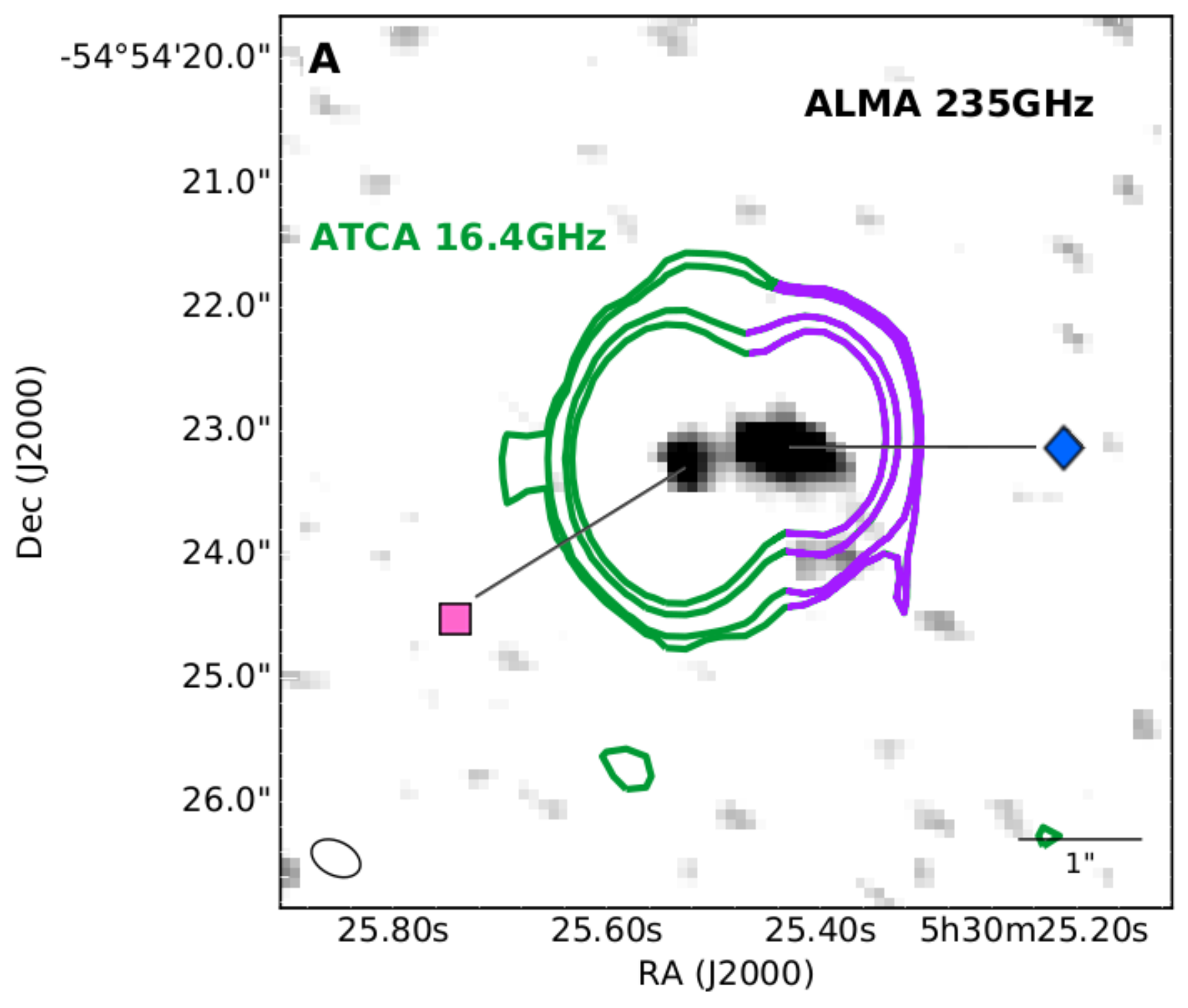} \\
\includegraphics[scale=0.35]{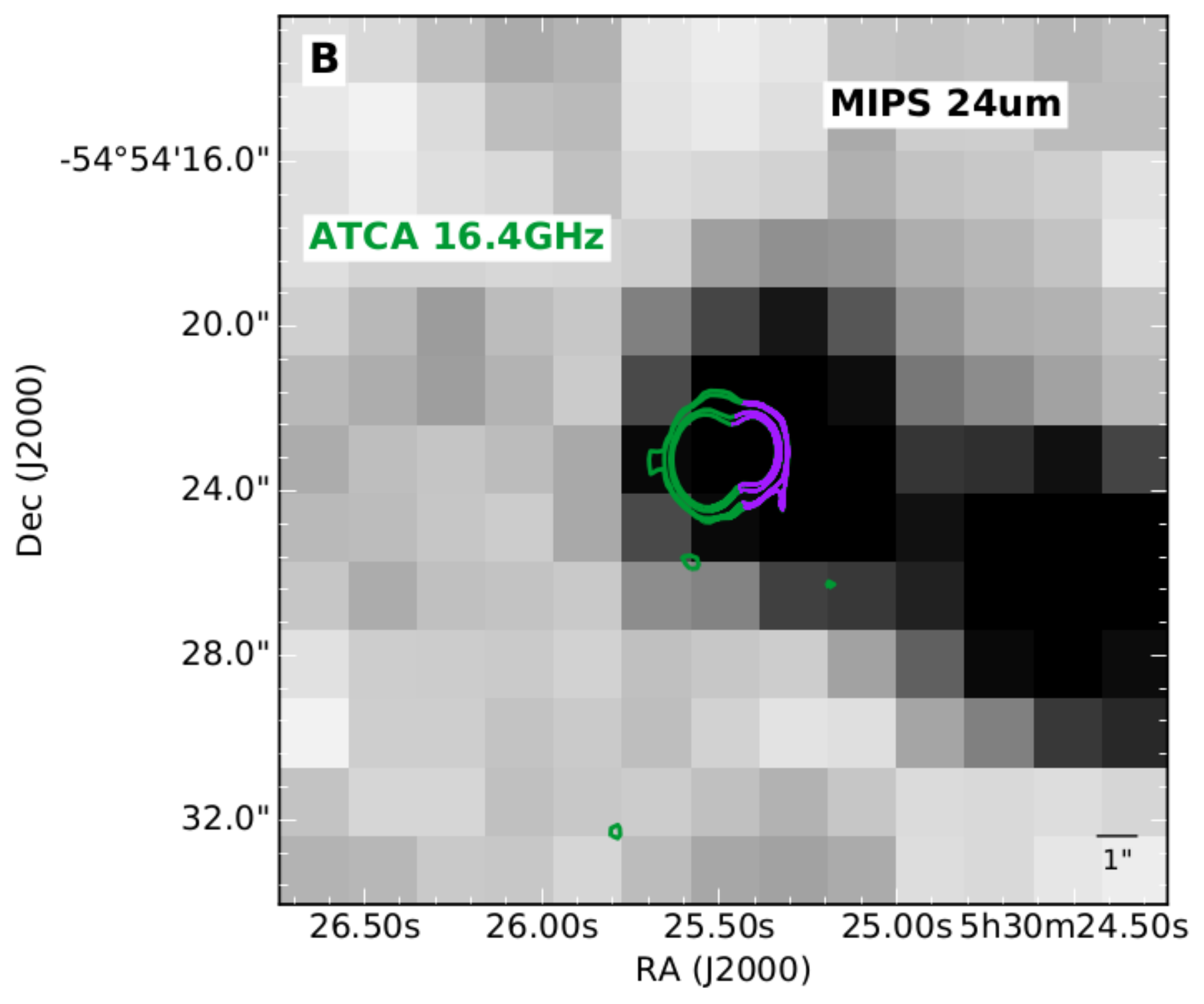}
\caption{\textit{Panel A:} continuum map of ALMA band 6 with overlaid ATCA 16.4\,GHz contours (levels are as Fig. \ref{map_0037}, $\sigma=55\,\mu$Jy). The blue and pink markers indicate the two ALMA detections and correspond to the same marker used in the SED (Fig. \ref{fig_0529}). The green and purple contours show the two components of the VLA data and correspond to the markers of the same color as in the SED fit. \textit{Panel B: } MIPS 24\,\mum\ continuum map, observe that the scale of the MIPS image is 5 times larger than panel A.}
\label{map_0529}
\end{figure}

PKS\,0529-549 has two continuum detections, both coinciding with the
two radio components (Fig. \ref{map_0529}). The SED fit is done with
five components, two synchrotron power-laws (eastern and western radio
components), two modified BB and one AGN component. The western radio
component (detected in ATCA 8640-18496 MHz) is assigned to the western
ALMA detection and fitted with an individual synchrotron power-law
and the same setup is also applied to the eastern radio and ALMA
components. The ATCA 7\,mm data do not resolve the individual components and
are only considered for fitting the total radio flux (the combination of
the western and eastern synchrotron power-law components). The eastern
ALMA detection is assigned to the first synchrotron component and
to a modified BB component, while the western ALMA detection is similarly assigned to the second synchrotron component and another modified BB component. The
SPIRE, PACS, MIPS and IRS data are fitted to the combination of the two
modified black bodies and an AGN component. The best solution finds that
the western ALMA detection is dominated by thermal dust emission and the
eastern component is pure synchrotron emission (Fig.~\ref{fig_0529}). 

\begin{table}
\begin{threeparttable}
\caption{Data for PKS\,0529-549 (z=2.58) }
\label{table_0529}
\centering
\begin{tabular}{lcc}
\toprule
Photometric band                   & Flux{[}mJy{]}   & Ref. \\
\midrule
\irs 		& 0.248  $\pm$     0.089		& A \\
\mips1 	& 0.966  $\pm$     0.040		& A \\
\mipssju 	& <4.11            				& A \\
\mipssex	& <74.1            				& A \\
\pacsg	& 8.3    $\pm$     4.0  		& B \\
\pacsr	& 31.9  $\pm$      9.0  		& B \\  
\spires	& 35.1   $\pm$     7.3    		& B \\
\spirem	& 43.8   $\pm$     8.3    		& B \\
\spirel	& 40.0   $\pm$     8.9    		& B \\
ALMA 6$^w$ 	& 1.33  $\pm$   0.16		& this paper\\
ALMA 6$^e$	& 0.37 $\pm$    0.074	& this paper\\
ATCA (18496 MHz)$^w$	& 23.4   $\pm$     1.5   	& C\\ 
ATCA (18496 MHz)$^e$	& 6.3  $\pm$       0.63    	& C\\ 
ATCA (16448 MHz)$^w$	& 26.3    $\pm$    1.3 	& C\\ 
ATCA (16448 MHz)$^e$	& 7.3   $\pm$      0.73    	& C\\   
ATCA (8640 MHz)$^w$	& 55.2  $\pm$  5.52    	& C\\ 
ATCA (8640 MHz)$^e$	& 20.7   $\pm$ 2.07    	& C\\ 
\atca		& 158     $\pm$        15.8 				& this paper\\ 
\bottomrule                          
\end{tabular}
     \begin{tablenotes}
      \small
      \item \textbf{Notes}  ($w$) West component, ($e$) East component.
      \item \textbf{References.} (A) \cite{DeBreuck2010}, (B) \cite{Drouart2014}, (C) \cite{Broderick2007}
    \end{tablenotes}
\end{threeparttable}
\end{table}


\clearpage
\newpage
\subsection{TN\,J0924-2201}
TN\,J0924-2201 has one single continuum detection which coincides with the
eastern radio component (Fig.~\ref{map_0924}). SED fitting with \mrmoose\
is done with four components, two synchrotron power laws (eastern and
western radio component), one modified BB and one AGN component.

\begin{figure}
	\includegraphics[scale=0.52]{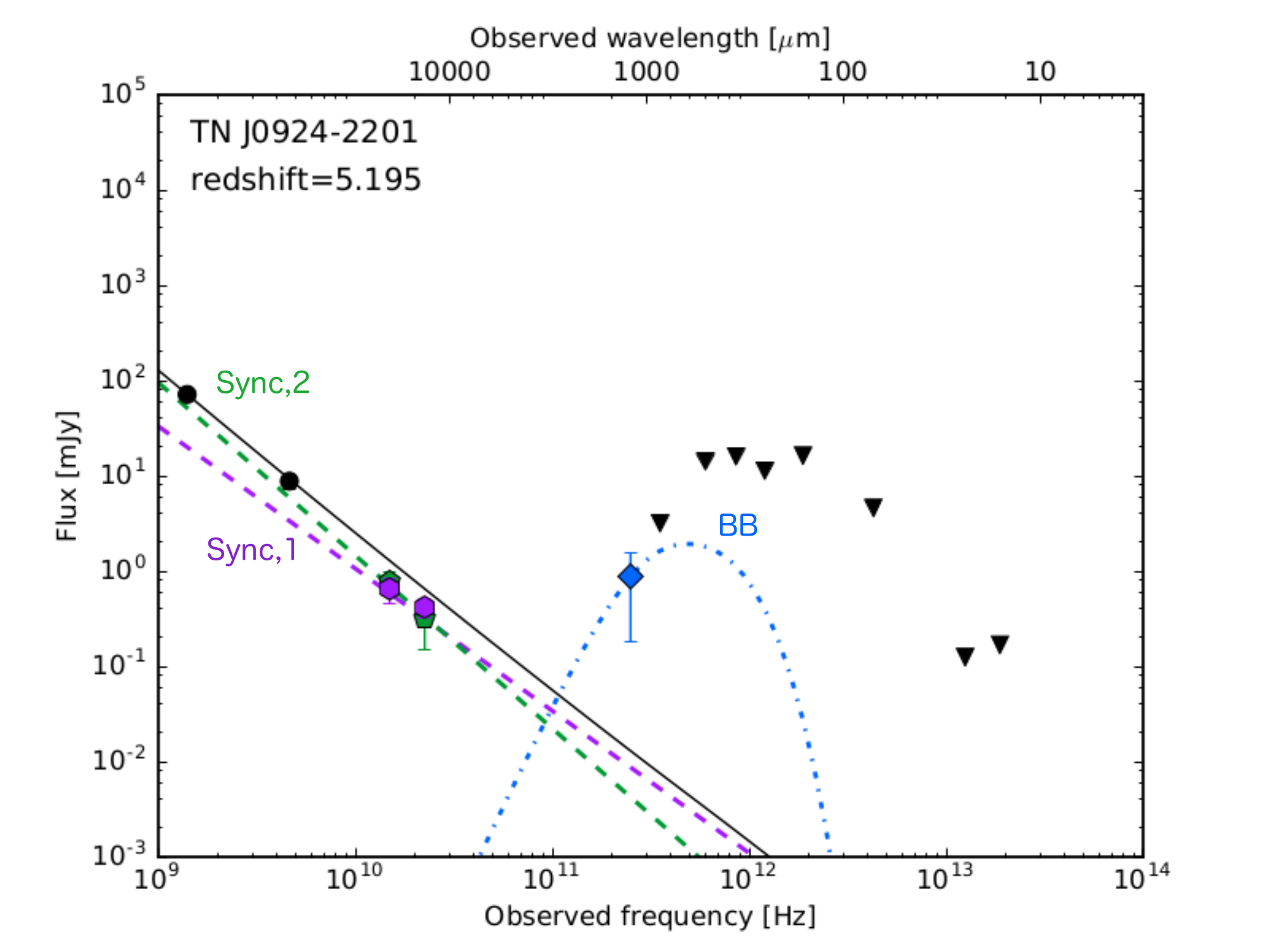}
\caption{SED of \textbf{TN\,J0924-2201}. Black solid line shows best
fit total model, black dashed line is the total synchrotron, the black
dash-dotted lines is the blackbody and the black dotted line indicates
the AGN component. The colored data points are sub-arcsec resolution
data and black ones indicate data of low resolution. The blue diamond
indicates the ALMA band 3 detection. Filled black circles indicate
detections (>$3\sigma$) and downward pointing triangles the $3\sigma$
upper limits (Table~\ref{table_0924}).} \label{fig_0924}
\end{figure}

\begin{figure}
	\centering
      	\includegraphics[scale=0.35]{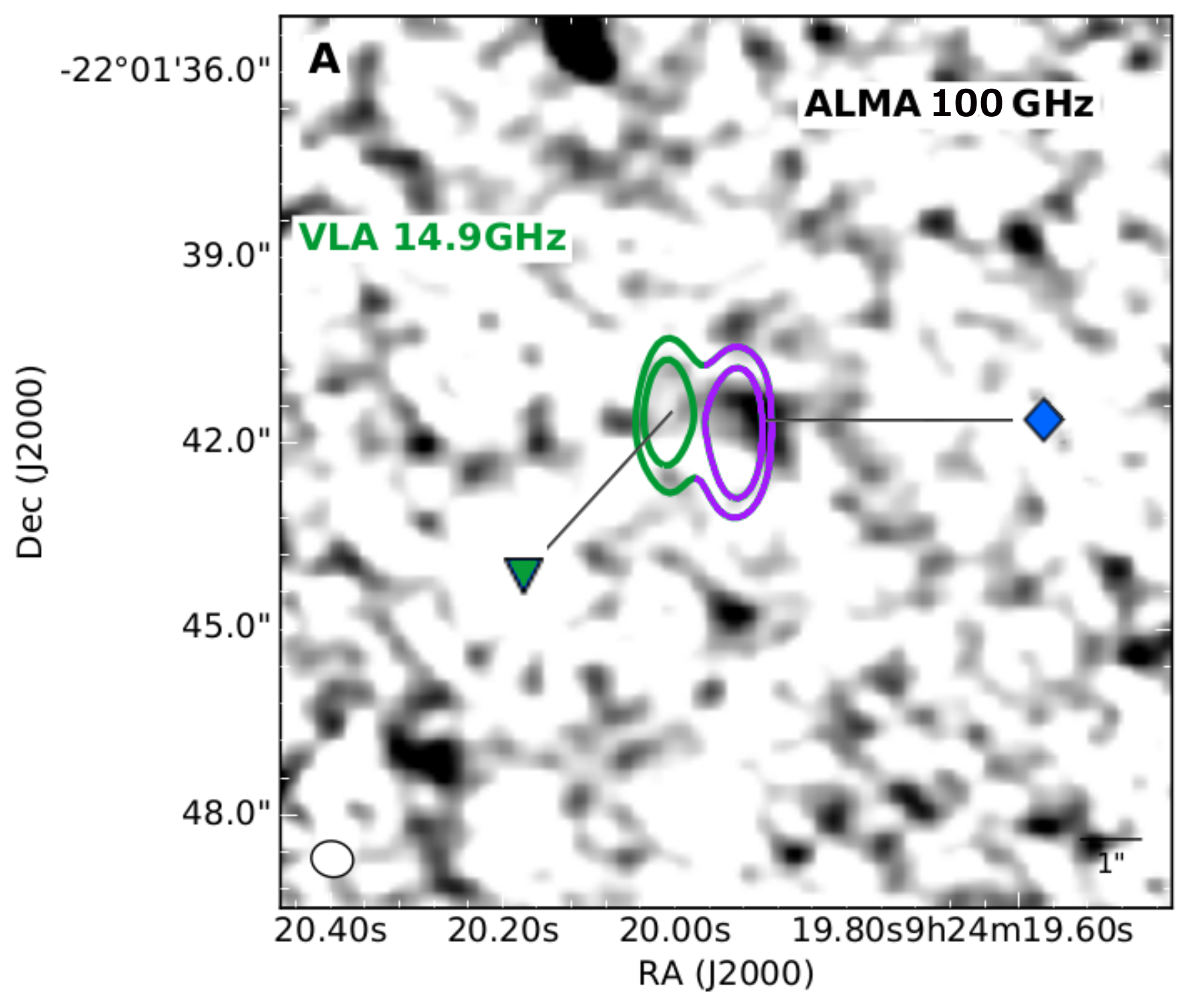}
\\
     	\includegraphics[scale=0.35]{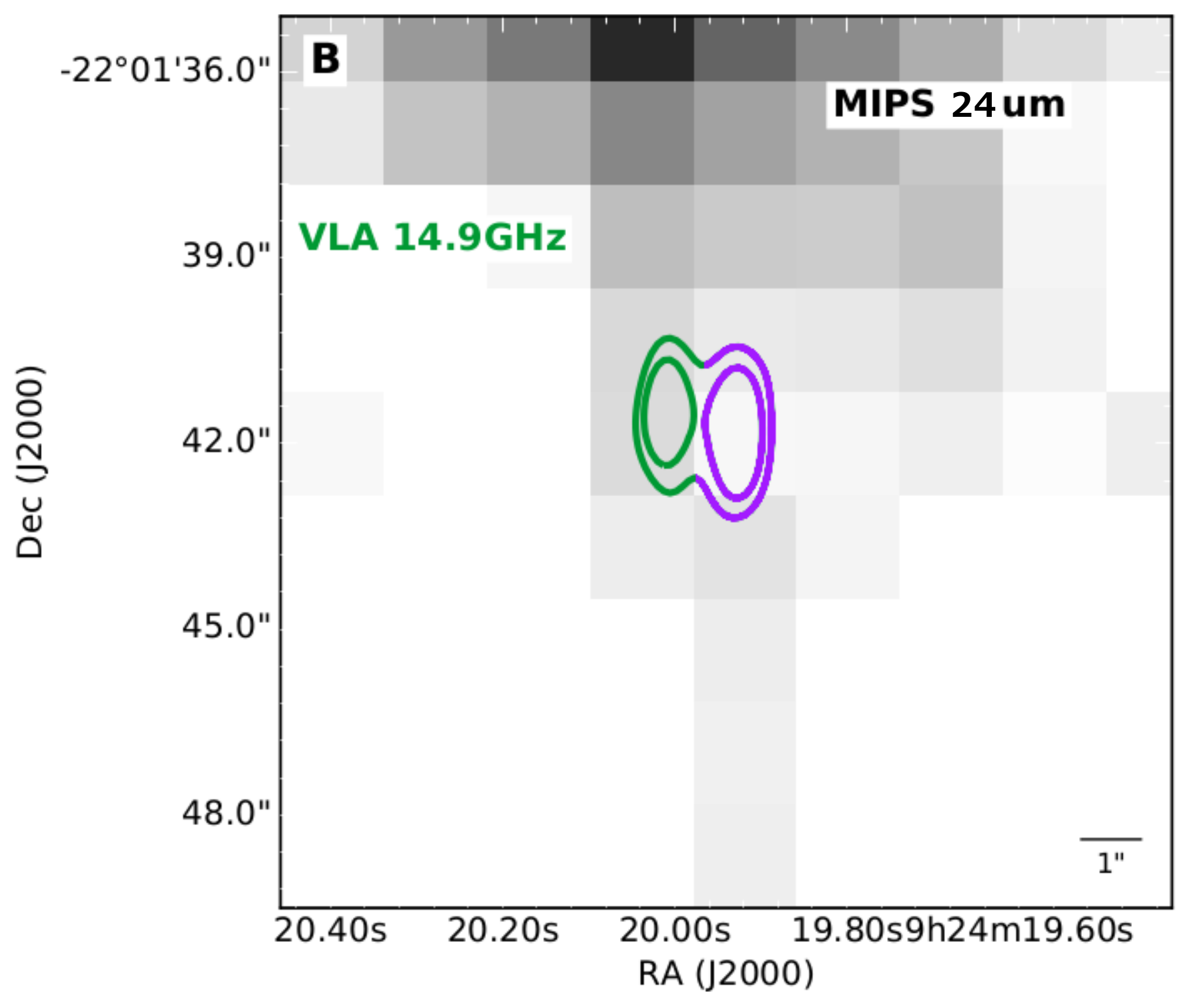}
 	\caption{ \textit{Panel A:} continuum map of ALMA band 3 with overlaid VLA U contours (levels are as Fig. \ref{map_0037}, $\sigma=85\,\mu$Jy). The blue marker indicates the ALMA detection and correspond to the same marker in the SED (Fig. \ref{fig_0924}). The green and purple contours show the two components of the VLA data and corresponds to the markers if the same color as in the SED. \textit{Panel B:} MIPS 24\,\mum\ continuum map.}
    	\label{map_0924}
\end{figure}

The western radio component (detected in VLA bands U and K) is assigned
to the ALMA detection and fitted with an individual synchrotron
power-law. The eastern radio component (detected in VLA bands U and
K) is only fitted with a synchrotron power-law Sync, 2). The VLA bands C and
L do not resolve the individual components and are only considered
for fitting the total radio flux (the combination of the western and
eastern synchrotron power-law components). The western ALMA detection
is assigned to the western synchrotron component (Sync, 1), and to a modified
BB component. The SCUBA, SPIRE, PACS, MIPS and IRS data is fitted to
the combination of the modified BB and a AGN component. The SED fit
only constrain the synchrotron models and the modified BB and AGN are
unconstrained. This is due to the fact that there is only one detection
in the FIR and it is not possible to constrain the modified BB with two
free parameters to only one data point. The solution for the two radio
lobes shows that the ALMA detection is likely completely dominated by
thermal dust emission. The plotted modified BB in Fig. \ref{fig_0924}
is scaled to the ALMA detection with slope and temperature fixed to
$\beta=2.5$ and T$=50$, like in the case where the upper-limit of the
\lirsf\, is calculated for the sources with ALMA non-detections. 

\begin{table}
\begin{threeparttable}
\caption{Data for TN\,J0924-2201 (z=5.195) }
\label{table_0924}
\centering
\begin{tabular}{lcc}
\toprule
Photometric band                   & Flux{[}mJy{]}   & Ref. \\
\midrule
\irs     		&    <0.168		&   A\\
\mips1     		&    <0.125		&   A\\
\pacsb    		&    <4.6         		&   B\\
\pacsr    		&    <16.3      		&   B\\
\spires   		&    <11.4   		&   B\\
\spirem   		&     <16.1        		&   B\\
\spirel   		&     <14.3 		&   B\\
\scuba  		&     <3.2			&   C\\
ALMA 6$^e$      &    	<0.25				&   this paper\\
ALMA 6$^w$     &    	0.88$\pm$0.7		&   this paper\\
VLA U$^w$       	&  0.65$\pm$0.2$^1$ 			& this paper   \\
VLA K$^w$      	&  0.42$\pm0.1^1$  			&  this paper \\
VLA U$^e$      	&  0.78$\pm0.22^1$ 			&  this paper \\
VLA K$^e$      	&  0.32$\pm0.17^1$			&  this paper \\
VLA C	      	&  8.79$\pm1.44^1$			&  this paper \\
VLA L       		& 71.5$\pm$2.2			&   F\\                            
\bottomrule                          
\end{tabular}
     \begin{tablenotes}
      \small
      \item ($a$) Integrated flux extracted using AIPS with imfit 1 Gaussian component (e) eastern component, (w) western component.
      \item \textbf{References.} (A) \cite{DeBreuck2010}, (B) \cite{Drouart2014}, (C) \cite{Reuland2003} (F) \cite{Pentericci2000}.
    \end{tablenotes}
\end{threeparttable}
\end{table}


\clearpage
\newpage

\subsection{MRC\,0943-242}
\begin{figure}
\includegraphics[scale=0.52]{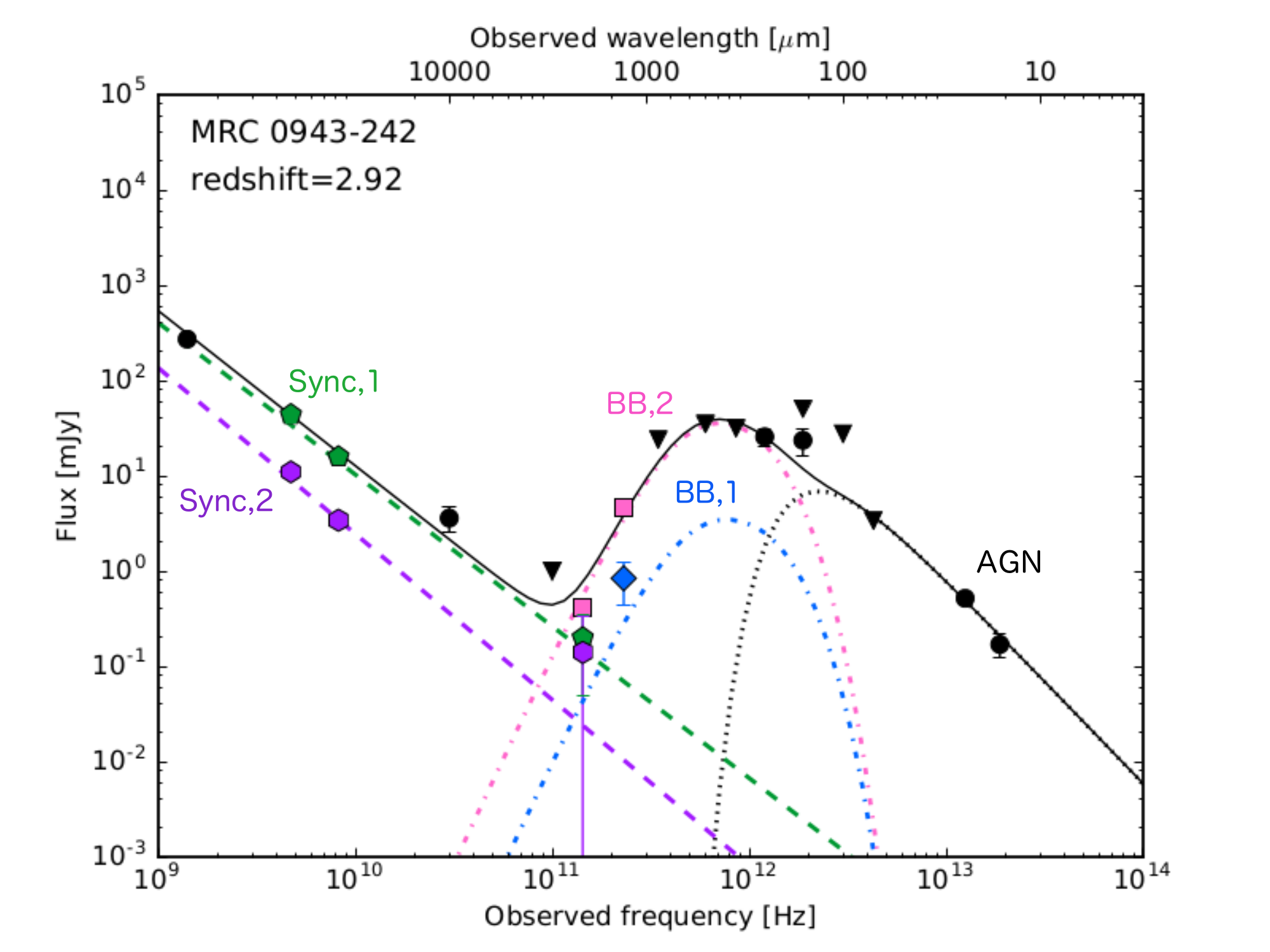}
\caption{SED of \textbf{MRC\,0943-242}. Black solid line shows best fit
total model, green and purple dashed line is north and south synchrotron
lobes, respectively. Black dashed-dotted line shows the blackbody
fitted associated to one of the ALMA detections and the black dotted
line indicates the AGN component. The colored data points are sub-arcsec
resolution data and black ones indicate data of low resolution. Green
pentagons are north synchrotron, purple hexagons are the south radio
component, the blue diamond shows the ALMA 6 host detection and magenta
squares indicates the total ALMA 4 and 6 flux of the 3 companions. Filled
black circles indicate detections (>$3\sigma$) and downward pointing
triangles the $3\sigma$ upper limits (Table~\ref{table_0943}).}
\label{fig_0943}
\end{figure}

\begin{figure}
	\centering
      	\includegraphics[scale=0.3]{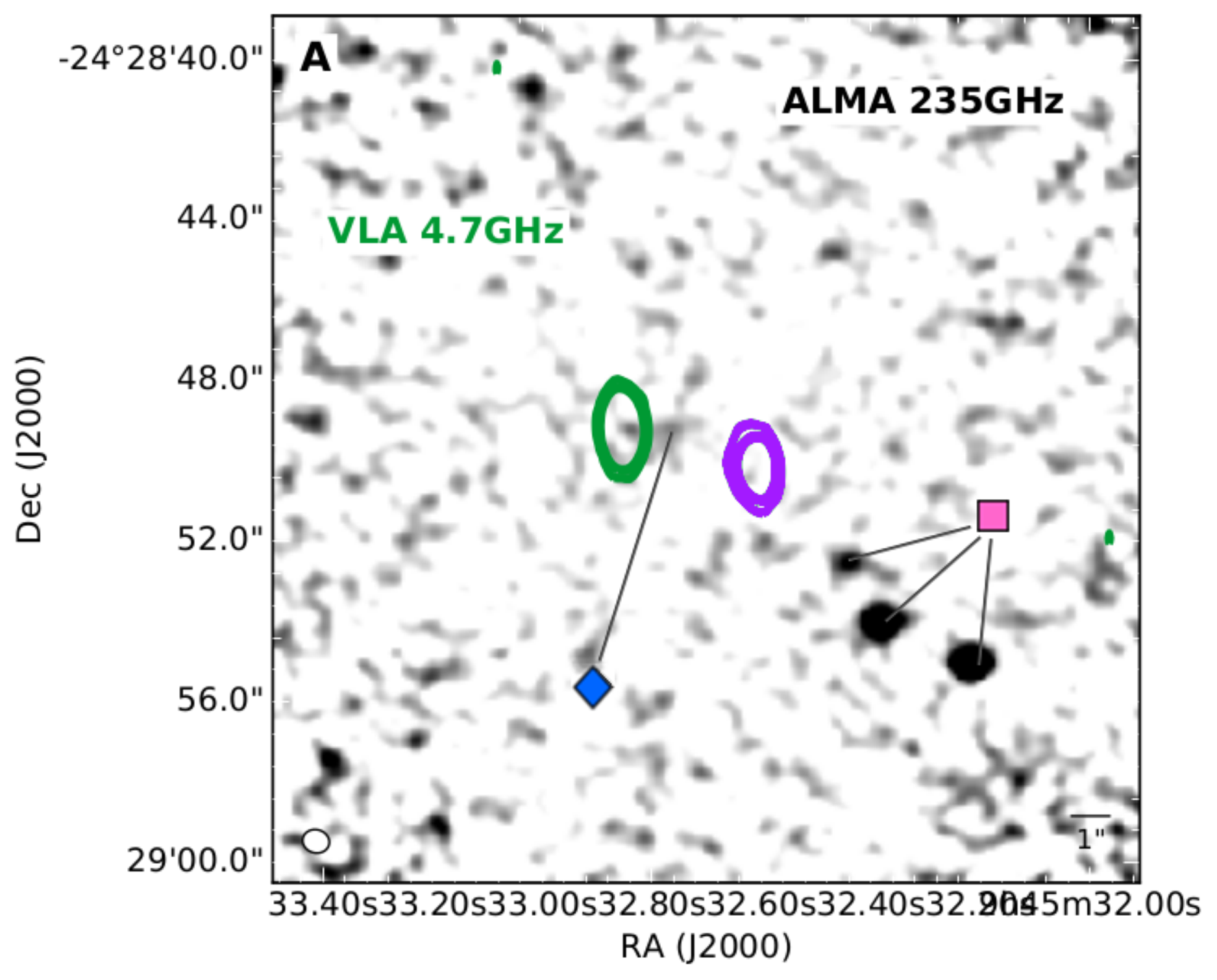}
\\
     	\includegraphics[scale=0.3]{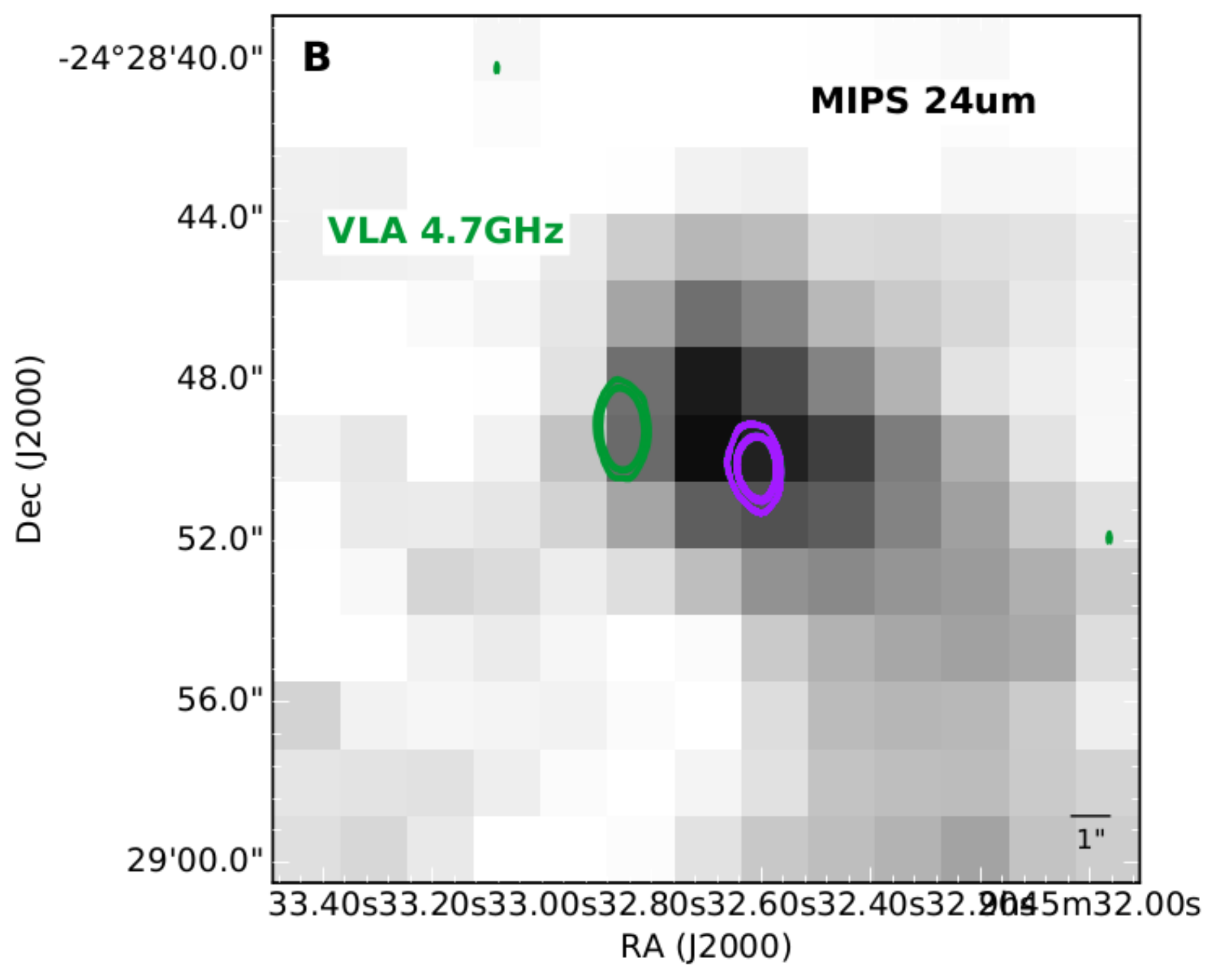}
 	\caption{ \textit{Panel A:} continuum map of ALMA band 6 (cycle 1 and 2) with overlaid VLA C contours (levels are as Fig. \ref{map_0037}, $\sigma=75\,\mu$Jy). Blue and pink markers indicate the two different components detected with ALMA detections and corresponds to the same data points in the SED fit. The green and purple contours show the two components if the VLA data and corresponds to the markers of the same colors as in the SED fit. \textit{Panel B: } MIPS 24\,\mum\ continuum map.}
    	\label{map_0943}
\end{figure}

MRC\,0943-242 has several continuum detections with ALMA band 6, one at
the host galaxy and three resolved companions (Fig. \ref{map_0943}). With
ALMA band 4 the three companions are also detected, as well as two
additional continuum components which coincide with the two radio
components seen with VLA. SED fitting with \mrmoose\ is done with five
models, two synchrotron power-laws (north and south component), two
modified BB (one for the host and one for the three companions grouped
together) and one AGN component. The VLA X, C and ALMA band 4 resolve
the synchrotron components and the two radio lobes are fitted with two
individual power-laws. The VLA L, ATCA 3\,mm and 7\,mm do not resolve
any individual components and are assigned to the combination of both
synchrotron models. The LABOCA, SPIRE, PACS, MIPS and IRS data are
assigned to both the modified BB of the host and companions as well as the
AGN component. The best fit model, shows that ALMA detection at the host
galaxy is dominated by thermal dust emission, and the three companions
are also consistent with emission from heated dust. The two synchrotron
components are consistent with a power-law from radio to ALMA band 4
(Figure \ref{fig_0943}). One reason why the PACS 100\,\mum\, is not well
fitted is because either the temperature of the modified BB must be higher
which is outside the parameter space or because the fixed $\nu_{cut}$ does
not allow the AGN component to account fully for the observed flux. 

\begin{table}
\begin{threeparttable}
\caption{Data for MRC\,0943-242 (z=2.933) }
\label{table_0943}
\centering
\begin{tabular}{lcc}
\toprule
Photometric band                   & Flux{[}mJy{]}   & Ref. \\
\midrule
\irs     			&0.170$\pm$ 0.048  		& A\\
\mips1     			&0.518 $\pm$0.040 		& A\\
MIPS (70\mum)     	&<3.390           		& A\\
MIPS (160\mum)    	&<50.900         		& A\\
\pacsg				&<27.6          			& B\\
\pacsr    			&23.6 $\pm$7.7   		& B\\
\spires   			&25.7 $\pm$5.2   		& B\\
\spirem   			&<31.7$^a$          		& B \\
\spirel   			&<35.2$^a$      			& B\\
\laboca  			&<24        				& B\\
ALMA 6$^h$  		&0.84 $\pm$ 0.4   		&this paper\\
ALMA 6$^c$  		&40.72$\pm$ 0.47 		&this paper\\
ALMA 4$^c$  		&0.41 $\pm$ 0.04 		&this paper\\
ALMA 4$^n$  		&0.21 $\pm$0.15   		&this paper\\
ALMA 4$^s$  		&0.14 $\pm$0.07  		&this paper\\
ATCA (3mm)    	&<1    					& this paper \\
ATCA (7mm) 		&3.6 $\pm$1.08  			& C \\
VLA X $^n$    	&15.86 $\pm$1.5$^b$   	& D\\
VLA X $^s$   	&3.42 $\pm$0.3 $^b$  	& D\\
VLA C $^n$    	&44 $\pm$4.3   			& D\\
VLA C $^s$  		&11 $\pm$1.1   			& D\\
VLA L       		&272.1 $\pm$27.2  		& E\\
\bottomrule                          
\end{tabular}
     \begin{tablenotes}
      \small
      \item \textbf{Notes}  ($h$) AGN host,  ($c$) companions, ($n$) north synchrotron lobe, ($n$) south synchrotron lobe, ($a$) changed to upper limits because of contamination sources seen in MIPS image, ($b$) flux obtained from \cite{Carilli1997} using the listed I$_{4.7}$ and $\alpha^{8.2}_{4.7}$ values of the north and south hot spot.
      \item \textbf{References.} (A) \cite{DeBreuck2010}, (B) \cite{Drouart2014}, (C) \cite{Emonts2011} (D) \cite{Carilli1997}, (E) \cite{Condon1998}.
    \end{tablenotes}
\end{threeparttable}
\end{table}


\clearpage
\newpage
\subsection{MRC\,1017-220}
\begin{figure}
\includegraphics[scale=0.52]{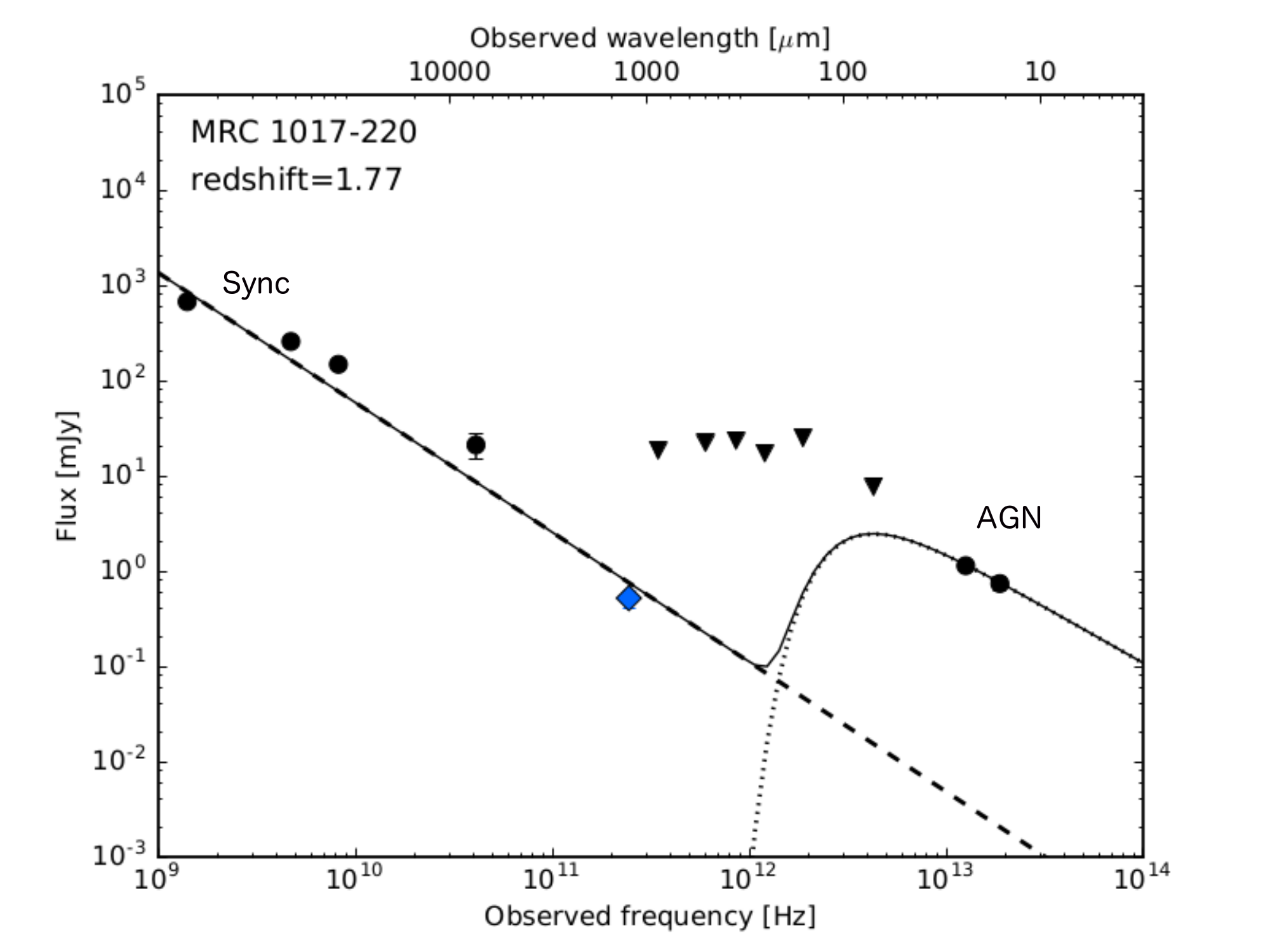}
\caption{SED of \textbf{MRC\,1017-220}. Black solid line shows best
fit total model, black dashed line is the total synchrotron and the
black dotted line indicates the AGN component. The colored data point
is sub-arcsec resolution data and black ones indicate data of low
resolution. The blue diamond indicates the ALMA band 6 detection. Filled
black circles indicate detections (>$3\sigma$) and downward pointing triangles
	the $3\sigma$ upper limits (Table~\ref{table_1017}).
}

\label{fig_1017}
\end{figure}

\begin{figure}
	\centering
      	\includegraphics[scale=0.35]{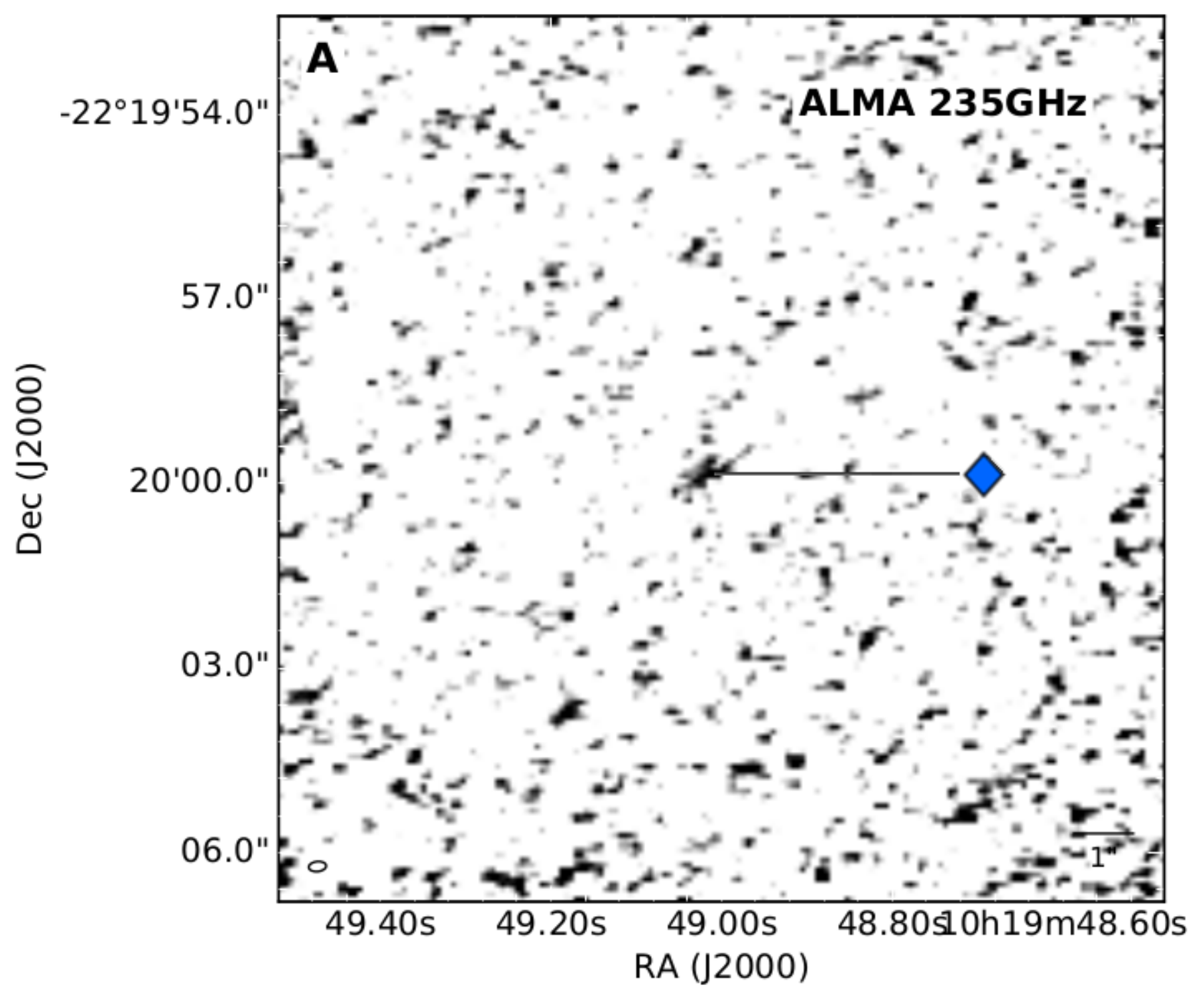}
\\
     	\includegraphics[scale=0.35]{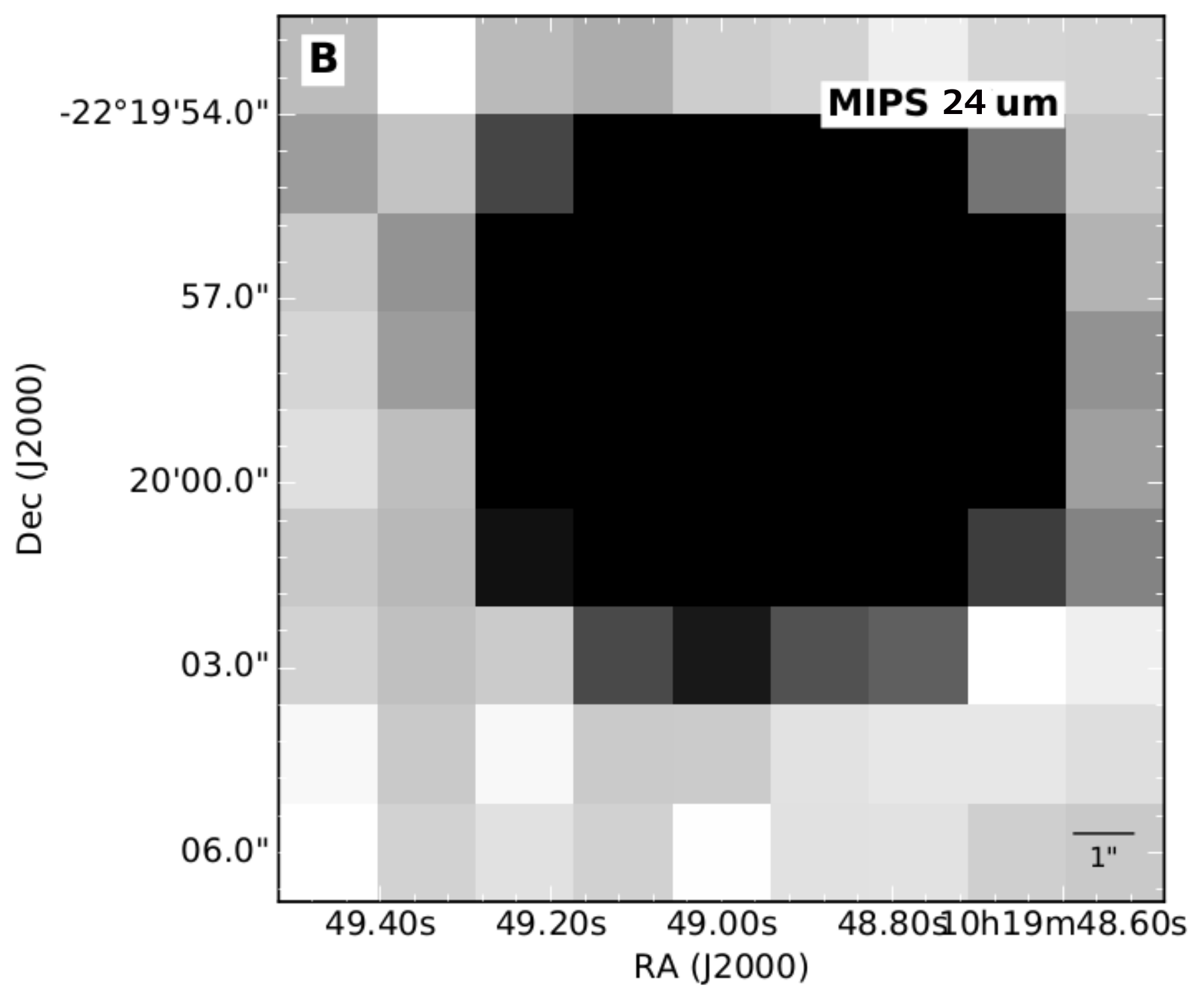}
 	\caption{\textit{Panel A:} continuum map of ALMA band 6, VLA maps not accessible, but the source is unresolved as shown in \cite{Pentericci2000}. The blue diamond indicate the ALMA component and correspond to the same data point as Fig. \ref{fig_1017}. \textit{Panel B: } MIPS 24\,\mum\ continuum map}
    	\label{map_1017}
\end{figure}

MRC\,1017-220 has one single continuum detection with ALMA,
which coincides with the unresolved synchrotron component (Fig.~\ref{map_1017}). SED
fitting with \mrmoose\ is done with three components, one synchrotron, one modified BB and one
AGN component. The VLA L, C, X and ATCA 7\,mm bands are assigned to the
synchrotron component and the ALMA detection is assigned to both the synchrotron power-law and a modified BB. The LABOCA SPIRE, PACS, MIPS and IRS points are assigned to the combination of the
modified BB and an AGN component. The best fit solution gives a dominating synchrotron
contribution to the ALMA detection (Fig. \ref{fig_1017}), due to the
many upper limit in the FIR. But studying the SED it seems likely that
the synchrotron is turning over at high frequencies but without a data
point at an intermediate frequency between the radio and ALMA data, it
is not possible to make a final conclusion whether the ALMA detection
is synchrotron dominated or not. 

\begin{table}
\begin{threeparttable}
\caption{Data for MRC\,1017-220 (z=1.77) }
\label{table_1017}
\centering
\begin{tabular}{lcc}
\toprule
Photometric band                   & Flux{[}mJy{]}   & Ref. \\
\midrule
\irs     	&   0.740$\pm$0.120   	&  	A \\  
\mips1   	&   1.140$\pm$0.030   	&   A\\ 
\pacsb    	&   <7.7        	 		&    B\\
\pacsr   	&   <25.1        		&    B\\
\spires  	&   <17.4        		&    B\\
\spirem  	&   <23.6        		&    B\\
\spirel  	&   <22.4        		&    B\\
\laboca 		&   <18.6       	 		&    B\\
ALMA 6 		&   0.52$\pm$0.11    	&    this paper\\
ATCA (7\,mm)  &	21.1$\pm$6.33    	&    this paper \\
VLA X      	&   148$\pm$14.8    		&    C\\
VLA C      	&  257$\pm$25.7    		&     C\\
VLA L      	&  673.2$\pm$67.3    	&     D\\
\bottomrule                          
\end{tabular}
     \begin{tablenotes}
      \small
      \item \textbf{References.} (A) \cite{DeBreuck2010}, (B) \cite{Drouart2014}, (C) \cite{Pentericci2000}, (D) \cite{Condon1998}.
    \end{tablenotes}
\end{threeparttable}
\end{table}


\clearpage
\newpage
\subsection{4C\,03.24}
4C\,03.24 two continuum detections in ALMA band 3, one that
coincides with the north synchrotron lobe and one that covers both the
radio core and hot spot of the south lobe (Fig.~\ref{map_4C0324}). The
SED fitting with \mrmoose\ is done with six components, three synchrotron (core, northern and southern radio component),
two modified BB (northern and southern) and one AGN component. The VLA L detection does not resolve
the individual components and are only considered for fitting
the total radio flux (the combination of the western, eastern and core
synchrotron power-law components). The C and X bands resolves the three components are there assigned to individual
synchrotron models. The northern ALMA detection is assigned to the northern
synchrotron power-law and one modified BB and the southern ALMA detection is assigned to a combination of the southern and core synchrotron components, as well as a modified BB. The SCUBA,
SPIRE, PACS, MIPS and IRS are fitted to the combination of the two modified BB and a AGN
component. The best fit model finds that the
north ALMA detection is dominated by thermal dust emission and the southern ALMA detection is dominated by synchrotron emission from the south radio lobe. 

\begin{figure}
	\includegraphics[scale=0.52]{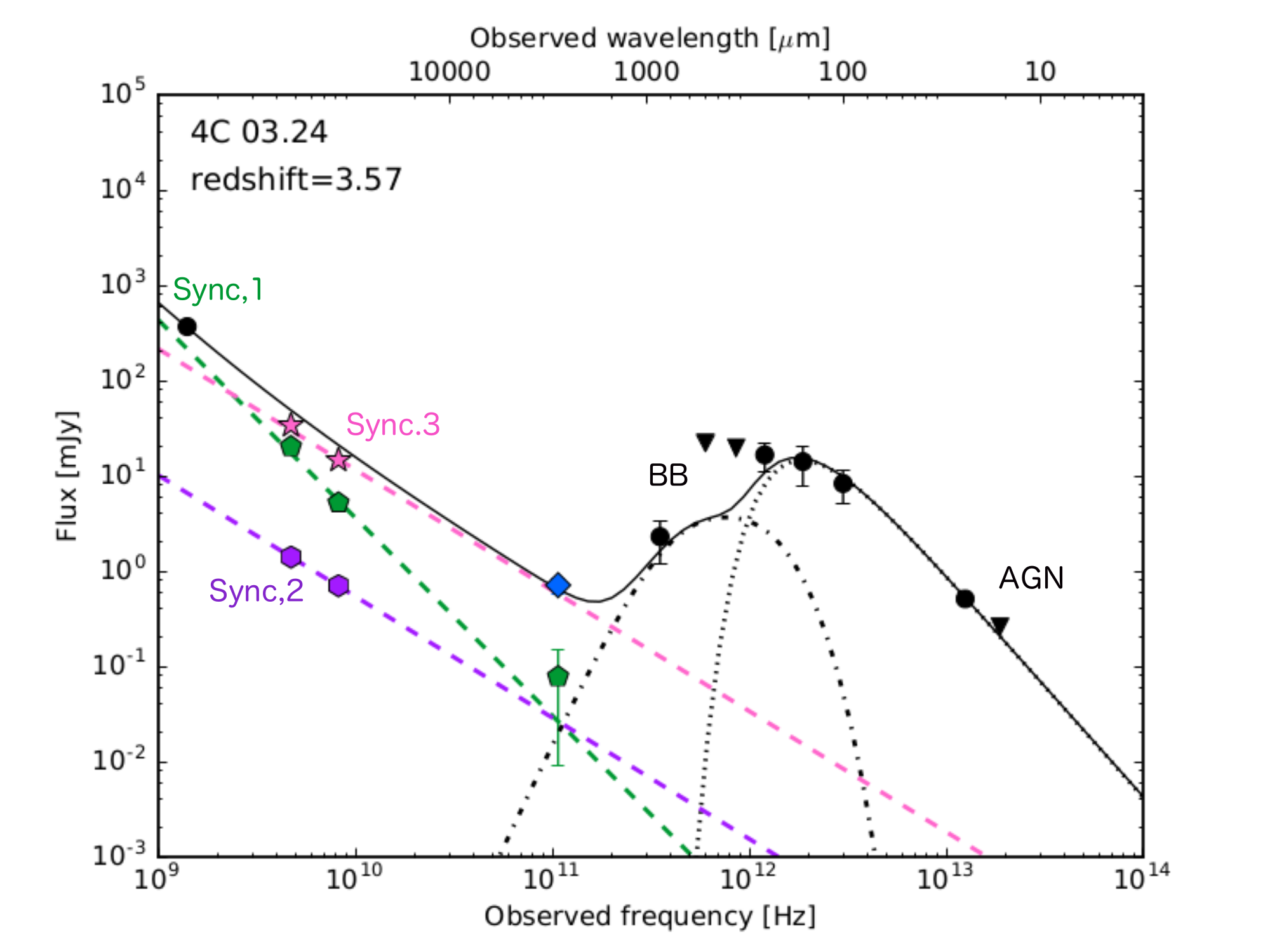}
\caption{SED of \textbf{4C\,03.24}. Black solid line shows best fit
total model, pink, green and purple dashed line is the north, core
and south synchrotron component, respectively, the black dash-dotted
lines is the blackbody and the black dotted line indicates the AGN
component. The colored data points are sub-arcsec resolution data and
black ones indicate data of low resolution. Green pentagons are the
north synchrotron lobe, purple hexagrams are the radio core and the pink
stars corresponds to the southern lobe. The blue diamond indicates the
ALMA band 3 detection spatially coincident with the host galaxy, radio
core and southern hotspot. Filled black circles indicate detections
(>$3\sigma$) and downward pointing triangles the $3\sigma$ upper limits (Table~\ref{table_4C0324}). }

	\label{fig_4C0324}
\end{figure}

\begin{figure}
	\centering
      	\includegraphics[scale=0.35]{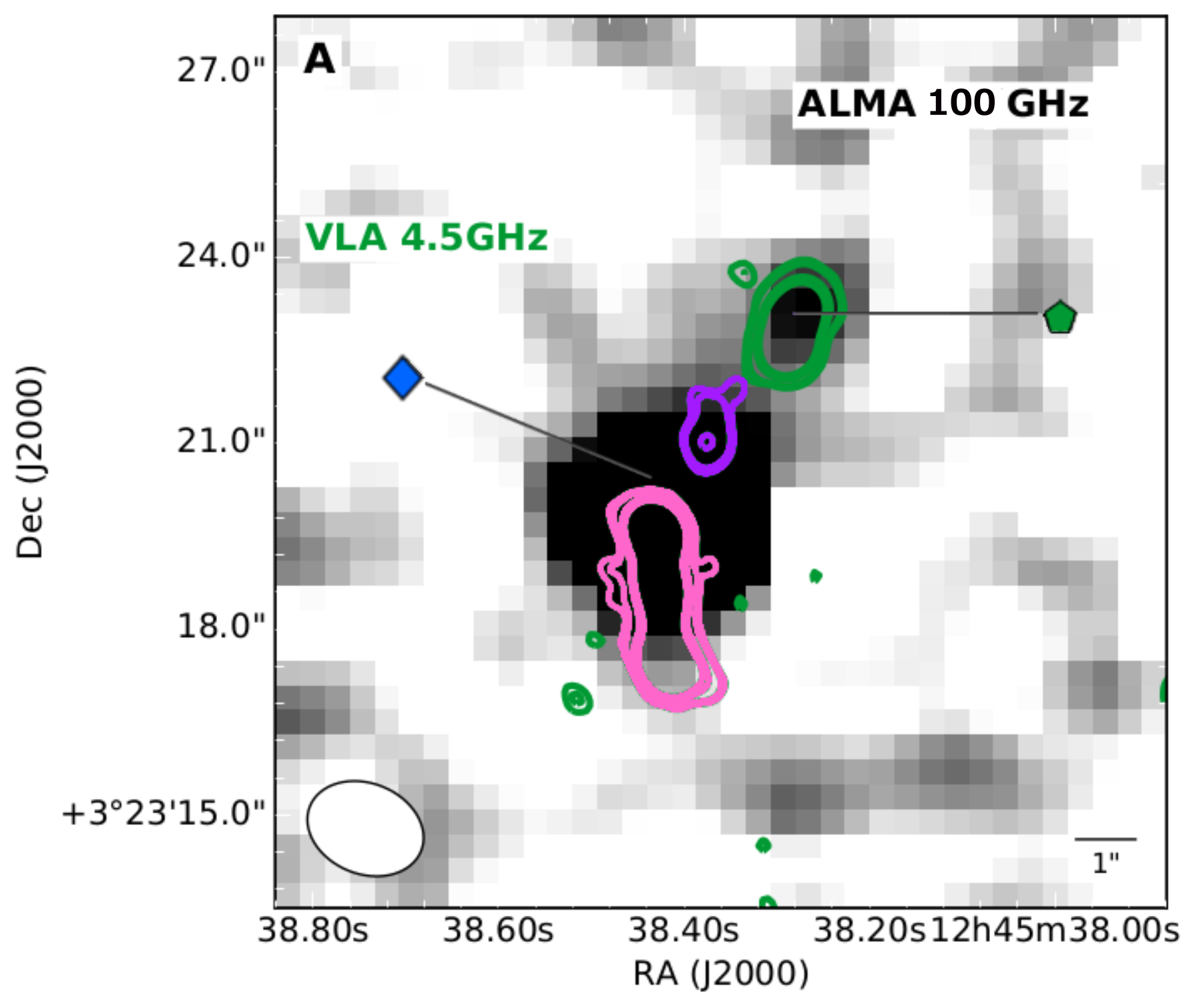}
\\
     	\includegraphics[scale=0.35]{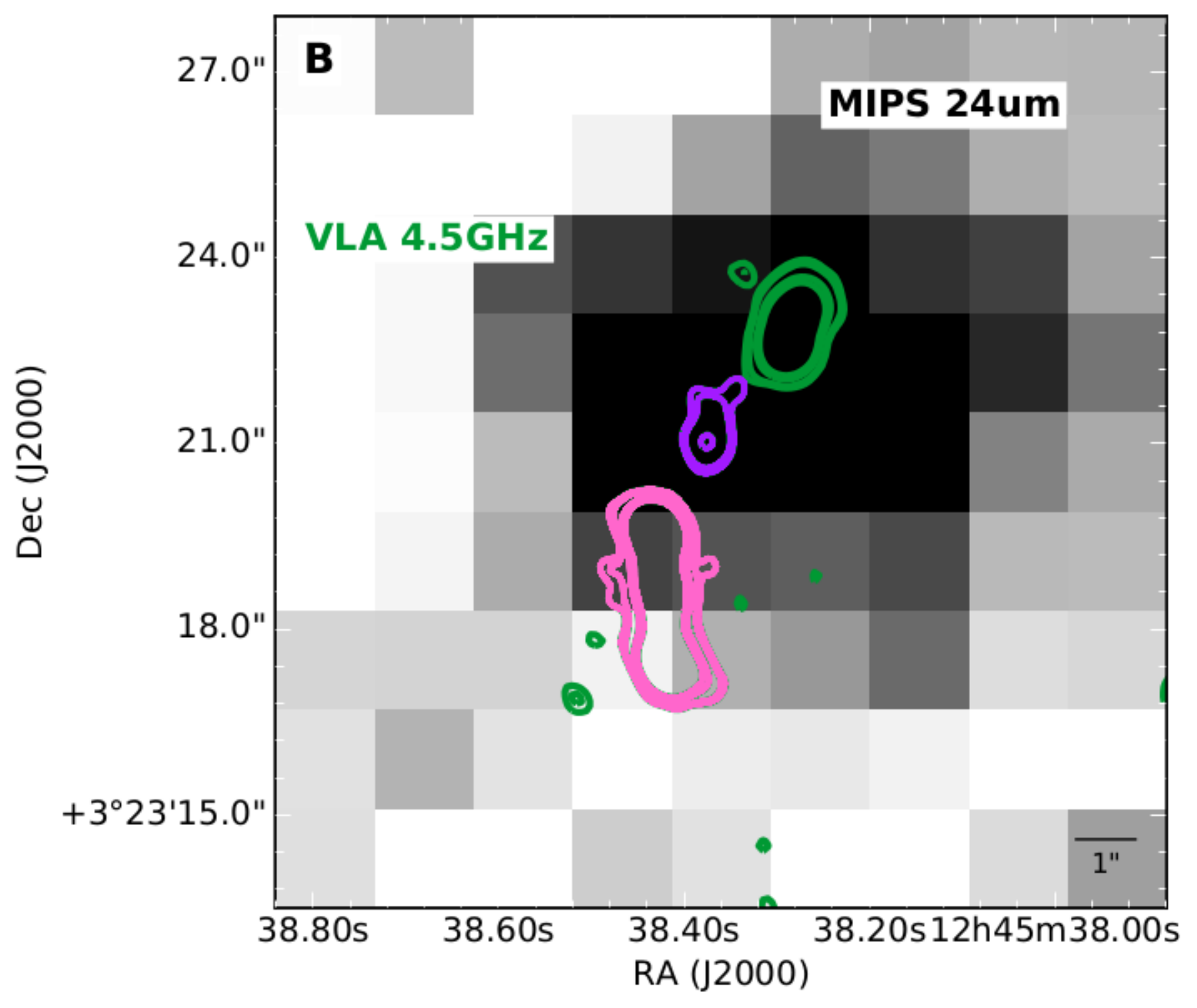}
 	\caption{\textit{Panel A:} continuum map of ALMA band 3 with overlaid VLA C contours (levels as Fig. \ref{map_0037}, $\sigma=50\,\mu$Jy). The blue and green markers indicate the two different ALMA detections and correspond to the same data points as Fig. \ref{fig_4C0324}. The green, purple and pink contours show the three components of the VLA data and correspond to the markers of the same colors as in the SED fit. \textit{Panel B: } MIPS 24\,\mum\ continuum map.}
    	\label{map_4C0324}
\end{figure}

\begin{table}
\begin{threeparttable}
\caption{Data for 4C\,03.24 (z=3.57) }
\label{table_4C0324}
\centering
\begin{tabular}{lcc}
\toprule
Photometric band                   & Flux{[}mJy{]}   & Ref. \\
\midrule
\irs     	&  <0.260			  	&  A \\  
\mips1   	&  0.511$\pm$0.043   	&   A\\ 
\pacsg    	&  8.3$\pm$3.1  		&    B\\
\pacsr   	&  14.1$\pm$6.3		&    B\\
\spires  	&  16.6$\pm$5.6		&    B\\
\spirem  	&  <19.6       			&    B\\
\spirel  	&  <22.2        			&    B\\
\scuba 	&  2.3$\pm$1.1	 		&    C\\
ALMA 3$^h$ 	&  0.08$\pm$0.07		&    this paper\\
ALMA 3$^s$ 	&  0.71$\pm$0.14 		&    this paper\\
VLA X$^n$& 5.2 $\pm$0.52    		&    D\\
VLA C$^n$& 20.3 $\pm$2.03    		&     D\\
VLA X$^c$&  0.7$\pm$0.07    		&    D\\
VLA C$^c$&  1.4$\pm$0.14    		&     D\\
VLA X$^s$& 14.6$\pm$1.5   		&    D\\
VLA C$^s$&  33.8$\pm$3.4    		&     D\\
VLA L      	 &  368.2$\pm$11.1  		&     E\\
\bottomrule                          
\end{tabular}
     \begin{tablenotes}
      \small
      \item \textbf{References.} (A) \cite{DeBreuck2010}, (B) \cite{Drouart2014}, (C) , \cite{Archibald2001}, (D) \cite{vanOjik1996}, (E) \cite{Condon1998}.
    \end{tablenotes}
\end{threeparttable}
\end{table}


\clearpage
\newpage
\subsection{TN\,J1338-1942}
TN\,J1338-1942 has one continuum detection with ALMA, which coincides with the northern radio component (Fig.~\ref{map_1338}). SED fitting with \mrmoose\ is done with four components, two synchrotron (northern and southern radio component), one modified BB and one AGN
component. The northern radio component (detected in VLA bands C and X) is assigned to
the northern ALMA detection and fitted with an individual synchrotron power-law. The southern radio component (detected in VLA bands C and X) is also fitted with an individual power-law and assigned to the southern ALMA upper limit. The VLA L data do not resolve
the individual components and are only considered for fitting
the total radio flux (the combination of the northern and southern
synchrotron power-law components). The northern ALMA detection is assigned to the norther synchrotron component and to a modified BB component. The MAMBO SCUBA, SPIRE,
PACS, MIPS and IRS are fitted to the combination of the modified BB and a AGN component. The best
fit is where the ALMA flux is a combination of the
northern synchrotron lobe and dust emission. 

\begin{figure}
	\includegraphics[scale=0.52]{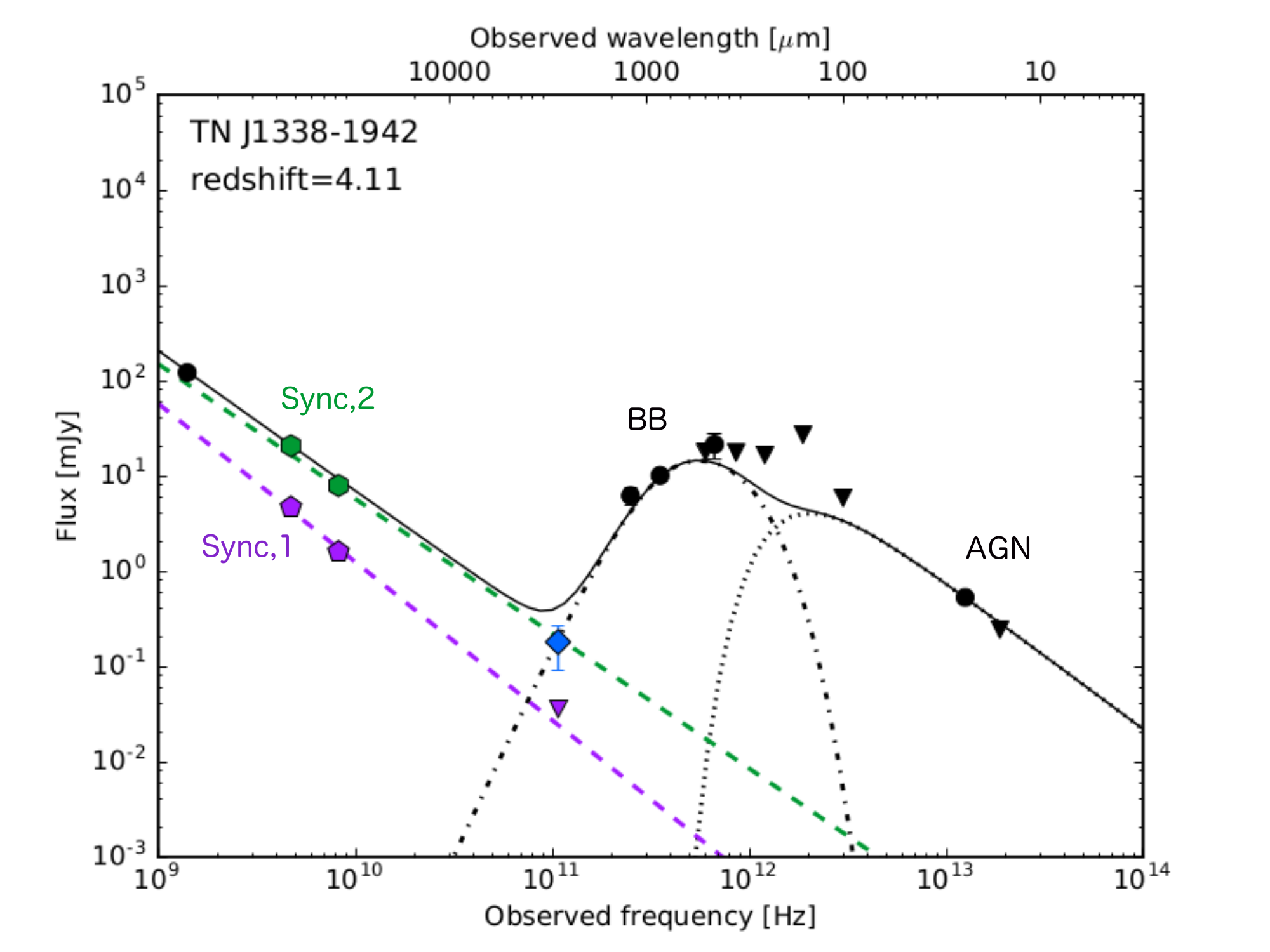}
\caption{SED of \textbf{TN\,J1338-1942}. Black solid line shows best
fit total model, green and purple dashed line is the north and south
synchrotron component, respectively. The black dash-dotted lines represents the
blackbody and the black dotted line indicates the AGN component. The
colored data points are sub-arcsec resolution data and black ones indicate
data with low spatial resolution. The green hexagons and purple pentagons are radio
data of the northern and southern synchrotron component, respectively. The
blue diamond indicates the ALMA band 3 detection. Filled black circles
indicate detections (>$3\sigma$) and downward pointing triangles the $3\sigma$ upper limits (Table~\ref{table_1338}).
}

	\label{fig_1338}
\end{figure}

\begin{figure}
	\centering
      	\includegraphics[scale=0.35]{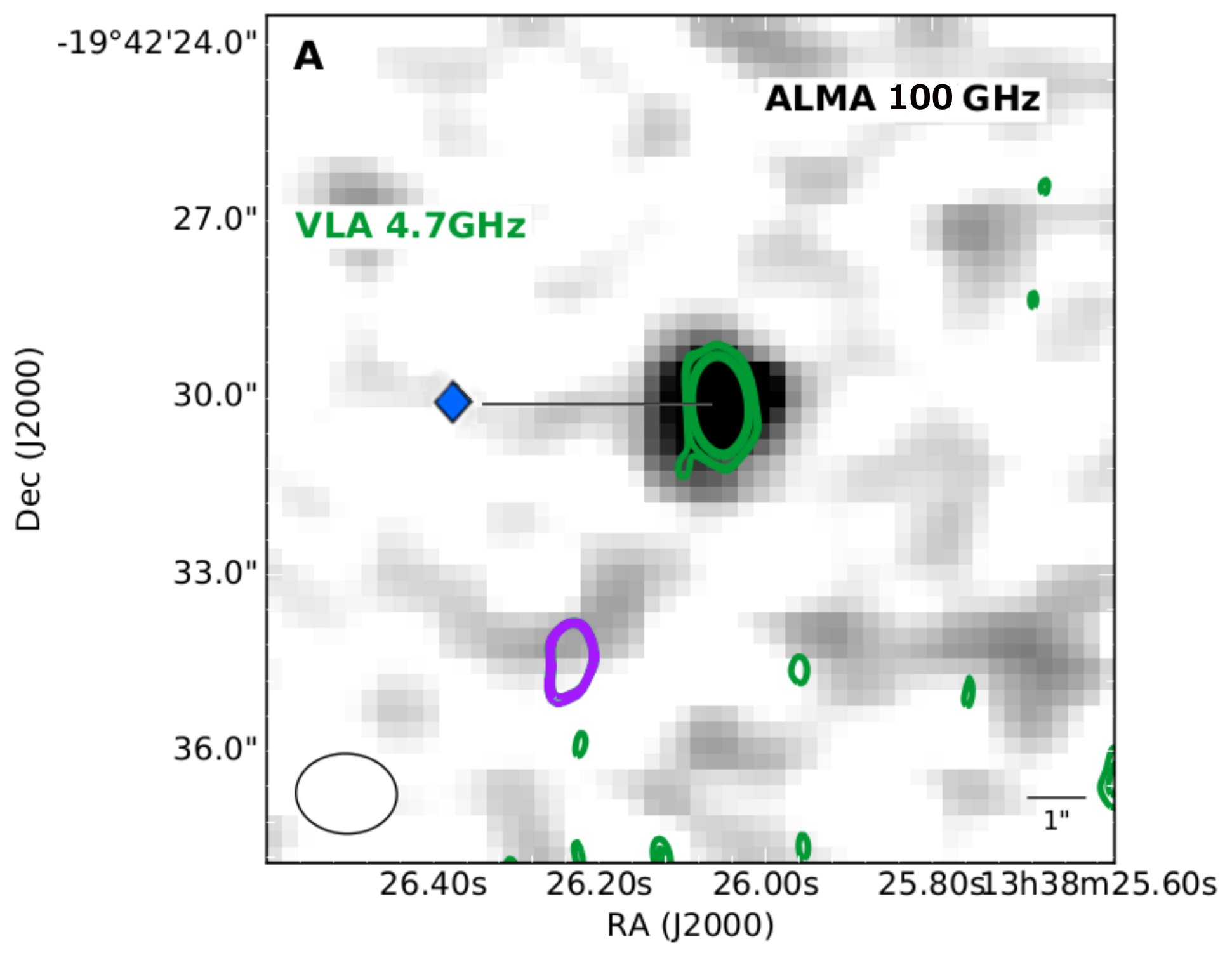}
\\
     	\includegraphics[scale=0.35]{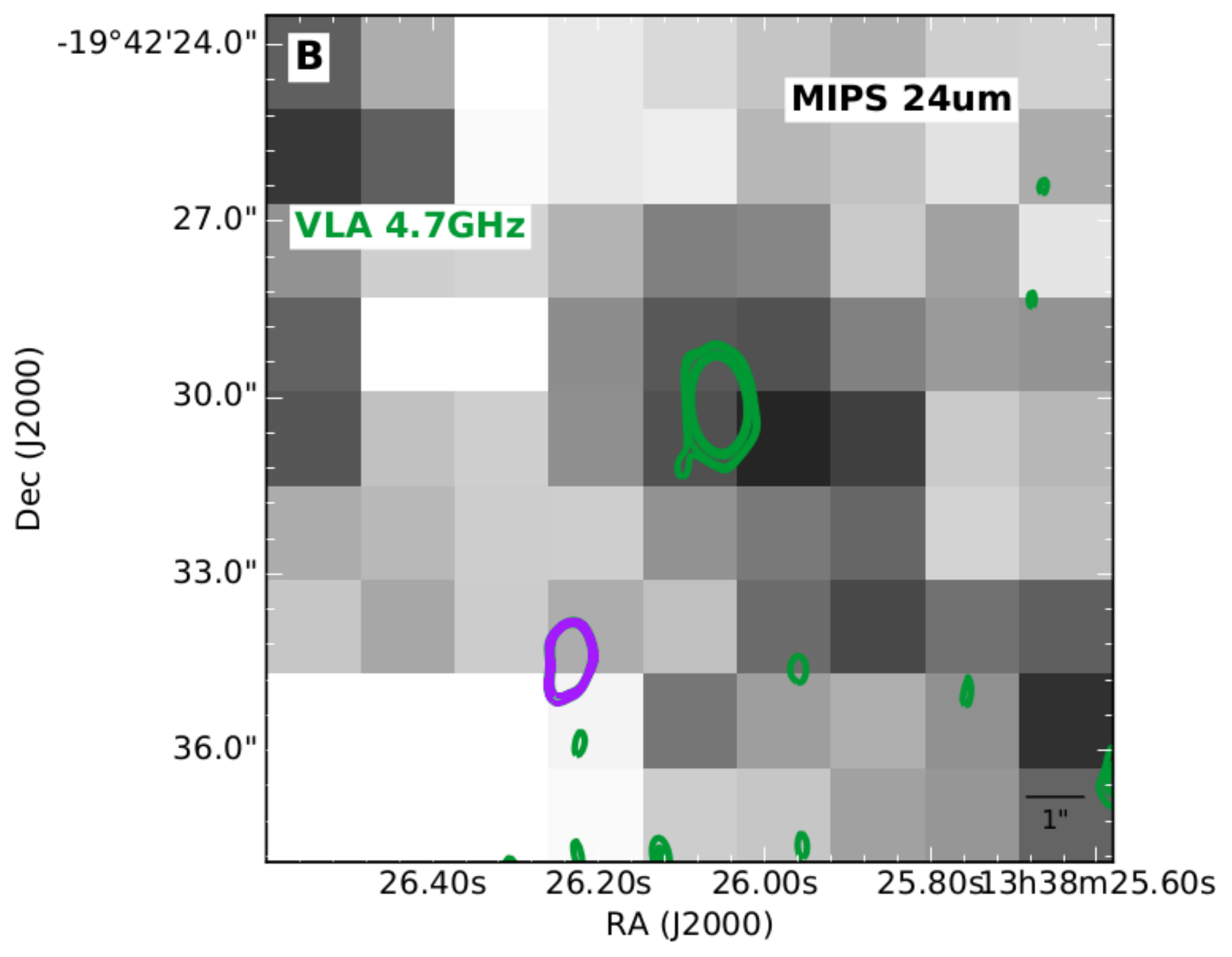}
 	\caption{\textit{Panel A:} continuum map of ALMA band 3 with plotted VLA C contours (levels are as Fig. \ref{map_0037}, $\sigma=55\,\mu$Jy) The blue diamond indicates the ALMA detection and is the same marker used in Fig. \ref{fig_1338}. The green and purple contours show the north and south radio component with the same color coding as in the SED plot. \textit{Panel B:} MIPS 24\,\mum\ continuum map. }
    	\label{map_1338}
\end{figure}

\begin{table}
\begin{threeparttable}
\caption{Data for TN\,J1338-1942 (z=4.110) }
\label{table_1338}
\centering
\begin{tabular}{lcc}
\toprule
Photometric band                   & Flux{[}mJy{]}   & Ref. \\
\midrule
\irs     	&  <0.226		  		&  A \\  
\mips1   	&  0.384$\pm$0.178   	&   B\\ 
\pacsg    	&  <5.9 				&    C\\
\pacsr   	&  <27.1				&    C\\
\spires  	&  <16.6				&    C\\
\spirem  	&  <17.6       			&    C\\
\spirel  	&  <18.0        			&    C\\
\scuba 	&  10.1$\pm$1.3	 		&    D\\
\scubas	&  21.4$\pm$6.4			& 	D\\
MAMBO	& 6.2$\pm$1.2				& 	D\\
ALMA 3 	&  0.18$\pm$0.09			&    this paper\\
ALMA 3 	&  <0.036 				&    this paper\\
VLA X$^s$& 1.6 $\pm$0.16    		&    E\\
VLA C$^s$& 4.7 $\pm$0.47    		&    E\\
VLA X$^n$&  7.9$\pm$0.79    		&    E\\
VLA C$^n$&  20.6$\pm$2.06    		&    E\\
VLA L      	 &  121.4$\pm$4.3  		&    F\\
\bottomrule                          
\end{tabular}
     \begin{tablenotes}
      \small
      \item \textbf{References.} (A) \cite{DeBreuck2010},(B) \cite{Capak2013} (C) \cite{Drouart2014}, (D)  \cite{DeBreuck2004}, (E) \cite{Pentericci2000}, (F) \cite{Condon1998}.
    \end{tablenotes}
\end{threeparttable}
\end{table}


\clearpage
\newpage
\subsection{TN\,J2007-1316}
\begin{figure}
	\includegraphics[scale=0.52]{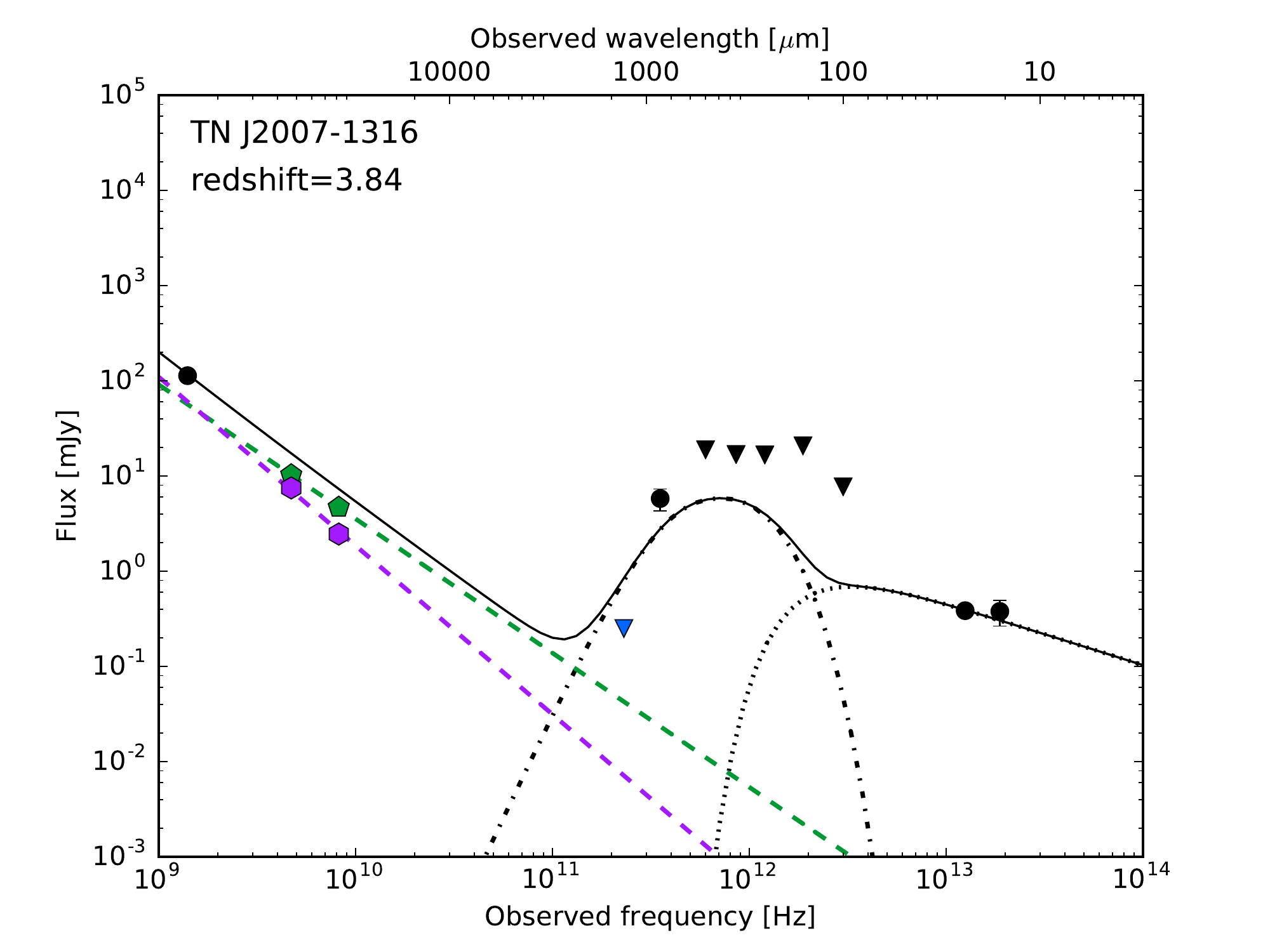}
\caption{SED of \textbf{TN\,J2007-1316}. Black solid line shows best
fit total model, green and purple dashed line are northern and southern
synchrotron lobes, respectively. The black dotted line represents the
AGN component. The colored data points indicate sub-arcsec resolution
data and black ones indicate data with low spatial resolution. Green
pentagons are northern synchrotron emission, purple hexagons are from
the southern radio component and the blue triangle indicates the ALMA
4 $3\sigma$ upper limit. Filled black circles indicate detections
(>$3\sigma$) and downward pointing triangles the
	$3\sigma$ upper limits (Table~\ref{table_2007}).
}

	\label{fig_2007}
\end{figure}

\begin{figure}
	\centering
      	\includegraphics[scale=0.35]{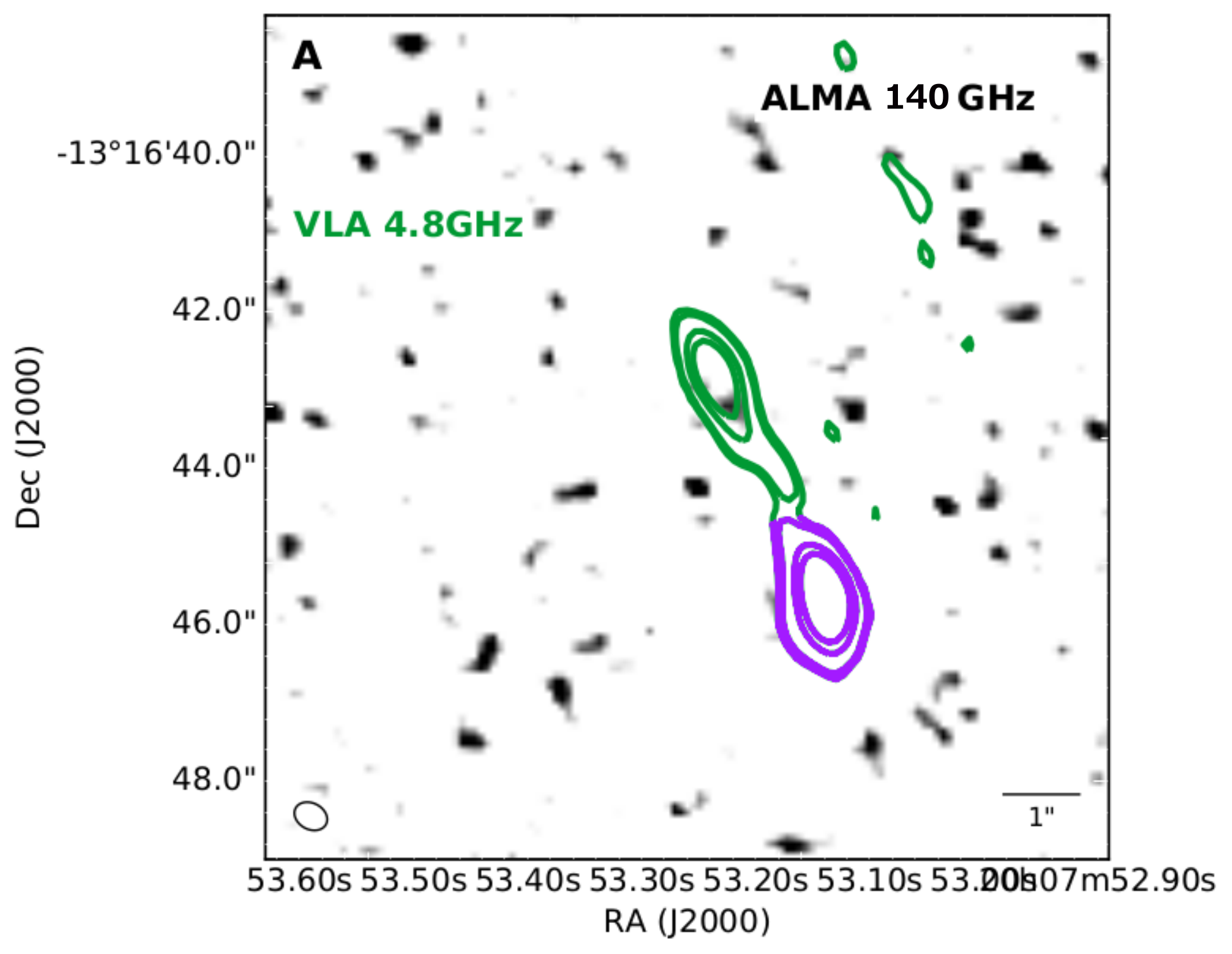}
\\
     	\includegraphics[scale=0.35]{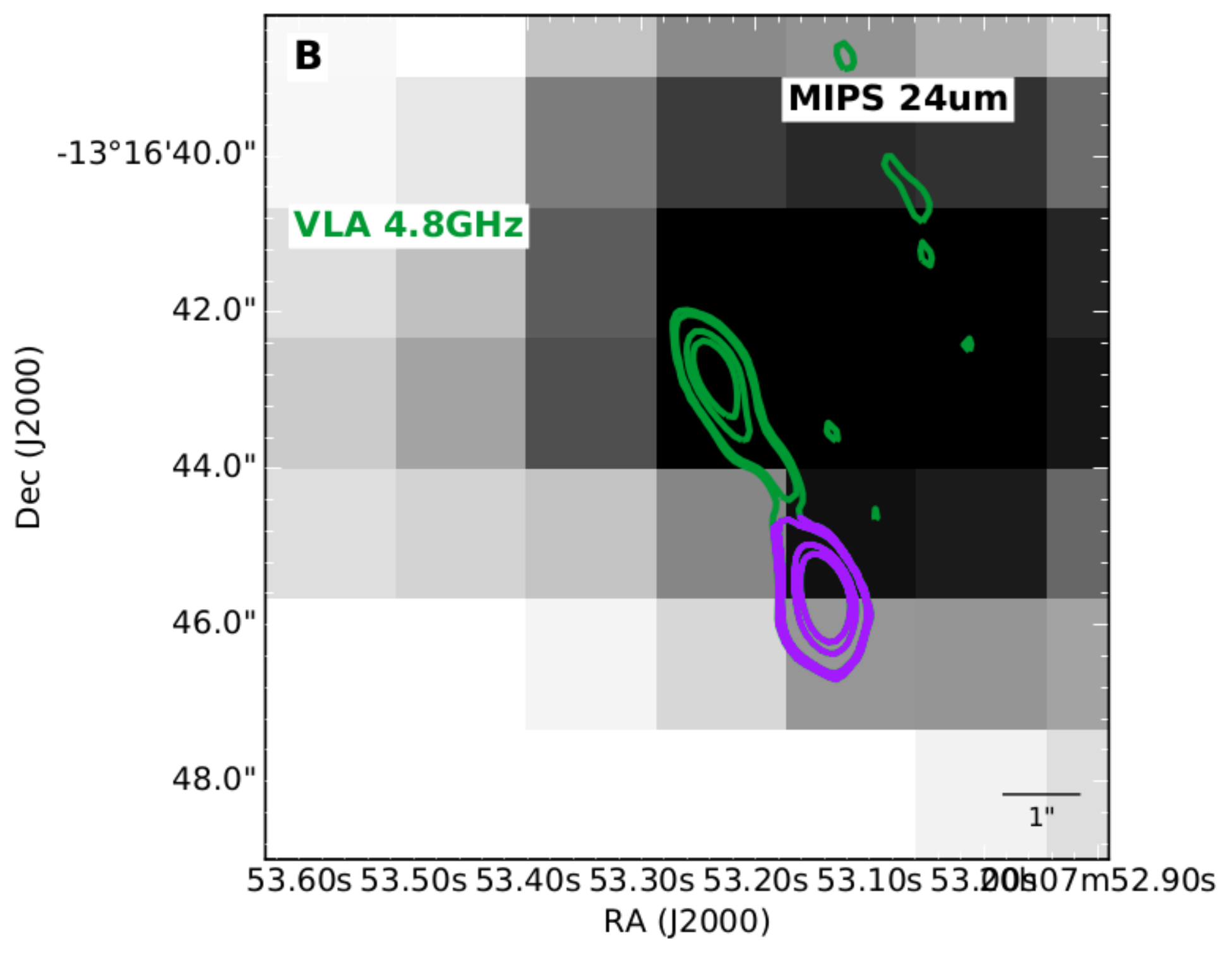}
 	\caption{\textit{Panel A:} continuum map of ALMA band 4 with overlaid VLA C contours (levels are as Fig. \ref{map_0037}, $\sigma=12\,\mu$Jy). The green and purple contours show the two components of the VLA data and corresponds to the markers of the same colors as in the SED fit. \textit{Panel B: } MIPS 24\,\mum\ continuum map.}
    	\label{map_2007}
\end{figure}

TN\,J2007-1316 has no ALMA continuum detection (Fig.~\ref{map_2007}). SED fitting with \mrmoose\ is done with four components,
two synchrotron (northern and southern radio component), one modified BB and one AGN component. The northern and southern radio components (detected in VLA bands C and X) are each assigned to an individual synchrotron power-law. The VLA L band do not resolve individual components and is only considered for fitting the total radio flux (the combination of the north and south synchrotron power-law). The ALMA upper limit at the AGN host location
is assigned to both synchrotron components and the modified BB. The SPIRE,
PACS, MIPS and IRS data a fitted to the combination of the modified BB and a AGN component. The best fit model only constrains the synchrotron and AGN components, while the modified BB remains unconstrained because there are only upper limits in the FIR
(Fig.~\ref{fig_2007}). 

\begin{table}
\begin{threeparttable}
\caption{Data for TN\,J2007-1316 (z=3.840) }
\label{table_2007}
\centering
\begin{tabular}{lcc}
\toprule
Photometric band                   & Flux{[}mJy{]}   & Ref. \\
\midrule
\irs     	&      0.378$\pm$0.113  		& A\\
\mips1     	&      0.385$\pm$0.040  		&A\\
\pacsg    	&      <7.7        				&B\\
\pacsr    	&      <20.8       				&B\\
\spires   	&      <16.7       				&B\\
\spirem   	&      <16.8$^a$        			&B\\
\spirel   	&      <18.9$^a$        			&B\\
ALMA 4  		&      <0.25       				&this paper\\
VLA X$^s$    	&      4.71$\pm$0.3$^b$    	&A\\
VLA X$^n$    	&      2.46$\pm$0.38$^b$	&A\\
VLA C$^s$    	&      10.30$\pm$1.03$^b$	&A\\
VLA C$^n$   	&      7.49$\pm$0.7$^b$ 		&A\\
VLA L     		&      113.2$\pm$13.    		&C\\
\bottomrule                          
\end{tabular}
     \begin{tablenotes}
      \small
      \item \textbf{Notes}  ($s$) South synchrotron lobe, ($n$) north synchrotron lobe, ($a$) changed to upper limits because of foreground object contamination ($b$) flux estimated using AIPS from original radio map, convolved to the resolution of the VLA C band.
      \item \textbf{References.} (A) \cite{DeBreuck2010}, (B) \cite{Drouart2014}, (C)  \cite{Condon1998}.
    \end{tablenotes}
\end{threeparttable}
\end{table}


\clearpage
\newpage
\subsection{MRC\,2025-218}
\begin{figure}
	\includegraphics[scale=0.52]{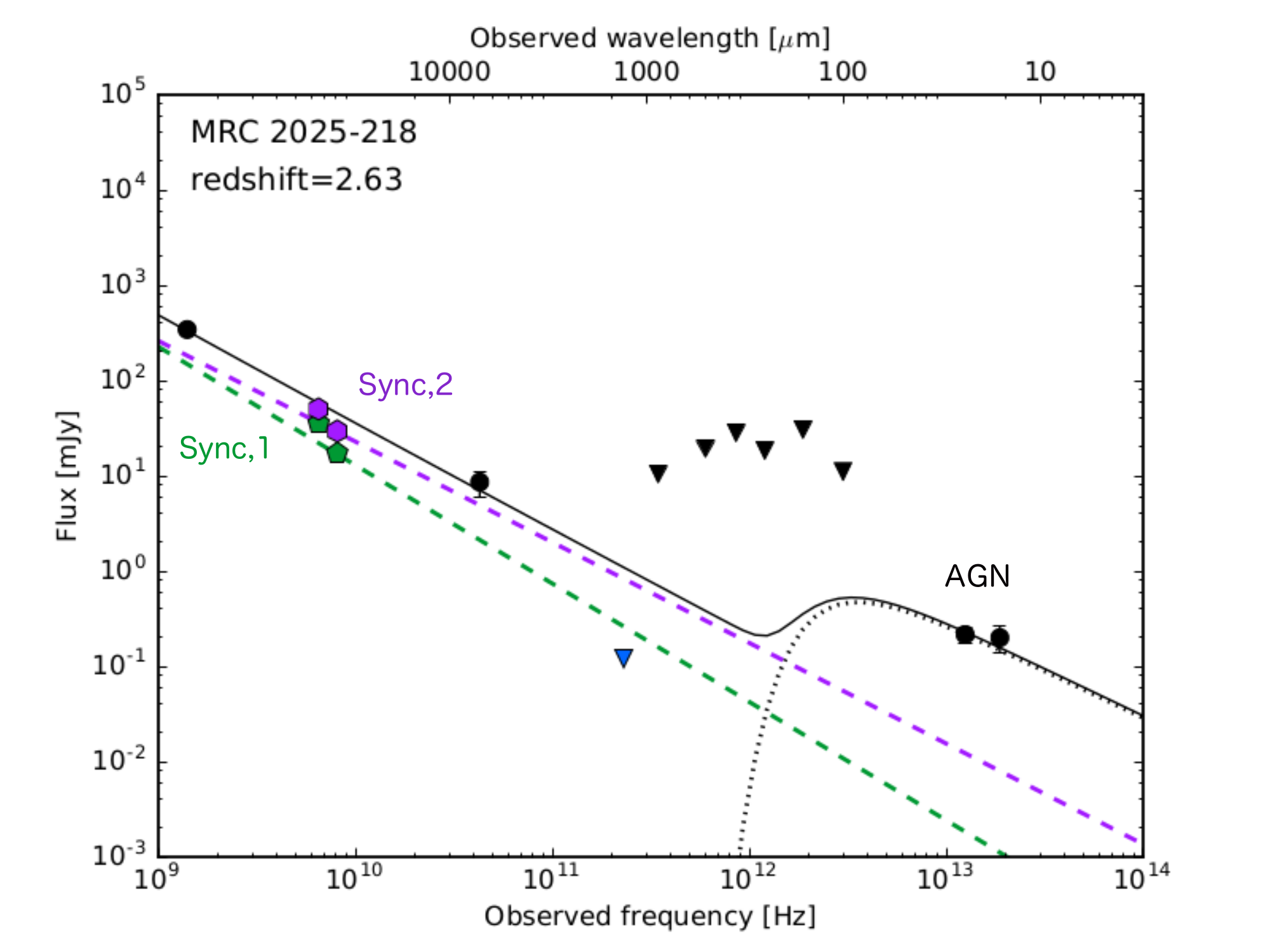}
	\caption{SED of \textbf{MRC\,2025-218}. Black solid line shows
	best fit total model, green and purple dashed lines are northern
	and southern synchrotron lobes, respectively and the black dotted
	line indicates the AGN component. The colored data points are
	sub-arcsec resolution data and black ones indicate data with low
	spatial resolution. Green pentagons are for the northern radio component, purple
	hexagons are for the southern radio component, and the blue triangle indicates
	the ALMA 6 $3\sigma$ upper limit. Filled black circles indicate
	detections (>$3\sigma$) and downward pointing triangles the
	$3\sigma$ upper limits (Table~\ref{table_2025}).
	}

	\label{fig_2025}
\end{figure}

\begin{figure}
	\centering
      	\includegraphics[scale=0.35]{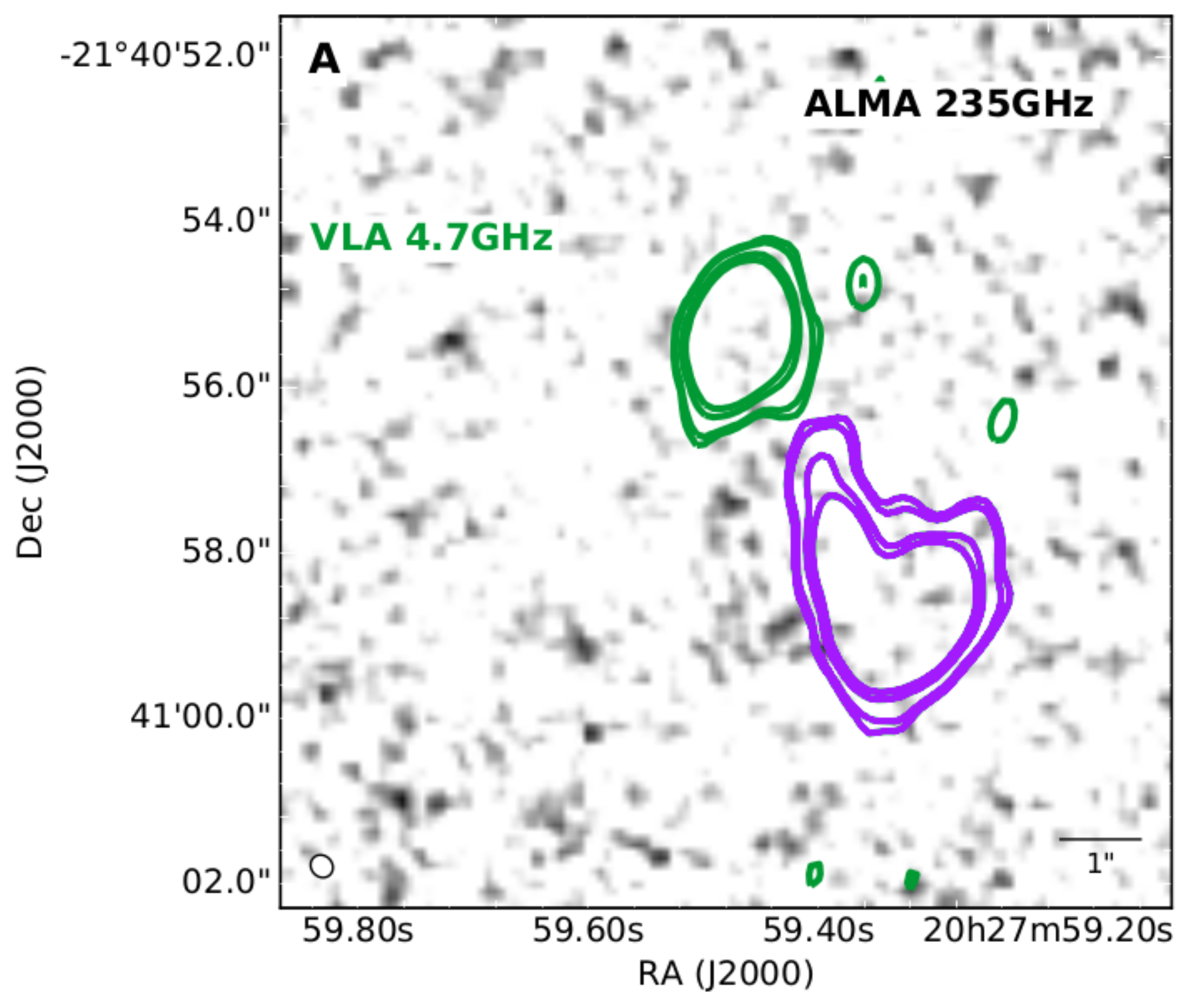}
\\
     	\includegraphics[scale=0.35]{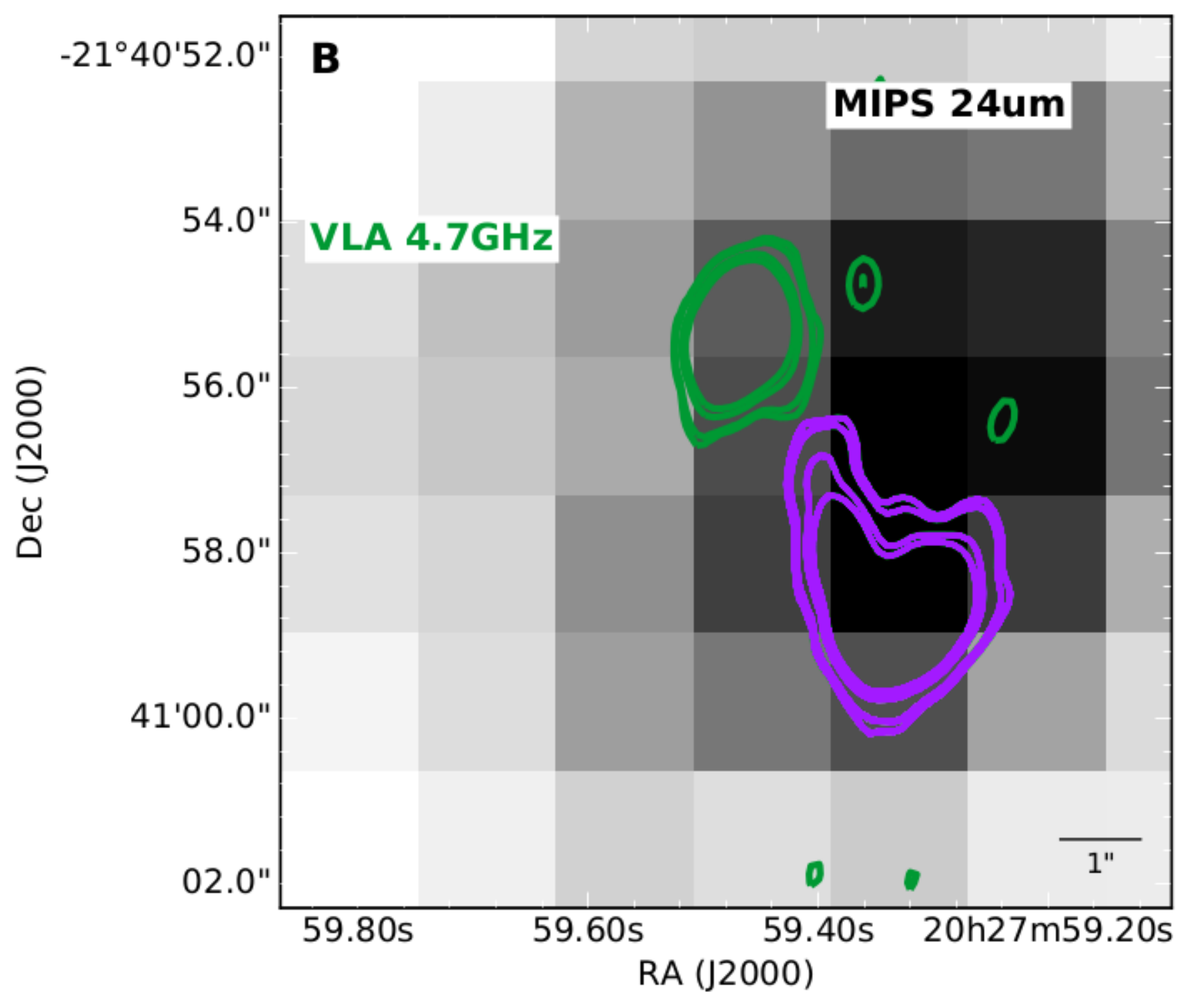}
 	\caption{ \textit{Panel A:} continuum map of ALMA band 6, plotted VLA C contours (levels are as Fig. \ref{map_0037}, $\sigma=57\,\mu$Jy). The green and purple contours show the two components of the VLA data and correspond to the makers of the same colors as in the SED fit. \textit{Panel B: } MIPS 24\,\mum\ continuum map.}
    	\label{map_2025}
\end{figure}

MRC\,2025-218 has no detection continuum detection (Fig. \ref{map_2025}, neither the host nor the synchrotron are detected. SED fitting with \mrmoose\ is done with four components, two synchrotron (north and south radio component), one modified BB and one AGN component. The north radio component (detected in VLA bands C and X) is assigned to an individual synchrotron power-law, the same set up is used for the southern radio component. The VLA L and ATCA 7\,mm to does not resolve any individual components and are only considered for fitting the total synchrotron power-law (the combination of the north and south power-laws). The ALMA upper limit at the AGN host location is associated to both synchrotron components and the modified BB. The LABOCA, SPIRE, PACS, MIPS and IRS data are fitted to combination of the modified BB and a AGN component. The best fit model only constrains the synchrotron and AGN parts, the modified BB are unconstrained because there are only upper limits in the FIR, see Fig. \ref{fig_2025}. 

\begin{table}
\begin{threeparttable}
\caption{Data for MRC\,2025-218 (z=2.630) }
\label{table_2025}
\centering
\begin{tabular}{lcc}
\toprule
Photometric band                   & Flux{[}mJy{]}   & Ref. \\
\midrule
\irs    	&       0.2$\pm$0.062  	&  A\\
\mips1   	&       0.216$\pm$0.043  	&  A\\
\pacsg  	&       <11.1          		&  B\\
\pacsr  	&       <30.7              		&  B\\
\spires 	&       <18.5              		&  B\\
\spirem 	&       <28.4              		&  B\\
\spirel 	&       <19.5              		&  B\\
\laboca	&       <10.5              		&  B\\
ALMA 6 	&       <0.12              		&  this paper\\
ATCA (7mm)   &       8.6$\pm$2.58   	&  this paper \\
VLA X$^n$     &       17.35$\pm$1.7$^a$	&  C\\
VLA X$^s$     &       29.59$\pm$2.9$^a$	&  C\\
VLA C$^n$     &       35.46$\pm$3.5$^a$	&  C\\
VLA C$^s$     &       50.36$\pm$5.0$^a$	&  C\\
VLA L     &       343.2$\pm$34     	&  D\\
\bottomrule                          
\end{tabular}
     \begin{tablenotes}
      \small
      \item \textbf{Notes}  ($1$) North synchrotron lobe, ($s$) south synchrotron lobe, ($a$) flux estimated using AIPS from original radio map, convolved to the resolution of the VLA C band.
      \item \textbf{References.} (A) \cite{DeBreuck2010}, (B) \cite{Drouart2014}, (C) \cite{Carilli1997}, (D)\cite{Condon1998}.
    \end{tablenotes}
\end{threeparttable}
\end{table}


\clearpage
\newpage
\subsection{MRC\,2048-272}
MRC\,2048-272 has no continuum detection with ALMA (Fig.~\ref{map_2048}). SED fitting with \mrmoose\ is done with four components, two
synchrotron (north and south radio component), one modified BB and one AGN component. The northern and southern radio component (detected in VLA bands C and X) are each assigned to an individual synchrotron power-law. The VLA L and ATCA 7\,mm bands do not resolve any individual components and are only considered for fitting the total radio flux (the combination of the northern and southern synchrotron power-law components). The ALMA upper
limit at the AGN host location is associated to both synchrotron components
and the modified BB. The LABOCA, SPIRE, PACS, MIPS and IRS data are fitted to the combination of the modified BB and AGN component. The best fit model only constrains
the synchrotron, and both the AGN the modified BB models are unconstrained
 because the FIR through mid-infrared data are only upper limits. 

\begin{figure}
	\includegraphics[scale=0.52]{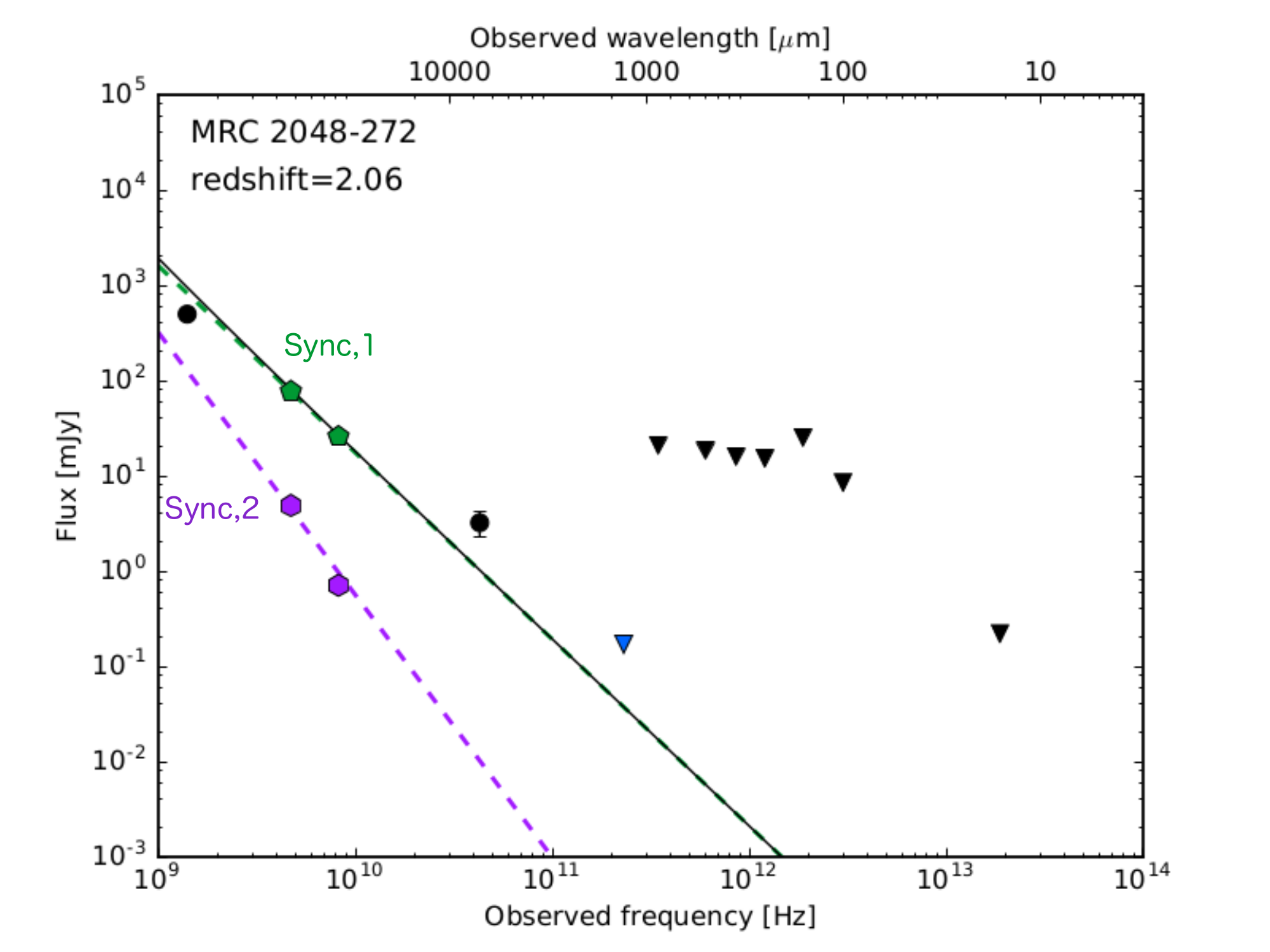}
\caption{SED of \textbf{MRC\,2048-272}. The black solid line shows the best fit
total model, green and purple dashed lines are the northern and southern
synchrotron lobes. Neither the blackbody nor the AGN component are fitted
to the upper limits. The colored data points are sub-arcsec resolution
data and black ones indicate data of low spatial resolution. Green
pentagons are northern synchrotron, purple hexagons are the southern
synchrotron component and the blue triangle shows the ALMA 6 $3\sigma$
upper limit. Filled black circles indicate detections (>$3\sigma$)
	and downward pointing triangles the $3\sigma$ upper limits (Table~\ref{table_2048}).}

\label{fig_2048}
\end{figure}

\begin{figure}
	\centering
      	\includegraphics[scale=0.35]{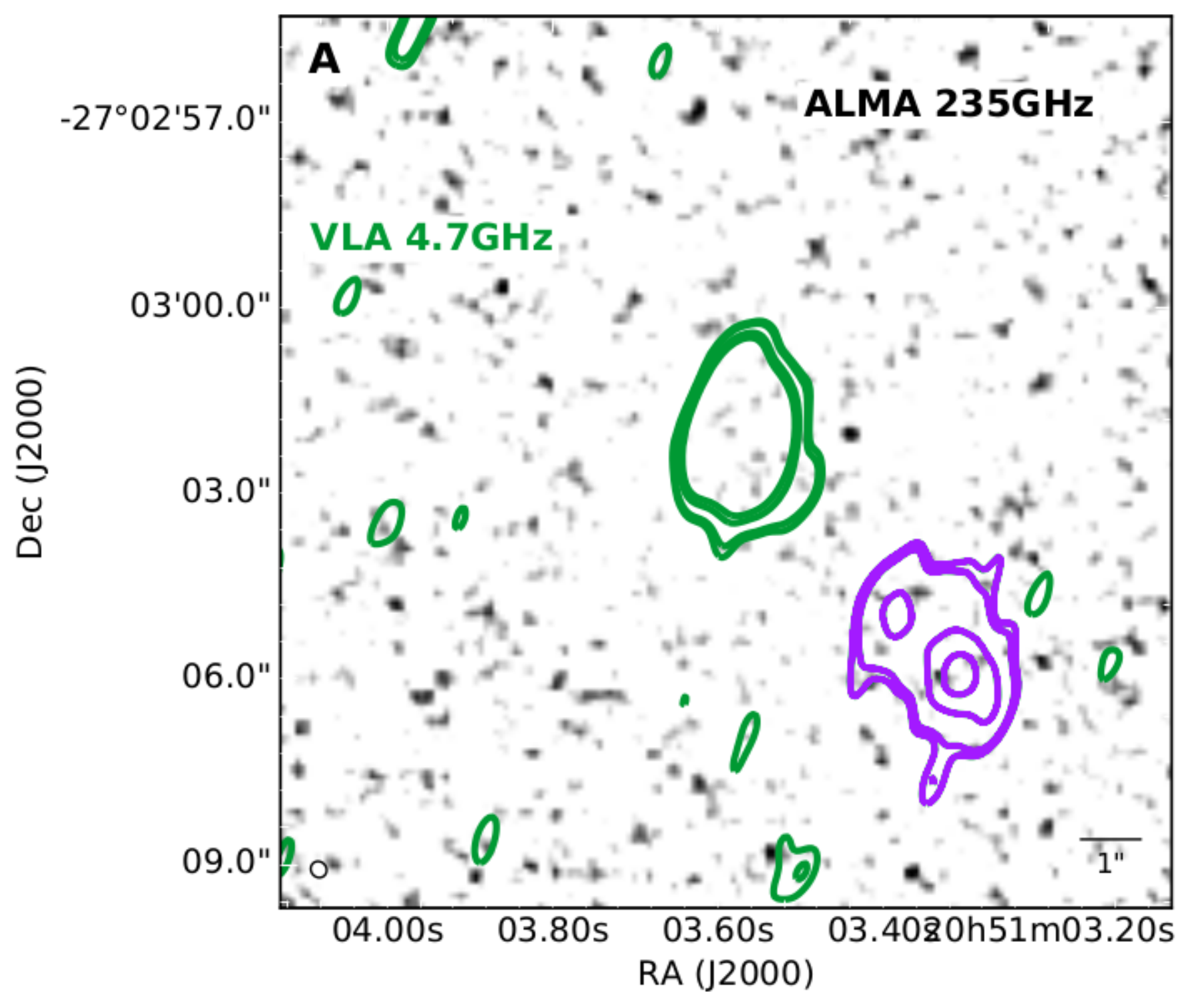}
\\
     	\includegraphics[scale=0.35]{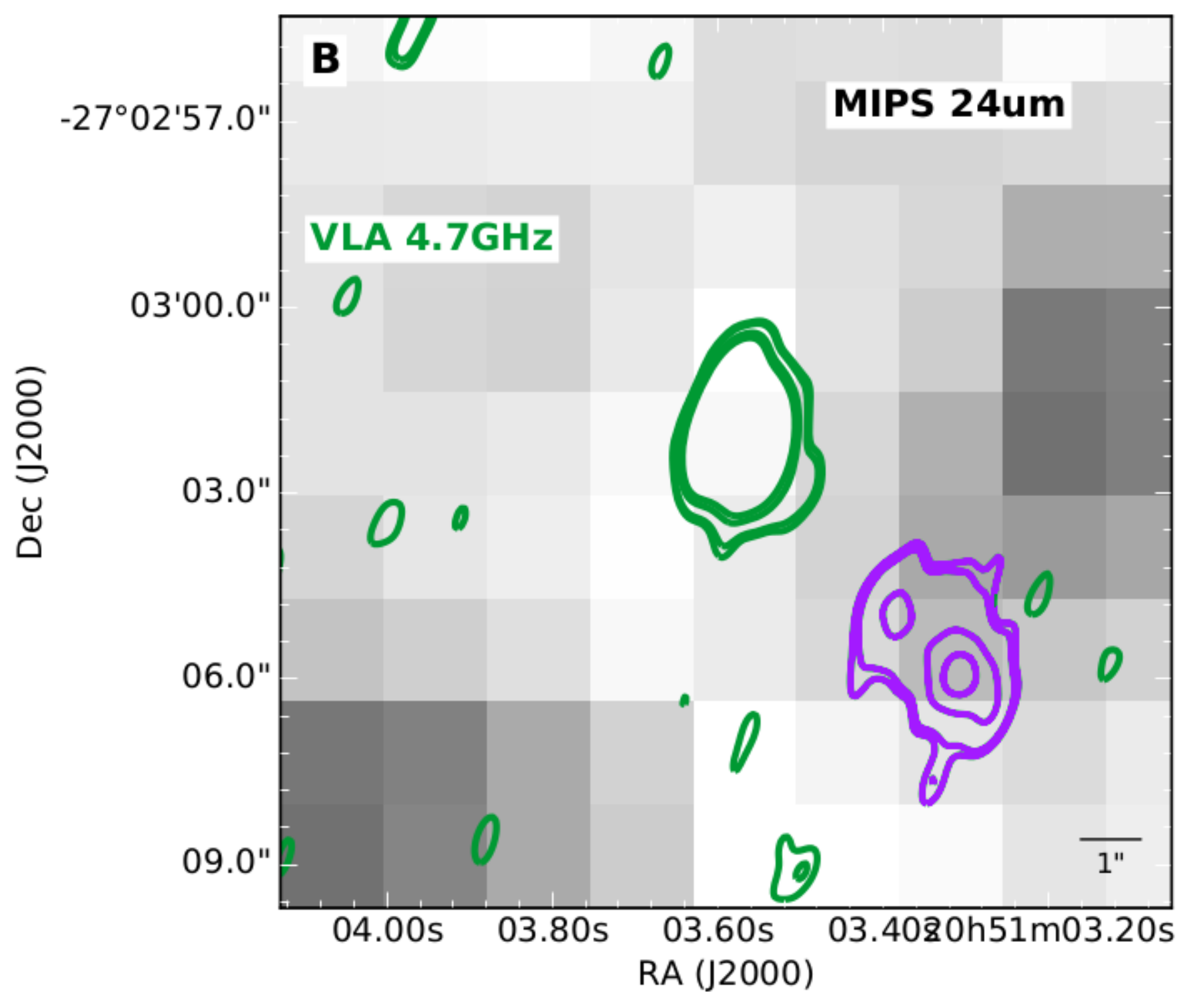}
 	\caption{ \textit{Panel A:} continuum map of ALMA band 6 with overlaid VLA C contours (levels as Fig. \ref{map_0037},$\sigma=53\,\mu$Jy). The green and purple contours show the two components of the VLA data and corresponds to the markers of the same colors as in the SED fit. \textit{Panel B: } MIPS 24\,\mum\ continuum map.}
    	\label{map_2048}
\end{figure}

\begin{table}
\begin{threeparttable}
\caption{Data for MRC\,2048-272 (z=2.060) }
\label{table_2048}
\centering
\begin{tabular}{lcc}
\toprule
Photometric band                   & Flux{[}mJy{]}   & Ref. \\
\midrule
\irs     	&    <0.22            				&   A\\  
\pacsg   	&    <8.5             				&	B\\     
\pacsr   	&    <25.3            				&     B\\
\spires  	&    <15.3           				&     B\\
\spirem  	&    <16.0            				&     B\\
\spirel  	&    <18.6            				&     B\\
\laboca 	&    <21.0           				&     B\\
ALMA 6 	&    <0.17            				&     this paper\\
ATCA (7\,mm)    &    3.22$\pm$0.96$^a$  		&     this paper \\
VLA X$^n$      	&    25.81$\pm$0.04$^a$	&     C\\
VLA X$^s$      	&    0.71$\pm$0.07$^a$	&    C\\ 
VLA C$^n$      	&    77.06$\pm$0.10$^a$ 	&    C\\
VLA C$^s$    	&    4.90$\pm$0.15  		&     C\\
VLA L      		&   498.1$\pm$49			&   D\\
\bottomrule                          
\end{tabular}
     \begin{tablenotes}
      \small
      \item \textbf{Notes} ($n$) North lobe, ($s$) South lobe, ($a$) flux estimated using AIPS from original radio map, convolved to the resolution of the VLA C band.
      \item \textbf{References.} (A) \cite{DeBreuck2010}, (B) \cite{Drouart2014}, (C) \cite{Carilli1997}, (D)\cite{Condon1998}.
    \end{tablenotes}
\end{threeparttable}
\end{table}


\clearpage
\newpage
\subsection{MRC\,2104-242}
\begin{figure}
\includegraphics[scale=0.52]{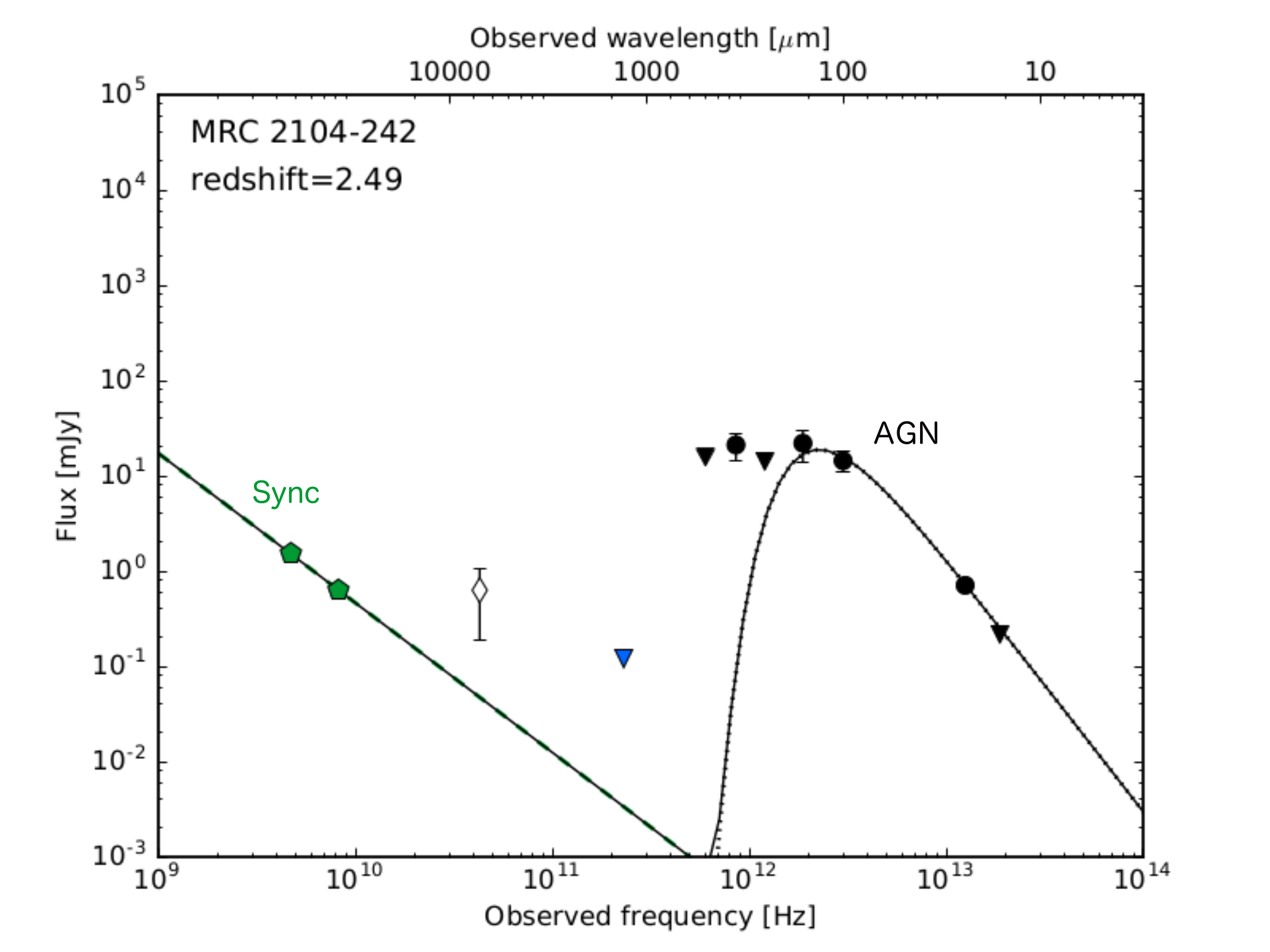}
\caption{SED of \textbf{MRC\,2104-242}. Black solid line shows best fit
total model, green dashed line is the radio core and the black dotted line
is the AGN component. The colored data points are sub-arcsec resolution
data and black ones indicate data of low resolution. Green pentagons
represent the synchrotron core and the blue triangle indicates the ALMA 6 $3\sigma$
upper limit. Filled black circles indicate detections (>$3\sigma$)
	and downward pointing triangles the $3\sigma$ upper limits (Table~\ref{table_2104}). The open diamond indicates available ATCA data but is only plotted as a reference and not used in the SED fit.}

\label{fig_2104}
\end{figure}

\begin{figure}
	\centering
      	\includegraphics[scale=0.35]{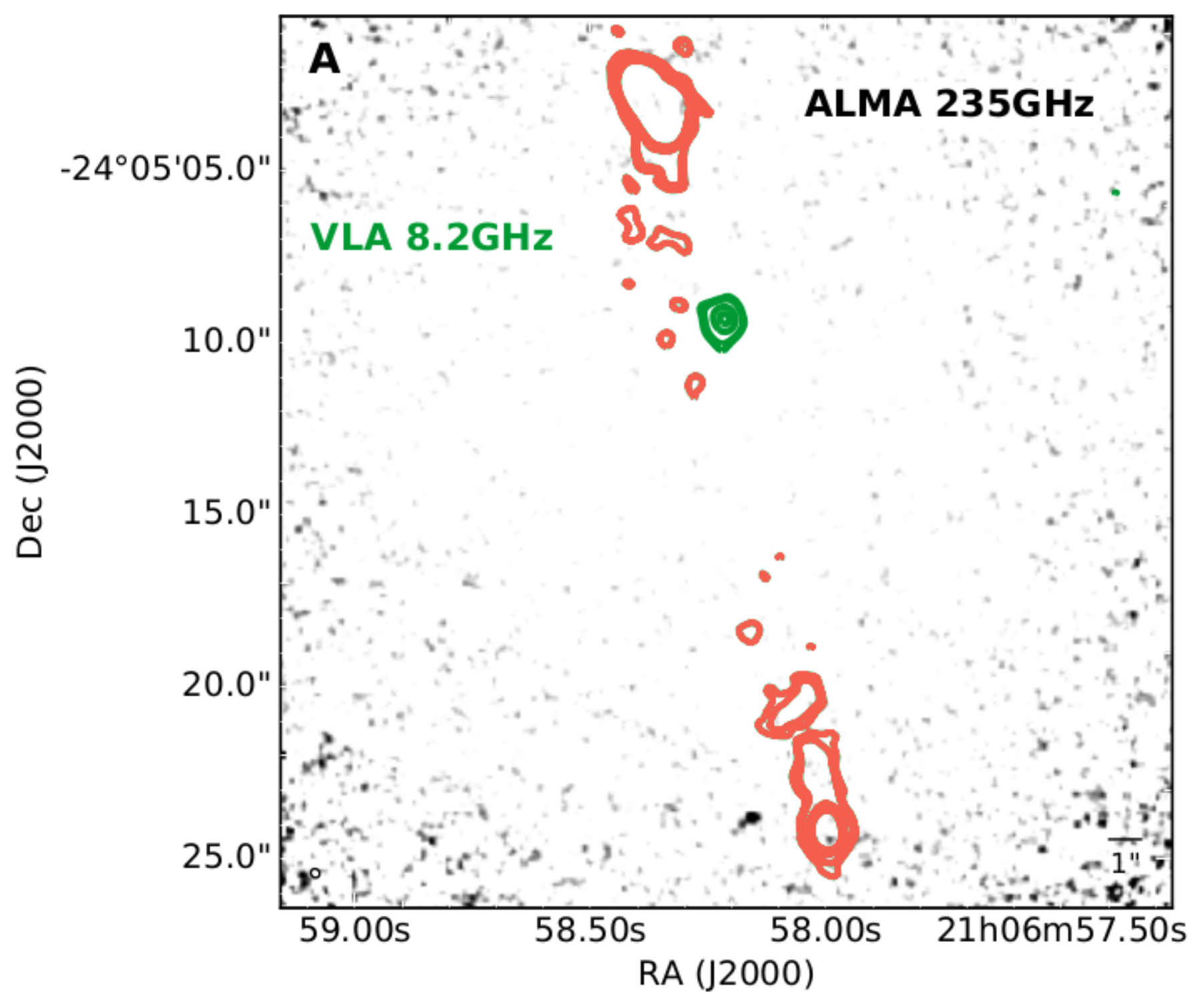}
\\
     	\includegraphics[scale=0.35]{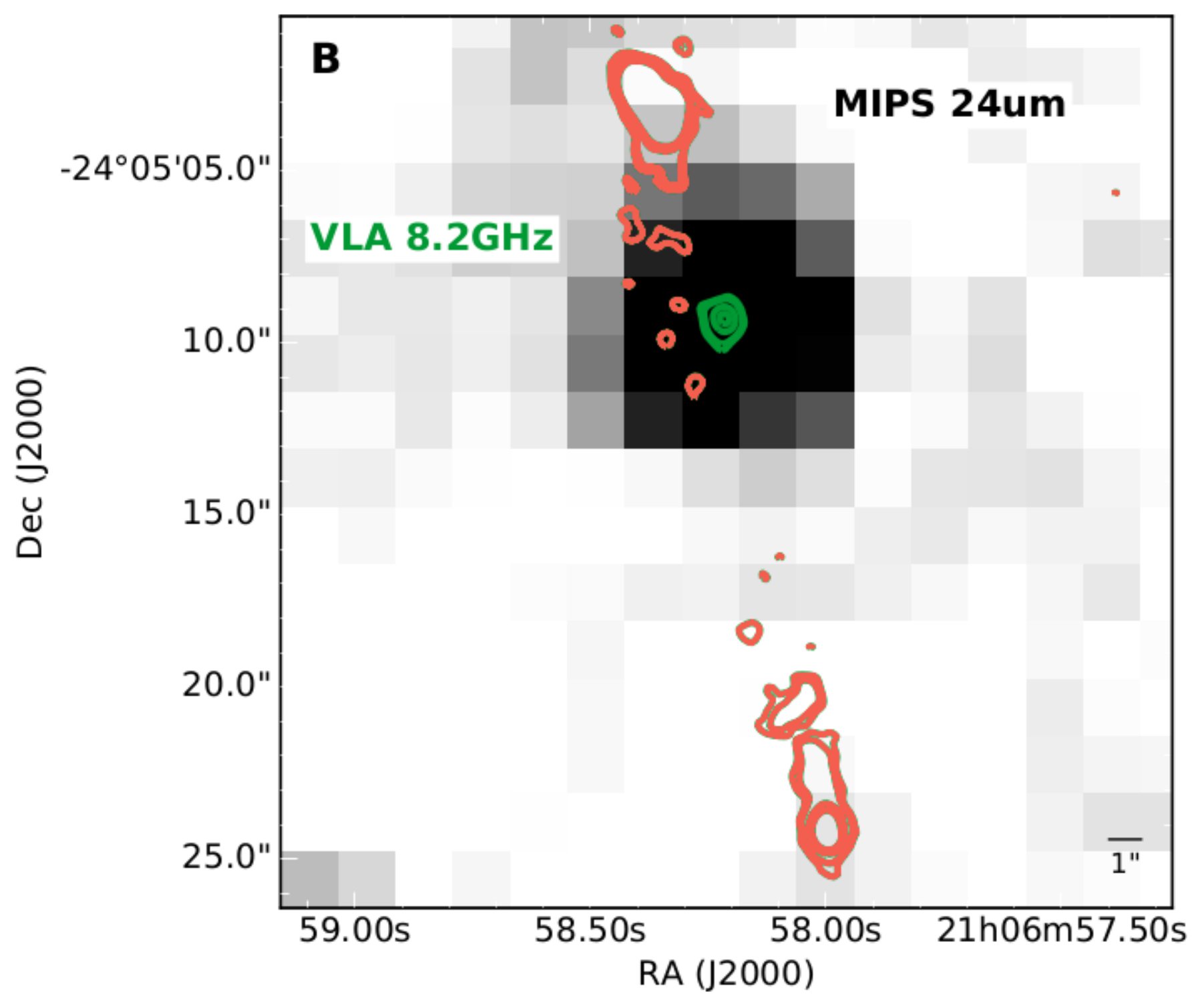}
 	\caption{\textit{Panel A:} continuum map of ALMA band 6 with overlaid VLA C contours (levels are as Fig. \ref{map_0037}, $\sigma=22\,\mu$Jy). The green contours show the two components of the VLA data and correspond to the markers of the same colors as in the SED fit (Fig. \ref{fig_2104}). The red contours are excluded in the fit. \textit{Panel B: } MIPS 24\,\mum\ continuum map.}
    	\label{map_2104}
\end{figure}

MRC\,2104-242 has no continuum detection with ALMA (Fig. \ref{map_2104}). SED fitting with \mrmoose\ is done with three components,
one synchrotron (only the core, the northern and southern radio components are excluded in the fit), one modified BB and one
AGN component. The  VLA C and X band data are assigned to the sychrotron
power law of the radio core. The ALMA upper limit at the AGN host location is associated
to both synchrotron components and the modified BB. The SPIRE, PACS, MIPS and IRS
data are fitted to the combination of the modified BB and a AGN component. The best fit model
only constrains the synchrotron and AGN component, the modified BB models are
unconstrained due to the FIR data consisting only of upper limits (Fig. \ref{fig_2104}).

\begin{table}
\begin{threeparttable}
\caption{Data for MRC\,2104-242 (z=2.491) }
\label{table_2104}
\centering
\begin{tabular}{lcc}
\toprule
Photometric band                   & Flux{[}mJy{]}   & Ref. \\
\midrule
\irs      		&      <0.217     			&A\\
\mips1     		&      0.709$\pm$0.048  	&A\\
\pacsg    		&      14.4$\pm$3.5    	&B\\
\pacsr    		&      22.0$\pm$8.4    	&B\\
\spires   		&      <14.2$^a$      		&B\\
\spirem   		&      21.1$\pm$6.6    	&B\\
\spirel   		&      <15.8       			&B\\
ALMA 6  			&      <0.12        		&this paper\\ 
VLA X$^c$       	&      0.63$\pm$0.03$^b$ 	&C\\
VLA C$^c$       	&     1.53$\pm$0.14$^b$	&C\\                              
\midrule
ATCA (7mm)$^*$	&	0.64  $\pm$0.44		& this paper\\
\bottomrule                          
\end{tabular}
     \begin{tablenotes}
      \small
      \item \textbf{Notes} ($c$) Radio core, ($a$) changed to upper limit because of contaminating forground object ($b$) flux estimated using AIPS from original radio map, convolved to the resolution of the VLA C band (*) data not used in fitting.
      \item \textbf{References.} (A) \cite{DeBreuck2010}, (B) \cite{Drouart2014}, (C) \cite{Pentericci2000}.
    \end{tablenotes}
\end{threeparttable}
\end{table}


\clearpage
\newpage
\subsection{4C\,23.56}
\begin{figure}
	\includegraphics[scale=0.52]{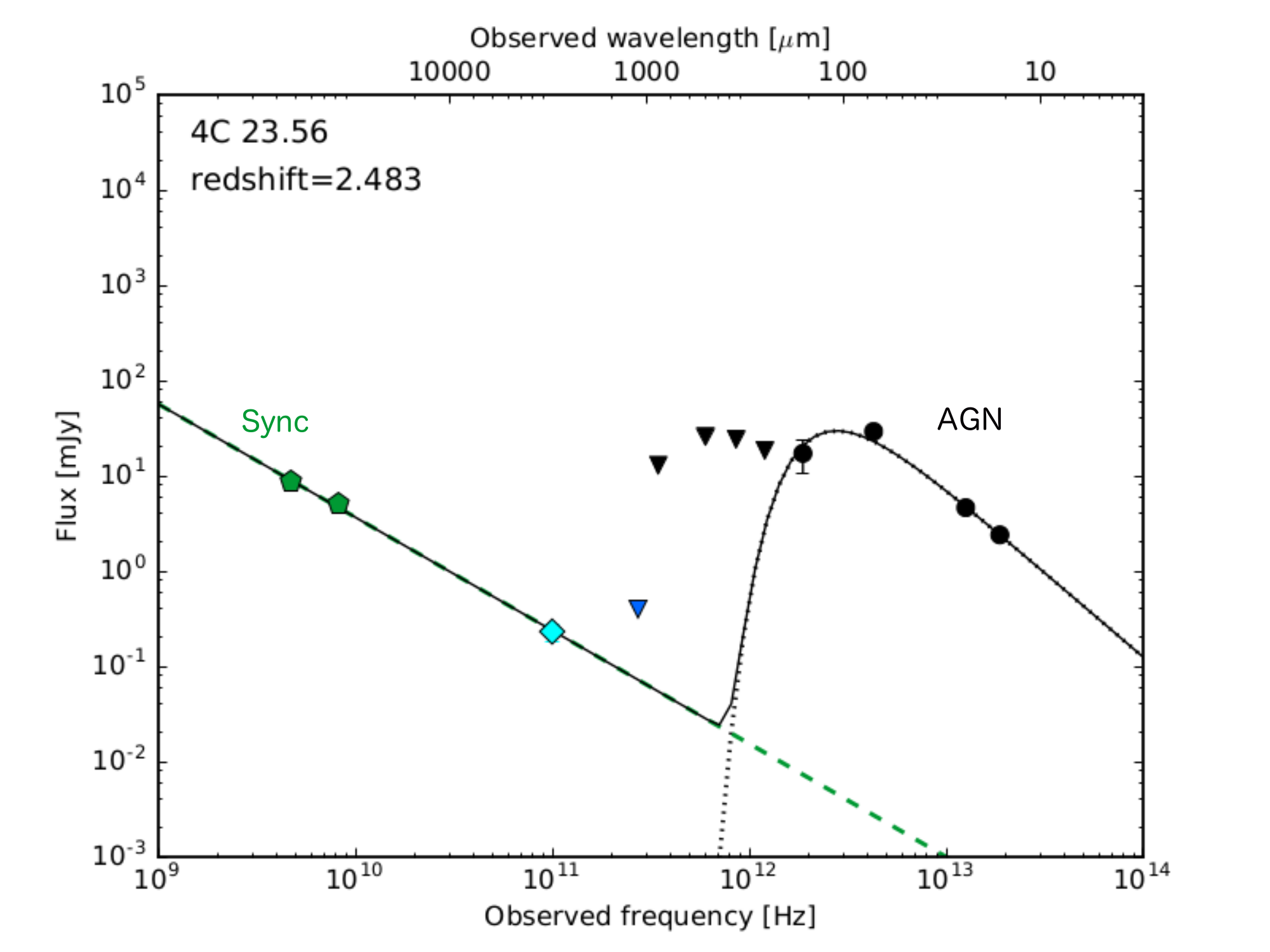}
\caption{SED of \textbf{4C\,23.56}. Black solid line shows best fit total
model, green dashed line is synchrotron power law for the radio core,
dotted line indicates the AGN component. The colored data points
are sub-arcsec resolution data and black ones indicate data with
low spatial resolution. Green pentagons are emission from the radio core, cyan diamond from
ALMA band 3, blue triangle indicates the ALMA band 3 3$\sigma$  upper
limit. Filled black circles indicate detections (>$3\sigma$), downward
	pointing triangles the $3\sigma$ upper limits (Table~\ref{table_0037}).
}
	\label{fig_2356}
\end{figure}

\begin{figure}
	\centering
      	\includegraphics[scale=0.35]{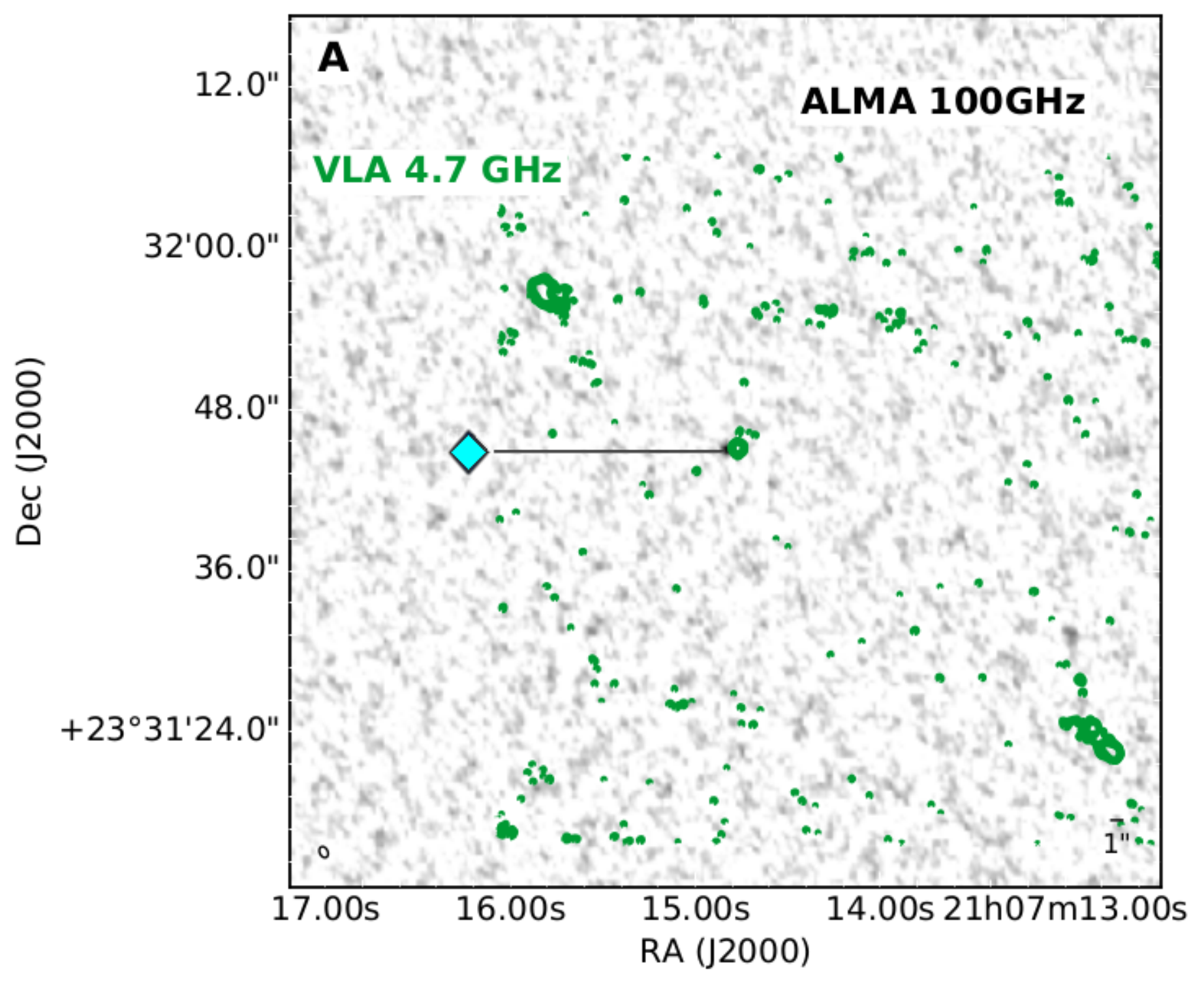}
\\
     	\includegraphics[scale=0.35]{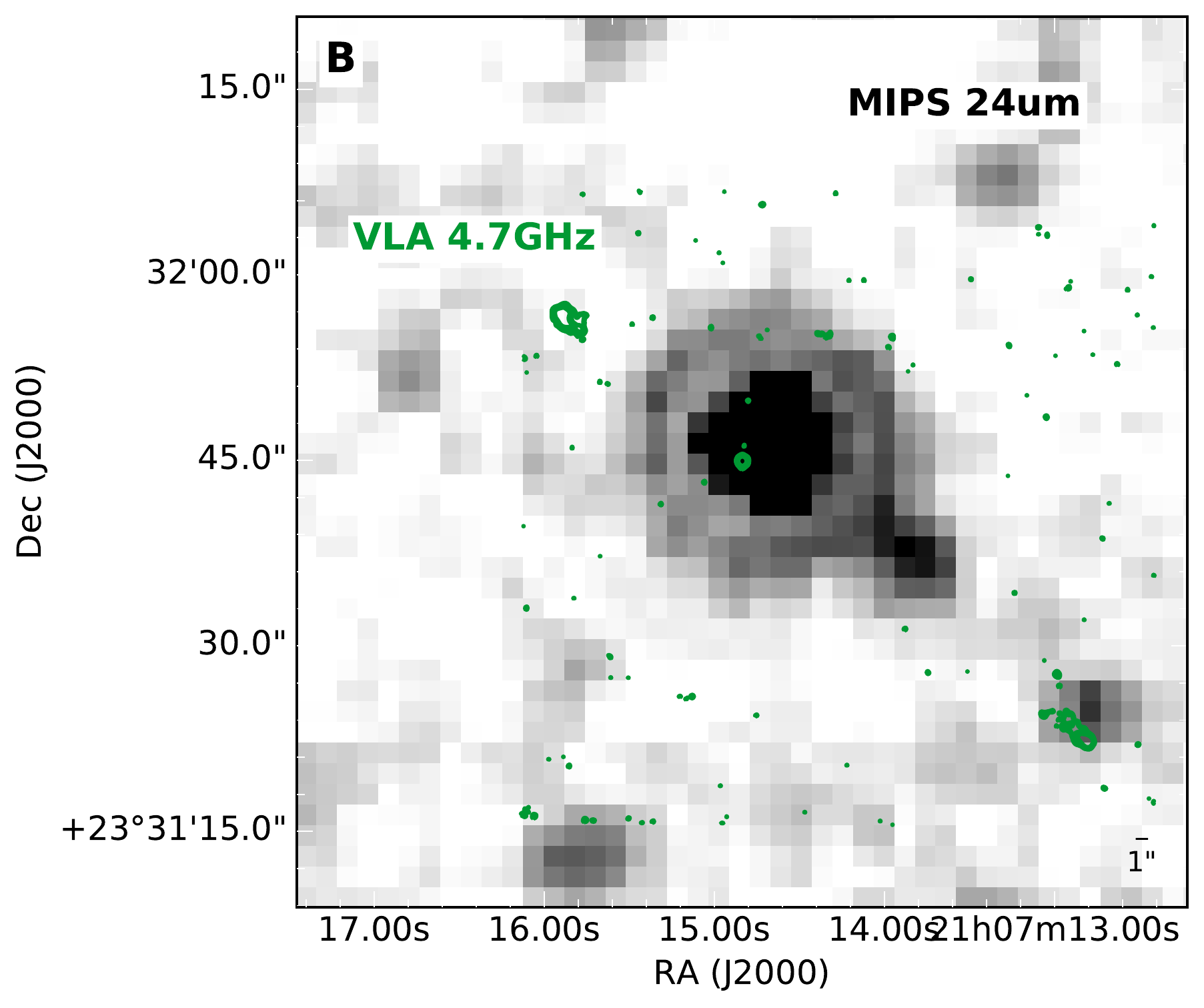}
 	\caption{ \textit{Panel A:} continuum map of ALMA band 3 with overlaid VLA C contours (levels are as Fig. \ref{map_0037}, $\sigma=67\,\mu$Jy). The cyan marker indicate the ALMA detection and corresponds to the marker use in the SED fit (Fig. \ref{fig_2356}). The green contours show the VLA data and only the is used in the SED fit, the lobes are excluded in the fit. \textit{Panel B:} MIPS 24\,\mum\ continuum map.}
    	\label{map_2356}
\end{figure}

4C\,23.56 has no continuum detection in ALMA band 6, but is detected in band
3. The detection coincides with the radio core (Fig. \ref{map_2356}). SED fitting with \mrmoose\ is done with three components, one synchrotron power-law
(of the core, the north and south lobe are excluded for the fit), one modified BB and one AGN component. The
VLA C and X band fluxes are assigned to the synchrotron power-law. The ALMA band 3 detection is assigned to both the synchrotron model and modified BB and the same set up is used for the ALMA band 6 upper limit. The LABOCA, SPIRE, PACS, MIPS and IRS are assigned
to the combination of the modified BB and a AGN component. The best fit model constrains the synchrotron and gives that the ALMA band 3 detection is dominated by synchrotron emission. The
modified BB is unconstrained (Fig. \ref{fig_2356}) due to the fact that all the FIR data points are upper limits. This 
source has a low SFR and the IR emission is dominated by the AGN. 4C\,23.56 was use to fix the $\nu_{cut}$=33\,\mum\ for the
model describing the AGN heated dust, because the SF contribution in
the FIR is very weak. 

\begin{table}
\centering
\begin{threeparttable}
\caption{Data for 4C\,23.56  (z=2.483) }
\label{table_4C23}
\begin{tabular}{lcc}
\toprule
Photometric band                   & Flux{[}mJy{]}   & Ref. \\
\midrule
\irs  		&      2.40   $\pm$     0.090    	&A\\
\mips1     	&      4.630  $\pm$     0.040    	&A\\
\pacsb     	&      29.2  $\pm$      3.2      	&B \\
\pacsr    	&      17.2    $\pm$    6.8      	&B \\
\spires   	&      <18.5              			&B \\
\spirem   	&      < 24.2              			&B \\
\spirel   	&      <25.9             			&B \\
\laboca  	&      <12.9              			& C\\
ALMA 6    	&  	<0.4$^a$        			& D \\
ALMA 3    &      0.23   $\pm$     0.05$^a$& this paper \\
VLA X$^c$   &      5.11  $\pm$      0.51 $^b$& E\\
VLA C$^c$  &      8.77   $\pm$      0.87 $^b$& E\\
\bottomrule                          
\end{tabular}
     \begin{tablenotes}
      \small
      \item \textbf{Notes} ($c$) Radio core, ($a$) Flux estimated using AIPS for primary beam corrected images. ($b$) flux estimated using AIPS from original radio map, convolved to the resolution of the VLA C band.
      \item \textbf{References.} (A) \cite{DeBreuck2010}, (B) \cite{Drouart2014}, (C) \cite{Archibald2001}, (D) \cite{Lee2017}, (E)  \cite{Carilli1997} 
    \end{tablenotes}
\end{threeparttable}
\end{table}


\clearpage
\newpage
\subsection{4C\,19.71}
4C\,19.71 has three detected continuum components in ALMA band 3, one coincides
with the AGN host galaxy and two coincide with the northern and southern
synchrotron lobes (Fig. \ref{map_1971}). SED fitting with \mrmoose\ is done with four components, two synchrotron (northern and southern radio components), one modified BB and
one AGN component. The northern radio
component (detected in VLA bands C and X) is assigned to
the northern ALMA detection and fitted to a individual synchrtoron component, the same setup is also applied
to the southern radio and ALMA components. The VLA L detection does not resolve any individual components and are only considered for fitting the the total radio flux (the combination of the northern and southern synchrtorn power-laws). The ALMA detection at the host location is assigned to only the
modified BB and the SCUBA, SPIRE, PACS, MIPS and IRS are fitted to the
combination of the modified BB and a AGN component. The best fit model constrains the two synchrotron components and they are fitted out to ALMA band 3. The ALMA detection
at the host is dominated by thermal dust emission. 

\begin{figure}
	\includegraphics[scale=0.52]{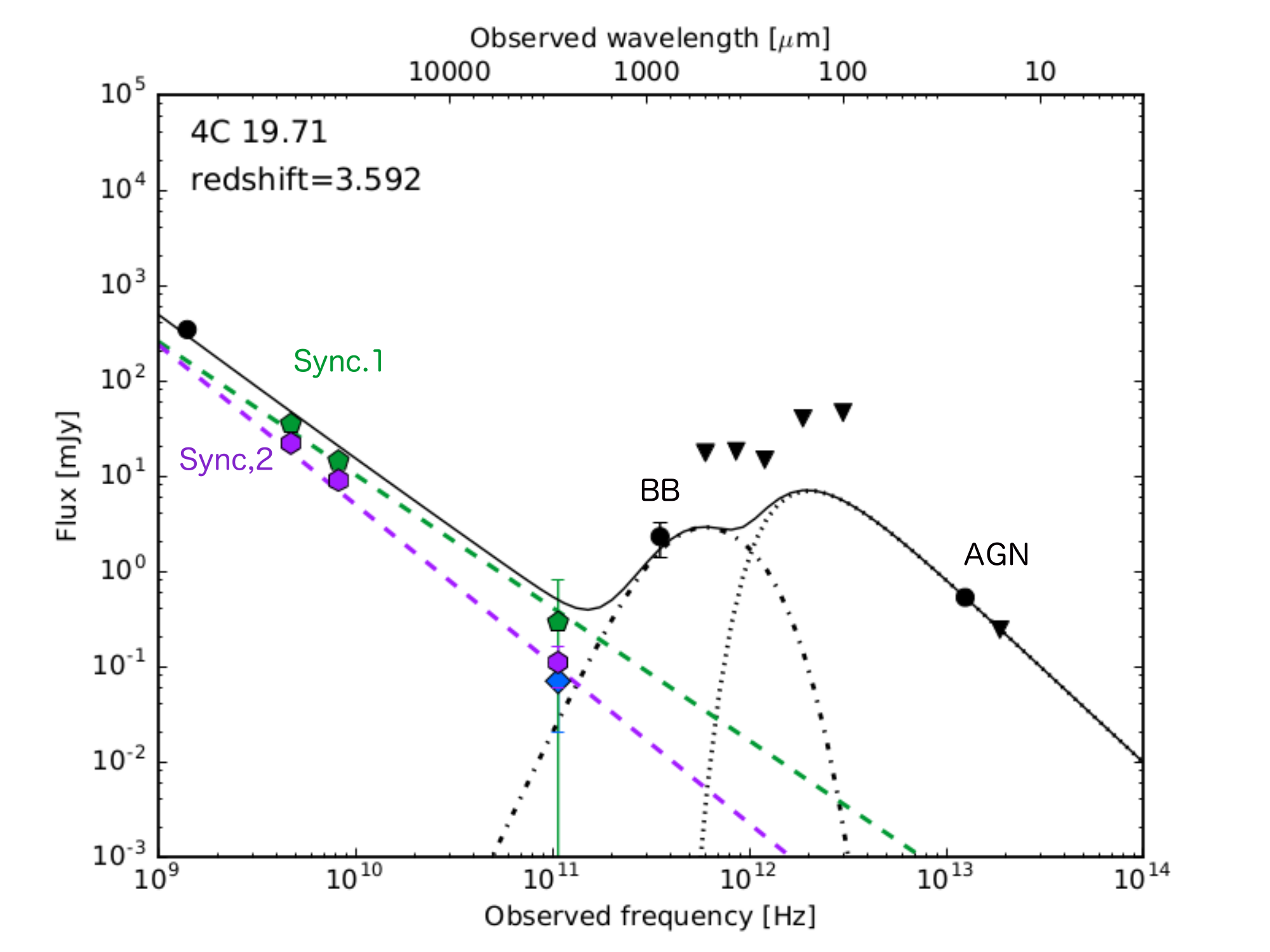}
	\caption{SED of \textbf{4C\,19.71}. Black solid line shows best fit total model, black dashed line is the total synchrotron, the black dash-dotted lines is the blackbody and the black dotted line indicates the AGN component. The colored data points are sub-arcsec resolution data and black ones indicate data of low resolution. The blue diamond indicates the ALMA band 3 detection. Filled black circles indicate detections (>$3\sigma$) and downward pointing triangles the $3\sigma$ upper limits (Table~\ref{table_1971}).}
	\label{fig_1971}
\end{figure}

\begin{figure}
	\centering
      	\includegraphics[scale=0.35]{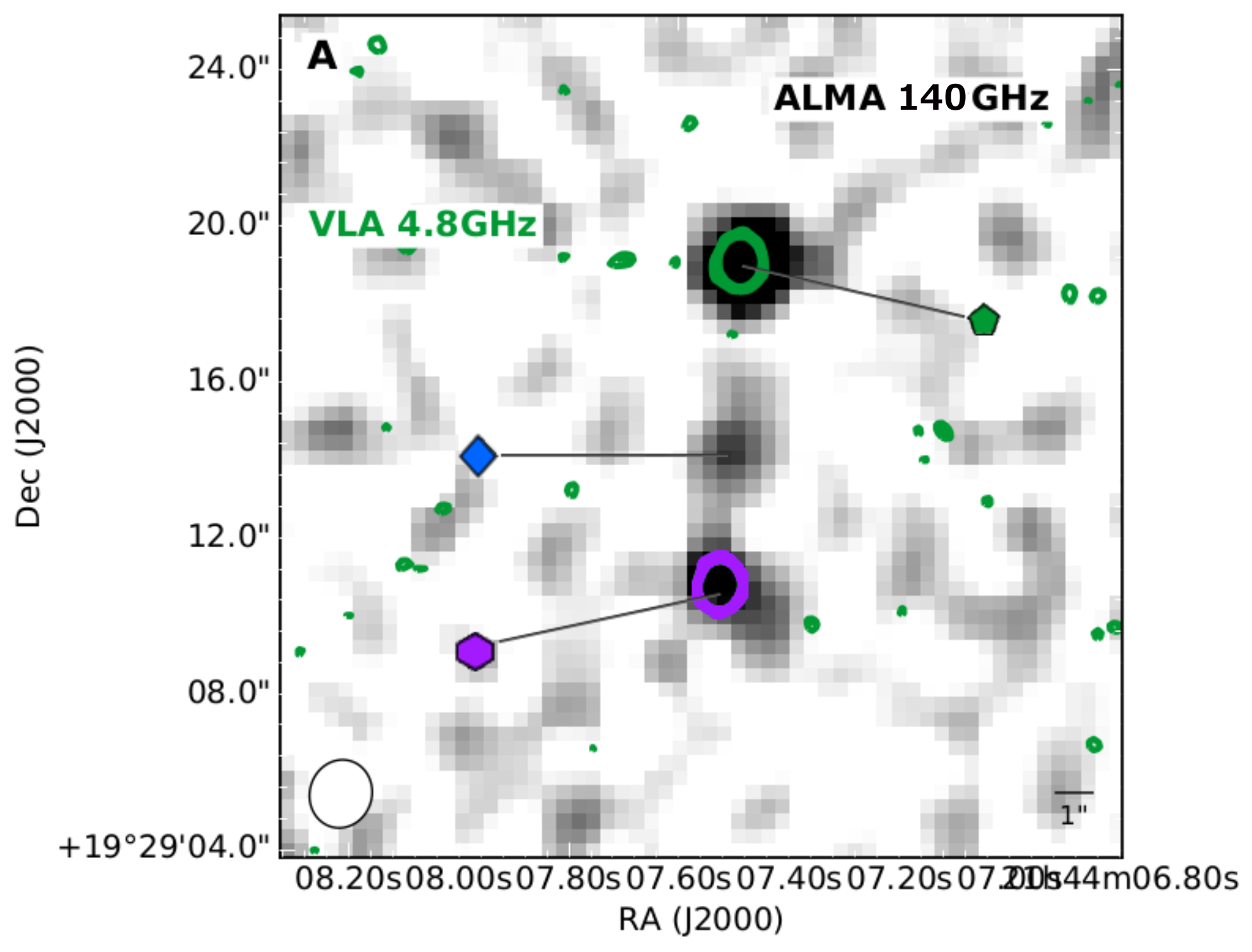}
\\
     	\includegraphics[scale=0.35]{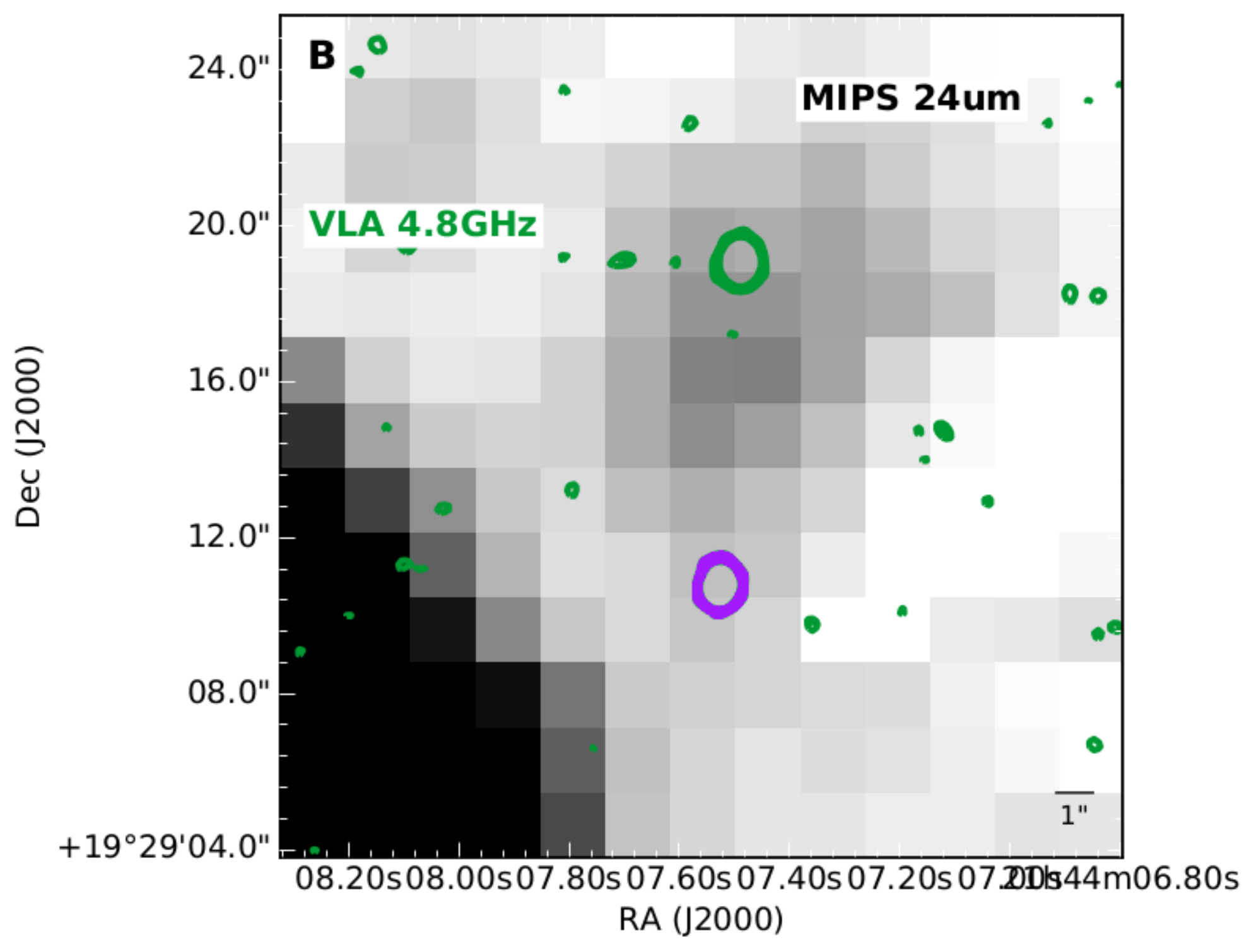}
 	\caption{\textit{Panel A:} continuum map of ALMA band 3 with overlaid VLA C contours (levels are as Fig. \ref{map_0037}, $\sigma=45\,\mu$Jy). The blue, green and purple markers indicate the three different components detected with ALMA and correspond to the same data point as Fig. \ref{fig_1971}. The green and purple contours show the two components of the VLA data and correspond to the markers of the same colors as in the SED fit. \textit{Panel B:} MIPS 24\,\mum\ continuum map.}
    	\label{map_1971}
\end{figure}

\begin{table}
\begin{threeparttable}
\caption{Data for 4C\,19.71 (z=3.592) }
\label{table_1971}
\centering
\begin{tabular}{lcc}
\toprule
Photometric band                   & Flux{[}mJy{]}   & Ref. \\
\midrule
\irs     		&    <0.244			&   A\\
\mips1     		&    0.529$\pm$0.033	&   A\\
\pacsg    		&    <46.1         		&   B\\
\pacsr    		&    <40.4      		&   B\\
\spires   		&    <14.8   		&   B\\
\spirem   		&     <18.1        		&   B\\
\spirel   		&     <17.5 		&   B\\
\scubas	 	&	12$\pm$13	&   C\\
\scuba  		&     2.3$\pm$0.9	&   C\\
ALMA 3$^c$      &    	0.07$\pm$0.05$^1$		&   this paper\\
ALMA 3$^n$     &    	0.29$\pm$0.53$^1$		&   this paper\\
ALMA 3$^s$     	&    	0.11$\pm0.05^1$		&   this paper\\
VLA X$^n$       	&   14.36 		&   D\\
VLA C$^n$      	&  34.99$\pm3.5$  	&   D\\
VLA X$^s$      	&   9.02$\pm0.9$ 		&   D\\
VLA C$^s$      	&  21.98$\pm2.2$	&   D\\
VLA L       		& 343.2$\pm$10.3	&   E\\                            
\bottomrule                          
\end{tabular}
     \begin{tablenotes}
      \small
      \item ($1$) Integrated flux extracted using AIPS with imfit 1 Gaussian component
      \item \textbf{References.} (A) \cite{DeBreuck2010}, (B) \cite{Drouart2014}, (C) \citep{Reuland2004} (D) \cite{Reuland2003} (E) \cite{Pentericci2000}.
    \end{tablenotes}
\end{threeparttable}
\end{table}


\clearpage
\newpage
\subsection{MRC\,2224-273}
MRC\,2224-273 has a single continuum detection which coincides with the compact
synchrotron emission (Fig. \ref{map_2224}). SED fitting with \mrmoose\ is done with three
components, one synchrotron power-law, one modified BB and one AGN component. The VLA X, C and L data are assigned to the synchrotron power-law, the ALMA detection is assigned to both synchrotron component
and a modified BB. The LABOCA, SPIRE, PACS, MIPS and IRS are fitted to the combination of the
modified BB and a AGN component. The best fit model gives that the ALMA detection is dominated by thermal dust emission. Without a
flux at intermediate radio frequencies between the VLA and ALMA data, it is not possible to constrain whether or not the synchrotron is steeping. In any case, the synchroton emission does not significantly contribute to the emission detected by ALMA. 

\begin{figure}
	\includegraphics[scale=0.52]{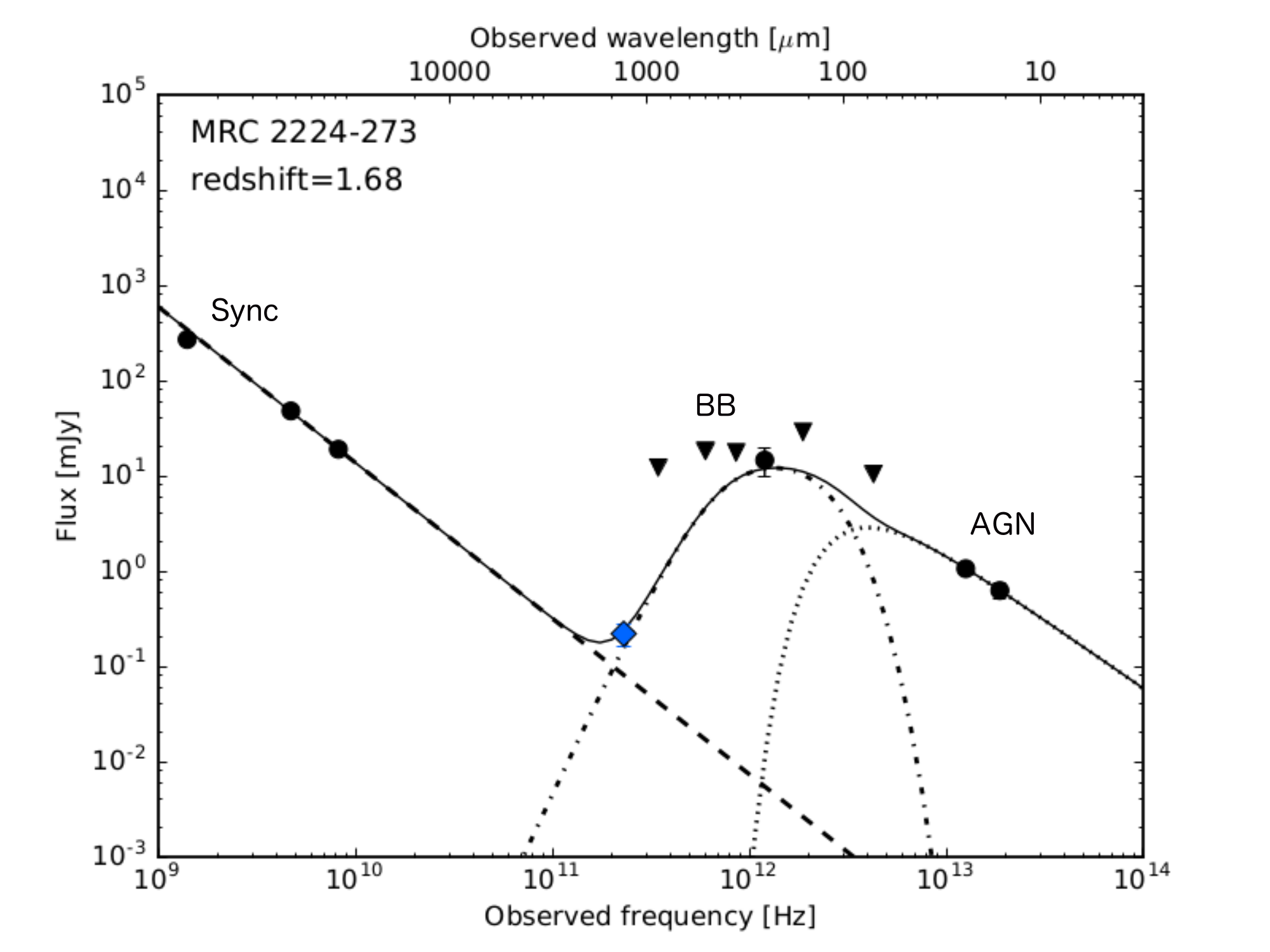}
	\caption{SED of \textbf{MRC\,2224-273}. Black solid line shows best fit total model, black dashed line is the total synchrotron, the black dash-dotted lines is the blackbody and the black dotted line indicates the AGN component. The colored data points are sub-arcsec resolution data and black ones indicate data of low resolution. The blue diamond indicates the ALMA band 6 detection. Filled black circles indicate detections (>$3\sigma$) and downward pointing triangles the $3\sigma$ upper limits (Table~\ref{table_2224}).}

	\label{fig_2224}
\end{figure}

\begin{figure}
	\centering
      	\includegraphics[scale=0.35]{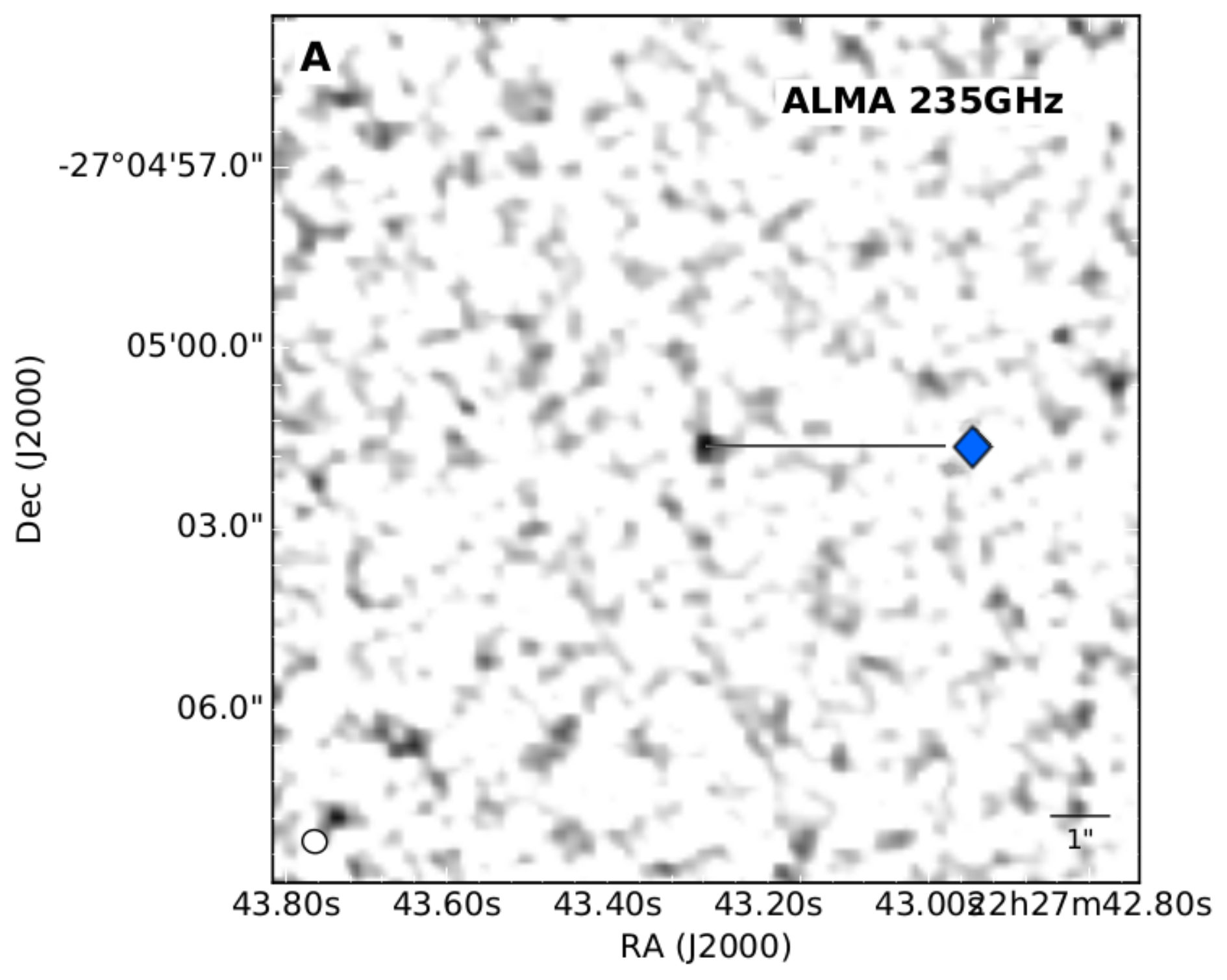}
\\
     	\includegraphics[scale=0.35]{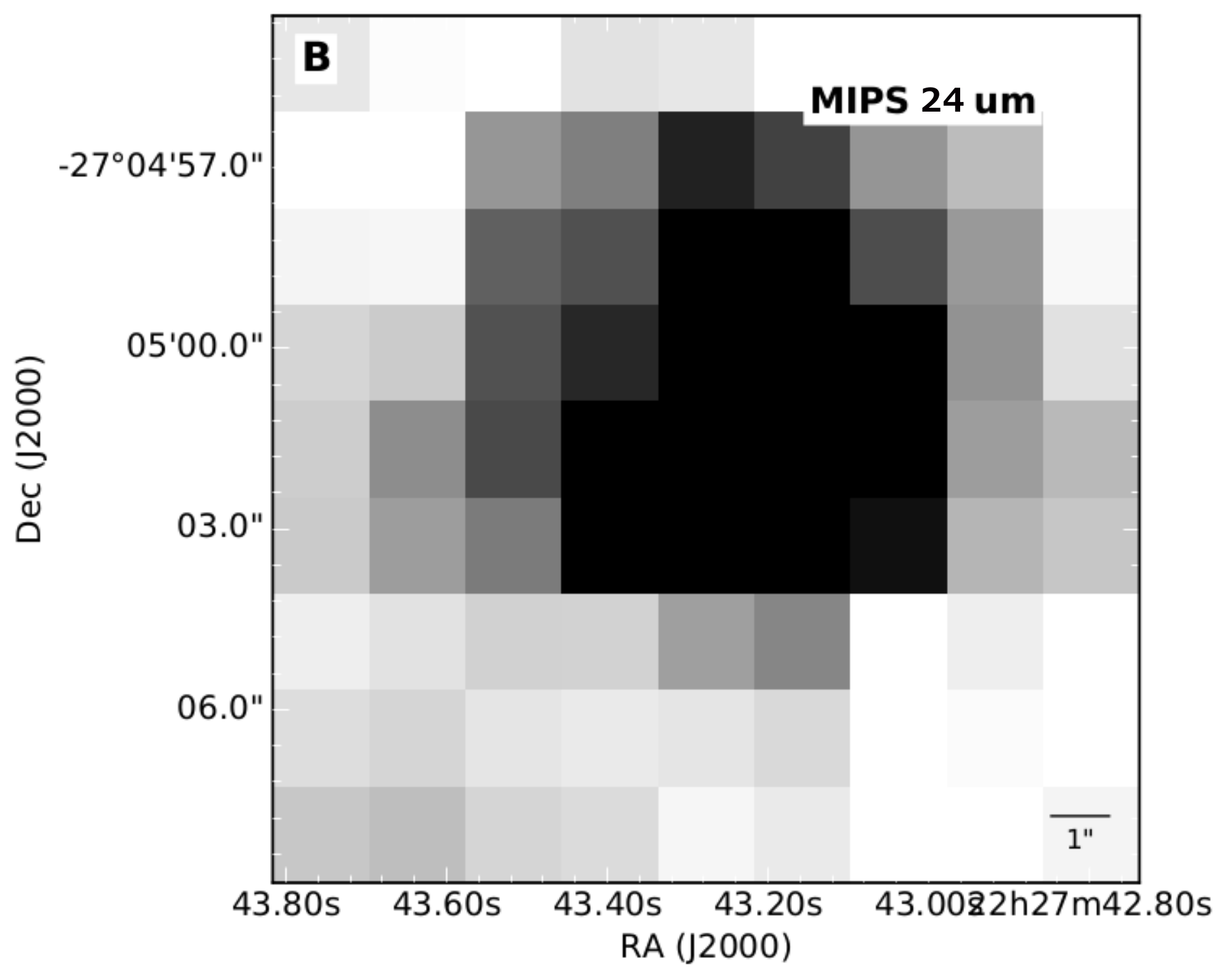}
 	\caption{\textit{Panel A:} continuum map of ALMA band 6. The blue marker indicates the ALMA dectection and corresponds to same data point in Fig. \ref{fig_2224}. VLA maps not accessible, but the source is unresolved, as shown in \cite{Pentericci2000}. \textit{Panel B: } MIPS 24\,\mum\ continuum map.}
    	\label{map_2224}
\end{figure}

\begin{table}[ht]
\begin{threeparttable}
\caption{Data for MRC\,2224-273 (z=1.68) }
\label{table_2224}
\centering
\begin{tabular}{lcc}
\toprule
Photometric band                   & Flux{[}mJy{]}   & Ref. \\
\midrule
\irs     		&    0.625$\pm$0.117  	&   A\\
\mips1     		&    1.06$\pm$0.04   		&   A\\
\pacsb    		&    <10.5              		&   B\\
\pacsr    		&    <28.9           		&   B\\
\spires   		&    14.6$\pm$4.8    		&   B\\
\spirem   		&    <17.6            		&   B\\
\spirel   		&    <18.3           		&   B\\
\laboca  		&    <12.3             		&   B\\
ALMA 6      	&    0.22$\pm$0.06   		&   this paper\\
VLA X       	&    19$\pm$0.022  		&   C\\
VLA C       	&    48$\pm$0.56   		&   C\\
VLA L       		&    269.9$\pm$26    		&   D\\                            
\bottomrule                          
\end{tabular}
     \begin{tablenotes}
      \small
      \item \textbf{References.} (A) \cite{DeBreuck2010}, (B) \cite{Drouart2014}, (C) \cite{Pentericci2000}, (D)\cite{Condon1998}.
    \end{tablenotes}
\end{threeparttable}
\end{table}


\end{appendix}

\end{document}